\documentclass[11pt,a4paper,twoside]{mydiss}
\usepackage{hyperref}

\usepackage{mathrsfs}
\usepackage{graphicx}
\usepackage{amsmath} 
\usepackage{amsthm}
\usepackage{amssymb}

\usepackage{fancyhdr}
\renewcommand{\chaptermark}[1]{\markboth{#1}{}}
\renewcommand{\sectionmark}[1]{\markright{\thesection\ #1}}
\def\xp{\sigma^{++}}
\def\xm{\sigma^{--}}

\def\O{\mathcal{O}}
\def\Hol{\mathrm{Hol}}

\def\del{\partial_{++}}
\def\dem{\partial_{--}}

\newcommand{\delr}{\raise.3ex\hbox{$\stackrel{\leftarrow}{\partial }$}}
\newcommand{\dell}{\raise.3ex\hbox{$\stackrel{\rightarrow}{\partial}$}}

\newcommand{\dr}{\raise.3ex\hbox{$\stackrel{\leftarrow}{\delta }$}}
\newcommand{\dl}{\raise.3ex\hbox{$\stackrel{\rightarrow}{\delta}$}}

\def\Bra#1{\left<#1\right|} 
\def\Ket#1{\left|#1\right>} 
{\catcode`\|=\active 
  \gdef\Braket#1{\left<\mathcode `\|"8000\let|\bravert {#1}\right>}} 
\def\bravert{\egroup\,\vrule\, \bgroup} 

 \def\L{\mathcal{L}}

\def\Theta{\overline{\theta}}

\def\D{\overline{d}}
\def\S{\mathcal{S}}
\def\M{\mathcal{M}}
\def\N{\mathcal{N}}

\def\half{\frac{1}{2}}

 \def\G{\mathcal{G}}
\def\L{\mathcal{L}}
\def\D{\mathcal{D}}
\def\B{\mathcal{B}}

 \def\balpha{{\overline{\alpha}}}
 \def\bbeta{{\overline{\beta}}}
 \def\bgamma{{\overline{\gamma}}}
 \def\bdelta{{\overline{\delta}}}

 \def\bmu{{\overline{\mu}}}

 \def\btau{{\overline{\tau}}}

 \def\bg{{\overline{g}}}
 
 \def\bM{{\overline{M}}}
 \def\bL{{\widehat{L}}}
 \def\bB{{\overline{B}}}
 \def\bw{{\overline{w}}}
 \def\bOmega{\overline{\Omega}}
 
 \def\bpartial{{\overline{\partial}}}
 
\def\gh{\mathrm{gh}}
\def\BRST{\mathrm{BRST}}

\def\FP{\mathrm{FP}}
\def\YM{\mathrm{YM}}
\def\GF{\mathrm{GF}}
\def\top{\mathrm{top}}
\def\Oc{\overline{c}}
 
\def\ext{\mathrm{ext}}

 \def\wa{\tilde{a}}
 \def\wA{\tilde{A}}
 
 \def\sig{\sigma}
 
\def\bbR{\mathbb{R}}
\def\bbC{\mathbb{C}}
\def\bbH{\mathbb{H}}
\def\bbZ{\mathbb{Z}}

\def\calA{\mathcal{A}}
\def\calB{\mathcal{B}}

\def\calR{\mathcal{R}}
\def\calK{\mathcal{K}}
\def\calP{\mathcal{P}}
\def\calQ{\mathcal{Q}}
\def\calG{\mathcal{G}}
\def\calH{\mathcal{H}}
\def\calN{\mathcal{N}}
 
 \def\kah{K\"{a}hler }
 \def\hkah{hyper-K\"{a}hler }
 \def\Hkah{Hyper-K\"{a}hler }
 
  \def\deppi{\frac{\delta S_{\top}}{\delta \pi_{+}}}
 \def\dempi{\frac{\delta S_{\top}}{\delta \pi_{-}}}

\newcommand{\parp}{\partial_{+}}
\newcommand{\parm}{\partial_{-}}

\newcommand{\psib}{\overline{\psi}}
\newcommand{\nub}{\overline{\nu}}
\newcommand{\mub}{\overline{\mu}}
\newcommand{\alphab}{\overline{\alpha}}
\newcommand{\betab}{\overline{\beta}}

\newlength{\dblsetlen}
\def\doubleset#1#2{{%
\def\doit#1#2{%
\settowidth{\dblsetlen}{\hbox{$\cstyle #1$}}%
\raise#2\hbox{$\cstyle #1$}%
\hskip-\dblsetlen%
\raise-#2\hbox{$\cstyle #1$}}%
\mathchoice%
{\def\cstyle{\displaystyle}\doit#1#2}%
{\def\cstyle{\textstyle}\doit#1#2}%
{\def\cstyle{\scriptstyle}\doit#1#2}%
{\def\cstyle{\scriptscriptstyle}\doit#1#2}}}

 \def\dem{\partial_{--}}
 \def\dep{\partial_{++}}
  \def\sdem{\partial_{-}}
 \def\sdep{\partial_{+}}
  \def\hp{h^{++}}
 \def\hm{h^{--}}
 \def\ahp{h^*_{++}}
 \def\ahm{h^*_{--}}

\newcommand{\pip}[1]{\pi_{+ #1}}
\newcommand{\pim}[1]{\pi_{- #1}}
\newcommand{\apip}[1]{\pi_*^{+ #1}}
\newcommand{\apim}[1]{\pi_*^{- #1}}
\def\Tpl{\mathcal{T}_{++}}
\def\Tmi{\mathcal{T}_{--}}

\def\tN{N}

\def\half{\frac{1}{2}}

\def\pl{ \: + }
\def\mi{ \: - }

\newfont{\bbbold}{msbm10 scaled \magstep1}

\def\bbA{\mbox{\bbbold A}}

\def\cR{{\cal R}}

\newfont{\goth}{eufm10 scaled \magstep1}

\def\ge{\mbox{\goth e}}

\def\gg{\mbox{\goth g}}

\def\go{\mbox{\goth o}}

\def\gR{\mbox{\goth R}}
\def\gs{\mbox{\goth s}}

\def\gu{\mbox{\goth u}}

\def\a{\alpha}
\def\b{\beta}
\def\c{\gamma}\def\C{\Gamma}
\def\d{\delta}\def\D{\Delta}
\def\e{\epsilon}
\def\f{\phi}

\def\F{\Psi} 

\def\vf{\varphi}
\def\h{\eta}

\def\n{\nu}
\def\p{\pi}
\def\P{\Pi}

\def\s{\sigma}\def\S{\Sigma}

\def\th{\theta}

\def\be{\begin{equation}}\def\ee{\end{equation}}
\def\bea{\begin{eqnarray}}\def\eea{\end{eqnarray}}
\def\barr{\begin{array}}\def\earr{\end{array}}

\def\o{\omega}\def\O{\Omega}
\def\del{\partial}

\def\xz{\times}

\def\nab{\nabla}

\def\hE{\hat{E}}\def\hR{\hat{R}}

\def\tnab{\tilde{\nabla}}

\def\hE{\widehat{E}}

\def\hL{\widehat{L}}
\def\hR{\widehat{R}}

\def\ghh{\widehat{\h}}
\def\ghl{\widehat{\l}}

\def\hnab{\widehat{\nab}}

\def\pl{{(+)}}\def\mi{{(-)}} \def\plmi{{(\pm)}}
\let\la=\label

\def\nn{\nonumber}
\def\bd{\begin{document}}
\def\ed{\end{document}}
\def\ba{\begin{array}}
\def\ea{\end{array}}
\def\bea{\begin{eqnarray}}
\def\eea{\end{eqnarray}}
\def\ft#1#2{{\textstyle{{\scriptstyle #1}\over {\scriptstyle #2}}}}
\def\fft#1#2{{#1 \over #2}}
\newcommand{\eq}[1]{(\ref{#1})}
\newcommand{\w}[1]{\\[0.#1cm]}
\def\eqs#1#2{(\ref{#1}-\ref{#2})}
\def\det{{\rm det\,}}
\def\tr{{\rm tr}}

\begin{document}

\begin{titlepage}

\vspace*{3cm}

\begin{center}
\Huge Special Holonomy and \\
Two-Dimensional \\
Supersymmetric $\sigma$-Models\\
\end{center}
\vspace{1cm}

\begin{center}
\Large
Vid Stojevic
\end{center}

\vspace{10cm}
\begin{center}
\large
Submitted for the Degree of Doctor of Philosophy \\
Department of Mathematics \\ 
King's College \\ 
University of London \\
2006
\end{center}

\end{titlepage}

\pagestyle{empty}

\cleardoublepage




\begin{center}\LARGE\bfseries{\sc Abstract}\end{center}

$d=2$ $\sigma$-models describing superstrings propagating on manifolds of special holonomy
are characterized by symmetries related to covariantly constant forms that these manifolds hold, which are generally non-linear and close in a field dependent sense. The thesis explores various aspects of the special holonomy symmetries (SHS).

Cohomological equations are set up that  enable a calculation of potential quantum anomalies in the SHS. It turns out that it's necessary to linearize the algebras by treating composite currents as generators of additional symmetries. Surprisingly we find that for most cases the linearization involves a finite number of composite currents, with the exception of $SU(3)$ and $G_2$ which seem to allow no finite linearization. We extend the analysis to cases with torsion, and work out the implications of generalized Nijenhuis forms.

SHS are analyzed on boundaries of open strings. Branes wrapping calibrated cycles of special holonomy manifolds are related, from the $\sigma$-model point of view, to the preservation of special holonomy symmetries on string boundaries. Some technical results are obtained, and specific cases with torsion and non-vanishing gauge fields are worked out.

Classical $W$-string actions obtained by gauging the SHS are analyzed. These are of interest both as a gauge theories and for calculating operator product expansions. For many cases there is an obstruction to obtaining the BRST operator due to relations between SHS that are implied by Jacobi identities. We relate the problem to extra gauge symmetries that exist when only a subalgebra of the SHS on the same special holonomy background is gauged. Such gauge symmetries are infinitely reducible and don't imply conserved currents. We propose a solution which avoids the Jacobi problem by gauging a subalgebra of SHS with complex conserved currents that involves spectral flow generators in one direction only, unlike the reducible algebra which involves both.

\pagestyle{empty}
\cleardoublepage




\begin{center}\LARGE\bfseries{\sc Statement of Original Work}\end{center}

Chapters 1 to 4 explain the background to the original work, which is contained in Chapters 5 to 8. The only exception to this is the unconventional gauge fixing procedure for the bosonic string in \ref{section:bosonic_string_theory} and the related observations in \ref{section:aside_on_bi_hamiltonian}, which to my knowledge are original contributions. Chapters \ref{chapter:algebras_of_general_L_symmetries}, \ref{chapter:algebras_on_manifolds_of_sh}, and \ref{chapter:W-strings} contain my personal work, except \ref{section:spechol_discussion} which was done in collaboration with my supervisor Paul Howe. Chapter \ref{chapter:boundary} contains work done in collaboration with Paul Howe and Ulf Lindstr\"{o}m, with each of us contributing roughly an equal amount.

The above doesn't account for the important help I received through personal communication. This is hopefully rectified in the Acknowledgements.

\newpage
\pagestyle{empty}{} \vspace*{1cm}

\tableofcontents



\chapter{Introduction}


The biggest challenge that physics has faced in the recent decades is to make sense of gravity at the quantum level.  It is difficult to assess how much progress has been made so far by the two main attempts, string theory and loop quantum gravity. Each certainly has its problems. The problem they both share is that   at present they are unable to make any testable physical predictions. Loop quantum gravity has an infinite number of undetermined coefficients in the Hamiltonian constraints \cite{Nicolai:2006id}, and is generally unable to reproduce  even the most basic qualities of the continuity of space on the large scale. String theory is formulated in a background dependent  way, and at present provides no mechanism for singling out the background that describes our universe. Some estimates \cite{Douglas:2003um} of the number of vacua in which the standard model can be realized are of the stupendously high order of $10^{500}$, which has recently led many string theorists to resort to the anthropic principle. However, arguments have also been put forward that this number is really vastly lower \cite{Vafa:2005ui}, and that most of the effective field theories we see through our perturbative understanding of string theory are in fact inconsistent. When compared to string theory, loop quantum gravity has the disadvantage that it's derived in a rather contrived way, and while there certainly are some correct ideas within the formalism, progress is slow. On the other hand, progress in string theory is being made at a  fast pace, but admittedly not always in a direction directly relevant for physics.
 
On can argue that any attempt to get a handle on quantum gravity using perturbative techniques seems to lead to string theory. General relativity is not perturbatively renormalizable for $D>2$, and since the uncontrollable ultraviolet divergences can be related to the point particle description of gravitons it is natural to try and work with relativistic extended objects instead. These can be described by the $\sigma$-model  action
\begin{equation}
\label{eq:renormalizability_of_sigma}
S = \int d^D \sig G_{ij}(\phi) \gamma^{\mu \nu} \frac{\partial \phi^i}{\partial \sigma^\mu}  \frac{ \partial \phi^j}{\partial \sigma^\nu} \ ,
\end{equation}
where $\sigma^\nu$ are the coordinates and $\gamma^{\mu \nu}$ a metric on the $D$-dimensional worldvolume traversed by a $(D-1)$-dimensional object moving through a target space manifold with coordinates $\phi^i$. The target space metric can be expanded in Riemann normal coordinates as 
\begin{equation}
G_{ij}(\phi) = \delta_{ij} + R_{ijkl}(\phi^i_0) \phi^{k} \phi^{l} + \cdots
\end{equation}
around some point $\phi^i_0$. Since $[\phi] = \frac{D-2}{2}$, we have that $[R_{ijkl}] = D-2$. So unless the target space is flat, (\ref{eq:renormalizability_of_sigma}) is renormalizable only for $D=2$, when it describes a string. 
 

Of course, this is only the beginning of the story. The ingredients that fix string theory as a virtually unique  theory are supersymmetry and conformal invariance. Supersymmetry is related to eliminating the tachyonic lowest energy states present in the bosonic string, while conformal invariance is related to unitarity of string scattering amplitudes. Technically, requiring conformal invariance is tantamount to requiring the $\beta$-function of the $\sigma$-model to vanish. One of the many miracles of string theory is that, at zeroth order in the $\sigma$-model expansion parameter $\alpha'$, this constrains the background in which a superstring can consistently propagate to obey Einstein's equations, with corrections due to the string being an extended object at higher orders. Furthermore, there are various fields other than the metric that feature in string theory: the two-form $b_{ij}$, which couples to the fundamental string itself, and various higher dimensional forms related to the existence of branes.  Branes play a role in a whole series of further miracles, especially in the context of dualities. Among the most significant  of these is the AdS/CFT correspondence, which in its original formulation \cite{Maldacena:1997re} states that string theory on  $\mathrm{AdS}_5\times \mathrm{S}^5$ is equivalent to the $N=4$ supersymmetric Yang-Mills theory (a conformally invariant theory) on the conformal compactification of four-dimensional Minkowski space, which is the boundary of $\mathrm{AdS}_5$. 

The thesis is for the most part more old fashioned, in that it is largely concerned with supersymmetric backgrounds that were studied before the advent of branes, M-theory, and AdS/CFT. This includes the simplest physically desirable background, in which six dimensions are compact and four large. The requirement of supersymmetry puts very stringent requirements on the nature of the small spaces:  if the only non-zero field is the metric they have to be six-dimensional Calabi-Yau manifolds. When $\alpha'$ corrections are included the metric is deformed from the Ricci-flat one, but the manifold remains Calabi-Yau in the topological sense. For compactifications to other dimensions other manifolds become relevant, but it turns out they are always of the special holonomy type, meaning that their holonomy groups are contained in $SU(n)$, $Sp(n)$, $G_2$, or $Spin(7)$.

The main aim of the thesis is to study the preservation of target space supersymmetry on special holonomy backgrounds in the presence of stringy corrections. There exist a number of  different types of $\sigma$-models describing string theory, each with its own advantages and disadvantages, and its own manifestation of spacetime supersymmetry. The formulations are remarkably different, given that they are ultimately equivalent. I will work in the Neveu-Schwarz (NS)  formalism, in which the $\sigma$-model base space is a supermanifold and the target space an ordinary manifold. Specifically, I will only consider the case when the base space has $(1,1)$ supersymmetry. The other possibilities are:
 \begin{itemize}
 \item The Green-Schwarz (GS)  formalism \cite{Green:1980zg}, in which the target space is a supermanifold. The advantage of the GS formalism is that spacetime supersymmetry is manifest.  However, it can only be quantized in the light-cone gauge; a covariant quantization is not  possible, leading to problems of infinite reducibility.
 \item The Berkovits formalism \cite{Berkovits:2000fe}, which builds on the GS formalism and uses various tricks to obtain a covariant formulation with manifest spacetime supersymmetry. The manner of obtaining the BRST operator is not the usual one, i.e. it is not obtained by gauge fixing an action that couples to a worldsheet metric, and the underlying conformal symmetry is not completely manifest.
 \item A further formalism, based on the GS string, that involves an infinite tower of ghosts has been proposed very recently \cite{Lee:2006pa}.
\end{itemize}

 
The NS formalism has the advantage that it is relatively straightforward to quantize, and the disadvantage that target space supersymmetry is not manifest. Rather, it is realized indirectly via worldsheet current algebras.  Any conformally invariant theory with $(1,1)$ worldsheet supersymmetry on a Calabi-Yau target space has the infinite-dimensional N=2  current algebra:\footnote{See, for example, the review by Greene \cite{Greene:1996cy} for details.}
\begin{align}
\label{eq:N2_abstract_algebra}
& [L_m , L_m ] = (m-n) L_{m+n} + \frac{c}{12} m (m^2 -1 ) \delta_{(m+n),0} \ ,  \\ \nonumber
& [J_m, J_n ] = \frac{c}{3} m \delta_{(m+n), 0}  \ , \\ \nonumber
& [L_n, J_m ] = -m J_{m+n}  \ , \\ \nonumber
& [L_n,  G^{\pm}_{m \pm a} ] = \left( \frac{n}{2} - (m \pm a) \right) G^{\pm }_{m+n \pm a} \ , \\ \nonumber
& [J_n, G^{\pm}_{m \pm a} ] = \pm G^{\pm}_{m+n \pm a} \ , \\ \nonumber
& \{ G^+_{n+a} G^-_{m-a} \} = 2 L_{m+n} + (\n-m +2a) J_{n+m} + \frac{c}{3} \left( (n+a)^2 - \frac{1}{4} \right) \delta_{m+n, 0} \ ,
\end{align}
where $L_m$ are modes of the conformal current, $J_n$ are modes of the $U(1)$ current, and $G^{\pm}_m$ are modes associated with two worldsheet supersymmetry currents. $c$ is the central charge, and is related to the dimension of the target space by $D=\frac{2}{3} c$. The $N=2$ algebra is invariant under the spectral flow,
 \begin{align}
& L_n \rightarrow L_n + \eta J_n + \frac{c}{6} \eta^2 \delta_{n,0} \ , \\ \nonumber
& J_n \rightarrow J_n + \frac{c}{3} \eta \delta_{n,0} \ , \\ \nonumber
& G^{\pm}_r \rightarrow  G^{\pm}_{r \pm \eta} \ ,
 \end{align}
which means that making the above replacements in (\ref{eq:N2_abstract_algebra}) leaves the commutation relations invariant. The spectral flow by $\eta = \pm \frac{1}{2}$ takes us from the NS sector, which describes target space bosons, to the R sector, which describes fermions (or vice versa, depending on the initial boundary conditions on the currents), and gives the description of spacetime supersymmetry for Calabi-Yau target spaces.  In addition, the $N=2$ algebra can be extended by currents related to the holomorphic and antiholomorphic $\frac{D}{2}$ forms that exist on CY manifolds \cite{Odake:1988bh}. These extra currents can be interpreted as generators of spectral flow by $\pm 1$.  Analogous extended algebras have been defined for all manifolds of special holonomy relevant for string compactifications \cite{Howe:1991ic, Figueroa-O'Farrill:1996hm}.

It was thought in the early days \cite{Alvarez-Gaume:1985ww} that a $\sigma$-model with $N=2$  symmetry on a Ricci flat CY manifold is conformally invariant to all orders in $\alpha'$. However, this was too hopeful, and it was soon found that counterterms are needed at four loop \cite{Grisaru:1986px} which deform the metric  from its Ricci flat solution. Corrections to supersymmetry transformations and equations of motion have been studied extensively since then in the context of supergravity, both in the context of effective actions of string theory, and M-theory (see \cite{Lu:2005im, Lu:2003ze} and references therein). However, so far there is no direct proof that supersymmetry is preserved when all the $\alpha'$ corrections are present, although some arguments are very convincing.
 
This thesis contributes towards analyzing the preservation of spacetime supersymmetry in the presence of stringy corrections by calculating potential anomalies in the special holonomy algebras. It is extremely difficult to evaluate special holonomy $\sigma$-model path integrals directly. In fact, it is impossible to do so for specific manifolds, since there are no explicit expressions for Ricci-flat metrics on special holonomy manifolds. The explicit evaluation can be avoided by using methods of algebraic renormalization and obtaining cohomological equations that enable the calculation of potential anomalies. The important advantage of this approach is that it can potentially provide constraints on $\alpha'$ corrections valid to all orders.
 
Algebras described in the abstract CFT language, such as (\ref{eq:N2_abstract_algebra}), start from the assumption that  the model under consideration is well defined, the symmetries in question are not broken by quantum effects, and so on. When working with the algebras in the context of a particular non-linear $\sigma$-model there are all kinds of issues one must confront before contact with the abstract CFT expressions can be made. The special holonomy algebras are particularly problematic, because they are generally non-linear in the $\sigma$-model fields and close only in a field dependent manner. It turns out that it's not possible to write down cohomological equations for algebras that close with field-dependent structure functions. One way to obtain such equations is to linearize the algebras, by treating composite currents as generators of additional symmetries. Alternatively, one can consider subalgebras that close linearly. However, to understand the full algebras the former approach seems to be the only possibility. It turns out that all the special holonomy algebras can be linearized in a finite number of steps except the ones of most physical interest: CY3 and $G_2$.  For these the only possibility seems to be to analyze various linear subalgebras. 
 
In addition, the thesis deals with a a list of other topics related to special holonomy algebras, and I'll briefly explain them as I summarize the contents:
\begin{itemize}
\item Chapter \ref{chapter:geometric_background} provides the geometric background. This includes a discussion of complex differential geometry, manifolds of special holonomy, $G$-structures, and calibrations.
\item Chapter \ref{ch:antifield_formalism} describes the antifield formalism, which is a natural framework  for analyzing symmetries at the quantum level and writing down cohomological equations for potential anomalies. I discuss  gauge symmetries,  methods of gauge fixing, as well as methods for working with local and conformal-type symmetries. The latter are symmetries whose transformation parameters depend on some but not all of the coordinates of the base space,  which is the case for symmetries related to current algebras.  I also give a description of quantum theories with background fields, which is relevant when one actually attempts to evaluate the path integral of the $\sigma$-model, since a quantum-background split of the fields needs to be performed. It is also relevant for expressing operator product expansions in the context of a non-trivial $\sigma$-model.
\item In Chapter \ref{chapter:2d_sigma_models} bosonic and $(1,1)$ $\sigma$-models are described in detail. A bi-Hamiltonian formulation of bosonic string theory is presented. When momenta are integrated out the extended action expressing the usual BRST symmetries of the gauge fixed string action is recovered. As an original result, I show that a different extended action can be obtained for which the antifields of the $b$ field parameterize the background metric, so one has a description of residual symmetries after gauge fixing that involves an arbitrary power of antifields.
\item Chapter \ref{chapter:algebras_of_general_L_symmetries} is concerned with the algebra of $L$-type symmetries of the NS $\sigma$-model, possibly with torsion, which are by definition obtained from covariantly constant forms in the target space. The special holonomy symmetries are of this general type.
\item In the first part of Chapter \ref{chapter:algebras_on_manifolds_of_sh} the full linearized special holonomy algebras are calculated, when this is possible. I show in detail what happens for the problematic cases of CY3 and $G_2$. In the second part  torsionfull cases are analyzed. The classification of special holonomy manifolds is no longer valid, but one can still consider manifolds with the same structure groups and covariantly constant tensors. The Nijenhuis form can be generalized to any pair of covariantly constant forms. We analyze the structure groups from the special holonomy list in the setting when the generalized Nijenhuis forms don't vanish. Since they are covariantly constant, an important question is whether the holonomy groups are further constrained by their presence, or whether it's possible to construct them using the original special holonomy forms. In the $SU(\frac{D}{2})$ case only the almost complex Nijenhuis three-form restricts the geometry further. Interestingly, the structure group of \emph{any} almost complex six-dimensional manifold is automatically reduced to $SU(3)$. The generalized Nijenhuis forms also imply the reduction of the structure group for the almost symplectic case, and almost quaternionic \kah case for $D>8$.
\item  Chapter \ref{chapter:boundary} is basically the content of the paper \cite{howe-2006-0601} published in colaboration with my supervisor Paul Howe and Ulf Lindstr\"{o}m.  It involves an analysis of the special holonomy symmetries of the $(1,1)$ $\sigma$-model describing open strings, i.e. having a worldsheet with boundaries. Preserving these on the boundaries implies that open strings end on calibrated submanifolds of the special holonomy target spaces, and gives a description of branes wrapping calibrated cycles. Some technical results are obtained, and we analyze various cases with non-vanishing torsion, in particular cases with two $G$-structures covariantly constant with respect to different torsionfull connections. We also examine scenarios with gauge fields living on the branes. 
\item  In Chapter \ref{chapter:W-strings} I discuss $W$-strings obtained by gauging special holonomy algebras. It turns out that for many algebras there is an obstruction to obtaining the $W$-string BRST operator due to relations between symmetry generators that are implied by the Jacobi identities. This problem is explored in detail, and related to the existence of extra gauge symmetries that exist when only a subset of the full algebra is gauged on the same special holonomy manifold. Possible ways of avoiding the problem are discussed. A resolution that is possible in many cases is to work with subalgebras involving complex currents; these involve spectral flow generators in one direction only. Gauging the special holonomy symmetries is also a necessary step for analyzing the operator product expansion between the currents, except that in this setting one treats the  ghosts and gauge fields as background rather that quantum fields. 
\end{itemize} 
 

\chapter{Geometric Background}
\label{chapter:geometric_background}


In this chapter I give an introduction to geometry  leading to the classification of Riemannian manifolds using their holonomy groups. The main aim is to describe manifolds of special holonomy, as well as more general manifolds with non-vanishing torsion. These are the relevant target spaces for the supersymmetric $\sigma$-models that feature in Chapters \ref{chapter:2d_sigma_models}-\ref{chapter:W-strings}. I will also introduce the idea of calibrations and calibrated submanifolds.    When branes are present in special holonomy backgrounds the condition for preserving some fraction of space-time supersymmetry requires that the branes wrap calibrated submanifolds. In Chapter \ref{chapter:boundary} this will be discussed from the $\sigma$-model perspective.

I will attempt to give a somewhat comprehensive overview of the more basic concepts that lead up to the above topics, but will not attempt  a self-contained presentation.   A very good resource  for topology and differential geometry for physicists is \cite{Nakahara:1990th}.  \cite{Isham:1999rh}  is also meant for physicists, but  is a lot more mathematical. The lectures by Candelas \cite{Candelas:1987is} were  an especially useful resource for me.

\section{Differential manifolds}
\label{section:differential_manifolds}

\begin{figure}
\includegraphics[scale=0.4]{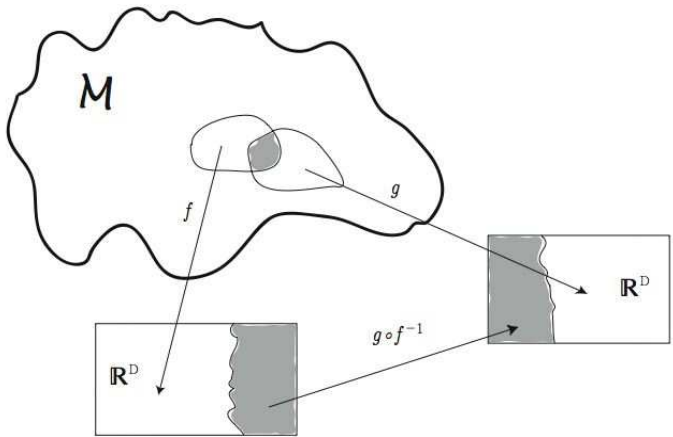}
\caption{A schematic drawing of a manifold $\mathcal{M}$ with a part of an atlas covering it.}
\label{figure:manifold}
\end{figure}

A $D$-dimensional \emph{topological manifold} $\M$ is a space that locally looks like an open set in $\mathbb{R}^D$, where $\bbR^D$ is assigned its usual topology.\footnote{The definition does not allow for spaces with a boundary, since a neighborhood of a boundary does not look like an open subset of $\mathbb{R}^D$. A \emph{manifold with a boundary} is defined to be locally either like $\mathbb{R}^D$ or like $\mathbb{H}^D$, 
\begin{equation}
\bbH^D = ( x^1, \cdots , x^D )  \ \ \ \mathrm{such \ that} \ \ \ x^D \geq 0 \ .
\end{equation}}
$\M$ is described by covering it with charts of $\mathbb{R}^D$, which are used as local coordinates, while requiring that on overlaps the maps between charts be continuous (see Figure \ref{figure:manifold}). Then continuous functions can be defined consistently on the whole of $\M$.  Some particular covering of the manifold is called an \emph{atlas}. One atlas is equivalent to any other one provided that the maps relating them are continuous.  

On a $C^\infty$-\emph{differential manifold}  the maps between charts are required to be differentiable infinitely many times. It's also possible to work with manifolds that can only hold functions which are differentiable a finite number of times, but I won't be making this generalization. The \emph{differential structure} of a manifold is the equivalence class of atlases related by differentiable maps.\footnote{This is an important definition, since it's not generally true that a  topological manifold  admits a unique differential structure.   The first case was discovered in 1956 by  John Milnor, who showed that on a seven-dimensional sphere it is possible to define  a differential structure inequivalent to the  one inherited from its embedding in $\mathbb{R}^8$. Even more remarkably, in 1983 Simon Donaldson discovered that only $\mathbb{R}^4$ among the Euclidian n-dimensional manifolds admits an infinite number of differential structures  \cite{Donaldson:1986mr}.}
  
In the rest of this section  I'll describe   various  structures that can exist on a differential manifold. I'll also describe fiber bundles.  In particular, the introduction of the metric will be left to the following section.
 
 \subsection{Fiber bundles}
   A \emph{fiber bundle} is built up from two manifolds, the \emph{base space}\footnote{Not to be confused with the usage of this term in the context of $\sigma$-models. See Chapter \ref{chapter:2d_sigma_models}.} $\M$  and the \emph{total space} $E$,  together with a continuous map $\pi: E \rightarrow \M$. 
   A fiber $F$ above a point $x \in \M$  is defined as $F = \pi^{-1}(x)$.\footnote{This is an inverse set map, and as such it is one-to-many. It's defined as the set of points $y$ in E for which $\pi(y) = x$.}  The total space is required to look like $U \times F$ locally, where $U$ is some open set on $\M$. Globally the structure can have all kinds of twists.  A \emph{section} $s$ of a bundle is a map  $s: \M \rightarrow E$. 
  
  While fiber bundles can have many other uses, here I'm taking the point of view that $\M$ is the manifold of interest, and we're trying to study it by introducing a structure $E$ over it.   
The particular type of fiber bundle that we'll be making use of is a \emph{vector bundle}, where the fiber is required  to have  the structure of a vector space. Vector bundles that exist naturally over a differential manifold are the tangent bundle $T \M$ and the cotangent bundle $T^* \M$, as well as all the tensor bundles constructed from these. Tensor fields are sections of tensor bundles.
 
 \subsection{$n$-forms}
 \label{subsection:nforms}

In this section I attempt to show why $n$-forms  are useful  for probing the topological properties of $\M$.  

 Consider two vectors  at some point on $\M$. If they are linearly independent they define a plane element. It's not  possible to assign an area to the plane element without an inner product, and thus a metric, but there is a topological notion of plane elements being different simply because they form a continuous space. This space is given by the set of $(2,0)$-tensors antisymmetric in their indices that are obtained by taking the \emph{wedge product} (or \emph{Grassmann product}) of two vectors $v$ and $w$:
\begin{equation}
v \wedge w := \frac{1}{2}(v \otimes w - w \otimes v) \ .
\end{equation}
Of course, there are antisymmetric $(2,0)$-tensors that can't be obtained by wedging two vectors, in which case they don't have such a direct geometrical interpretation. The information about the orientation of the plane is also encoded in the Grassmann product since $v \wedge w =  - w \wedge v$. Thus, $v\wedge w$ can define one orientation and $w \wedge v$ the other. Extending this to the whole of $\M$,   an antisymmetric $(2,0)$ tensor field obtained by wedging two vector fields defines a plane element. Similarly, a totally antisymmetric tensor field of type $(n,0)$ obtained by wedging $n$ vector fields defines an oriented $n$-dimensional volume element. 

It turns out that it's more useful to work with $(0,n)$-type totally antisymmetric  tensor fields, or  \emph{n-forms}. Forms contain exactly the same information as totally antisymmetric vectors, but in an inverted way. A cotangent vector, which is also a 1-form, defines a $(D - 1)$ volume element, while a $(D-1)$-form defines a line element.  A $(D-2)$-form obtained by wedging $(D-2)$ $1$-forms will describe a plane element, and so on. 

Locally an $n$-form $\alpha$ has the expansion
\begin{align}
\alpha = \frac{1}{n!} \alpha_{i_1 ... i_n} dx^{i_1} \wedge \cdots \wedge dx^{i_n}  \ .
\end{align}
Forms can be combined using the wedge product: 
\begin{align}
\alpha \wedge \beta = \frac{1}{m! n!} \alpha_{i_1 ... i_n} \beta_{j_1 ... j_m} dx^{i_1} \wedge \cdots \wedge dx^{i_n} \wedge dx^{j_1} \wedge \cdots \wedge dx^{j_m}  \ .
\end{align}
There is  a   natural differential operation called the \emph{exterior derivative} $d$ that maps an $n$-form to an $(n+1)$-form,
\begin{equation}
\label{eq:exterior_derivative}
d \alpha = \frac{1}{n!} \alpha_{i_1 ... i_n ,k} dx^k \wedge dx^{i_1} \wedge \cdots \wedge dx^{i_n}  \ ,
\end{equation}
where  
\begin{equation}
\alpha_{i_1 ... i_n , k} := \frac{\partial}{\partial x^k}  \alpha_{i_1 ... i_n} \ ,
\end{equation}
and the subscript "$,k$" is a shorthand for a partial derivative with respect to $x^k$. $d$ is a linear operator that obeys
\begin{equation}
d(\alpha \wedge \beta) = (d \alpha) \wedge \beta + (-1)^n \alpha \wedge d \beta \ ,
\end{equation}
and is also nilpotent. The latter property is easy to see from the definition (\ref{eq:exterior_derivative}),
\begin{equation}
d^2 \alpha = \frac{1}{n!} \alpha_{i_1 ... i_n ,kl} dx^l \wedge dx^k \wedge dx^{i_1} \wedge \cdots \wedge dx^{i_n} \equiv 0 \ ,
\end{equation}
since partial derivatives commute.

 On a $D$-dimensional manifold a $D$-form is locally determined by a function (for this reason it is  useful to think of functions as $0$-forms). However, under a change of coordinates a $D$-form transforms like a density,
\begin{equation}
f' dx'^1 \wedge \cdots \wedge dx'^D = (\det T) f  dx^1 \wedge \cdots  \wedge dx^D \ ,
\end{equation}
where $T^i_j$ is the Jacobian matrix
\begin{equation}
T^i_j = \frac{\partial x'^i}{\partial x^j} \ .
\end{equation}
If $\det{T}$ is everywhere positive a $D$-form transforms in the same way as an integral under a change of variables, which means that one can integrate it over $\M$.\footnote{There are quite a  few technicalities involved in defining integration over a manifold. For a nice explanation see \cite{Baez:1995sj}.}  A wedge product of $D$ vectors, on the other hand, can't be integrated. If it's possible to find a set of charts such that $\det{T} > 0$, $\M$ is said to be \emph{orientable} and we can define a nowhere vanishing $D$-form that characterizes the volume of $\M$ called a \emph{volume form}. A nowhere vanishing $n$-form for $n<D$ can be integrated over an oriented $n$-dimensional submanifold of $\M$.

Topological information can be extracted by studying  $n$-forms that obey
\begin{equation}
d \alpha_n = 0 \ .
\end{equation}
 These are called \emph{closed}, and the vector space of all such forms is denoted $Z^n$. Due to the exterior derivative being nilpotent, an $n$-form can be closed simply because it's of the form
 \begin{equation}
 \alpha_n = d \beta_{n-1} \ .
 \end{equation}
Such forms are called \emph{exact}, and form a vector space $B^n$. It is not generally true that  every closed form is exact, but  it's not hard to show that \emph{locally} they can be expressed as $\alpha = d \beta$.\footnote{This result is know as the Poincar\'{e} lemma.} Therefore, an obstruction to a closed form being exact is a topological one, and the objects of interest are closed forms under the equivalence relation
 \begin{equation}
 \alpha^1_n \sim \alpha^2_n  \ \ \ \ \  \mathrm{if} \ \ \ \ \ \ \alpha^1_n - \alpha^2_n = d\beta_{n-1} \ .
 \end{equation}
 The space of all closed forms modulo the above relation, 
 \begin{equation}
 \label{eq:real_cohom_def}
 H^n = \frac{Z^n}{B^n} \ ,
 \end{equation}
 is called the \emph{n-th deRham cohomology group}.  $H^n$ is Abelian, and in fact  it has more structure, being a vector space. The term "group" is used due to other cohomology theories, which can involve groups that are not vector spaces. The dimension of the $H^n$ is the \emph{n-th Betti number}:
 \begin{equation}
 \label{eq:betti_number}
 b_n = \mathrm{dim} H^n \ .
 \end{equation}
The \emph{Euler characteristic}  topological invariant $\xi$ is related to the Betti numbers:
 \begin{equation}
 \label{eq:euler_characteristic}
 \xi  = \sum_n (-1)^n b_n \ .
 \end{equation}


What role do these groups play in the topology? In rough terms, they are related to integrals of $n$-forms over compact submanifolds $\N$ of $\M$. Making use of Stokes' theorem,
\begin{equation}
\label{eq:stokes_theorem}
\int_\N \alpha_n = \int_{\N} d \beta_{n-1} = \int_{\partial \N} \beta_{n-1} = 0 \ ,
\end{equation}
we see that an integral of an exact form over $\N$ is zero, since $\N$ has no boundary. Conversly, it can also be shown that a form is exact if and only if its integral over all compact submanifolds vanishes. So if we can find some $\N$ such that the integral of $\alpha_n$ doesn't vanish, $\alpha_n$ can't be exact.

Taking for simplicity a closed 1-form $\alpha$, we can attempt to show that it's exact by constructing a function $\beta$ such that $\alpha= d \beta$.  This can be done as follows,
\begin{equation}
\beta = \int _y^x \partial_i \beta dx^i \ ,
\end{equation}
 but only if the integral is independent of the starting point $y$. Due to Stokes' theorem, this will necessarily be the case when the difference between two paths, which is a circle, is a boundary of some open subset of $\M$. Such a boundary can be found when the circle is contractible to a point. If non-contractible circles exist the construction will not always work, and it is possible to have closed forms that are not exact. That such forms always exist is true, but harder to show.
  
As an example we can think of a torus, which has two classes of non-contractible circles that wrap the two holes of the torus. Given a closed $1$-form that is not exact, it follows from the result stated after (\ref{eq:stokes_theorem}) that integrating it over circles in at least one of these classes gives non-vanishing results. In fact, it is not hard to see that
\begin{equation}
H^1 \cong \bbR \oplus \bbR \ ,
\end{equation}
since (modulo exact forms) there is only one linearly independent closed 1-form with non-vanishing integrals over circles belonging to one of the above classes, and vanishing integrals over circles belonging to the other.   Closed $0$-forms are constant functions, and none of them are exact since there are no $-1$-forms. Thus, $H^0 \cong \bbR$. In general $H^0$ just counts the connected components of the manifold. We'll see later that $H^0 \cong H^D$, so in fact $H^2 \cong \bbR$ for a torus.  

 
  In summary, I've briefly introduced forms and the theory of cohomology, and touched on a relation between cohomology and the enumeration of submanifolds of $\M$ that are not boundaries of open sets.  The latter formalism is known as \emph{homology}. I've not attempted to explain it in any detail\footnote{I refer the reader to Chapter 3 in \cite{Nakahara:1990th} for a nice treatment of the subject.}, but in a precise sense it is a theory dual  to cohomology. Cohomology is significantly more powerful since it involves calculus.
 
 \subsection{Almost complex and complex structures}
 \label{subsec:almost_cpx_structures}
 
 A \emph{complex manifold} of $D/2$ complex dimensions, which is also a real manifold of $D$ dimensions, is required to look like $\mathbb{C}^{\frac{D}{2}}$ locally. In order for holomorphic functions to make sense globally, the transition functions between the charts  are themselves required to be holomorphic. 
 
  
On a complex manifold the tangent vector spaces are naturally complex. A real tangent vector is written as a sum of holomorphic and antiholomorphic  parts,
\begin{equation}
V^{\mathbb{R}} = 
V^{\alpha} \frac{\partial}{\partial z^\alpha} + V^{\balpha } \frac{\partial}{\partial z^\balpha }  \ ,
\end{equation}
with $V^{\balpha }$   the complex conjugate of $V^{\alpha }$.  More general vectors are of the form
\begin{equation}
V^{\mathbb{C}} = V^{\alpha} \frac{\partial}{\partial z^\alpha} + W^{\balpha } \frac{\partial}{\partial z^\balpha }  \ , 
\end{equation}
 with $W^\balpha \neq V^\balpha$, and will have  complex components when expressed in a real basis. 
 

Since a complex manifold is also a real manifold, it should be possible to define it without reference to complex charts. It can be shown that a real $D$-dimensional manifold is tantamount to a  complex manifold of $D/2$ dimensions if it admits a $(1,1)$ tensor $I^i_{\ j}$ that squares to $-1$,
\begin{equation}
I^i_{\ j} I^k_{\ i} = - \delta^k_j \ ,
\end{equation}
such that the \emph{Nijenhuis tensor}, 
\begin{equation}
\label{eq:Nijenhuis_tensor_geom_ch}
N^i_{\ lm} :=  I^k_{\ [l}  I^i_{\ m],k} + I^i_{\ k}  I^k_{\ [l,m]} \ , 
\end{equation}
vanishes.\footnote{For a proof of this statement see \cite{Candelas:1987is}.} The vanishing of the Nijenhuis tensor implies that it's possible to find an chart $\bbC^{\frac{D}{2}}$ such that
\begin{equation}
\label{eq:cpx_structure_in_homor_coords}
I^{\alpha}_{\ \beta} = i \delta^{\alpha}_{\beta}  \ \ \ , \ \ \ I^{\balpha}_{\ \bbeta} = -i \delta^{\balpha}_{\bbeta} \ ,
\end{equation}
and that such charts can be patched across $\M$ so that $I^i_{\ j}$ really defines a \emph{complex structure}. If the Nijenhuis tensor doesn't vanish, a globally defined tensor $I^i_{\ j}$ that squares to $-1$ is referred to as an \emph{almost complex structure}. 


On a complex manifold tensors are naturally decomposed into terms containing different numbers of  holomorphic and antiholomorphic components. In real coordinates a grading is provided by the projection tensors\footnote{Meaning that $P^2 = 1, Q^2 = 1, PQ = QP = 0$, which enables one to perform a grading consistently.}:
\begin{equation}
P^i_{\ j} = \frac{1}{2} ( \delta^i_j - i I^i_{ \ j}) \ \ \ , \ \ \  Q^i_{\ j} = \frac{1}{2} ( \delta^i_j + i I^i_{ \ j}) \ .
\end{equation}
In the complex case one can write equations such as
\begin{align}
& K_{\balpha \bbeta } dx^{\balpha} \otimes dx^{\bbeta} = P^k_{\ i} P^l_{\ j} K_{kl} dx^i \otimes dx^j \ \ \ ,  \\ \nonumber
&  K_{\alpha \bbeta} dx^{\alpha} \otimes dx^{\bbeta} = Q^k_{\ i} P^l_{\ j} K_{kl} dx^{i} \otimes dx^{j}\ \ \  , \ \ \ \mathrm{etc. \ ,}
\end{align}
for some tensor $K_{ij}$, but in the almost complex case these equations hold only in reference to a single point. It is always legitimate to perform algebraic manipulations using holomorphic coordinates, but in the almost complex case one is not allowed to make conclusions about differential operations. For example, if the only non-vanishing component of a two form $\alpha$ is $\alpha_{\alpha \bbeta}$ one would conclude naively that $d \alpha$ can't have a part pure in its indices, $d \alpha_{\alpha \beta \gamma}$. In the almost complex case this is not necessarily true.

Let us now concentrate on the complex case. A $(p+q)$-form with $p$ holomorphic and $q$ antiholomorphic indices will be written as a $(p,q)$-form, or $\alpha^{(p,q)}$, with $p,q <  \frac{D}{2}$. For example, a general $3$-form $\alpha$ is decomposed as
\begin{equation}
\alpha = \alpha^{(3,0)} + \alpha^{(2,1)} + \alpha^{(1,2)} + \alpha^{(0,3)} \ .
\end{equation}
The exterior derivative can be decomposed in the same way, $d \alpha = d \alpha^{(4,0)} +  d \alpha^{(3,1)} + \cdots$. Furthermore, one can define an operator $\partial$ that maps a $(p,q)$-form into a $(p+1, q)$-form, and an operator $\bpartial$ that maps a $(p,q)$-form into a $(p, q+1)$-form.  These are called \emph{Dolbeault operators} and act in the expected manner:
\begin{align}
\label{eq:dolbeaut_action}
\partial (\alpha_{\alpha \bbeta} dx^\alpha \wedge dx^\bbeta) = \alpha_{\alpha \bbeta, \gamma} dx^\gamma \wedge dx^\alpha \wedge dx^\bbeta \ , \\ \nonumber
\bpartial  (\alpha_{\alpha \bbeta} dx^\alpha \wedge dx^\bbeta) = \alpha_{\alpha \bbeta, \bgamma}  dx^\bgamma \wedge dx^\alpha \wedge dx^\bbeta \ .
\end{align}
They obey $\partial \bpartial + \bpartial \partial = 0$ and are separately nilpotent, so on a complex manifold a more refined cohomology than  (\ref{eq:real_cohom_def}) exists:
\begin{equation}
H_{\bpartial}^{(p,q)} \equiv \frac{Z_{\bpartial}^{(p,q)}}{B_{\bpartial}^{(p,q)} }  \ .
\end{equation}
The Betti numbers (\ref{eq:betti_number}) are now denoted as $b_{p,q}$. 

\subsection{Symplectic manifolds}
\label{subsection:symplectic_manifolds}
  
  The  \emph{symplectic} structure is defined by a 2-form, 
\begin{equation}
\label{eq:symplectic_structure}
S = S_{ij} dx^i \wedge dx^j \ ,
\end{equation}
that is non-degenerate, meaning that $\det(S) \neq 0$. So it's possible to define an inverse matrix $S^{ij}$ such that
\begin{equation}
S^{ij} S_{jk} = \delta^i_k \ .
\end{equation}

The symplectic structure features in classical Hamiltonian mechanics and in canonical quantization.
 As will be discussed in Chapter \ref{ch:antifield_formalism}, in the context of covariant quantization it is natural to work on a special type of a  \emph{Grassmannian manifold}, or supermanifold, on  which to every bosonic coordinate we also append a fermionic one. On such  manifolds the natural structure is an  antisymplectic one.

\subsection{Connections and holonomy groups}
\label{subsection:connections_holonomy_groups}

The structure that allows us to compare tangent vectors at different points on $\M$,  or more generally, compare vectors in different fibers lying in some vector bundle over $\M$, is called a \emph{connection}. 


Consider  $\M$ as the base space of a vector bundle $E \rightarrow \M$, with fiber $E_x$ over the point $x \in \M$. Denoting the space of sections of $E$ as $\Gamma(E)$, a connection $\nabla$ on $\M$ assigns to each vector field $v \in \mathrm{Vect} ( \M )$ a map from $\Gamma(E)$ to itself satisfying: 
\begin{eqnarray}
\label{eq:connection_def}
\nabla_v( s+s' ) & = & \nabla_v s + \nabla_v s'  \ , \\ \nonumber
\nabla_{v} (f s) & = & v(f) s + f \nabla_v (s) \ , \\ \nonumber 
\nabla_{(v + v')} s & = & \nabla_v s + \nabla_{v'} s   \ , \\ \nonumber
\nabla_{fv} s & = & f \nabla_{v} s \ ,
\end{eqnarray}
for all $f \in C^{\infty} (\M)$, $s, s' \in \Gamma(E)$, and $v, v' \in \mathrm{Vect} (\M )$. $\nabla_v (s)$ is also referred to as the \emph{covariant derivative} of $s$ along $v$.

Given a a map  $\gamma : [0,T] \rightarrow \M$, we can pull back the connection on $E \rightarrow \M$ to a connection on $E \rightarrow [0,T]$. A \emph{constant section} $s(\gamma(t))$ is required to satisfy
\begin{equation}
\label{eq:parallel_transport}
\nabla_{\frac{\partial \gamma (t)}{\partial t}} s(\gamma (t)) = 0 \ ,
\end{equation}
where $t \in [0, T]$. It defines  parallel transport along $\gamma$. The \emph{holonomy} of a connection along a smooth path $\gamma$, $H (\gamma, \nabla)$, is defined as the linear map between fibers over different points of $\M$ obtained by parallel transporting along $\gamma$ from $\gamma(0)$ to $\gamma(T)$. This naturally extends to  piecewise smooth maps, $H(\gamma_3, \nabla) := H(\gamma_2, \nabla) H(\gamma_1, \nabla)$, where  $\gamma_3$ is constructed from the smooth components $\gamma_1$ and $\gamma_2$.


Taking $\gamma$ to be a loop based at $x \in \M$, i.e. $\gamma(0) = \gamma(T)=x$, parallel transport defines an automorphism of the vector space at $E_x$. The set of automorphisms obtained from all possible loops based at $x$ form a subgroup of $\mathrm{GL}(\mathrm{dim}(E), \mathbb{R})$, known as the \emph{holonomy group} $\Hol_x (\nabla)$. One can drop the reference to a particular point, since given two points $x$ and $y$, and a path $\gamma'$ from $x$ to $y$
\begin{equation}
\Hol_y (\nabla) = H(\gamma', \nabla) Hol_x (\nabla) H(\gamma', \nabla)^{-1} \ .
\end{equation}
Once we have identified $E_x$ and $E_y$ with $\mathbb{R}^{{\mathrm{dim}(E)}}$, the holonomy groups based at different points can be thought of as subgroups of $\mathrm{GL}(\mathrm{dim}(E), \bbR)$ related by conjugation, and are clearly isomorphic. Thus the holonomy group is a topological property of $\M$, independent of reference to a particular point.

In local coordinates the covariant derivative of a section $s$ along a vector field $v$ is written as
\begin{equation}
\label{eq:cov_derivative_def1}
\nabla_v (s) = v^i(\partial_{i} s^a + A^a_{i b} s^b) e_a \ ,
\end{equation}
where $e_a$ are the basis vectors of the fibers. I've used letters from the beginning of the alphabet for the fiber indices, and letters from the middle of the alphabet for the tangent indices of the base space. For each value of the index $a$, $e_a(x)$ is locally a  non-vanishing section of the fiber bundle. For many bundles there are no nowhere vanishing global sections. The \emph{connection coefficients} $A^a_{i b}$ define the action of the covariant derivative  on the basis vectors of the fibers along the basis vector fields of the tangent space,
\begin{equation}
\nabla_{\partial_i} e_a \equiv \nabla_i e_a = A^{b}_{i a} e_b \ .
\end{equation}
The expression for the covariant derivative of a generic section (\ref{eq:cov_derivative_def1}) follows from the properties (\ref{eq:connection_def}). In practice it is convenient to work with derivatives along the basis vector fields,
\begin{equation}
\nabla_i s^a = \partial_{i} s^a + A^a_{i b} s^b \ ,
\end{equation} 
since a covariant derivative along a generic vector field is obtained simply by contracting with $v^i$.

A natural object that can be constructed from a connection is the \emph{curvature}:
\begin{equation}
\label{eq:riemann_tensor_def}
F (U, V, s) = (\nabla_U \nabla_V - \nabla_V \nabla_U  - \nabla_{[U,V]})s \ ,
\end{equation}
where the notation indicates that we can think of it acting on two vector fields $U$ and $V$ and a section $s$. The result is of course another section.

When the total space is the tangent bundle, all the indices on the connection coefficients are the same:
\begin{equation}
\nabla_i \partial_j = \Gamma^k_{\ ij} \partial_k  \ .
\end{equation}
One can also construct general tensor bundles over $\M$. The covariant derivative of a $(p,q)$ type tensor is given by
\begin{equation}
\nabla_i T^{j_1 ... j_p}_{k_1 ... k_q} = \partial_i T^{j_1 ... j_p}_{k_1 ... k_q} 
+ \Gamma^{j_1}_{\ i l} T^{l j_2 ... j_p}_{k_1 ... k_q} + \cdots - \Gamma^l_{\ i k_1}  T^{j_1 ... j_p}_{l k_2 ... k_q} - \cdots \ .
\end{equation}
 and can be used to define the parallel transport of a $(p,q)$-type tensor.  If $\M$ has additional structure (for example, a metric, an (almost) complex, or a symplectic structure), it is natural to restrict the connection  by demanding that these structures be covariantly constant. Such constraints are generally too weak to define the connection uniquely. 

The curvature of a connection on a tangent bundle is a tensor of type $(1,3)$ and is conventionally denoted as $R$. When $\M$ is a Riemannian manifold $R$ is the Riemann tensor (one normally also assumes that the connection preserves the metric). $R$ acts on three vector fields, and the result is also a vector field. In components:
\begin{align}
\label{eq:riemann_tensor_def_compon}
& R (U, V, W)  = R^i_{\ jkl}U^j V^k W^l \ ,  \\ \nonumber
& R^i_{ \ jkl} = \Gamma^i_{\ lj,k} - \Gamma^i_{\ kj,l} + \Gamma^m_{\ lj} \Gamma^i_{\ km} 
- \Gamma^m_{\ kj} \Gamma^i_{\ lm} \ .
\end{align}
From a connection on the tangent bundle we can also construct a type $(1,2)$ tensor called the \emph{torsion}: 
\begin{align}
\label{eq:torsion}
& T(U, V) =  \nabla_U V - \nabla_V U - [U,V]  \  , \\ \nonumber
& T^i_{ \ jk} = \Gamma^i_{\ jk} - \Gamma^i_{\ kj} \ .
\end{align}
For a general vector bundle torsion has no meaning.

\begin{figure}
\includegraphics[scale=0.49]{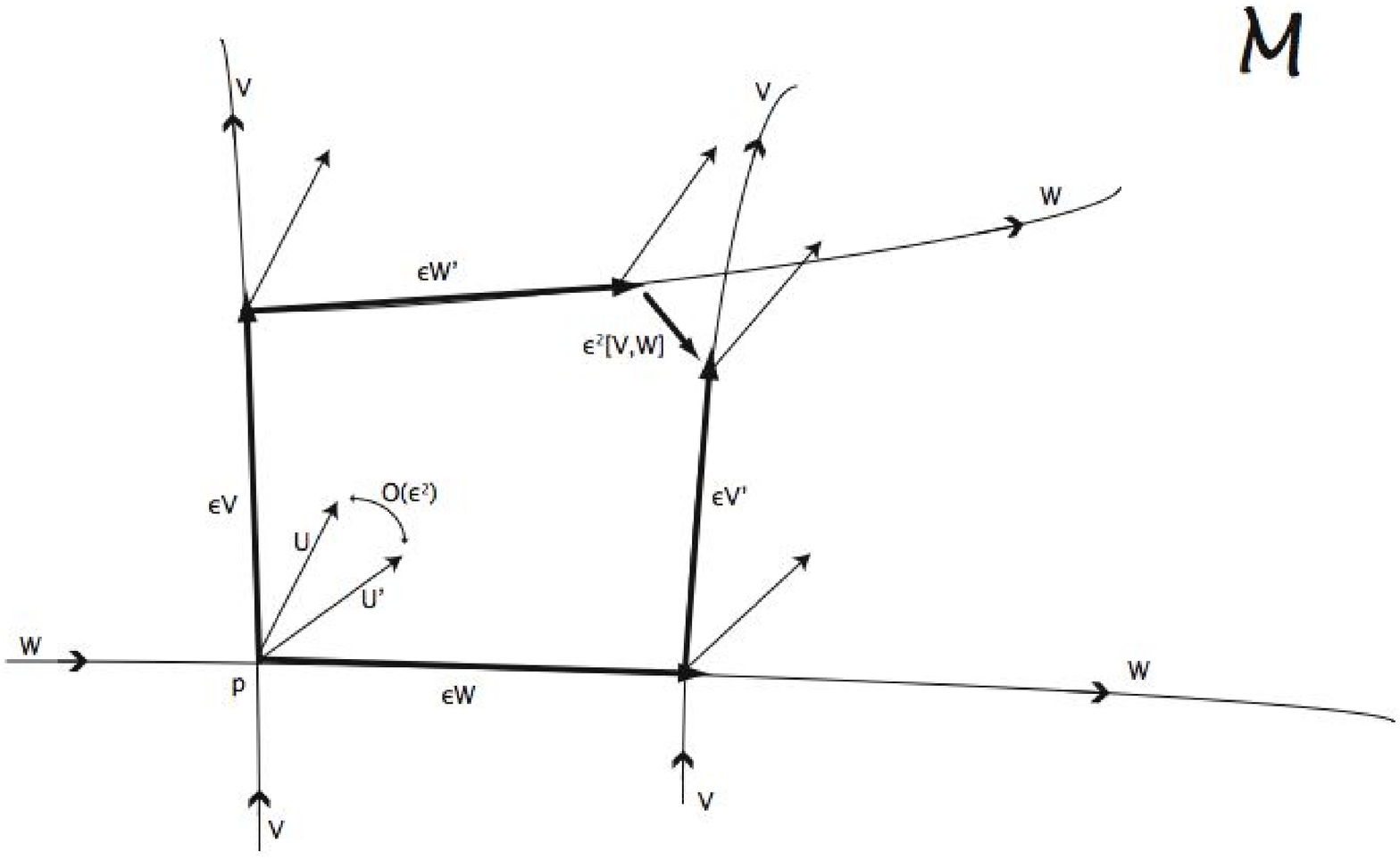}
\caption{The infinitesimal effect of the Lie derivative  $\L_V W$, shown by the darker arrows, is of order $\epsilon^2$. Parallel transporting a vector $U$ using a connection along an infinitesimal  loop defined by the dark arrows, with the $\epsilon^2$ gap bridged by the Lie derivative, induces a linear transformation such that $U^i - U^{'i} = \epsilon^2 R^i_{ \ jkl}U^j V^k W^l $. }
\label{figure:lie_derivative}
\end{figure}

The definition of the curvature tensor  (\ref{eq:riemann_tensor_def}) involves the Lie bracket. I'll remind the reader of its geometric meaning, since it is important for understanding the infinitesimal meaning of the curvature and torsion tensors. The Lie bracket is given by
\begin{equation}
\L_V W \equiv [ V, W ] :=  V^i \partial_i (W^j \partial_j) - W^i \partial_i (V^j \partial_j) 
=  (V^i \partial_i W^j  - W^i \partial_i V^j )\partial_j  \ .
\end{equation}
We call $\L_V W$ the Lie derivative of the vector field $W$ along the vector field $V$. Infinitesimally,  $\L_V W$ is the vector field obtained by dragging $W$ along $V$ and $V$ along $W$, and taking the difference. If the vectors $V$ and $W$ are of order $\epsilon$, the Lie derivative will be of order $\epsilon^2$ (see Figure \ref{figure:lie_derivative}). Such parallel transport is possible given two predefined vector fields; we can't compare arbitrary tangent vectors at different points on $\M$ using the Lie derivative.   

The infinitesimal meaning of the curvature tensor can also be understood with the help of Figure \ref{figure:lie_derivative}. We now use the connection to parallel transport some third vector  $U$ along  $V$ and then along $W$, and vice versa. The gap between the two paths is of order $\epsilon^2$ and it prevents us from comparing the vectors since their difference is of the same order.  The part of (\ref{eq:riemann_tensor_def}) containing the Lie bracket  compensates for this gap, so that the vectors  can be subtracted. In the end, $\epsilon^2 R (U, V, W)$ is the result of this subtraction. The same result is obtained by parallel transporting a  single vector along the loop, and comparing the result with the original vector, which is the situation depicted in  Figure \ref{figure:lie_derivative}.

It follows that the curvature tensor gives us infinitesimal elements of the holonomy group:
\begin{equation}
\label{eq:holonomy_generator1}
U^i - U^{'i} =  \delta U^i = \epsilon^2 R^i_{ \ jkl}U ^j V^k W^l = \delta a^{kl} R^i_{ \ jkl}U^j  \ .
\end{equation}
So  $\delta a^{kl} R^i_{ \ jkl}$ are matrices of the Lie algebra of the holonomy group; $i$ and $j$ are the matrix indices, and $k$ and $l$ form a basis of the Lie algebra. Alternatively,
\begin{equation}
\delta U^i = \int_S dx^k \wedge dx^l R^i_{ \ jkl}U^j 
\end{equation}
when transported along a small closed curve that bounds the surface $S$.
 On a manifold that's simply connected\footnote{Meaning that every loop can be contracted to a point.} the Lie algebra determines the holonomy group straightforwardly. For manifolds that are not simply connected one has to be more careful (see \ref{section:spechol_manifolds}).

The holonomy group of an unrestricted connection is contained in $GL(D, \bbR)$. If there is a  covariantly constant vector field we obtain the integrability condition
\begin{equation}
[ \nabla_k, \nabla_l ] V^i =  R^i_{ \ jkl}V^j = 0  \ ,
\end{equation}
and a similar expression for a general tensor 
\begin{equation}
\label{eq:general_holonomy_expression}
[ \nabla_k, \nabla_l ] T = 0 \ .
\end{equation}
If there is a nowhere vanishing invariant tensor on $\M$ the holonomy group is reduced to a proper subgroup of $GL(D, \bbR)$. In terms of the special tensors introduced so far, the covariant constancy of the almost complex structure and the volume form will reduce the holonomy group to $SL(D, \bbR)$ and $SL(D_{\bbC}, \bbC)$ respectively. When the holonomy group lies in a subgroup $G$ of $GL(D, \bbR)$, the manifold is said to have \emph{G-structure}.

Torsion is best understood geometrically by taking two vector fields and transporting one along the other, but now using the connection rather than the Lie derivative. The result is that the gap between vectors - the analogue of $\epsilon^2 [V,W]$ in Figure \ref{figure:lie_derivative} - will be of order $\epsilon^2$ in the presence of torsion, but only of order $\epsilon^3$ when torsion vanishes.

\section{Riemannian manifolds leading to the Calabi-Yau}  
  
Any differential manifold admits a metric, which is a non-degenerate symmetric $(0,2)$ type tensor:
 \begin{equation}
 G = G_{ij} dx^i \otimes dx^j \ .
 \end{equation} 
A \emph{Riemannian manifold} is the pair $(\M, G)$.  The presence of a metric  has the following implications:
 \begin{itemize}
 \item  The inner product can be defined, since there is a natural map between tangent and cotangent vectors, i.e. we can raise, lower, and contract indices. On a Riemannian manifold the inner product is required to be positive definite. Otherwise the manifold is semi-Riemannian.
 \item On an orientable manifold there exists a canonical volume form,
  \begin{equation}
  \label{eq:metric_vol_form}
  Vol = \sqrt{ G} dx^1 \wedge \cdots \wedge dx^D \ ,
  \end{equation}
  where $G$  stands for the determinant of the metric.
  \item Using $Vol$ one can define the \emph{Hodge $*$-operator}  that maps $n$-forms to $(D-n)$-forms,
  \begin{equation}
  \label{eq:hodge_operator}
  *(dx^{i_1}  \wedge \cdots  \wedge dx^{i_n}) = \frac{1}{(D-n)!} \sqrt{G} G^{i_1 j_1} \cdots G^{i_n j_n} \epsilon_{j_1 \cdots j_n j_{n+1} \cdots j_D}  dx^{j_{n+1}} \wedge \cdots \wedge dx^{j_D}   \ ,
  \end{equation}
   where  $\epsilon_{j_1 \cdots j_D }$ is the Levi-Civita alternating symbol. It provides us with an inner product between $n$-forms,
   \begin{equation}
   \label{eq:form_inner_product}
   (\alpha, \beta ) := \int_\M \alpha \wedge *\beta \ .
   \end{equation}
  \item Demanding that the metric be covariantly constant defines the connection up to torsion,
  \begin{equation}
  \label{eq:riemannian_connection}
  \Gamma^i_{ \ jk} = \Gamma^{i(l.c.)}_{ \ jk} + \frac{1}{2}(T^{ \ i}_{j \ k} + T^{ \ i}_{k \ j} + T^i_{ \ jk}) \ ,
  \end{equation}
  where $ \Gamma^{i(l.c.)}_{jk}$ are the Levi-Civita coefficients:
  \begin{equation}
  \label{eq:LC_coefficients}
   \Gamma^{(l.c.)}_{ijk} = \frac{1}{2} \left( G_{ij,k} + G_{ik,j} - G_{jk,i} \right) \ .
  \end{equation}
  The holonomy group of a connection that preserves the metric is contained in $O(D, \bbR)$. On an orientable manifold on which $Vol$  (\ref{eq:metric_vol_form}) is preserved the holonomy group is contained in $SO(D, \bbR)$.
  Requiring that the torsion vanishes gives us the unique Levi-Civita connection. As we'll see in Chapter \ref{chapter:2d_sigma_models},  $T_{ijk}$ is automatically totally antisymmetric in the $\sigma$-model context, in which case the first two terms in the bracket in  (\ref{eq:riemannian_connection}) cancel. 
 \end{itemize}
  
  The inner product (\ref{eq:form_inner_product}) allows us to define the adjoint of the exterior derivative, $d^\dagger$, by requiring that
\begin{equation}
(\alpha, d \beta) = (d^\dagger \alpha, \beta)
\end{equation}
for some $n$-form $\alpha$ and $(n-1)$-form $\beta$. Clearly, $d^\dagger$ maps $n$-forms to $(n-1)$-forms. The explicit expression is:
  \begin{equation}
  d^\dagger \alpha := (-1)^{n(D-n+1)} * d * \alpha = - \frac{1}{(n-1)!} \nabla ^k \alpha_{k i_2 ... i_{n-1}} dx^{i_2} \wedge \cdots \wedge dx^{i_{n-1}} \ .
  \end{equation}
  It's not hard to see that $(d^\dagger)^2=0$, and thus one can define \emph{co-closed} forms to be annihilated by $d^\dagger$, and \emph{co-exact} ones to be of the form $\alpha = d^\dagger \beta$.

The \emph{Laplacian} is defined as:
\begin{equation}
\Delta  = (d + d^\dagger)^2 = d d^\dagger + d^\dagger d \ .
\end{equation}
A form annihilated by $\Delta$ is called \emph{harmonic}, and it follows straightforwardly that a harmonic form must be both closed and co-closed. A result known as \emph{Hodge's theorem} states that any form can be decomposed uniquely into harmonic, closed, and co-closed parts,
\begin{equation}
\alpha = \tau + d\beta + d^\dagger \gamma \ ,
\end{equation}
  where $\tau$ is the harmonic part. A general closed form can be written as
  \begin{equation}
  \alpha = \tau + d\beta  \ ,
  \end{equation}
so once a metric is fixed each member of the cohomology group is uniquely represented by a harmonic form. Changing the metric will change the harmonic form, but the difference must be exact.
 
 Now we are in a position to prove powerful theorems quite easily. For example, from the above results  it's easy to see that the harmonic condition implies the following:
 \begin{equation}
 d \tau = d^\dagger \tau = 0 \longrightarrow * d \tau = (-1)^{n(-D-n+1)} d^\dagger * \tau \ .
 \end{equation}
 Thus, if $\tau$ is harmonic $*\tau$ is as well, and the spaces $H^n$ and $H^{(D-n)}$ have to be isomorphic. This theorem is known as Poincar\'{e} duality.\footnote{One immediate consequence is that the Euler characteristic  (\ref{eq:euler_characteristic}) for an odd dimensional manifold vanishes.}
 
 I will not make use of more results than those  presented here, but it is the start of quite a wonderful subject which explores the relation between cohomology and homology mentioned at the end of  \ref{subsection:nforms}. The idea is that we can make precise the notion of an $n$-dimensional submanifold $\N$ of $\M$ being dual to a $(D-n)$-form $\alpha$ by requiring that
 \begin{equation}
 \int_\N \beta = \int_\M \alpha \wedge \beta
 \end{equation}
  for all closed $n$-forms $\beta$.  I refer the reader to \cite{Candelas:1987is, Eguchi:1980jx}.

 \subsection{(Almost) Hermitian manifolds}
 \label{subsection:hermitian_manfiolds}
  
  When working on an (almost) complex manifold it is natural to require the (almost) complex structure to be compatible with the metric, in the sense that
  \begin{equation}
  \label{eq:hermitian_cond}
  I^i_{\ j} I^k_{ \ l} G_{ik} = G_{jl} \ .
  \end{equation}
  Then $\M$ is an \emph{(almost) Hermitian manifold}, which is a restriction on the metric rather than $\M$, since a manifold with a non-Hermitian metric $G_{ij}$ also admits the Hermitian metric  \cite{Candelas:1987is}
  \begin{equation}
  G'_{ij} = G_{ij} +   I^i_{\ j} I^k_{ \ l} G_{ik} \ .
  \end{equation}
  Condition   (\ref{eq:hermitian_cond}) is equivalent to the requirement that in holomorphic coordinates the metric has only mixed components ($G_{\alpha \bbeta}$ and $G_{\bbeta \alpha}$), or to the requirement
  \begin{equation}
  I_{ij} = - I_{ji}  \ .
  \end{equation}
 Thus on an (almost) Hermitian manifold there is a natural real 2-form $\omega$ called the \emph{K\"{a}hler form}\footnote{Sometimes this term is reserved for the case when $\M$ is complex.}:
  \begin{equation}
  \label{eq:def_kahler_form}
  \omega = \frac{1}{2} I_{ij} dx^i \wedge dx^j = G_{\alpha \bbeta} dx^\alpha \wedge dx^\bbeta  \ .
  \end{equation}
  
Using the K\"{a}hler  form a nowhere vanishing real $D_{\bbC}$-form can be constructed as
  \begin{equation}
  \label{eq:kahler_vol_form}
  \underbrace{\omega \wedge \cdots \wedge \omega}_{D_{\bbC} \ \mathrm{times}} \ ,
  \end{equation}
 where  $D_{\bbC} \equiv D/2$, and we can conclude that any (almost) complex manifold is orientable. 

Poincar\'{e} duality for complex manifolds tells us that  $b_{p,q} = b_{(D_{\bbC} - p), (D_{\bbC}- q)}$. It is customary to write the Betti numbers in an array known as the \emph{Hodge diamond}, for example,
  \begin{equation}
  \label{eq:hodge_diamond_2d}
  {\arraycolsep=2pt
\begin{array}{ccccc} & & b_{0,0} & &\\
& b_{1,0}& & b_{0,1} &\\
b_{2,0} & & b_{1,1} & & b_{0,2}\\
& b_{2,1}& & b_{1,2} &\\
& & b_{2,2} & &
\end{array}  \ 
}
  \end{equation}
   for $D_{\bbC}=2$:
 Due to Poincar\'{e} duality the Hodge diamond is symmetric about the horizontal axis.
  
  A connection on a Hermitian manifold is not fully specified by requiring that the complex structure and the metric be covariantly constant. However, it's not necessary to require that the torsion vanish to obtain a unique connection, instead we can restrict the torsion to be pure in its lower indices. The unique \emph{Hermitian connection} is given by
 \begin{equation}
 \label{eq:hermitian_connection}
 \Gamma^{\alpha}_{ \ \beta \gamma} = G^{\alpha \btau} \partial_\beta G _{\gamma \btau} \ \ \ \ \mathrm{and \ c.c.} \ ,
 \end{equation}
  and the only non-vanishing components of the Riemann tensor are
  \begin{equation}
  \label{eq:hermitian_riemann_tensor}
  R^{\balpha}_{ \  \bgamma \beta \bdelta} = \partial_\beta \Gamma^\balpha_{ \  \bgamma \bdelta}
  \ \ \ \ \ \mathrm{and \ c.c.} \ .
  \end{equation}
It is clear from the index structure of $R$ that  the holonomy group is reduced to $U(D_{\bbC})$, and it's not hard to see that the same is true for any connection  that preserves an (almost) complex structure and an (almost) Hermitian metric. 
  
  From a Hermitian connection a non-trivial member of $H^2$ can be constructed as
  \begin{align}
  \label{eq:ricci_form}
  \calR =&  \frac{1}{4} R^i_{ \ jkl} I^{j}_{\ i} dx^k \wedge dx^l 
  = iR^{\alpha}_{ \ \alpha \mu \bbeta} dx^\mu \wedge dx^\bbeta \\ \nonumber
  = & i \partial \bpartial \ln \sqrt{G} \ .
  \end{align}
  The last line follows from (\ref{eq:hermitian_riemann_tensor}) and indicates that $\calR$ is closed, since $\partial \bpartial =  - \frac{1}{2} d( \partial - \bpartial)$ (see from (\ref{eq:dolbeaut_action})). However,  $\ln \sqrt{G}$ is not a scalar so the form is not necessarily exact. 
  
$\calR$ is related to a special member of $H^2$ known as the \emph{first Chern class}, $c_1$, by
\begin{equation}
\label{eq:first_chern_class}
c_1 = \frac{1}{2 \pi} \left[  \calR\right ]  \ .
\end{equation}
The factor of $1/2 \pi$ is a convention related to normalization of integrals over $\M$, and is chosen in such a way that $c_1$ is an \emph{integral class}, meaning that  integrating it over $\M$ gives an integer.

  \subsection{\kah  and Calabi-Yau manifolds}
  \label{subsection:kah_and_CY}
  
  A \emph{ \kah manifold} is a Hermitian manifold for which the \kah form is closed:
  \begin{equation}
  \label{eq:closed_kahler_form}
  d \omega = 0 \ .
  \end{equation}
  Thus on a \kah manifold $\omega$ defines a symplectic structure (\ref{eq:symplectic_structure}). An equivalent condition is that complex structure be covariantly constant with respect to the Levi-Civita connection:
  \begin{equation}
  \nabla^{(l.c.)}_\mu I = 0 \ .
  \end{equation}
  It follows directly from (\ref{eq:closed_kahler_form}) that 
  \begin{equation}
  \label{eq:kah_metric_relations}
  G_{\alpha \bbeta, \bgamma} =   G_{\alpha \bgamma, \bbeta}  \ \ \  , \ \ \ 
  G_{\balpha \beta, \gamma} =   G_{\balpha \gamma, \beta}  \ .
  \end{equation}
  So on a \kah manifold the Hermitian connection is equivalent to the Levi-Civita connection, since it is symmetric in its lower indices. The covariant constancy of the complex structure implies that $\omega$ is also covariantly constant, from which it can be concluded that it's co-closed. Thus, $\omega$ is harmonic. 
  
  On a \kah manifold the cohomology groups  of $\partial$ and $\bpartial$ are the same. This can be seen most easily by showing that 
  \begin{equation}
  \Delta_{\partial} = \Delta_{\bpartial} = \frac{1}{2} \Delta_{d} \ .
  \end{equation}
  It follows that $b_{p,q} = b_{q,p}$, so the Hodge diamond (\ref{eq:hodge_diamond_2d}) is symmetric about its vertical axis.
  
By invoking Stokes' theorem it can be shown that if $\omega$ were exact the integral of the volume form would be zero, at least for compact manifolds without a boundary. Thus, at least for these cases, $\omega$ can't be exact.  This means that locally  $\omega$  can be expressed in terms of the \emph{\kah potential} $\calK$ as
\begin{equation}
G_{\alpha \bbeta} = \partial_\alpha \partial_\bbeta \calK \ ,
\end{equation}
  but $\calK$ doesn't transform as a scalar. Any \kah metric can be deformed by adding an exact form to $\omega$. The class of $\omega$ in $H^2$ is referred to as the \emph{\kah class}.
  
    The Riemann tensor now has an extra symmetry
  \begin{equation}
  R^{\balpha}_{ \ [\bbeta \bgamma] \tau} = 0 \ \ \ , \  \ \  R^{\alpha}_{ \ [\beta \gamma] \btau}  =0 \ ,
  \end{equation}
  so that the components of the Ricci form (\ref{eq:ricci_form}) coincide with those of the Ricci tensor 
  \begin{equation}
  \label{eq:ricci_tensor}
  R_{ij} = R^k_{\ ikj} \ .
  \end{equation}
  Thus a Ricci-flat \kah manifold has the property that its first Chern class $c_1$ (\ref{eq:first_chern_class}) vanishes.  
  
  In general, the holonomy group of the \kah manifold is not further reduced from $U(D_{\bbC})$, which is already the holonomy group of the more general Hermitian connection. However, if the Ricci form is zero the holonomy group is reduced to $SU(D_{\bbC})$. This is true because
  \begin{equation}
  U(D_{\bbC}) \cong U(1) \otimes SU(D_{\bbC}) \ ,
  \end{equation}
  and the $U(1)$ part of the transformation is generated by the trace of the Lie algebra generators, $R^i_{ \ ijk}$. When the manifold is Ricci-flat, we are left only with the $SU(D_{\bbC})$ part.

A \emph{Calabi-Yau manifold} is defined as a \kah manifold with vanishing $c_1$. We've just seen that if the metric is Ricci-flat, $c_1$ must vanish. It is far from obvious that such a metric can be found on all \kah manifolds for which $c_1 = 0$.  This conjecture was made by Calabi in 1957 \cite{Calabi:1957}, and proved only in 1977 by Yau \cite{Yau:1977ms}. The proof refers to compact Calabi-Yau manifolds,  and shows that if $c_1=0$ there is a Ricci-flat metric in each \kah class. String theorists are  interested in the compact cases in the context of compactification, but  non-compact Calabi-Yau manifolds do play a role in the AdS/CFT correspondence  \cite{Lu:2003gp}.
  
  
  The $SU(D_{\bbC})$ structure implies that $\M$ should have a nowhere vanishing covariantly constant $(D_{\bbC}, 0 )$-form $\Omega$, and a $(0, D_{\bbC} )$-form $\bOmega$. Showing that a manifold which admits such a form is Ricci-flat is not difficult. Locally $\Omega$ can be written as
  \begin{equation}
  \label{eq:holomorphic_form}
  \Omega = f(x) \epsilon_{\mu_1 ... \mu_{D_{\bbC}}} dx^{\mu_1} \wedge \cdots \wedge dx^{\mu_{D_{\bbC}}}
  \end{equation}
  for some holomorphic function $f$. One can show that the determinant of the metric can be written as
  \begin{equation}
  \label{eq:G_cov_constant_form}
  \sqrt{G} = \frac{ |f|^2}{ || \Omega ||^2} \ ,
  \end{equation}
  where 
  \begin{equation}
   || \Omega ||^2 = \frac{1}{D_{\bbC}!} \Omega_{\mu_1 ... \mu_{D_{\bbC}}} \Omega^{\mu_1 ... \mu_{D_{\bbC}}}
  \end{equation}
  is normalized appropriately. The point is that the right hand side of (\ref{eq:G_cov_constant_form}) is a scalar, and therefore the Ricci form (\ref{eq:ricci_form})
   has to be exact. The proof of the converse is more elaborate and I refer the reader to \cite{Candelas:1987is}.
   
  In Chapter \ref{chapter:algebras_on_manifolds_of_sh} I will be making use of the fact that one can choose coordinates such that $\sqrt{G}$ and $f$ are constants.  The solution to the Ricci-flatness condition is locally\footnote{The proof of the Calabi conjecture involves showing that this solution is locally compatible  with the \kah condition, and that the only global obstruction is that $c_1$ vanishes.}
  \begin{equation}
  \ln \det G = h(x^\alpha) + \overline{h} (x^{\balpha})
  \end{equation}
   and one can always make a holomorphic transformation of coordinates such that $h$, and thus $\det G$, are constant. The function $f$ in (\ref{eq:holomorphic_form}) is related to $h$ by a constant multiple \cite{Candelas:1987is}.\footnote{Using this fact it's easy to show that the $\Omega$ is covariantly constant precisely for the Ricci-flat metric.}
  
  It follows from the existence of the holomorphic form that $b_{\dim_\bbC(\M), 0} = b_{0, \dim_\bbC(\M)} = 1$. If these Betti numbers were more than one, there would be more than one $SU(D_{\bbC})$ structure on $\M$, and $\M$ would no longer be a generic Calabi-Yau. Further relations for the Betti numbers  can be derived \cite{Candelas:1987is}. It turns out that they are completely determined for $D_{\bbC}=2$, and that  there is a unique differential manifold with $SU(2)$ holonomy called K3. For $D_{\bbC}=3$  there are two undetermined Betti numbers: $b_{1,1}$ and $b_{2,1}$.
  
 It should be noted that it is not possible to find the Ricci-flat metrics explicitly, which is one of the reasons why the proof of the Calabi conjecture is so difficult. The equations one has to solve are non-linear, and the  only way to tackle the problem is to resort to numerical methods. For the K3 case Ricci-flat metrics have been computed, and were used to explicitly calculate geometric quantities of interest \cite{headrick-2005-22}. There has been some progress very recently \cite{donaldson-2005-} on the Calabi-Yau three-fold, which was thought to be computationally extremely demanding.  The implication for string theory is that the information about Calabi-Yau manifolds is very restricted, because one can do virtually no explicit computations.

\section{Manifolds of special holonomy}
\label{section:spechol_manifolds}

Possible holonomy groups of the Levi-Civita connection on simply connected Riemannian manifolds, with irreducible and non-symmetric metrics, have been classified in 1955 by Berger \cite{Berger:1955}. They are known as manifolds of \emph{special holonomy}. For a modern review see  \cite{Joyce:2000}, or for a shorter introduction \cite{Joyce:2001xt}. The possible holonomy groups and the  associated manifolds are as follows:
\begin{enumerate}
\item Non-orientable Riemannian: $O(D)$. The only nowhere-vanishing covariantly constant tensor is the metric.
\item Orientable Riemannian: $SO(D)$. The volume form obtained from the metric  (\ref{eq:metric_vol_form}) is also covariantly constant. 
\item  K\"ahler: $U(\frac{D}{2} ) \subset SO(D)$ for $\frac{D}{2} \geq 2$. The additional covariantly constant form is  the \kah form $\omega$.
\item  Calabi-Yau:  $SU(\frac{D}{2}) \subset U(\frac{D}{2})$ for $\frac{D}{2} \geq 2$. The additional covariantly constant forms are  the $(\frac{D}{2},0)$-form $\Omega$, and the $(0, \frac{D}{2})$-form $\bOmega$.
\item Hyperk\"{a}hler:  $Sp(\frac{D}{4}) \subset SU(\frac{D}{2})$, for $D \geq 8$ and a multiple of 4. There are three covariantly constant complex structures $I_r$, with $r = \{1 ,2, 3 \}$, which obey the algebra of quaternions: $I_r I_s = - \delta_{rs} + \epsilon_{rst} I_t$. With respect to each $I_r$ one can construct a $(0, \frac{D}{2})$- and $(\frac{D}{2}, 0)$-form by taking appropriate combinations of wedge products of the $w_r$s.
\item Quaternionic K\"ahler: $Sp(\frac{D}{4}) \cdot Sp(1) \subset SO (D)$, for $ D \geq 8$ and a multiple of 4. Locally one has three complex structures, like in the Hyperk\"ahler case, but they are not covariantly constant. The nowhere vanishing covariantly constant form is given by $ \omega_r \omega_r$. 
\item $G_2$ manifold in $D=7$. In addition to the $SO(7)$ tensors, there is a covariantly constant three form $\varphi$, and its Hodge dual $*\varphi$
\item $Spin(7)$ manifold in $D=8$.  In addition to the $SO(8)$ tensors, there is a covariantly constant self-dual four-form $\Psi$, known as the Cayley form.
\end{enumerate}
The $G_2$ and $Spin(7)$ manifolds are known as manifolds of \emph{exceptional holonomy}, since they involve the exceptional Lie groups. The invariant forms associated with these groups are given explicitly in Chapter \ref{chapter:algebras_on_manifolds_of_sh} (see also \ref{section:target_space_geometry} for useful ways to express them).

For non-reducible metrics, holonomy groups that are products of the listed groups need to be considered.\footnote{This does not necessarily mean that $\M$ is a product manifold $( \calP \times \calQ, G \times G')$, where $G'$ is a metric on $\calQ$, but only that it is locally of this form.  Globally there may be twists, like those in a bundle.} 

A symmetric space is one for which there is a map $s_p: \M \rightarrow N$ for every $p \in \M$, such that $(s_p)^2$ is the identity map and $p$ is an isolated fixed point. It turns out that such spaces are of the form $\calG / \calH$, where $\calG$ is  a Lie group and $\calH$ a connected Lie subgroup of $\calG$. They can be classified using  Cartan's classification of Lie groups. 

A further point of interest is that the Riemann tensor is covariantly constant  if and only if the metric is locally symmetric, and thus for special holonomy groups 
\begin{equation}
\label{eq:cov_constant_riemann}
\nabla R \neq 0 \ .
\end{equation}
To obtain the Berger classification  one needs to consider possible subgroups of $SO(D)$ by looking at the equation  (\ref{eq:general_holonomy_expression}) with various invariant tensors associated with these subgroups.  The special holonomy classification comes after various restrictions: those imposed by his starting assumptions, the  Bianchi identity, equation (\ref{eq:cov_constant_riemann}), as well as other more technical restrictions. Berger also gave a finite list of torsion-free non-metric (affine) connections. It was thought that this list exhausted the classification, but more recently it has been shown that this is not the case \cite{chi-1996-126}.

The restriction of $\M$ being simply connected means that the local holonomy group is the same as the global one; since every loop is contractible to the identity, the holonomy group must be connected. If the condition that $\M$ is simply connected is relaxed, we have to allow for the possibility that the holonomy group is not connected. The connected components of these groups are still be those on the special holonomy list.

Of direct interest to string theory are those manifolds in the list that are Ricci-flat:
\begin{equation}
SU(n), \ \ \  Sp(n), \ \ \  G_2, \ \ \ Spin(7) \ .
\end{equation}
This classification of manifolds with these groups as \emph{local} holonomy groups has been done in \cite{McInnes:1990ph} (see also \cite{moroianu-1999-} and \cite{botvinnik-1999-}). Relaxing only the condition that $\M$ is simply connected, the  Ricci-flat   possibilities are as follows:\footnote{If  $\calH_1$ and $\calH_2$ are subgroups of $\calG$,  $\calH_1 \cdot \calH_2$  is the set of all elements $h_1 h_2$; $h_1 \in \calH_1$, $h_2 \in \calH_2$. It is a subgroup if $h_1 h_2 = h_2 h_1$.}
\begin{enumerate}
\item For any $D \geq 4: SO(D), O(D)$.
\item For any odd  $\frac{D}{2} \geq 3: SU(\frac{D}{2}), SU(\frac{D}{2}) \rtimes \bbZ_2$.
\item For any even $\frac{D}{2} \geq 4: \bbZ_{\frac{D}{2}} \cdot SU(\frac{D}{2}), SU(\frac{D}{2}) \rtimes \bbZ_2, (\bbZ_{\frac{D}{2}} \cdot SU(\frac{D}{2})) \rtimes \bbZ_2$.
\item For any even $k$ in $D = 4k: \bbZ_r \times Sp(k)$, where $r$ divides $k+1$ and $r \neq 1$.
\item  For any odd $k$ in $D = 4k: \bbZ_r \times Sp(k)$ for $r$ odd, $\bbZ_{2r} \cdot Sp(k)$ for $r$ even, or  $Q_{4r} \cdot Sp(k)$, $B_{4r} \cdot Sp(k)$ where $r$ divides $k+1$ and $r \neq1$, and $r=6,12$ or 40 for $B_{4r}$.
\item For $D=7$: $G_2$ or $\bbZ_2 \times G_2$.
\item For $D=8: Spin(7)$.
\end{enumerate}
It is of interest that the only case when the local holonomy group always coincides with the global one is $Spin(7)$, and that odd dimensional manifolds with local holonomy  $SU(\frac{D}{2})$  have the same global holonomy only if they are simply connected. It should also be noted that the case $\bbZ_2 \times G_2$ can only occur for non-orientable manifolds.

A much deeper construction of the Berger classification is obtained by restricting the holonomy group to preserve a structure on $\M$ related to one of the  normed division algebras $\bbA$: real $\bbR$, complex $\bbC$, quaternionionic $\mathbb{H}$, and octonionic $\mathbb{O}$  \cite{leung-2003-}. There is also a natural  way to generalize the notion of  orientation. The possible holonomies are precisely those from the Berger list,  as summarized by the table in Figure \ref{figure:normed_algebra_berger}.
\begin{figure}
\centering
\begin{tabular}{| c | c | c |} \hline
$\bbA$ & $Hol$ preserving $\bbA$ & $Hol$ preserving $\bbA$ + $\bbA$-orientation \\ \hline \hline
$\bbR$ & Riemannian   &  Oriented Riemannian \\ \hline
$\bbC$ & \kah & Calabi-Yau \\ \hline
$\mathbb{H}$ & Quaternionic \kah & Hyperk\"{a}hler \\ \hline
$\mathbb{O}$ & $Spin(7)$ &  $G_2$ \\ \hline
\end{tabular} \label{figure:normed_algebra_berger}
\caption{Holonomy groups preserving normed division algebras $\bbA$, and a generalization of orientation which I'll call $\bbA$-orientation, are those from the Berger list.}
\end{figure}

For torsionfull connections the Berger classification no longer applies. One can still study manifolds with $G$-structure (see the end of \ref{subsection:connections_holonomy_groups}) involving the same groups as those in the Berger list. This turns out to be a useful thing to do, especially in the context of string theory where these remain important as more general backgrounds preserving supersymmetry - ones that include torsion as well as higher order forms related to the presence of branes.  For recent progress in the classification of such solutions see \cite{Gillard:2004xq, MacConamhna:2004fb}, and also \cite{Duff:2004jj} for a review. From the point of view of the  normed algebra construction the Berger list remains important since this construction doesn't depend on the connection being Levi-Civita.

\section{Calibrations}
\label{section:calibrations}

 In this section we begin to explore the structure of a manifold by studying a particular subset of its submanifolds.  The object that defines these submanifolds is a \emph{calibration} \cite{Harvey:1982xk}, which is an $n$-form $\phi$ with the following properties
\begin{align}
\label{eq:calibration_def}
& d \phi = 0 \ , \\ \nonumber
& \phi |_{\xi_n} \leq Vol |_{\xi_n} \ ,
\end{align}
for all tangent $n$-planes $\xi_n$ on $\M$, where $Vol$ is derived from the metric (\ref{eq:metric_vol_form}).  An $n$-cycle, which is a submanifold $\calN \in \M$ that is an element of the $n$-th homology group, is said to be calibrated by $\phi$ if 
\begin{equation}
\label{eq:calibrated_form}
\phi |_\calN = Vol |_\calN  \ .
\end{equation}
In this case $\calN$ is said to be a \emph{calibrated submanifold}. If we pick a different element $\calN'$ in the homology class of $\calN$, meaning that the difference between the two is a boundary of some $(n+1)$-dimensional submanifold $\calB$ ($\calN - \calN' = \calB$), then
\begin{align}
Vol (\calN) = \int_{\calN} \phi  = \int_\calB d \phi + \int_{\calN'} \phi = \int_{\calN'} \phi \leq Vol(\calN')  \ .
\end{align}
The second equality uses Stokes' theorem,  the third is due to $d \phi = 0$, and the inequality is due to the second line in (\ref{eq:calibration_def}). Thus $\calN$ has the property that it is volume minimizing in its homology class. 

The covariantly constant forms of special holonomy manifolds are all calibrations. In Chapter \ref{chapter:boundary} we'll meet the following calibrations and calibrated submanifolds:
\begin{enumerate}
\item $U(\frac{D}{2})$: The calibrating forms on a \kah manifold are $\omega^p$, which includes $Vol$ for $p=\frac{D}{2}$ (see (\ref{eq:kahler_vol_form})). The calibrated cycles are complex submanifolds of $\M$.
\item $SU(\frac{D}{2})$: In addition to the above, on a Calabi-Yau manifold we have the calibrating forms $\cos \theta \  \mathrm{Re}( \Omega) + \sin \theta \  \mathrm{Im}( \Omega )$  parameterized by a constant $\theta$, $0 \leq \theta < 2\pi$. They calibrate Lagrangian submanifolds of $\M$. 
\item $G_2$: $\varphi$ calibrates associative three dimensional submanifolds, while $*\varphi$  calibrates co-associative four dimensional ones.
\item $Spin(7)$: $\Psi$ calibrates four dimensional Cayley cycles.
\end{enumerate}

A Lagrangian submanifold is defined by requiring the symplectic form (see \ref{subsection:symplectic_manifolds}) to vanish when restricted to it. When $\theta=0$ the calibrated submanifold is called \emph{special Lagrangian}. In this case the calibrating form is  $\mathrm{Re} (\Omega)$, and  $\mathrm{Im}(\Omega)$ vanishes when restricted to $\calN$.  For other values of $\theta$ an appropriate linear combination of $\mathrm{Re}(\Omega)$ and $\mathrm{Im}(\Omega)$ will vanish.  Special Lagrangian submanifolds have been studied in great detail; in addition to \cite{Harvey:1982xk} see also  the lectures by Joyce \cite{Joyce:2001nm} and Hitchin \cite{Hitchin:1999fh}.  

To study the deformations of the calibrated cycles it is necessary to understand the properties of $T \M$ restricted to $\calN$.\footnote{Section 4.2 of \cite{Gauntlett:2003di} is a reference for the next few paragraphs.}   The natural decomposition is in terms of the vector bundles normal and tangent  to $\calN$
\begin{equation}
T\M |_{\calN} = T \calN  \oplus \nu \ ,
\end{equation}
where $\nu$ consists of all vector fields in $T \M |_{\calN}$ normal to $T \calN$. The subspace of $\nu$ that defines deformations through the space of calibrated cycles, i.e. the moduli space of $\calN$,  is of special interest.

For special Lagrangian submanifolds $T \calN$ and $\nu$ are isomorphic. This is the case because $I_{ij}|_{\calN}$ vanishes, so for any vector field $V^j$ on $\calN$, the one-form given by $I_{ij} V^j$ is orthogonal to all vector fields on $\calN$, and  $I^i_{ \ j} V^j$ provides a one-to-one map from $T \calN$ to $\nu$. In addition, it turns out that  the the one-form $I_{ij} V^j$ defines a deformation through the space of special Lagrangian submanifolds if and only if it's harmonic \cite{mclean98deformation}. Thus, the dimension of the moduli space is given by $b_1(\calN)$.

In the case of $G_2$, co-associative cycles also have the property that $\nu$ is isomorphic to $T \calN$. It turns out that $\nu$ is isomorphic to the space of anti-self-dual forms on $\calN$, while the moduli space is given by the subspace of closed forms (because they are anti-self-dual, it also follows that such forms have to be harmonic).  

For associative cycles in $G_2$ and $Spin(7)$ manifolds the isomorphism between $\nu$ and $T\calN$ no longer holds. For the $G_2$ case, $\nu$ is isomorphic to $S \otimes V$, where $S$ is the spin bundle of $\calN$ and $V$ a rank two $SU(2)$ bundle. For $Spin(7)$, $\nu$ is isomorphic to $S_- \otimes V$, where $S_-$ is the bundle of spinors of negative chirality. The moduli space is given by the kernel of the Dirac operators on the spaces  $S \otimes V$ and  $S_- \otimes V$. Both of these cases are discussed in detail in \cite{Harvey:1982xk}.

Hyperk\"{a}hler manifolds contain cycles calibrated by the three \kah forms,  but they also contain Lagrangian cycles. These have been studied in the mathematics \cite{hitchin-1999-3, leung-lagrangian}, but not as much in the physics literature. In \cite{Gauntlett:1997pk} they come up in supergravity solutions that contain branes, but to my knowledge have not been analyzed from the point of view of calibrations. Calibrated cycles of quaternionic  \kah manifolds have not been explored much. They are defined in \cite{leung-2003-}, which otherwise also gives a nice interpretation of calibrations in the context of normed algebras.

In string theory  calibrations are related to brane solutions preserving supersymmetry. This can be explored in a number of ways: 
\begin{enumerate}
\item In the  $\sigma$-model setting open strings ending on calibrated cycles describe branes that preserve supersymmetry. 
\item Superstring theory can be studied at tree level by ten dimensional supergravity theories, and calibrations feature in brane solutions that preserve some fraction of supersymmetry. Of course, via dualities, $M$-theory becomes relevant.
\item  Branes can also be analyzed  using supersymmetric extensions of  the Nambu-Goto action (see the introduction to Chapter \ref{chapter:2d_sigma_models}). From this point of view, one can  directly see that such solutions minimize volume. It's also natural to introduce the notion of a generalized calibration \cite{Gutowski:1999tu},  when torsion and/or other forms related to brane backgrounds are present. Since the supersymmetric Nambu-Goto action is deformed by the presence of these fields, the generalized calibration will no longer minimize the volume, which is manifested by dropping the requirement in (\ref{eq:calibration_def}) that the calibrating form be closed. 
\end{enumerate}
Calibrations are  explored further  in the setting ({\it i}) in Chapter \ref{chapter:boundary}, as are generalized calibrations,  but to a more limited extent.


\chapter{Antifield formalism}
\label{ch:antifield_formalism}


The path integral calculates time ordered expectation values of field operators in the vacuum.\footnote{In this chapter I will assume a basic knowledge of quantum field theory. For a detailed exposition to perturbative QFT the reader is referred to one of a large number of textbooks. A selection in order of increasing abstraction from nasty details is \cite{Ryder:1985wq, Bailin:1986wt, Ramond:1989yd, Piguet:1995er}, while \cite{Peskin:1995ev} is a very comprehensive textbook.  For a detailed explanation of renormalization and the renormalization group flow the reader is referred to the Witten and D. Gross lectures in the first volume of \cite{Deligne:1999qp}, or the Callan and Gross lectures in the much older text \cite{Balian:1976vq}.} Schematically,
\begin{equation}
\label{eq:path_integral}
 \int [ d \phi ] F(\phi(x)) \exp \left( \frac{i}{\hbar}  W[\phi(x)]  \right)  = \Bra{0} T \{ F(\phi(x)) \}\Ket{0}  \ ,
\end{equation} 
where $F$ is some function of the fields $\phi$, in general non-local, $T$ denotes time ordering, $\Ket{0}$ is the vacuum state of the theory, and $x$ stands for coordinates on the Lorentzian manifold on which the fields live. In the spirit of $\sigma$-models, I'll refer to this manifold as the base space (see Chapter \ref{chapter:2d_sigma_models}). 

Free theories are relatively simple, but for interacting theories (with which this thesis is concerned) one must have a way of handling infinities that occur when one tries to define the path integral. To have a chance of obtaining finite amplitudes, $W$ in (\ref{eq:path_integral}) has to be formally  infinite, differing from the classical action $S$ by an infinite renormalization of the fields, masses, and coupling constants.  Operators which contain more than a single power of $\phi$ at the same spacetime point, called \emph{composite operators}, must themselves be renormalized. For renormalizable theories the procedure of absorbing all infinities into the parameters of the theory works, yielding finite expectation values, but one is still forced to implement a regularization procedure to handle infinite expressions at intermediate steps. Regularization generally breaks classical symmetries, and these are not always  restored after the infinite renormalization, when the limit in which the theory no longer depends on the regulators is taken. The breaking of scale is particularly important; it is encoded in the $\beta$-function and related to the dependence of the coupling constants on the energy scale. The renormalization group studies this dependence and is crucial for checking whether the perturbative evaluation of the QFT is consistent.  In the context of string theory, the $\sigma$-models studied in this thesis are required to be conformally invariant at the quantum level, which is equivalent to the requirement that the $\beta$-function vanishes.

The antifield formalism is a general setting both for gauge fixing and for analyzing symmetries of quantum field theories.   In \ref{section:quant_of_interest_in_qft} I will define the basic objects of interest in a QFT.  In \ref{section:local_and_global_sym}  the infinitesimal symmetries of classical field theories, and their algebras, are explored in a general framework.  A symmetry can either be global, local (gauge), or of conformal-type (where the transformation parameter depends on some but not all coordinates of the base space). The presence of gauge symmetries in the classical action prevents the evaluation of the path integral, and a gauge fixing procedure must be implemented in order to make sense of the QFT. How this works in the antifield formalism is outlined in  \ref{section:faddeev_popov}, \ref{section:BRST_quantization}, and \ref{section:BV}.  The gauge fixed action is no longer gauge invariant, but it is invariant under a residual global symmetry called the BRST symmetry, or, depending on the type of gauge algebra, a more general BV symmetry. The presence of a BRST/BV symmetry, or other global or conformal-type symmetries, implies relations between the Feynman diagrams in the perturbative evaluation known as Ward identities. How to obtain \emph{naive} Ward identities in the antifield formalism for BRST/BV symmetries is explained in \ref{section:BRST_quantization} and \ref{section:BV},  and for general global or conformal type symmetries in  \ref{sec-global_symmetries}. The adjective 'naive' refers to the assumption that a regulator which preserves the classical symmetries can be found. Furthermore, the action in a path integral generally involves both quantum fields and other classical background fields (this can happen, for example, because we are evaluating the path integral around some non-trivial classical background), so one needs to understand what role the background fields play in the Ward identities. This is also discussed in \ref{sec-global_symmetries}.  Of course, the naive assumptions about classical symmetries making it through the renormalization procedure may not actually hold, in which case one speaks of an anomaly in the Ward identities. In \ref{sec:anomalies} I discuss cohomological equations that enable all possible anomalies related to a classical symmetry to be computed. These equations are very naturally expressed in the antifield formalism, and are relatively easy to handle because they involve only local expressions. The main benefit is that an explicit evaluation of the path integral is avoided.  

\section{Basic objects of interest in the path integral approach to QFT}
\label{section:quant_of_interest_in_qft}

The \emph{generating functional} $Z[J]$ contains the information about all the scattering amplitudes of the theory:
\begin{eqnarray}
\label{eq:generating_functional1}
Z[J] & = & \int [ d \phi ] \exp \left( \frac{i}{\hbar}  W[\phi] + \frac{i}{\hbar} \int dx \  \phi^i(x) J_i(x) \right) \\ \nonumber
& = & \left< 0 | 0 \right>_J \ ,
\end{eqnarray}
where  $J_i$ are classical sources for the quantum fields $\phi^i$. As indicated in the second line, $Z[J]$ calculates the vacuum-to-vacuum amplitude in the presence of the sources.
By expanding the exponential containing $J$,  $Z[J]$ is expressed in terms of the $N$-point \emph{Green functions} as
\begin{equation}
\label{eq:generating_functional2}
Z[J]  =  \sum^{\infty}_{N=0} \frac{(i/  \hbar )^N}{N!} \int dx_1 ... dx_N J_{i_1}(x_1) ... J_{i_N}(x_N)
 G^{i_1 ... i_N}( x_1, \ ... \ , x_N) \ .
\end{equation}
Each of these Green functions is evaluated  by summing over all Feynman diagrams with $N$ external legs. It is obtained by setting $J=0$ after functionally differentiating $Z[J]$ with respect to $J$ $N$ times, 
\begin{equation}
G^{i_1 ...   i_N}( x_1, \  ... \ , x_N) =  \Bra{ 0 } T ( \phi^{i_1} (x_1) ...  \phi^{i_N} (x_N) ) \Ket{0} \ ,
\end{equation}
and is related to the scattering of $N$ particles. 

The generating functional for \emph{connected} Green functions, $Z_{conn}[J]$, is defined by
\begin{equation}
\label{eq:generating_connected}
Z[J] = \exp \left(  \frac{i}{ \hbar} Z_{conn} [J]  \right) \ .
\end{equation} 
One expands $Z_{conn} [J]$ in powers of $J$  as in  (\ref{eq:generating_functional2}), except that the evaluation of each connected $N$-point Green function involves a sum only over Feynman diagrams that are connected.\footnote{The connected Green functions are related to scattering amplitudes when there are interactions between all the particles involved. The disconnected ones are related to all scattering scenarios, even when no interaction takes place.}

An important quantity is the expectation value of a field in the presence of sources, referred to as the \emph{classical field}:
\begin{equation}
 \label{eq:classical_field}
\phi_{(c)}^i (x)[J] \equiv \frac{\delta Z_{conn}[J]}{\delta J_i(x)} = \Bra{0} \phi^i(x) \Ket{0}_J \ .
\end{equation}
The classical fields are used to define the \emph{effective action}:
\begin{align}
 \label{eq:effective_action1}
\Gamma [\phi_{(c)}] & \equiv   Z_{conn} [J] - \int dx  J_i(x) \phi_{(c) }^i(x) \\ \nonumber
& = -i \ln{Z} - \int dx   J_i(x) \phi_{(c) }^i(x) \ .
\end{align}
By functionally differentiating $\Gamma$ with respect to $J$, but not differentiating $\phi_{(c)}$ itself, we see that it indeed depends only on  $\phi_{(c)}$. Or stated more accurately, its dependence on  $J$ enters only through $\phi_{(c)}^i $. The expansion in powers of $\phi_{(c)}$ is in terms of one-particle-irreducible  (1PI)  $N$-point functions, $\Gamma^{(N)}$:
\begin{equation}
 \label{eq:effective_action2}
\Gamma[\phi_{(c)}] = \sum^{\infty}_{N=0} \frac{(i/  \hbar )^N}{N!} \int dx_1 ... dx_N \Gamma^{(N) i_1 ... i_N}(x_1, \ ... \  ,x_N)  \phi_{(c) i_1} (x_1) ... \phi_{(c)i_N} (x_N)   \ .
\end{equation}
These are expressed in terms of Feynman diagrams that can not be separated in two by cutting any of the internal lines, and with all the propagators corresponding to external lines divided out.  Because the relations between $Z$, $Z_{conn}$, and $\Gamma$ are invertible, the 1PI diagrams are the fundamental building blocks of quantum field theory. The term 'effective action' is appropriate since in the tree approximation $\Gamma[\phi_{(c)}]$ is the classical action, with $n$-loop diagrams contributing non-local corrections proportional to powers of $(\hbar)^n$. In fact, one can use the effective action itself  to evaluate the generating functional, but with the prescription that one only takes tree diagrams into account in the evaluation.

\section{Local, global, and conformal-type symmetries}
\label{section:local_and_global_sym}

Here is a good point to start using the deWitt convention, which is often useful when talking about field theories in general terms. It is generally more confusing than useful when talking about a particular theory, and I will refrain from doing so. The basic idea is to generalize the Einstein summation convention so that repeated indices imply integration over spacetime, in addition to summation over discrete indices. For example, one would write the free single boson action $S = \int dx \partial^\mu \phi \partial_\mu \phi$ as 
\begin{equation}
\label{eq:free_action2}
S(\phi) =  \phi^i K_{ij} \phi^j \equiv \int dx  dy  \phi(x) \partial^\mu \partial_{\mu} \delta(x-y) \phi(y) \ ,
\end{equation}
so $K_{ij} \equiv \partial^\mu \partial_{\mu} \delta(x-y)$, and the summation over $i$ and $j$ stands just for integration. In the above case it makes no difference whether the partial derivatives are with respect to $x$ or $y$, but in general this must be specified.
In the deWitt notation the dependence of objects on fields may still be indicated, but the dependence on spacetime is contained entirely in the indices. The number of free indices in an expression is equal to one more than the number of $\delta$-functions left in the usual notation, after all possible integrations have been performed. As in the example (\ref{eq:free_action2}), no free indices means that the final expression is integrated. 

Also, from now I will be working in units in which $\hbar  = 1$.

A theory possesses local (or gauge) symmetries if its action is invariant under the transformations
\begin{align}
& \phi^i (x) \rightarrow \phi^i (x) + \delta \phi^i(x)  \ \ \ , \ \ \ \delta \phi^i(x) = \int d y \ \varepsilon^A(y) R^i_A (\phi(x), \delta(x, y))  = \varepsilon^A R^i_A \ , \label{eq:variation1} \\
\label{eq:variation_of_action}
& \frac{\dr S(\phi)}{\delta \phi^i} \varepsilon^A R^i_A  \phi^i \equiv  \sum_{A} \int dx \ \frac{\dr S(\phi)}{\delta \phi^i(x)} \delta_{\varepsilon^A (x)} \phi^i (x) =0 \ ,
\end{align} 
where I have written out explicitly what the deWitt notation implies. The parameters $\varepsilon^A(x)$ are unconstrained functions over spacetime. They have the same statistics as the fields for even, and opposite statistics for odd symmetries.  The objects $R^i_A$ are generally field dependent.\footnote{If this is not the case, the symmetry is automatically Abelian. For example, for the $U(1)$ transformation of a gauge field, $\delta_{\Lambda} A_{\mu} = \partial_{\mu} \Lambda$.} Because $\phi^i$  stands for a general set of fields that can include both fermions and bosons, it is necessary to specify a direction for the functional derivatives in (\ref{eq:variation_of_action}).

The classical action can  possess continuous global symmetries, and also conformal-type symmetries\footnote{Conformal and superconformal symmetries of $D=2$ $\sigma$-models are of this type (see Chapter \ref{chapter:2d_sigma_models}), but in this chapter I'll keep the discussion general.}
\begin{align}
 \label{eq:variation2}
& \phi^i (x) \rightarrow \phi^i (x) + \delta \phi^i(x) \ , \\ \nonumber
& \delta \phi^i(x) = \varepsilon^A R^i_A: =  \left\{ \begin{array}{ll}
				\sum_A \varepsilon^A R^i_A(\phi) & \mbox{for global} \\
				\sum_A \int d \overline{y} \ \varepsilon^A ( \overline{y}) R^i_A (\phi(x), \delta(\overline{x}, \overline{y}))  
								& \mbox{for conformal-type}
				\end{array}
				\right. \ .
\end{align}
 For global symmetries the parameters $\varepsilon^A$ are independent of the coordinates of the base space,  while for conformal-type symmetries they depend on a subset $\overline{x}$ of the coordinates $x$. The objects $R^i_A$  are again generally functions of fields. In the conformal-type case they contain $\delta$-functions, but only over the subspace of coordinates $\overline{x}$.

The statement that the  transformations form an algebra is that a graded commutator doesn't generate any new symmetries except ones that are graded antisymmetric in the equations of motion,
\begin{align}
\label{eq:graded_commutator}
 \left( \frac{\dr R^i_A}{\delta \phi^j} R^j_{B} - (-1)^{\epsilon_A \epsilon_B} \frac{\dr R^i_B}{\delta \phi^j} {R^j_A} \right)  \varepsilon^B \varepsilon^A = 
 \left( R^i_C N^C_{AB} - \frac{\dr S}{\delta \phi^j} E^{ji}_{AB} \right) \varepsilon^B \varepsilon^A \ ,
\end{align}
where  $\epsilon_A$ stands for the parity of the parameter $\varepsilon^A$ ($1$ for fermions and $0$ for bosons).    $N^{C}_{AB}$  are the structure functions, and in general they are field dependent. The non-closure functions $E^{ji}_{AB}$ are required to obey $E^{ji}_{AB} = - (-1)^{\epsilon_i \epsilon_j} E^{ij}_{AB}$ in order for the second term to be a symmetry. In this case
\begin{equation}
\frac{\dr S}{\delta \phi^i} \frac{\dr S}{\delta \phi^j} E^{ji}_{AB} \equiv 0 \ ,
\end{equation}
due to a (graded) symmetric-antisymmetric contraction of the $i,j$ indices. Algebras that close up to such terms proportional to the equations of motion are said to be  \emph{open}. Obviously, they are tied to a particular action, which is not true for the \emph{closed} algebras.   Lie algebras are closed algebras with structure functions that are not field dependent.

To evaluate the path integral perturbatively one first has to calculate the free propagator, which is essentially the inverse of the operator that enters the kinetic term (for example, $K_{ij}$ in (\ref{eq:free_action2})). The infinite-dimensional matrix 
\begin{equation}
\label{eq:Hessian}
H_{ij} := \frac{\dl}{\delta \phi^i} \frac{\dr S}{\delta \phi^j} 
\end{equation}
is referred to as the \emph{Hessian} of the action. The matrix whose inverse gives the free propagator is  the Hessian evaluated at $\phi^i=0$. More generally, quantum field theories can be expanded around other solutions, $\phi^i = \phi^i_0$, to the equations of motion. The set of all such solutions forms a surface in the space of fields referred to as the \emph{stationary surface}.\footnote{Alternatively, the fields $\phi^i_0$ are said to be \emph{on-shell}.}  Therefore, the matrix that we generally wish to invert is the Hessian evaluated at some point on the stationary surface: $H_{ij}(\phi_0)$.

When the action has symmetries, $\varepsilon^A R^i_A(\phi_0)$ are eigenvectors of $H_{ij}(\phi_0)$ with zero eigenvalue\footnote{They are also said to be \emph{zero modes} of $H_{ij}$.},
\begin{equation}
\label{eq:null_vector_of_hessian}
H_{ij}(\phi_0) \varepsilon^A R^i_A(\phi_0) = 0 \ ,
\end{equation}
 which can be seen by functionally differentiating (\ref{eq:variation_of_action}). It follows that
 \begin{equation}
\left. \frac{\dr S}{\delta \phi^j} \right|_{\phi^i_0 + \varepsilon^A R^i_A(\phi^i_0)} = 0
 \end{equation}
 for $\varepsilon^A$ small. So, close to some solution to equations of motion there exist other solutions. The space of solutions  in the vicinity of $\phi_0$ is finite-dimensional if $\varepsilon^A$ are global parameters, but is infinite-dimensional for gauge and conformal-type  symmetries.
 
Solutions related by global symmetries are physically distinct.\footnote{According to Noether's theorem, for every global or conformal-type symmetry there is a conservation equation. This can be seen by making the parameters $\varepsilon^A$  local, so that the transformations are no longer  symmetries of the action, but instead 
 \begin{equation} 
 \label{eq:noethers_theorem}
 \delta S = \int dx J_A \partial_{\mu} \varepsilon^A(x)  \ .
 \end{equation} 
 However, any infinitesimal transformation is a symmetry on the stationary surface  (this is the definition of the stationary surface!). So after integrating the right hand side of (\ref{eq:noethers_theorem}) by parts, we can conclude that $\partial_{\mu} J = 0$ on-shell.}  Local symmetries, on the other hand, are related to some redundancy in the formulation of the theory, and solutions differing by them are for all purposes physically the same. For example, in electrodynamics  the field strength can be written in terms of a one-form $A$ as $F = dA$. The gauge freedom lies in the fact that  $A$ can be changed by an exact form, $A \rightarrow A+ d\Lambda$, without affecting the physics.
 
Up to this point we have assumed that the path integral sums over all fields, regardless whether they are related by symmetries or not. For global symmetries this is desirable, because we do want to sum over all solutions that are physically distinct, but for gauge symmetries it causes severe problems.   In the   perturbative evaluation of the path integral we are, roughly speaking, picking some solution to the equations of motion and functionally integrating over nearby field configurations. When there are symmetries these nearby configurations can be divided into those that remain on-shell, and those that are perpendicular to the stationary surface:
 \begin{equation}
 \label{eq:sep_of_perpendicular_modes}
 \phi^i \approx \phi^i_0 + \varepsilon^A R^i_A + \phi^i_{\bot} = \phi^i_0 + \delta \phi^i \ ,
 \end{equation}
where $\phi^i_{\bot}$ are the perpendicular transformations, while the nearby on-shell configurations are parameterized by $\varepsilon^A$. Expanding the action around some stationary point $\phi^i = \phi_0^i$, 
\begin{equation}
S(\phi) = S(\phi_0) +  (\delta \phi^i - \phi_0^i) \left. \frac{ \delta S}{ \delta \phi^i} \right|_{\phi_0} 
+  (\delta \phi^i - \phi_0^i) (\delta \phi^j - \phi_0^j)  \left. \frac{ \delta^2 S}{ \delta \phi^i \delta \phi^j } \right|_{\phi_0}  + \cdots \ ,
\end{equation}
where I assume temporarily that the fields are bosonic, the path integral can be expressed in terms of the modes (\ref{eq:sep_of_perpendicular_modes}).\footnote{See for example \cite{Balian:1976vq} for an explanation how the path integral is evaluated in this approximation. The expansion here is sufficient for an evaluation to two-loops.} Using (\ref{eq:null_vector_of_hessian}) we get:
 \begin{align}
 & \int [ d \phi ] e^{ i S(\phi) } \approx \\ \nonumber 
 & \exp{i \left( S(\phi_0) +  \phi_0^i\phi_0^j  \left. \frac{ \delta^2 S}{ \delta \phi^i \delta \phi^j } \right|_{\phi_0} \right) }
  \int [d\phi^i_{\bot}] [d \varepsilon^A] \exp{i \left(  \phi_{\bot}^i \phi_{\bot}^j  \left. \frac{ \delta^2 S}{ \delta \phi^i \delta \phi^j } \right|_{\phi_0} \right) } \ .
 \end{align}
 The first factor is numerical, and can be absorbed in the normalization. The important point is that the integral over $\varepsilon^A$ factorizes from the rest. For local symmetries $\varepsilon^A$ has a functional freedom, but because it doesn't feature in the exponent it has no propagator. For this reason we can't make sense of the quantum theory. To do so we must ignore the integral over $\varepsilon^A$, and integrate only over a subset of the space of  fields whose elements are not related by gauge transformations. Methods of achieving this will be explored in the next sections.
 
How does the existence of a null eigenvector of the on-shell Hessian (\ref{eq:null_vector_of_hessian}) relate to the existence of a propagator?  In the finite-dimensional case, if a matrix has a zero eigenvector its determinant must vanish and then it can't be inverted. However,  (\ref{eq:null_vector_of_hessian})  is an infinite-dimensional matrix equation, and we must distinguish between eigenvectors related to global, conformal-type, and local symmetries. For global symmetries  (\ref{eq:null_vector_of_hessian}) states that there is a finite-dimensional subspace of the the space of fields on which $H_{ij} |_{\phi_0}$ is not invertible.   For conformal-type symmetries this space is infinite-dimensional, but its functional freedom is less than that of the fields. In either case, the impact is that the symmetries carry over into the quantum realm as Ward identities. Since only gauge symmetries prevent the evaluation of a propagator,  we can decide whether a particular action can be used in a path integral simply by thinking of the Hessian as a finite-dimensional matrix, with a single entry in the matrix for each field. The existence of the propagator is ensured if the Hessian  has maximal rank in this finite-dimensional sense.

\section{Faddeev-Popov gauge fixing of Yang-Mills and  BRST symmetry}
\label{section:faddeev_popov}

The four-dimensional Yang-Mills action coupled to massive fermions is given by:
\begin{equation}
\label{eq:yang_mills_with_ferm} 
S_{\YM} = \int d^4x \left( \overline{\mathbf{\Psi}} (i \gamma^{\mu} \mathbf{D}_{\mu} - m \mathbf{1}) \mathbf{\Psi} -  
\frac{1}{4} F_{\mu \nu}^{a}F^{\mu \nu}_{a} \right) \ .
\end{equation}
The gauge fields $\mathbf{A_{\mu}} \equiv A_{\mu}^{a} \mathbf{T}_a$ are Lie algebra valued forms\footnote{I will denote a basis of the Lie algebra in a fundamental representation of a compact Lie group by $\mathbf{T}_a$.  Matrices and vectors will be denoted by the bold font, their indices will always be suppressed. The generators  satisfy $[\mathbf{T}_a, \mathbf{T}_b] = f^{c}_{ \ ab} \mathbf{T}_c $ and the Jacobi identity, $f^a_{ \ [bc} f_{|a|de]} \equiv 0$. They are normalized so that $\mathrm{Tr}( \mathbf{T}_a \mathbf{T}_b) = \frac{1}{2} \delta_{ab}$. The field strength $\mathbf{F}_{\mu \nu} = -i g^{-1} [ \mathbf{D}_{\mu}, \mathbf{D}_{\nu} ] = F^a_{\mu \nu} \mathbf{T}_a$ also takes values in the Lie algebra, and because of the normalization we have $\mathrm{Tr}( \mathbf{F}_{\mu \nu} \mathbf{F}^{\mu \nu} )= \frac{1}{2} F_{\mu \nu}^{a}F^{\mu \nu}_{a}$.},  $\mathbf{D}_{\mu} := \mathbf{1} \partial_{\mu} - ig \mathbf{A}_{\mu}$, and $\mathbf{\Psi}$ takes values in the vector space holding the fundamental representation of the gauge group. The spinor indices of  $\mathbf{\Psi}$ have also been suppressed. 

Action (\ref{eq:yang_mills_with_ferm}) is invariant under the gauge transformations
\begin{equation}
 \label{BRST_gauge}
\delta_{\Lambda} \mathbf{\Psi} = - i g \mathbf{\Lambda} \mathbf{\Psi}  \ \ \ ,  \ \ \ 
\delta_{\Lambda} \mathbf{A}_{\mu} = (\partial_{\mu} \Lambda^c + \Lambda^a f_{ \ ab}^c A^b_\mu ) \mathbf{T}_c \ , 
\end{equation}
where $\mathbf{\Lambda} \equiv \Lambda^a \mathbf{T}_a $ are local parameters taking values in the Lie algebra. 
It is not possible to invert the on-shell Hessian for the gauge fields.\footnote{The on-shell Hessian for the gauge fields is given by 
\begin{equation}
\left. \frac{\delta^2 S_{\YM}}{\delta A^a_{\mu} \delta A^b_{\nu}} \right|_{A^a_{\mu} = 0} = \delta_{ab}(g_{\mu \nu} \partial^2 - \partial_{\mu} \partial_{\nu}) \equiv   \delta_{ab} K_{\mu \nu} .
\end{equation}
 The easiest way to see that this matrix is not invertible is to notice that $K_{\mu \nu} K^{\nu}_{ \ \eta} = \partial^2 K_{\mu \eta} \propto K_{\mu \eta}$, so $K_{\mu \nu}$ is a projection operator.} By looking at the case of imposing a constraint on an integral of finite dimension $n$ and then taking the limit $n \rightarrow \infty$ (see for example \cite{Bailin:1986wt}), one can argue that the path integral over gauge fields not related by gauge transformations is given by
\begin{equation}
Z[J]  =  \int [d A^a_{\mu} ]  \mathrm{det} 
\left| \frac{\delta G_b' }{\delta \Lambda^c } \right|
\delta(G_b) \exp i \left( S_{\YM} + 
\int d^4 x  J^{\mu}_{a} A_{\mu}^{a} \right) \ , \label{fadeev1} 
\end{equation}
where $\delta(G_b)$ defines the choice of a gauge slice, and $G_b'$ is the variation of $G_b$ under the transformations (\ref{BRST_gauge}). Formally,
\begin{equation}
 \int [d \Lambda^c ]   \mathrm{det} 
\left| \frac{\delta G_b'}{\delta \Lambda^c } \right|  
\delta(G_b) = 1 \ ,
\end{equation}
so trying to integrate over $\Lambda^a$ in  (\ref{fadeev1}) takes us back to the original ill-defined path integral.

It is useful to re-express (\ref{fadeev1}) so that all the terms are in the exponent. Without going into details  (see eg. \cite{Bailin:1986wt}), the main feature is that one needs to introduce two Lie algebra valued Grassmann fields,  $\mathbf{c} \equiv c^{a} \mathbf{T}_{a}$ and $\overline{\mathbf{c}} \equiv \overline{c}^a \mathbf{T}_{a}$. These are called the ghost and anti-ghost fields.  They don't obey the spin-statistics theorem and are thus not considered to be physical fields. The final final form of the gauge fixed action depends on the choice of $G_a$, but using any covariant gauge choice\footnote{The gauge fixing functions $G_a$ come in either of the two forms, $G_{a} = \partial_{\mu} A_{a \mu} - f_a(x)= 0$,  or  $G_{a} = t^{\mu} A_{a \mu} - f_a (x) = 0$, where the gauge choice is described by an arbitrary function $f_a (x)$, and $t^{\mu}$  a vector of norm 1. The former corresponds to the usual covariant gauge choices, while the latter is a non-covariant gauge choice, whose merit is that the ghosts decouple from the gauge fields. This indicates that ghosts can be avoided, but at the price of not having a covariant formulation.} the generating functional is given by
\begin{equation}
\label{fadeev2}
Z[J]  =  \int [d  A^a_{\mu} ]   [ d \overline{c}^a ] [ d c^a ]  
\exp i \left( S_{\FP} + 
\int d^4 x  J^{\mu}_{a} A_{\mu}^{a} \right) \ . 
\end{equation}
$S_{\FP}$ is the Faddeev-Popov\footnote{Named after the people who first performed this kind of gauge fixing \cite{Faddeev:1967fc, Faddeev:1969su}.}  action:
\begin{equation}
\label{eq:faddeev_popov_action1}
S_{\FP} = S_{\YM} + \int d^4 x  \left( -  \frac{1}{2 \xi} (\partial_{\mu} A^{ a \mu})^{2} +
\partial^{\mu} \overline{c}_{a}(\partial_{\mu} c^a + 
g f_{ \ bc}^{a} c^{b} A^{c}_\mu ) \right) \ , 
\end{equation}
where $\xi$ is an arbitrary constant. 

The gauge fixed action $S_{\FP}$ is no longer invariant under the local transformations (\ref{BRST_gauge}), but it does possess a global fermionic symmetry called the \emph{BRST symmetry}.\footnote{This acronym stands for the names of the people who discovered this symmetry: Becchi, Rouet, and Stora   \cite{Becchi:1975nq}, and independently Tyutin \cite{Tyutin:1975qk, Iofa:1976je}.}  On the matter and gauge fields it acts as
\begin{align}
& \delta_{\BRST} \mathbf{\Psi} = - i  \theta g \mathbf{c \Psi}   \ , \\ \nonumber
& \delta_{\BRST} \mathbf{A}_{\mu} =  \theta (\partial_{\mu} c^c + c^a f_{ \ ab}^c A^b_\mu ) \mathbf{T}_c  \ ,
\end{align}
and on the ghosts as
\begin{align}
\label{eq:brst_ghost_antighost} 
& \delta_{\BRST} c^{a} = - \theta g \frac{1}{2} f^{a}_{ \ bc} c^{b} c^{c}  \ , \\ \nonumber
& \delta_{\BRST} \overline{c}^a = - \frac{\theta}{\xi } \partial^{\mu} A^a_{\mu} \ ,
\end{align}
where $\theta$ is a global fermionic parameter. For the transformations on the matter fields we have essentially replaced the bosonic parameters by ghosts multiplied by $\theta$. However, the ghosts are quantum fields, and  $\delta_{\BRST}$ is a non-linear symmetry. It is a remnant of the original gauge symmetries in the gauge fixed action. 

The action $S_{\FP}$ has a second symmetry called the anti-BRST symmetry. On the matter and gauge fields it acts in the same way as the BRST symmetry, but with the replacement $c \rightarrow \overline{c}$. The transformations of the ghosts and anti-ghosts are given by
\begin{align}
& \delta_{\overline{\BRST}} c^a =   \overline{\theta} \left( \frac{1}{\xi } \partial^{\mu} A^a_{\mu} 
-  g  f^{a}_{ \ bc} \overline{c}^{b} c^{c} \right) \ , \\ \nonumber
& \delta_{\overline{\BRST}} \overline{c}^a =  - \overline{\theta} g \frac{1}{2} f^{a}_{ \ bc} \overline{c}^{b} \overline{c}^{c}  \ .
\end{align}

The number of gauge symmetries is equal to the number of generators of the Lie algebra, while $\delta_{\BRST}$ and $\delta_{\overline{\BRST}}$  are only two symmetry transformations, parameterized by $\theta$ and $\overline{\theta}$.  The general relation is that for each closed \emph{algebra} there is a single BRST and a single anti-BRST symmetry after gauge fixing.

The BRST transformation is nilpotent when acting on the fermions, the gauge fields, and the  ghosts, but it is only proportional to the equations of motion for $\mathbf{\overline{c}}$. An analogous remark is true for the anti-BRST symmetry. A fully nilpotent transformation can be obtained by introducing auxiliary Lie algebra valued scalar fields: $\mathbf{b} = b^{a} \mathbf{T}_{a}$.\footnote{These are sometimes referred to as Nakanishi-Lautrup fields.}, modifying $\delta_{\BRST}$ and $\delta_{\overline{\BRST}}$  as
\begin{equation}
\delta_{\overline{\BRST}} c^a  = - \overline{\theta} g \frac{1}{2} f^{a}_{ \ bc} \overline{c}^{b} c^{c}  - \overline{\theta} b^a  \ \ \  , \ \ \ 
\delta_{\BRST} \overline{c}^a =  \theta b^a  \ ,
\end{equation}
 and letting $\mathbf{b}$ transform as:
\begin{equation}
\delta_{\BRST} b^a = 0 \textrm{ \ \ \ , \ \ \ } \delta_{\overline{\BRST}} b^a = - \overline{\theta} g \frac{1}{2} f^{a}_{ \ bc} \overline{c}^b b^c \ .
\end{equation}
In this form,
\begin{equation}
\label{eq:BRST_aBRST_nilpotence}
(\delta_{\BRST} )^{2} = \delta_{\BRST} \delta_{\overline{\BRST}} + \delta_{\overline{\BRST}} \delta_{\BRST} = (\delta_{\overline{\BRST}})^2 = 0 \ .
\end{equation}
As can be inferred from the above transformations, the scaling dimension of $b$ is $2$, so it can't possess a renormalizable kinetic term. The action invariant under the above transformations is
\begin{equation}
S_{\FP \mathrm{'}} = S_{\YM} + \int d^4 x \ \left( \frac{\xi}{2} (b^a)^2 +  b_a \partial^{\mu} A^{ a }_{\mu} +
\partial^{\mu} \overline{c}_{a}(\partial_{\mu} c^a + 
g f_{ \ bc}^{a} c^{b} A^{c}_\mu ) \right)  \ .
\end{equation}
It is equivalent to $S_{\FP}$ (\ref{eq:faddeev_popov_action1}) after eliminating the auxiliary $b^a$ fields using their equations of motion.

In conclusion, all possible gauge fixed actions that are obtained by the Faddeev-Popov procedure share the feature that they possess nilpotent BRST and anti-BRST symmetries, although for off-shell nilpotence auxiliary  fields must be introduced. Since only the gauge fixed action is relevant for the quantum theory, it is the nilpotent symmetries  that are fundamental. BRST quantization is concerned with obtaining a nilpotent gauge symmetry from a classical gauge algebra, and then utilizing it to construct a gauge fixed action. This will be the subject of the next section.  It turns out that it's sufficient to consider only the BRST symmetry to obtain the most general gauge fixed action. The gauge fixing procedure can always be extended to include the anti-BRST symmetry, the benefit of which is that the introduction of the $b^a$ and $\overline{c}^a$ fields is more natural, whereas if one considers only the BRST symmetry these fields must be introduced by hand. Otherwise, the BRST-anti-BRST procedure seems to hold little practical benefit. One interesting application has been the superfield formulation of Yang-Mills theory \cite{Bonora:1980pt}. It may be that in the future, due to its larger covariance, this procedure will be seen to be more fundamental. The reader is referred to \cite{Baulieu:1983tg} for a review article.

\section{BRST quantization}
\label{section:BRST_quantization}

In this section we show how to construct a nilpotent BRST operator starting from an irreducible set of  gauge transformations $\varepsilon^A R^i_A$ that form a closed algebra, and how to use this operator to construct a gauge fixed action. The naive Ward identities related to the BRST symmetry are also derived.  The methods of this section will not work for open or reducible algebras, for which the full BV procedure (discussed in \ref{section:BV})  must be employed.

The first step to obtaining a nilpotent symmetry is to promote the transformation parameters $\varepsilon^{A}$  to ghosts $c^{A}$. Then in place of the original transformations there is instead a single BRST transformation
\begin{equation}
\label{eq:BRST_on_matter1}
\delta_{\mathrm{BRST}} \phi^i = \theta c^{A} R^i_{A}   \ ,
\end{equation}
with $\theta$ a fermionic  parameter. 

It is useful to drop the constant parameter, and think of the BRST transformation as changing the parity of the field:
\begin{equation}
\label{eq:BRST_on_matter2}
\delta_{\BRST} \phi^i = c^{A} R^i_{A}   \ .
\end{equation}
We'll also assign a grading, called the \emph{ghost number} and denoted by $\gh (\mathrm{field})$, so that  $\gh(\phi) = 0$, $\gh (c^A) = 1$, and $\gh (\delta_{\BRST} )  = 1$.

Up to a caveat involving the Jacobi identity, $\delta_{\mathrm{BRST}}$ is nilpotent if it is extended  to act on the ghosts as\footnote{I am allowing for the possibility that the structure functions depend on fields, although the only examples I'm aware of are \emph{open} $W$-string algebras, in which case it's not possible to use BRST anyway (see Chapter \ref{chapter:W-strings}).}
\begin{equation}
\delta_{\BRST} c^{A} = N^{A}_{BC} (\phi)  c^{B} c^{C} \ .
\end{equation}
  When acting on  $\phi^i$,  $\delta_{\mathrm{BRST}}$ is nilpotent due to the closure of the algebra (see (\ref{eq:graded_commutator})):
\begin{equation}
\delta_{\BRST}^2\phi^i =  \frac{\dr c^A R^i_A}{\delta \phi^j} c^B R^j_B  + (-1)^{(\epsilon_i +1) \epsilon_C } R^i_C N^C_{AB} c^A c^B = 0 \ .
\end{equation}
 It's interesting to note that there are no complicated sign factors involved; the ghosts impose precisely the correct (anti)-symmetrization in the sum, which wouldn't be the case if one wrote the same equation using the usual transformation parameters. The choice of sign before the second term is conventional, and could have been absorbed in the structure functions.  The Jacobi identity can be written as
\begin{equation}
\label{eq:jacobi_brst_problem}
  \left( 2 c^C N^A_{BC} c^{D} c^{E} N^B_{DE} + c^{B}c^{C} N^{A}_{BC,k} c^D R^k_D \right) R^i_{A} \equiv 0 \ ,
\end{equation}
and $\delta_{\BRST}$ is nilpotent when acting of the ghosts if the object in the parentheses vanishes on its own. For most algebras this is the case, but in certain circumstances that I will come to later  (see Chapters  \ref{chapter:algebras_on_manifolds_of_sh} and \ref{chapter:W-strings})  the Jacobi identity is satisfied only after summing over the transformations, even for some cases when the structure functions are not field dependent. It doesn't seem to be possible to obtain a nilpotent symmetry from such algebras.

Next we wish to obtain a gauge fixed action by exploiting the nilpotence of the BRST operator. An action $S(\phi)$ invariant under the transformations $\delta \phi^i = \varepsilon^{A} R^i_{A}$ will automatically be BRST invariant, but it will not do in a path integral because of the non-invertibility of the on-shell Hessian. We also know that an action of the form 
\begin{equation}
S_{\GF} = S + \delta_{\BRST} \Psi
\end{equation}
will be gauge invariant due to the nilpotency of the BRST transformation. Here $\Psi$ is some integrated fermionic object local in the fields, referred to as a \emph{gauge fixing fermion}. The hope is that for an appropriate choice of $\Psi$ the action $S_{\GF}$ will have a well defined propagator.

However, as things stand we can't construct such a fermion. The classical action has ghost number zero, and this property must be preserved when gauge fixing terms are introduced. For $\gh ( \delta_{\BRST} \Psi) = 0$, the gauge fixing fermion must have ghost number $-1$. But because none of the fields have negative ghost number, it is not possible to construct such a fermion.  So we need to extend the field content by introducing the fields $\Oc^A$ and $b^A$, referred to as \emph{trivial pairs}, and let them  transform in the following (clearly nilpotent) manner under the BRST transformation:
\begin{align}
\label{eq:brst_aux_field_transformations}
\delta_{\BRST} \Oc^A = b^A \ \ \ , \ \ \  \delta_{\BRST} b^A = 0 \ .
\end{align}
We can then take $\gh (\Oc^A) = -1$, so that $\gh (b^A) = 0$.  These fields are generalizations of the $\Oc^a$ and $b^a$ fields that appeared in the Faddeev-Popov gauge fixing in the previous section.
 
 The introduction of the antighost $\gh (\Oc^A)$ and the auxiliary field $b^A$ in this way is a little bit \emph{ad hoc}. It would be quite natural if we constructed both a BRST and an anti-BRST transformation from the gauge algebra, since then the antighost would be the parameter of the anti-BRST symmetry, and the $b^A$ field would have to be introduced because of the nilpotency requirements (\ref{eq:BRST_aBRST_nilpotence}). If we write $\delta_i$, with $i = \{ \BRST, \overline{\BRST} \}$, the gauge-fixing part of the action would be  $\epsilon^{ij} \delta_{i} \delta_{j} \Psi$, where $\epsilon^{ij}$ is the Levi-Civita alternating symbol. Since this is the invariant tensor for the group $Sp(2)$, the BRST-anti-BRST gauge fixing is  $Sp(2)$ covariant. Although more symmetric, it doesn't bring much new since the most general gauge fixed action can be obtained by BRST methods alone.
 
 Understanding the implications of the BRST symmetry at the quantum level is more difficult. The problem is that now we need to worry about the invariance of a generating functional,
 \begin{equation}
 Z[J] = \int [d \Phi ] \exp{ i ( W_{\GF}( \Phi) +   \Phi^\alpha J_\alpha )  } \ ,
 \end{equation}
 where $W_{\GF}$ is the bare version of a gauge fixed action, rather than just the classical gauge fixed action. I'm taking $\Phi^\alpha$ to stand for all the fields of the theory: 
 \begin{equation}
 \Phi^\alpha  = \{ \phi^i, c^A, \Oc^A, b^A \} \ . 
 \end{equation}
 In order to talk about the BRST variation of $Z[J]$ one would like to vary $W_{\GF}$. This is 
a very unsettling thing to do, since $W_{\GF}$ is a regularized object that is infinite in the limit when the regulators are taken to zero. Ignoring this, and treating $W_{\GF}$ as if it were a classical action will yield certain naive relations between Feynman diagrams, referred to as \emph{naive Ward identities},  that are presumed to be true order by order in perturbation theory. Treating $W_{\GF}$ as if it were the classical action is automatically justified if it is possible to find a regularization for the classical action that respects the BRST symmetry. If it is not possible to do so, the naive relations may not hold, but it may still be possible to add finite counterterms to $W_{\GF}$ in order to recover them. When this is not possible, the classical BRST symmetry possesses a genuine quantum anomaly. Anomalies will  be discussed in \ref{sec:anomalies}, in this section we'll only concern ourselves with writing down naive Ward identities related to the BRST symmetry of a gauge fixed action.

Performing the change of variables $\Phi^\alpha \rightarrow \Phi'^\alpha = \Phi^\alpha+ \theta \delta_{\BRST} \Phi^\alpha$ in the path integral, we can conclude that
\begin{align}
\label{eq:BRST_ward_derivation1}
 Z[J] & = \int [d \Phi ] \exp{ i ( W_{\GF}( \Phi) + \  \Phi^\alpha J_\alpha )  } \\ \nonumber
  & = \int [d \Phi ]  \exp{ i ( W_{\GF}( \Phi) + ( \Phi^\alpha +  \theta \delta_{\BRST} \Phi^\alpha)J_\alpha )  }  \ ,
\end{align}
provided that $W_{\GF}$ is invariant under the BRST transformation, and that there is no contribution from the Jacobian.  Because $\theta^2=0$, the following relation should hold:
\begin{equation}
  \Bra{0}  \delta_{\BRST} \Phi^\alpha \Ket{0}_{J} J_{\alpha} =0 \ .
\end{equation}

Because of the non-linearity of the BRST symmetry, the expression above would involve an evaluation of expectation values of  composite operators, which involves evaluating Feynman diagrams with single insertions of these operators. However, we wish to be more systematic and include the possibility of arbitrary insertions, as is done for the fundamental fields themselves in $Z[J]$. That is, we wish to include all the  information about the BRST symmetries in some extended action. To do this, sources for the BRST transformations must be included, and the generating functional becomes
\begin{equation}
 Z[J, \Phi^*] = \int [d \Phi ] \exp{ i ( W_{\GF}( \Phi) + J_\alpha \Phi^\alpha +  \Phi^*_\alpha   \delta_{\BRST} \Phi^\alpha )  } \ .
\end{equation}
The sources $\Phi^*_\alpha = \{ \phi^*_i, c^*_A, \Oc^*_A, b^*_A  \} $, are referred to as the \emph{antifields} of $\Phi^\alpha$.\footnote{I've introduced an antifield for the $b^A$ fields, but because its BRST transformation is trivial  it's not really necessary to do so.} They  are BRST invariant have opposite spin, and obey opposite statistics to the original fields. They are assigned ghost numbers such that  the source terms have ghost number zero:
\begin{align}
& \gh (\phi^*_i) = -1 \  \ , \  \ \gh (c^*_A) = -2 \ \ , \  \ \gh ( \Oc^*_A) = 0 \ , \\  \nonumber
&  \mathrm{in} \ \mathrm{general:} \ \ \  \gh ( \Phi^*_\alpha) = - \gh (\Phi^\alpha) - 1 \ .
\end{align} 

Now the naive Ward identity can be written as
\begin{equation}
\label{eq:ward_identity1}
 J_{\alpha} \frac{\dl Z[J, \Phi^*]}{\delta \Phi^*_\alpha}  = 0 \ ,
\end{equation}
which can also be rewritten as
\begin{align}
\label{eq:BRST_ward_rewritten}
J_{i} \frac{ \dl Z }{ \delta \Phi^*_i} & =  -i \int [ d \Phi ]  \frac{\dl}{\delta \Phi^*_\alpha }  \left( \vphantom{\frac{1}{2}} \exp{ i ( W_{\ext} ) } \right) \frac{\dr}{\delta \Phi^\alpha} \left( \vphantom{\frac{1}{2}} \exp{ i (J_\beta \Phi^\beta)} \right) \\ \nonumber
& = i \int  [ d \Phi ]  \frac{\dl}{\delta \Phi^*_\alpha }  \frac{\dr}{\delta \Phi^\alpha}  \left( \vphantom{\frac{1}{2}}  \exp{ i ( W_{\ext} ) } \right)  \exp{ i (J_\beta \Phi^\beta)}  \\ \nonumber
& = -i  \int  [ d \Phi ] \left(  -i \Delta W_{\ext} + \frac{1}{2} ( W_{\ext}, W_{\ext}) \right) \exp{ i ( W_{\ext} + J_\beta \Phi^\beta) } \ ,
\end{align}
with
\begin{equation}
 W_{\ext}(\Phi, \Phi^*) := W_{\GF}(\Phi) +  \Phi^*_\alpha  c^A R^\alpha_A \ .
\end{equation}
In the last line of (\ref{eq:BRST_ward_rewritten}) the \emph{antibracket} is defined to act on any two objects $A$ and $B$ as
\begin{equation}
\label{eq:antibracket_def}
(A, B) :=   \frac{ \dr A} { \delta   \Phi^\alpha } \frac{\dl B}{ \delta \Phi^*_\alpha} - 
\frac{\dr A}{ \delta \Phi^*_\alpha }  \frac{ \dl B}{ \delta   \Phi^\alpha }  \ ,
\end{equation}
and the \emph{$\Delta$-operator} is given by
\begin{equation}
\label{eq:delta_operator}
\Delta  := (-1)^{\epsilon_\alpha +1} \frac{\dr}{\delta \Phi^*_\alpha} \frac{\dr}{\delta \Phi^\alpha}  \ .
\end{equation}
They are objects of  fundamental importance in quantum field theory. By reinstating factors of $\hbar$, one can see that the $\Delta$-operator factor is proportional to $\hbar$, and is therefore a quantum effect. In fact, $\Delta$ is not well defined when acting on local functionals\footnote{This is equivalent to the fact that it is not possible to define the square of the Dirac $\delta$-function, which is itself deeply related to infinities of quantum field theory. If there was a way to do so, it would not be necessary to introduce sources for non-linear symmetries. Instead, it would be enough to take any number of functional derivatives at the same spacetime point to get an expectation value for a non-linear operator.} \label{footnote:delta operator}, and is related to the Jacobian of the path integral measure under the BRST transformations and to the existence of anomalies.

The last line  of (\ref{eq:BRST_ward_rewritten}) states that the vacuum expectation value  of $-i \Delta W_{\ext} + \frac{1}{2} ( W_{\ext}, W_{\ext})$ in the presence of the background fields $\Phi^*_\alpha$ and $J_\alpha$ is zero. Since this must hold for arbitrary background field configurations, we have that
\begin{equation}
\label{eq:quantum_master_eq1}
-i \Delta W_{\ext} + \frac{1}{2} ( W_{\ext}, W_{\ext}) = 0 \ ,
\end{equation}
which is known as the \emph{quantum master equation}. The requirement that $W_{\ext}$ satisfies it is  a lot less restrictive than the requirement of BRST symmetry, because terms of arbitrary power in the antifields are allowed in the solution. We summarize this important point as:
\begin{itemize}
\item If the bare action $W_{\ext}$ satisfies the quantum master equation (\ref{eq:quantum_master_eq1}), \\ then the  Ward identity  $J_{\alpha} \frac{\dl  Z[J, \Phi^*]}{\delta \Phi^*_\alpha}  = 0$ is satisfied.
\end{itemize}
Most of the information about BV quantization is contained in the above statement, because for \emph{any} gauge symmetry, as long one has a solution to the quantum master equation, Ward identities can be written down. Similarly,  a solution to the \emph{classical master equation},
\begin{equation}
\label{eq:classical_master_equation}
(S_{\ext}, S_{\ext} ) = 0 \ ,
\end{equation}
 enables one to write down naive Ward identities. The BRST procedure is not applicable for open or reducible algebras, but there is hope of curing this in the more general framework. 

In perturbation theory the idea is to look for a soultion to the quantum master equation order by order in $\hbar$. In the  approximation $\hbar = 0$ one wants $S_{\ext}$ to satisfy the classical master equation. 
Let's see how the $\hbar = 0$  solution is obtained in the BRST case. As a first step we obtain the \emph{minimal solution},
\begin{equation}
\label{eq:BRST_solution_to_ME}
S_{\min} = S_{0} + \phi^*_i c^AR^i_A  + c^*_D N^D_{AB} c^A c^B \ ,
\end{equation}
where $S_0$ is the original gauge invariant action. The master equation is just the statement that $\delta_{\BRST}^2 = 0$, but because $S_0$ has no propagator $S_{\min}$ can't be used in a path integral. This is cured by adding a term $\delta_{\BRST} \Psi$. The extended gauge fixed action can be written as
\begin{equation}
 S_{\ext}(\Phi, \Phi^*) = S_{\min} (\Phi, \Phi^*) +  (\Psi, S_{\min}) + \overline{c}^*_A b^A \ ,
\end{equation}
where $S_{\min}$ is the minimal solution and I've also added a source for the BRST transformation of $\overline{c}^A$ (\ref{eq:brst_aux_field_transformations}).  In the case of an open or reducible algebra, when BV quantization is required,  the principle is still simple: as a starting point one has to find a solution to the classical master equation such that propagators exist.  It is not immediately obvious how to do this, and will be discussed in the next section.

 If $W_{\ext}$ satisfies the quantum master equation, it can be shown that the partition function is independent of the choice of gauge fermion, by which we mean that
\begin{equation}
Z_{\Psi}[J, \Phi^*] - Z_{\Psi '} [J, \Phi^*]=0 \ ,
\end{equation}
where $\Psi' = \Psi + d \Psi$ for some small deformation $d \Psi$. 
The condition for gauge independence can be written as
\begin{equation}
\left< 0 |  (d \Psi, W_{\min} +  \overline{c}^*_A b^A) | 0 \right>_{J, \Phi^*} = \left<  0 |  (d \Psi, W_{\ext} ) | 0 \right>_{J, \Phi^*} = 0 \ ,
\end{equation}
where $W_{\ext}$ involves the gauge fermion $\Psi$. The second equality follows because the term $(\Psi, W_{\min} )$ is independent of antifields.\footnote{At this point it becomes apparent that gauge fixing, as described in this section, can't work for a minimal solution non-linear in the antifields; one needs the full BV recipe in such cases.} Therefore,
\begin{align}
\label{eq:gauge_inv_wo_insertion}
& \int [ d \Phi ]  i  \frac{\dr  d \Psi}{\delta \Phi^\alpha} \frac{\dl W_{\ext}}{\delta \Phi^*_\alpha} \exp{ i ( W_{\ext} + J_{\alpha} \Phi^{\alpha})}  \\ \nonumber
& = \int [ d \Phi ]   \frac{\dr  d \Psi}{\delta \Phi^\alpha} \frac{\dl}{\delta \Phi^*_\alpha} \left( \vphantom{\frac{1}{2}} \exp{ i  W_{\ext} } \right) \exp{ i J_{\alpha} \Phi^{\alpha}}   = 0 \ .
\end{align}
After integrating by parts we get two terms which must be zero separately: 
\begin{equation}
\Bra0  d \Psi \left( -i \Delta W_{\ext} + \frac{1}{2} ( W_{\ext}, W_{\ext}) \right)  \Ket0_{J, \Phi^*} = 0 \ ,
\end{equation}
and the Ward identities for the action with the deformed gauge fermion,
\begin{equation}
\label{eq:ward_identitiy_deformed_fermion}
J_\alpha \frac{\dl Z_{\Psi '} [J, \Phi^*]}{ \delta \Phi^*_\alpha} = 0 \ .
\end{equation}
In conclusion, it's possible to move around the space of theories gauge fixed by different gauge fermions only if the extended quantum action  satisfies the quantum master equation, or equivalently  the Ward identities (\ref{eq:ward_identitiy_deformed_fermion}).

Physically relevant operators have to be gauge invariant. At the quantum level the condition is that the expectation value of  some field operator $\O(\Phi)$   is unchanged when the gauge fermion is deformed:
\begin{align}
\label{eq:gauge_inv_with_insertion}
& \int [d \Phi ]  \O(\Phi) \exp{ i ( W_{\ext} + J_{\alpha} \Phi^{\alpha})} - \int [d \Phi ]  \O(\Phi) \exp{ i ( W_{\ext}' + J_{\alpha} \Phi^{\alpha})} \\ \nonumber
& = \int [ d \Phi ]   \O(\Phi) \frac{\dr  d \Psi}{\delta \Phi^\alpha} \frac{\dl}{\delta \Phi^*_\alpha} \left\{ \exp{ i  W_{\ext} } \right\} \exp{ i J_{\alpha} \Phi^{\alpha}}   = 0 \ ,
\end{align}
where $W_{\ext}'$ involves the deformed gauge fermion $\Psi'$. We again obtain (\ref{eq:ward_identitiy_deformed_fermion}) after integrating by parts, but in addition the condition
\begin{equation}
\label{eq:sigma_operator}
\sigma \O(\Phi)\equiv  -i \Delta \O(\Phi) + ( \O(\Phi), W_{\ext})  = 0
\end{equation}
is required to hold. This is equivalent to the Ward identity
\begin{equation}
J_\alpha \frac{ \dl }{\delta \Phi^*_\alpha}  \int [d \Phi ]  \O(\Phi) \exp{ i ( W_{\ext} + J_{\alpha} \Phi^{\alpha})}  =0 \ .
\end{equation}

The $\sigma$-operator is the quantum version of the BRST operator. Formally it can be shown to be nilpotent if $W_{\ext}$ satisfies the quantum master equation. Classically physical observables are simply gauge invariant combinations of fields, but since ghost fields have been introduced we need to specify that they must have ghost number zero. At the quantum level one would expect that physical observables are classified by the cohomology classes of the $\sigma$-operator at ghost number zero. The problem with trying to understand these kinds of cohomology classes is  that the equations are extremely difficult to deal with, since they involve the singular $\Delta$-operator. In the light of this, it is a remarkable fact that when working with the effective action, the antibracket is the relevant object and not the $\sigma$-operator. 

The original definition of the effective action  (\ref{eq:effective_action1}) had only the $J_{\alpha}$ sources as background fields, but here we also have the antifields, so now
\begin{equation}
\Gamma_{\ext}[\Phi_{(c)}, \Phi^*] \equiv -i \ln{Z[J, \Phi^*]} - J_{\alpha} \Phi_{(c)}^\alpha \ ,
\end{equation}
with
\begin{equation}
\Phi_{(c)}^\alpha \equiv \frac{1}{i} \frac{\dl \ln{Z[J, \Phi^*]}}{\delta J_\alpha} =   \Bra{0} \Phi^\alpha  \Ket{0}_{J, \Phi^*}  \ .
\end{equation}
The classical fields are functions of $\Phi^*_\alpha$ as well as $J_\alpha$, but since $\Gamma_{\ext}[\Phi_{(c)}, \Phi^*] $ is independent of $J_\alpha$, it must be true that $J_\alpha$ can be written as a function of $\Phi^*_\alpha$ and $\Phi^\alpha_{(c)}$. Taking the functional derivative of $\Gamma_{\ext}[\Phi_{(c)}, \Phi^*]$, we see that
\begin{equation}
\frac{\dr \Gamma_{\ext}[\Phi_{(c)}, \Phi^*] }{\delta \Phi^\alpha_{(c)} }= -J_\alpha  \ .
\end{equation} 
This simple result comes about because terms involving the functional differentiation by  $J_\alpha$ cancel. Now it's easy to see that (\ref{eq:ward_identity1}) implies
\begin{equation}
\label{eq:ward_identity_effective_action}
( \Gamma_{\ext}[\Phi_{(c)}, \Phi^*] , \Gamma_{\ext}[\Phi_{(c)}, \Phi^*] )_{(c)} = 0 \ .
\end{equation}
The antibracket is defined as in (\ref{eq:antibracket_def}), except that now the differentiation is with respect to $\Phi^\alpha_{(c)}$ instead of $\Phi^\alpha$. So unlike $W_{\ext}$,  $\Gamma_{\ext}$ satisfies the classical master equation. The classical BRST operator is
\begin{equation}
\delta_{\BRST} \Phi^\alpha = ( \Phi^\alpha , S_{\ext} ) ,
\end{equation}
and one can think of $( \Phi^\alpha , \Gamma_{\ext} )$ as the full BRST operator that includes all quantum corrections. The non-local contributions to the full BRST operator are neatly encoded in $\Gamma_{\ext}$, and nilpotence is due to the (\ref{eq:ab_jacobi}) property of the antibracket.

The BRST gauge fixing of the Yang-Mills action  with fermions (\ref{eq:yang_mills_with_ferm}) is very simple. The minimal solution is
\begin{align}
S_{\min} =& S_{\YM} + \int d^4 x \left( -i g \mathbf{\Psi}^*  \mathbf{c} \mathbf{\Psi} +i g \overline{\mathbf{\Psi}}^* \mathbf{c} \overline{\mathbf{\Psi}}
 + \mathbf{A}^*_{\mu} \mathbf{D}^\mu \mathbf{c} + \frac{1}{2} c^*_a f^{a}_{ \ bc} c^b c^c \right) \ .
\end{align}
After introducing the trivial pairs $\{ b^a, c^a \}$, the choice of gauge fixing fermion
\begin{equation}
\Psi = \int d^4 x \overline{c}_a \left( \partial^{\mu} A^a_\mu - \frac{b^a}{2 \xi}  \right) 
\end{equation}
yields the action $S_{\FP}$ (\ref{eq:faddeev_popov_action1}).

\section{BV quantization}
\label{section:BV}

In the previous section it was observed that as long as an extended action is a solution to the master equation, naive Ward identities can be written down. Importantly, the antibracket (\ref{eq:antibracket_def}) plays a more fundamental role than the existence of a BRST symmetry. BV quantization\footnote{BV stands for I. A. Batalin and G. A. Vilkovisky, the people who originally developed this formalism    \cite{Batalin:1984jr, Batalin:1981jr}.} exploits this fact, and enables theories that are more general than what BRST quantization can handle to be gauge fixed, such as cases with reducible or open algebras.  

The first step in the procedure is to introduce an antifield $\phi^*_i$ for every field $\phi^i$ in a classical gauge invariant action $S_0$. Then, given a set of gauge generators $R^i_A$, we construct a minimal extended action starting as
\begin{equation}
S_{\min} = S_{0}(\phi) + \phi^*_i c^A R^i_A(\phi) + \cdots \ ,
\end{equation}
and fill in the dots so that $S_{\min}$ is a solution to the classical master equation (\ref{eq:classical_master_equation}).
If the transformations form a closed algebra  the solution will be just what it was in the BRST case, (\ref{eq:BRST_solution_to_ME}), but here we allow for more general solutions. It turns out  that it's possible to gauge fix any minimal solution, as long as it is \emph{proper}. Properness is the requirement that the on-shell Hessian (see (\ref{eq:Hessian})) of $S_{\min}$ has half its maximal rank,
\begin{equation}
H_{a b} \equiv \frac{ \dl \dr S_{\min}}{\delta \varphi^a \delta \varphi^b} \ \ \ , \ \ \  \mathrm{rank} (H_{a b} |_{\Sigma} )= N/2 \ ,
\end{equation}
where  $\varphi^a$ is a collective field that includes \emph{all} the fields that feature in $S_{\min}$: the original fields, the ghosts, as well as their antifields.  $N$ is the total number of these fields, and $\Sigma$ indicates that the Hessian is evaluated on the stationary surface.  The requirement that the on-shell Hessian of $S_{\min}$ has only half its maximal rank is related to the fact that the antifields are not integrated over in the path integral. Roughly speaking, properness requires that $S_{\min}$ has no further gauge symmetries other than those taken care of by the ghosts, up to those that are automatic simply because the field content has been doubled - we'll come back to this point later in the context of (\ref{eq:bv_symmetry}). In the next paragraphs I'll describe the properness condition in detail, and the method of gauge fixing $S_{\min}$, after which the connection between properness and the existence of a propagator in the gauge fixed action can easily be made.

The point on the stationary surface that is usually most convenient for checking the properness of some solution $S_{\min}$ to the master equation is  given by $\phi^*_i = c^A = c^*_A = 0$ and $\phi^i = \phi^i_0$, where $\phi^i_0$ is a solution to the equations of motion obtained from $S_0$.
 $S_{0}$ itself is trivially a solution of the master equation, since it contains no antifields. However,  if $S_0$ has gauge symmetries the rank of its Hessian evaluated at $\phi_0^i$ will be less than $N/2$. In this case the solution is not proper, and the propagator doesn't exist. Next, suppose we found a minimal solution involving a set of  generators  $R^i_A$ that closes, possibly up to equations of motion (in which case terms non-linear in the antifields are needed in $S_{\min}$). In this case there are additional terms in the on-shell Hessian of the form
\begin{equation}
\frac{ \dl \dr S_{\min}}{\delta c^A \delta \phi^*_i} |_{\phi_0, \  \phi^*_i = 0} = 2 R^i_A | _{\phi_0}  \ .
\end{equation}
It may be that not all the gauge symmetries of $S_0$ were included in $R^i_A$, in which case the rank of the Hessian is again less then $N/2$; the remedy is simply enough to include them.  If $R^i_A$ does include all the symmetries, and the rank of $H_{a b} |_{\Sigma}$ is still less than $N/2$,  the implication is that there are reducibility relations between the generators,
\begin{equation}
R^i_{A_0} Z^{A_0}_{A_1} |_{\Sigma} = 0 \ .
\end{equation}
I have introduced an integer label for the indices $A_i$, $i = \{ 0, 1 , \cdots \}$, that will differentiate between the original  gauge generators, which are labeled by the index $A_0$, reducibility coefficients $Z^{A_0}_{A_1}$, and any further reducibility coefficients
$Z^{A_i}_{A_{i+1}}, i> 0$, that imply that the reducibility coefficients $Z^{A_{i-1}}_{A_i}$ are not linearly independent:
\begin{equation}
Z^{A_{i-1}}_{A_i} Z^{A_i}_{A_{i+1}} |_{\Sigma} = 0 \ .
\end{equation}
The point is that for each of these relations we must introduce a new set of ghosts $c^{A_i}$ in order to obtain a proper solution. If this procedure terminates after introducing the ghosts $c^{A_i}$ for all $i$ up to some integer $l$, we have what is referred to as an $l$-level reducible theory.  It may be that this procedure never terminates, in which case  the  theory is infinitely reducible. The general form of a proper minimal solution for a reducible theory is
\begin{equation}
\label{eq:minimal_reducible_solution}
S_{\min} = S_0 + \phi_{i}^{*} c^A R^i_A + \sum_{k=0} c^{*}_{A_k} Z^{A_k}_{A_{k+1}} c^{A_{k+1}} + \cdots  \ ,
\end{equation}
where $\cdots$ stands terms quadratic and higher in the antifields, such as those that would be present for open algebras, or in general any terms determined by the master equation. Inability to find such terms means that the gauge algebra generators or the reducibility relations have not all been identified (see \cite{Henneaux:1992ig} for a proof of this statement). 

In subsection \ref{subsection:properness} I include some additional technical details about properness that are relevant for Chapter  \ref{chapter:W-strings}.

The next  step is to gauge fix $S_{\min}$, and to do this it's again necessary to construct a gauge fermion $\Psi$ with $\gh (\Psi) = -1$. The ghost number of a level $l$ ghost is
\begin{equation}
\gh (c^{A_l}) = l+1 \ ,
\end{equation}
and since the antibracket increases the ghost number by one, it's not possible to construct $\Psi$ using the fields in the minimal solution. As in the BRST case, it's necessary to introduce auxiliary fields. Determining the number of fields that must be introduced for reducible theories is a  technically difficult issue that I won't have need for in this thesis, so I refer the reader  to the original paper 
 \cite{Batalin:1984jr}, or to the review article \cite{Gomis:1994he}. Here I simply assume that $\Phi^\alpha$ has been extended to include all these fields, and that $\Psi$ has been chosen appropriately.

The generalization of $\delta_{\BRST}$ is
\begin{equation}
\label{eq:BV_symmetry}
\delta_{\mathrm{BV}} \Phi^\alpha := (\Phi^\alpha, S_{\min} ) \ ,
\end{equation} 
where $\Phi^\alpha$ now includes the gauge fields, and for an $l$-level reducible theory all the ghosts up to $c^{A_l}$, as well as all the necessary auxiliary fields.  The properties shared between the BRST and BV operators are nilpotence, which is due to the graded Jacobi identity (\ref{eq:ab_jacobi}), and that $\gh (\delta_{\mathrm{BV}}) = 1$. The difference is that $\delta_{\mathrm{BV}} \Phi^\alpha$ can depend on antifields. When this is the case we can't gauge fix simply by adding $\delta_{\mathrm{BV}} \Psi$ to the minimal solution, because an extended action of the form
\begin{equation}
\label{eq:BV_gauge_fixing}
S_{\ext} = S_{\min} +  \delta_{\mathrm{BV}} \Psi + \cdots \ ,
\end{equation}
where $\cdots$ stands for source terms for the auxiliary fields, will not obey the classical master equation.

The resolution to the problem is most easily obtained by appealing to the geometry that underlies the antifield formalism. Before explaining this, it is helpful to recall the description of the Hamiltonian formalism in terms symplectic geometry (see \ref{subsection:symplectic_manifolds}). The Hamiltonian formalism is defined on the \emph{phase space}, which is the cotangent bundle of configuration space.   For an $n$-dimensional configuration space with coordinates $x^i$, the phase space has $2n$ coordinates $y^a = (x^i, p_i)$. The momenta $p_i$ provide a natural symplectic structure on the phase space,
\begin{equation}
S = S_{ab} dy^a \wedge dy^b :=  d p_i \wedge dx^j \ ,
\end{equation}
  which enables us to define the Poisson bracket,
\begin{equation}
\{ A, B \} := \frac{1}{2}S^{ab} \frac{\partial A}{\partial y^a}  \frac{\partial B}{\partial y^b} = \frac{\partial A}{\partial p_i}\frac{\partial B}{\partial x_i} + A \leftrightarrow B \ ,
\end{equation}
where $S^{ab}$ is the inverse of $S_{ab}$.  Symmetries of the Hamiltonian formalism are given by \emph{canonical transformations}, which are those diffeomorphisms of phase space that preserve the symplectic structure, and therefore the Poisson bracket.

The antifield formalism  is defined over a space that has fields and antifields as its coordinates. Since half of these coordinates are fermionic, this space is not a conventional manifold but a supermanifold \cite{Cartier:2002zp}. It has additional structure related to the existence of an antibracket. This can be seen by writing the antibracket  in terms of the collective field $\varphi^a = \{ \Phi^\alpha , \Phi^{*}_{\alpha} \}$,
\begin{equation}
\label{eq:master_equation_in_collective_field}
(A, B) = \frac{ \dr A (\varphi  ) } { \delta   \varphi^{a} } \xi^{ab} \frac{\dl B (\varphi ) }{  \delta \varphi^{b} } \ ,
\end{equation}
where
\begin{equation}
\xi^{ab} = (\varphi^a, \varphi^b) = \left( \begin{array}{cc} 0 & \delta^i_j \\ - \delta^i_j & 0 \end{array} \right) \ .
\end{equation}
The coordinates $(\Phi_\alpha^*, \Phi^\alpha)$ can be thought of as Darboux coordinates for the \emph{antisymplectic} structure,  $\xi = \xi_{ab} \delta \varphi^a \wedge \delta \varphi^b$, where $\xi_{ab}$ is the inverse  of $\xi^{ab}$.\footnote{Clearly the Hamiltonian formalism can be extended to include fermionic degrees of freedom. The difference to antisymplectic geometry is that the coordinates $p_i$ and $x^i$ always have the same parity.} The functional notation $\delta \varphi^a$ is used because the space of fields is infinite-dimensional.\footnote{There is a lot more to say about the geometry behind the antifield formalism. For example, momenta can be understood in terms of the cotangent bundle over configuration space, but we haven't elucidated on the origin of antifields. This issue was originally addressed by Witten \cite{Witten:1990wb}. Subsequently Schwarz  has worked extensively on the geometry behind BV quantization (his initial paper is \cite{Schwarz:1992nx}).}

For $S_{\min}$ linear in antifields, schematically
\begin{equation}
S_{\min} = S_0 (\Phi) + \Phi^*_\alpha F^\alpha \ ,
\end{equation}
$S_{\ext}$ can be written as
\begin{equation}
\label{eq:canonical_trans_gauge_fixing}
S_{\ext} = S_0(\Phi) + \left( \Phi^*_\alpha + \frac{\dr \Psi (\Phi) }{\delta \Phi^\alpha} \right) F^\alpha \ ,
\end{equation}
where I have assumed that $S_{\min}$ includes the trivial pair sources. This is just a restatement of BRST quantization,
\begin{equation}
S_{\GF} = S_0 + \frac{\dr \Psi }{\delta \Phi^\alpha}  F^\alpha = S_0 + \delta_{\BRST} \Psi \ ,
\end{equation}
but the point is that one can think of gauge fixing as the transformation 
\begin{equation}
\label{eq:canonical_transformation1}
\Phi^*_\alpha \rightarrow \Phi^*_\alpha + \frac{\dr \Psi(\Phi)}{\delta \Phi^\alpha}  \equiv \Phi^*_\alpha + (\Psi(\Phi), \Phi^*_\alpha) \ .
\end{equation}
When the minimal solution is not linear in antifields, gauge fixing is still performed by transforming the antifields in this manner.  (\ref{eq:canonical_transformation1}) is an example of an infinitesimal canonical transformation, by which we mean that it leaves the antibracket and the antisymplectic structure invariant.\footnote{In detail, $c: \varphi \rightarrow \varphi'$ is a canonical transformation if the following diagram commutes:
\begin{equation}
\label{eq:canonical_transformation_def}
\begin{array}{ccc} 
A(\varphi), B(\varphi) & \stackrel{c}{\longrightarrow} &  A(\varphi'), B(\varphi')  \\
\downarrow &			&		\downarrow \\
( A(\varphi), B(\varphi) )_\varphi &  \stackrel{c}{\longrightarrow} & ( A(\varphi'), B(\varphi') )_{\varphi'}
\end{array} \ .
\end{equation}
By infinitesimal we mean that if the order of  $f$ is $\epsilon$, then the antibracket is preserved up to terms proportional to $\epsilon^2$.  Writing canonical transformations in their finite form is more involved; see \cite{Batalin:1984ss} and Appendix A in \cite{Troost:1989cu}.}
 It is a special kind of a more general infinitesimal transformation given by
\begin{equation}
\Phi^*_\alpha \rightarrow \Phi^*_\alpha + ( f(\Phi, \Phi^*) , \Phi^*_\alpha ) \ \ \ , \  \ \  
\Phi^\alpha \rightarrow \Phi^\alpha   +  ( f(\Phi, \Phi^* ), \Phi^\alpha ) \ ,
\end{equation}
where the fermion $f(\Phi, \Phi^*)$ depends on both fields and antifields. It seems, however, that a gauge fermion that depends only on fields is always sufficient. If $S_{\min}$ is not proper the gauge fixing procedure can't be performed successfully, irrespective of the choice of $\Psi$. The gauge symmetry of $S_{\min}$ that hasn't been taken care of by a ghost will remain a symmetry of the part of $S_{\ext}$  that is independent of antifields. 
 
Of course, one must check that gauge fixing via  (\ref{eq:canonical_transformation1})   is really  the correct thing to do. When showing that the theory is invariant under an infinitesimal deformation of the gauge fermion, we can proceed as in the BRST case. The second line in (\ref{eq:gauge_inv_with_insertion}) follows from the first because the change in the extended action due to a change $\Psi \rightarrow \Psi +  d\Psi$ is given by
\begin{equation}
\delta W_{\ext} = \frac{\dr  d \Psi}{\delta \Phi^\alpha} \frac{\dl W_{\ext}}{\delta \Phi^*_{\alpha}} \ .
\end{equation}
The naive Ward identities  (\ref{eq:ward_identity1}) and (\ref{eq:ward_identity_effective_action}) follow as in the last section, since the only requirement in their derivation is that $S_{\ext}$ has a propagator and is a solution to the classical master equation.  The statements made about physical observables and cohomology classes in the previous section remain valid here, except of course we now have $\delta_{\mathrm{BV}}$ instead of $\delta_{\BRST}$.
 
Whereas BV quantization has been successful in quantizing theories with finitely reducible and/or open algebras, infinitely reducible algebras have generally not been fully understood. Examples of infinite reducibility  can be found in algebras of gauge symmetries of Green-Schwarz type actions for particles, strings, and branes  \cite{Kallosh:1988zt, VanProeyen:1991mp, Bergshoeff:1997kr}, and string field theory actions \cite{Samuel:1987gf, Thorn:1988hm, Berkovits:2001nr} (the latter also features prominently in the BV review article \cite{Gomis:1994he}).  

Sometimes the problem of infinite reducibility can be solved by introducing an outside idea that doesn't follow  the BV recipe. One example is the pure spinor formalism of Berkovits \cite{Berkovits:2000fe}.   Another is what we have been dealing with in this section.  Namely, the minimal solution to the classical master equation $S_{\min}$ has infinitely reducible gauge symmetries. Writing the master equation using collective coordinates (\ref{eq:master_equation_in_collective_field}) and differentiating it with respect to $\varphi^a$ shows that
\begin{equation}
\label{eq:bv_symmetry}
\frac{ \dr S_{\min} (\varphi  ) } { \delta   \varphi^{a} } \mathcal{R}^a_c = 0  \ ,
\end{equation}
where
\begin{equation}
\mathcal{R}^a_c = \xi^{ab}  \frac{ \dr \dl S_{\min}(\varphi)}{ \delta \varphi^b \delta \varphi^c } \equiv \xi^{ab} H_{bc} \ ,
\end{equation}
and $H_{ab}$ is the Hessian of $S(\varphi)$. The gauge generators of this symmetry  are linearly dependent at the stationary point $\varphi_0$, which can be seen by differentiating  (\ref{eq:bv_symmetry}):
\begin{equation}
\mathcal{R}^a_b \mathcal{R}^b_c |_{ \Sigma} = 0 \label{eq:bv_nilpotency} \ .
\end{equation}
Because the symmetry generators are at the same time reducibility relations, the theory is infinitely reducible. The BV recipe itself is the 'outside idea' used to make sense of $S_{\min}$, but due to the infinite reducibility it is not possible to apply the BV recipe to gauge fix the symmetries (\ref{eq:bv_symmetry}).  There has been an attempt by Vilkovisky \cite{Vilkovisky:1999mk} at addressing the infinite reducibility of $S_{\min}$ by solving a master equation (i.e. writing a master equation for the solution to a master equation), with some interesting insights about the role played by a gauge fixing fermion that depends on both fields and antifields. Never the less, the dual function of antifields, as fields that one transforms to gauge fix and as sources for BRST/BV transformations, remains somewhat mysterious: in the latter role they are entirely classical entities, but in the former it's preferable to think of them as quantum fields, since one is transforming them into quantum fields.

\subsection{Some technicalities about the properness condition}
\label{subsection:properness}

The properness condition requires that the matrix $\mathcal{R}^a_b$ be of maximal rank.  From the null relation (\ref{eq:bv_nilpotency}) one can see that this is satisfied if all the null vectors of the on-shell Hessian are contained in itself,
\begin{equation}
\label{eq:hessian_cond1}
H_{ab} v^b |_{\Sigma} = 0 \ \ \rightarrow \ \ v^b \approx \xi^{bc} H_{cd} w^d \ ,
\end{equation}
where $\approx$ on the right hand side indicates that the equality holds only on shell. In terms of expansion (\ref{eq:minimal_reducible_solution}), the Hessian at the stationary point consists of the matrices $R^i_a$, $Z^{a_k}_{a_{k+1}}$, and $H_{ij}$, the Hessian of $S_0$ evaluated at $\phi_0$, and (\ref{eq:hessian_cond1}) reads:
\begin{align}
 \label{eq:hessian_components}
H_{ij} v^j |_{\Sigma} = 0 \ \ \ &  \longrightarrow  \ \ \ v^{j} \approx R^j_A w^A  \ ,\\  \nonumber
R^{i}_{A} v^A |_{\Sigma} = 0 \ \ \ & \longrightarrow  \ \ \   v^A \approx Z^A_{A_1} w^{A_1} \ , \\ \nonumber 
Z^{A_k}_{A_{k+1}} v^{A_{k+1}} |_{\Sigma} = 0 \ \ \ & \longrightarrow 
 \ \ \ v^{A_{k+1}} \approx Z^{A_{k+1}}_{A_{k+2}} w^{A_{k+2}} \ .
\end{align}

Taking the first relation in (\ref{eq:hessian_components}), $v^j$ can be expanded as
\begin{equation}
v^{j} = R^{j}_A w^A + v^{j k} \frac{\dl S_0}{\delta \phi^k} \ .
\end{equation}
Normally it is assumed that the matrix $v^{jk}$ is graded antisymmetric, but it is also possible that $v^{jk}$ has a graded \emph{symmetric} part.\footnote{For any reasonable action, a quantity that vanishes on-shell is proportional to the equations of motion \cite{Henneaux:1992ig}.} Then one has symmetries that are proportional to the equations of motion but can't be absorbed by adding terms quadratic in the antifields to the extended action.  One could try incorporating them into $R^j_A$, but this is problematic.  Namely, from the second relation in (\ref{eq:hessian_components}) we see that $R^i_Av^A$ is zero on-shell for \emph{any} vector $v^A$. Properness then requires that all of these vectors be included in the reducibility matrix $Z^A_{A_1}$, and thus on-shell vanishing symmetries are infinitely reducible. The condition that \emph{all} symmetries be included in $R^i_A$ is referred to as \emph{completeness}:
\begin{equation}
 \frac{\dl S_0}{\delta \Phi^i} T^{i} (\Phi) 
 \equiv y_i T^{i} (\Phi) = 0 \ \ \rightarrow \ \ T^i(\Phi) = R^i_A w^A + v^{ij} y_j \ ,
\end{equation}
for any $T^i$, where  $v^{ij}$ is now required to be graded antisymmetric.  When on-shell symmetries are present there is a difference between  completeness and properness. Every complete solution is proper, but a proper solution need not be complete. In particular, properness doesn't require that we include on-shell symmetries in $R^i_A$.  Due to the problems of infinite reducibility it seems that this indeed should not be done, and completeness should not be satisfied.

One must check that excluding on-shell symmetries is consistent, i.e. that they have no bearing on the existence of the propagator.  This is certainly the case for the examples that we'll meet in this thesis, which arise in the context of $W$-strings and will be discussed in Chapter  \ref{chapter:W-strings}. In fact, the set of symmetries which can be excluded from $R^i_A$ consists of symmetries that are either related to the  nilpotency of fermions, or to relations between the conserved currents; only the latter vanish on-shell. I'll refer to them generally as \emph{null symmetries}.  In Chapter \ref{chapter:W-strings} I'll discuss situations in which null symmetries are generated by the algebra of gauge symmetries. To close the algebra one is forced to include them in the set of generators, and it's then not possible to avoid the infinite reducibility issues.

So far I've concentrated on the first equation in (\ref{eq:hessian_components}). The property that generalizes completeness to all the equations in (\ref{eq:hessian_components})  is something called the acyclicity of the Koszul-Tate (KT) differential \cite{Henneaux:1992ig, Vandoren:1993bw, Vandoren:1996ku}. People are interested in it for various reasons, among the most important is the role it plays in proofs of existence and uniqueness (up to canonical transformations) of a solution to the classical master equation, given some reasonably behaved classical action as the starting point.\footnote{These proofs were initiated by Voronov and Tyutin \cite{Voronov:1982cp} and Batalin and Vilkovisky \cite{Batalin:1985qj}, and improved on by Henneaux \emph{et al}  \cite{Fisch:1989dq, Fisch:1989rp}.}  Generalizing its relation to completeness, properness doesn't imply the acyclicity  of the KT differential, and so BV doesn't require it when there is a difference between the two. Never the less, the acyclicity of the KT differential can be used to add the reducibility relations into $Z^{a_k}_{a_{k+1}}$ in a consistent manner \cite{Vandoren:1996ku, Thielemans:1995hn}. In more detail, once one includes the null symmetries in $R^i_A$, properness requires all reducibility relations to be included in  $Z^{a_k}_{a_{k+1}}$, an infinite number for each $k$, but the acyclicity of the KT differential allows one to consistently pick out a finite subset for each $k$. In this way the infinite reducibility problems can be controlled, but are not bypassed. 

\section{Global and conformal-type symmetries}
\label{sec-global_symmetries}

In this section we derive the naive Ward identities related to global and conformal-type symmetries of the classical action. I will discuss these on their own, not in conjunction with gauge symmetries. It is entirely possible to have global symmetries closing to gauge, but I won't have a need to consider such scenarios and refer the reader to \cite{Brandt:1996uv, Amorim:1998mm}; for an application in the context of the Wess-Zumino model see \cite{Howe:1990pz}.

If the classical action  is invariant under a certain set of global or conformal-type symmetries, $\delta \phi^i =  \varepsilon^L R^i_L$, varying $W$ in the generating functional (\ref{eq:generating_functional1})  yields
\begin{align}
\label{eq:ward_derivation}
Z[J]  	& =   \int [ d \phi ] \exp{ i ( W(\phi) + \phi^i J_i ) }  \\ \nonumber
       	& =  \int [ d \phi ] \exp{i ( W(\phi) +  (\phi^i+  \delta \phi^i ) J_i ) } \\ \nonumber
	& =   \Bra{0} 1 +  \  i J_i \varepsilon^L R^i_L(\phi) \Ket{0}_J \ .
\end{align}
 As before, the second line is equal to the first because $\phi^i \rightarrow \phi^i + \delta \phi^i$ is simply a change of variables in the path integral, if we assume that the Jacobian of the path integral  is one.   The third line is true because the parameters  $\varepsilon^L$ are taken to be infinitesimal, so only the first term in the expansion of the exponent is relevant. Thus
\begin{equation}
\label{eq:composite_op}
 \varepsilon^L  \Bra{0} J_i R^i_L(\phi) \Ket{0}_J =
\varepsilon^L \frac{\delta \Gamma_{\ext}}{\delta \phi_{(c)}^i(x)} \Bra{0} R^i_L(\phi) \Ket{0}_J = 0 \ .
\end{equation}
So naively the effective action is invariant under the variation by the vacuum expectation value of $\varepsilon^L R^i_L$ in the presence of the source $J_i$. This is the quantum equivalent of a classical symmetry.  

When $R^i_L$ is linear in the fields,
\begin{equation}
R^i_L(\phi) =   T^i_{(L)j}   \  \phi^j  \ ,
\end{equation}
we have
\begin{equation}
\varepsilon^L \Bra{0} R^i_L(\phi) \Ket{0}_J  = \varepsilon^L T^i_{(L)j} \phi^j_{(c)} \ .
\end{equation}
Then the Ward identity in terms of the effective action is
\begin{equation}
\ \frac{\delta \Gamma_{\ext}}{\delta \phi_{(c)}^i} \ \varepsilon^L T^i_{(L)j} \phi_{(c)}^j = 0 \ ,
\end{equation}
and in terms of the generating functional
\begin{equation}
 T^i_{(L)j} \frac{\delta Z}{\delta J_j} = 0 \ .
\end{equation}
So when the symmetry transformations are linear in the fields, the Ward identities can be understood as relations between Feynman diagrams that don't involve insertions of composite fields. Non-linear symmetries, on the other hand, do lead to expressions  involving explicit insertions of composite operators. It is desirable to avoid this by adding sources for the composite operators to the extended action, as we have done in the last section, as a first step to obtaining nilpotent operators related to the symmetries. This in turn will enable the calculation of potential anomalies via a cohomological analysis (see \ref{sec:anomalies}).

We extend (\ref{eq:generating_functional1})  to include sources for the symmetry transformations:
\begin{equation}
Z[J]  =  \int [ d \phi ] \exp{ i ( W(\phi) + \phi^*_i a^L R_L^i(\phi)+  \phi^i J_i   ) }  \ .
\end{equation}
Unlike the gauge fixing ghosts $c^A$,  the ghosts $a^L$ are classical background fields.  They are  like the usual parameters of transformations which we had in  (\ref{eq:variation2}) and (\ref{eq:ward_derivation}), except that they have opposite parity. For conformal-type symmetries they depend on the subset $\overline{x}$ of coordinates of the base space.  

Following the philosophy of the last section, we introduce antifields $a^*_L$ for the structure functions of the global symmetry $N^L_{MN}$ 
\begin{equation}
\label{eq:global_sym_ext_action}
S_{\ext} = S(\phi) + \phi^*_i a^{L} R^i_{L} +  a^*_{L} N^L_{MN}a^M a^N + \cdots \ ,
\end{equation}
and add any non-linear terms in the antifields necessary for $S_{\ext}$ to satisfy the classical master equation
\begin{equation}
(S_{\ext}, S_{\ext} ) =0 \ .
\end{equation}
Here the antibracket is defined with respect to the collective field $\Phi^{\alpha} := \{ \phi^i, a^{L} \}$ which now includes both the quantum fields $\phi^i$ and the background fields $a^L$. Then we can define an analogue of the BV transformation for global/conformal-type symmetries which I will refer to as $\delta_{BVG}$:
\begin{equation}
\label{eq:BVG}
\delta_{BVG} \Phi^\alpha = (\Phi^\alpha, S_{\ext}) \ .
\end{equation}

The major difference to the solution of the master equation in the BV quantization context is that  global ghosts are simply c-numbers (possibly fermionic). The part of the master equation they feature in involves usual partial derivatives,
\begin{align}
\frac {\delr S_{\ext}}{\partial a^L} \frac {\dell S_{\ext}}{\partial a^*_L} \ ,
\end{align}
where no integration is implied by repeated indices. Similarly, the part of the master equation with conformal-type ghosts contains integrals and functional derivatives involving the appropriate subset of coordinates.  For the rest of this section I'll use global ghosts to illustrate things, the generalization to conformal-type ghosts being straightforward.

Compared to algebras of gauge transformations, an important difference is that field dependent global/conformal-type algebras can not be made sense of in the context of a master equation, and the naive Ward identities necessarily involve explicit insertions of composite operators.  If we attempt a solution to the master equation of the form (\ref{eq:global_sym_ext_action}), the term proportional to $\phi^*_i$ in the master equation (which should describe the closure of the algebra) is
\begin{align}
& \int dx dy   \phi^*_i(x) a^L \frac{\dr R^i_L(\phi(x)) }{\delta \phi^j(y) } a^M R ^j_M(\phi(y))  \\ \nonumber
& + (-1)^{(\epsilon_L)(\epsilon_{\phi^i} + 1)} \int dx \ \phi^*_i(x) R^i_{L}(x)  \int dy \  N^L_{MN}(\phi(y)) \stackrel{?}{=} 0 \ ,
\end{align}
where the coordinate dependence has been written out in full. The first term clearly involves a single integral when fully evaluated, because of the $\delta$-function due to the functional derivative. By 'field dependent closure' we mean that it is equal to
\begin{equation}
\label{eq:global_field_dep_commutator}
 - (-1)^{(\epsilon_L)(\epsilon_{\phi^i} + 1)} \int dx  \phi^*_i(x) R^i_{L}(x)  N^L_{MN}(\phi(x)) \ .
\end{equation}
This is clearly not equal to the second term which involves a product of two integrals, and thus  (\ref{eq:global_sym_ext_action}) is not a solution to the master equation when the structure functions are field dependent.
  
In the BRST construction, when trying to construct a nilpotent operator we need global ghost to transform into something involving $N^L_{MN}(\phi(x))$, which is local. This is seriously problematic because we're treating objects with different functional freedom on the same footing. At the quantum level it is an order of magnitude worse because background fields are being transformed into quantum fields.  
 
We should also check whether it's possible to disregard the master equation as a starting point, and never-the-less obtain Ward identities that don't involve explicit insertions of composite operators. Performing a derivation as in (\ref{eq:ward_derivation}) for
  \begin{equation}
 Z[J, \phi^*, a , a^*]  =  \int [ d \phi ] \exp{ i ( W + \phi^*_i a^L R^i_L +J_i \phi^i   + J_{L} a^{L} ) } 
 \end{equation}
gives us the naive condition that 
 \begin{align}
 \Bra{0}  -i (-1)^{(\epsilon_L)(\epsilon_{\phi^i} + 1)}   \phi^*_i R^i_{L}  N^L_{MN}(\phi) + i J_i a^L R^i_L(\phi) \Ket{0}_{J, \phi^*, a, a^*}  = 0 \ .
 \end{align}
The second term can be rewritten as a functional derivative of $Z$ with respect to $\phi^*_i$, and is therefore not problematic. However, when the structure functions are field dependent it is not possible to add a local term to the extended action that would enable us to re-express the first term by differentiating $Z$ with respect to the ghosts or their antifields. 
 
We are thus forced to conclude that that there is no way to make sense of field dependent algebras  of global or conformal-type symmetries  without having to evaluate composite operators explicitly. The only way out is to include all the composite generators in $R^i_L$, and hope that we can get some control over the full Lie algebra generated by the transformations that close in a field dependent manner. If we are able to do so, it is simple enough to construct $W_{\ext}$, and follow the derivation of the naive Ward identities as in (\ref{eq:ward_derivation}), with the symmetry transformation $\delta_{\mathrm{BVG}}$. Performing a change of integration variable $ \phi^i \rightarrow \phi^i + \delta_{BVG} \phi^i$ in
\begin{equation}
\label{eq:gen_fucntional2}
Z[J, \phi^*, a , a^*]  =  \int [ d \phi ] \exp{ i ( W_{\ext} +J_i \phi^i   + J_{L} a^{L} ) }  \ 
\end{equation}
yields the naive Ward identities
\begin{align}
\label{eq:ward2}
 J_{i} \frac{ \dl Z }{ \delta \phi^*_i}  -
 \frac{\dell W_{\ext}}{\partial a^*_L} \left( \frac{ \delr Z}{\partial a^{L}} + J_{L} \right)=0 \ .
\end{align}
The second term can now be written as a functional derivative of $Z$  because $\frac{\dl W_{\ext} }{\delta a^*_{L}}$ depends just on background fields and can be taken outside the path integral, i.e. $\frac{\dl W_{\ext} }{\delta a^*_{L}} = \frac{\dl S_{\ext} }{\delta a^*_{L}}$.\footnote{Of course, $\exp{iJ_{L} a^{L}}$ can be taken out of the path integral as well. I have introduced this source term to avoid having to modify the definitions of the effective action (\ref{eq:effective_action1}) and the classical field (\ref{eq:classical_field}) (which is  simply $a^{L}$ in this case).}

The first term can be rewritten as
\begin{align}
J_{i} \frac{ \dl Z }{ \delta \phi^*_i} & =  -i \int [ d \phi ]  \frac{\dl}{\delta \phi^*_i }  \left\{ \exp{ i ( W_{\ext} ) } \right\} \frac{\dr}{\delta \phi^i} \left\{ \exp{ i (J_\alpha \Phi^\alpha)} \right\} \\ \nonumber
& = i \int  [ d \phi ]  \frac{\dl}{\delta \phi^*_i }  \frac{\dr}{\delta \phi^i}  \left\{ \exp{ i ( W_{\ext} ) } \right\} \left\{ \exp{ i (J_\alpha \Phi^\alpha)} \right\} \\ \nonumber
& = -i  \int  [ d \phi ] \left\{  -i \Delta_{\phi} W_{\ext} + \frac{1}{2} ( W_{\ext}, W_{\ext})_{\phi} \right\} \exp{ i ( W_{\ext} + J_\alpha \Phi^\alpha) } \ .
\end{align}
The $\phi$ subscript on the antibracket and the $\Delta$-operator indicates that these are to be evaluated using only the quantum fields. The second term in (\ref{eq:ward2}) can be written as
\begin{equation}
i  \int  [ d \phi ]  \left\{ \frac{\delr W_{\ext} }{\partial a^{L}} \frac{\dell W_{\ext} }{\partial a^*_{L}}  \right\} \exp{ i ( W_{\ext} + J_i \phi^i + J_{M} a^{M}) } \ .
\end{equation}
Therefore the Ward identity (\ref{eq:ward2}) is equivalent to the condition
\begin{equation}
\label{eq:ward3}
-i \Delta_{\phi} W_{\ext} + \frac{1}{2} ( W_{\ext}, W_{\ext})_{\Phi} =0 \ .
\end{equation}
As the subscript indicates, the antibracket involves all the fields.  

One can think of the BV symmetry (\ref{eq:BV_symmetry}) of the classical extended action (\ref{eq:BV_gauge_fixing}) as a particular nilpotent fermionic symmetry that is  non-linear in the fields (except possibly for the Abelian gauge groups). However, it is crucial to the gauge fixing procedure and for extracting the physical content of the theory, and the breaking of the BV symmetry at the quantum level renders the theory inconsistent. The breaking of some other global symmetry is simply a feature brought about by the quantization procedure and is not generally related to a physical inconsistency. 

\section{QFTs with background fields}
\label{sec-background_fields}

Background fields can play a role in quantum field theories other than just as sources for fields and transformations, or as parameter ghosts.  For example, starting from a classical action $S_{0}$ we may wish to  integrate only over  a subset of fields in the path integral, and keep the rest of the fields as background structures. This may happen when coupling a quantum field theory to a background metric. We could have, for example, a Yang-Mills theory living in curved space, or a $\sigma$-model coupled to a background worldsheet metric. Unlike the ghosts for local or conformal-type symmetries, such background fields generally depend on all the coordinates. 

A particular scenario that we'll meet in this thesis is the case of $d=2$ $\sigma$-models whose conformal-type symmetries have been gauged, but the ghosts and gauge fields are not integrated over in the path integral. The gauging procedure will be discussed in detail in Chapters \ref{chapter:2d_sigma_models} and \ref{chapter:W-strings}. As we'll see, such path integrals are useful for describing the OPE between the conserved currents related to the conformal-type symmetries in question. On the other hand, if all the fields are treated as quantum fields, the gauge fixing procedures of  \ref{section:BRST_quantization} and \ref{section:BV} need to be applied. When the conformal currents themselves are gauged we have a description of string theory, and for general conformal-type symmetries a W-string theory is obtained. 

At present we'll only need to know the field content
\begin{equation}
\label{eq:collective_field2}
\Phi^\alpha := \{ \phi^i, c^A, h^L\} \ ,
\end{equation}
where $c^A$ are the ghosts and $h^L$ the gauge fields\footnote{In fact, the index structure of the ghosts and gauge fields is the usually the same, except in certain pathological cases that we'll meet in Chapter \ref{chapter:W-strings}. Here the distinction will be kept.}, and will not need to consider their transformation properties in detail. The important property is that in the quantum analysis only $\phi^i$ are treated as quantum fields. Classically we have a gauge symmetry, but it doesn't cause problems at the quantum level because the gauge fields are not treated as quantum fields.\footnote{In more detail, for $d=2$ theories gauge fields don't have a propagator even when they are treated as quantum fields. Rather, their presence enforces certain constraints. However, it is true in all dimensions that a theory is well defined as long as gauge fields are not integrated over in the path integral.} $S_{\ext}$ is assumed to satisfy a classical master equation with respect to $\Phi^\alpha$ and the antifields $\Phi^*_\alpha :=  \{ \phi^*_i, c^*_A, h^*_L \}$ that describes the diffeomorphism invariance of $S_0$. $\delta_{BVG}$ is defined as before (\ref{eq:BVG}).  Previously, when other than the antifields the only background fields were just the global or conformal-type ghosts, it was inconsistent to have them transform to quantum fields because of their different functional freedoms. Now all the background fields are local and this is no longer a problem.

The important question is whether there are still obstructions to writing down naive Ward identities that don't involve composite operators. Making a change of variables in the path integral with $\delta_{BVG} \phi^i $  yields
\begin{equation}
\label{eq:ward5}
\int  [ d \phi ] \left\{  -i \Delta_{\phi} W_{\ext} + \frac{1}{2} ( W_{\ext}, W_{\ext}))_{\Phi} \right\} \exp{ i ( W_{\ext} + J_\alpha \Phi^\alpha) } = 0 \ .
\end{equation}
The bare action will satisfy the analogue of  (\ref{eq:ward3}), but  it is very inconvenient to analyze due to the $\Delta$-operator.  So we wish to work back and try to obtain something akin to (\ref{eq:ward2}). The part of the master equation containing quantum fields is not a problem, and we can basically rewrite this as the first term in (\ref{eq:ward2}). The part containing background fields,
\begin{equation}
\label{eq:background_part_of_me}
\Bra 0 \left( \frac{\dr W_{\ext}}{\delta h^L}  \frac{\dl W_{\ext}}{\delta h^*_L} + \frac{\dr W_{\ext}}{\delta c^A}  \frac{\dl W_{\ext}}{\delta c^*_A}  \right) \Ket 0_{h, c, \Phi^*} \ ,
\end{equation}
 can be problematic if, in either of the two terms above, both factors depend on quantum fields. Assuming that $\frac{\dl W_{\ext}}{\delta h^*_L}$  and $\frac{\dl W_{\ext}}{\delta c^*_A}$ are independent of quantum fields, the naive Ward identity
 \begin{align}
 \label{eq:ward4}
 J_{i} \frac{ \dl Z }{ \delta \phi^*_i}  -
 \frac{\dl W_{\ext}}{\delta h^*_L} \left( \frac{ \dr Z}{\delta  h^{L}} + J_{L} \right)
 -  \frac{\dl W_{\ext}}{\delta c^*_A} \left( \frac{ \dr Z}{\delta  c^{A}} + J_{A} \right) =0
\end{align}
will hold. If both factors do depend on quantum fields, it is not possible to write a naive Ward identity without involving composite operators that don't have sources in $S_{\ext}$. These observations were first made by Bastianelli  \cite{Bastianelli:1991yk}, and further analyzed in the antifield context by Falkenberg et al.  \cite{Falkenberg:1997ia, Falkenberg:1998ia}.

From a slightly different point of view, it would be desirable if the effective action satisfied
\begin{equation}
\label{eq:bf_me_for_effective_action}
 ( \Gamma_{\ext}, \Gamma_{\ext})_{\Phi^{\alpha}_{(c)}} = 0 \ ,
\end{equation}
 where  $\Phi^{\alpha}_{(c)} = \{ \phi^i_{(c)}, c^A, h^{L} \}$. Let us see what the obstruction to obtaining (\ref{eq:bf_me_for_effective_action}) is. The definitions are as follows. Starting from
\begin{equation}
Z[J_\alpha, \Phi^*_\alpha, c^A, h^L ]  =  \int [ d \phi ] \exp{ i ( W_{\ext}(\varphi) +J_i \phi^i   + J_{L} h^{L} +J_{A} c^{A}  ) }  \ ,
\end{equation}
the classical field is
\begin{equation}
\Phi^\alpha_{(c)} := \frac{-i}{Z} \frac{\dl Z}{\delta J_\alpha}   \ 
\end{equation}
(so clearly $h^{L}_{(c)} \equiv h^{L}$ $c^{A}_{(c)} \equiv c^{A}$). The effective action is 
\begin{equation}
\label{eq:def_of_effective_action2}
\Gamma_{\ext}[\Phi^{\alpha}_{(c)}, \Phi^*_{(c) \alpha}] = -i \ln{ Z[J, \Phi^*, c, h ] }- J_i\phi^i_{(c)} - J_A c^A -J_{L} h^{L} \ .
\end{equation}
It is generally non-local in $\phi^i_{(c)}$, but will remain local in the background fields. 

The first term in  (\ref{eq:ward4}) is obtained straightforwardly from (\ref{eq:bf_me_for_effective_action}) since 
\begin{equation}
\frac{\dr \Gamma_{\ext}}{\delta \phi^i_{(c)} } = -J_i \ \ \  \ \   \mathrm{and} \ \ \  \ \ 
\frac{\dr \Gamma_{\ext}}{\delta \phi^*_i }= \frac{-i}{Z} \frac{\dl Z}{\delta \phi^*_i } \ .
\end{equation}
One must be slightly careful when deriving this because in (\ref{eq:def_of_effective_action2}) $J^i$ is expressed as functional of  $\phi^i_{(c)}$ and of all the antifields. 

The second term in (\ref{eq:ward4}) does not follow from (\ref{eq:bf_me_for_effective_action}) in general. This is easy to see because
\begin{equation}
\label{eq:field_dep_eff_action_me}
\frac{\dr \Gamma_{\ext}}{\delta h^{L} } = \frac{-i}{Z} \frac{\dl Z}{\delta h^{L}}  \ \ \ \ \    \mathrm{and} \ \ \  \ \  
\frac{\dr \Gamma_{\ext}}{\delta h^*_{L} } = \frac{-i}{Z} \frac{\dl Z}{\delta h^*_{L}}  \ ,
\end{equation}
and similarly for $c^A$. The second term in (\ref{eq:field_dep_eff_action_me}) is simply $\frac{\dl W_{\ext}}{\delta h^*_L}$ when this expression is independent of quantum fields. So only when this is the case is (\ref{eq:bf_me_for_effective_action}) equivalent to (\ref{eq:ward4}).

What kind of a naive identity does the effective action (\ref{eq:def_of_effective_action2}) obey when both $\frac{\dl W_{\ext}}{\delta h^*_L}$  and $\frac{\dl W_{\ext}}{\delta c^*_A}$ contain quantum fields? The generating functional satisfies
\begin{equation}
 i J_{i} \frac{ \dl Z}{ \delta \phi^*_i}  = - \Bra 0  \left( \frac{\dr W_{\ext}}{\delta h^L}  \frac{\dl W_{\ext}}{\delta h^*_L} + \frac{\dr W_{\ext}}{\delta c^A}  \frac{\dl W_{\ext}}{\delta c^*_A} \right) \Ket 0_{h, c \Phi^*}  \ .
\end{equation}
From this it is not hard to show that the effective action satisfies
\begin{equation}
\frac{\dr \Gamma_{\ext}}{\delta \phi^i_{(c)} } \frac{\dr \Gamma_{\ext}}{\delta \phi^*_i } + \left(  \frac{\dr W_{\ext}}{\delta h^L}  \frac{\dl W_{\ext}}{\delta h^*_L} + \frac{\dr W_{\ext}}{\delta c^A}  \frac{\dl W_{\ext}}{\delta c^*_A} \right) \cdot \Gamma_{\ext} = 0 \ ,
\end{equation}
where the second term stands for a sum of 1PI  graphs (\ref{eq:effective_action2}), with single insertions of the composite operator in the parentheses.\footnote{For details on how this result is obtained the reader is referred to \cite{Bastianelli:1991yk}.} In \cite{Falkenberg:1997ia, Falkenberg:1998ia} naive Ward identities that don't necessitate insertions of composite operators have been obtained. However, these have some nasty non-local properties that make a cohomological analysis impossible.

\section{Anomalies}
\label{sec:anomalies}

In this section I will explain a method for calculating potential anomalies using cohomological techniques, by exploiting the nilpotency of operators that are available to us in the antifield formalism. The $\Delta$-operator (\ref{eq:delta_operator}) is nilpotent, but is not well defined when acting on local functionals. In addition, $W_{\ext}$ contains the bare action together with bare source terms, so the quantum master equation (\ref{eq:quantum_master_eq1}) only makes sense if the $\Delta$-operator is also regularized (see also the footnote on pg. \pageref{footnote:delta operator}). There is a series of papers \cite{Troost:1989cu, Diaz:1989nx, Hatsuda:1989qy, DeJonghe:1992gu, DeJonghe:1992xk} in which gauge theory anomalies are calculated in this context. However, because the whole calculation depends on a regularization scheme it has only been successfully applied to one loop. The more powerful approach is to work with the effective action, which is regulator independent. The relevant nilpotent operator, $\delta_{\mathrm{BV}}$ (\ref{eq:BV_symmetry}), is obtained directly from the antibracket and enables a cohomological analysis of anomalies that is independent of a particular regularization scheme.

If a classical symmetry is broken at the quantum level the effective action no longer satisfies the master equation (\ref{eq:ward_identity_effective_action}), but rather
\begin{equation}
(\Gamma_{\ext}, \Gamma_{\ext}) = \calA \cdot \Gamma_{\ext}\ ,
\end{equation}
where the \emph{anomaly} $\calA$ is a local integrated object which generally contains all the fields and antifields that enter the classical extended action. That it must be local is a general feature of renormalization theory \cite{Clark:1976ym}. Due to the Jacobi property of the antibracket (\ref{eq:ab_jacobi}), $\calA \cdot \Gamma_{\ext}$ obeys
\begin{equation}
\label{eq:nonperturb_anomaly_equation}
(\Gamma_{\ext}, \calA \cdot  \Gamma_{\ext} ) = 0 \ .
\end{equation} 
This is the Wess-Zumino consistency condition in its full generality. 

In perturbation theory the effective action is expanded as a series in $\hbar$:
\begin{equation}
\Gamma_{\ext} = S_{\ext} + \hbar M_1 + \hbar^2 M_2 + \cdots \ .
\end{equation}
Supposing that the lowest order contribution to the anomaly comes at order $\hbar^n$, then
\begin{equation}
(\Gamma_{\ext}, \Gamma_{\ext}) = \hbar^n  \calA_n + O(\hbar^{n+1}) \ ,
\end{equation}
where $\calA_n$ is a local integrated object. The anomaly is called \emph{spurious} if  $\calA_n$ can be written as
\begin{equation}
\label{eq:spurious_anomaly}
\calA_n = (S_{\ext}, \calB_n) \ ,
\end{equation}
because then the effective action can be modified as
\begin{equation}
\Gamma_{\ext}' = \Gamma_{\ext} -\frac{1}{2} \hbar^{n} \calB_n \ ,
\end{equation}
so that $\Gamma_{\ext}'$ is anomaly free to order $\hbar^{n+1}$. We have a \emph{genuine anomaly} if  $\calA_n$ is not of this form.

The problem of calculating possible genuine anomalies is a cohomological one. It follows from (\ref{eq:nonperturb_anomaly_equation}) that $\calA_n$ satisfies
\begin{equation}
\label{eq:perturb_anomaly}
(S_{\ext}, \calA_n) = 0 \ ,
\end{equation}
and we also have the equivalence relation
\begin{equation}
\calA_n \approx \calA_n  + (S_{\ext}, \calB_n ) \ .
\end{equation}
Since the antibracket increases the ghost number by one, $\gh (\calA_n) = 1$, and possible genuine anomalies are given by the cohomology classes of the antibracket at ghost number one. The approach is especially powerful since (\ref{eq:perturb_anomaly}) is a local equation, and thus one only needs to worry about the cohomology of local functionals. The theory is perturbatively anomaly free if this cohomology is trivial. Otherwise one has to conclude that the theory is anomalous, unless it is possible to argue that the coefficient of the anomaly vanishes. It should also be noted that the most general classical action compatible with a given set of linear symmetries\footnote{Non-linear symmetries must be handled by introducing sources for them, as explained in \ref{sec-global_symmetries}.}  (or equivalently the structure of all possible counterterms), is determined by the cohomology of local functionals at ghost number zero that respects these symmetries.

$\calA$ is expanded in the number of antifields. For example, for a gauge theory based on an extended action that describes a closed algebra:
\begin{equation}
\calA = c^A \calA_A + \phi^*_i c^A c^B \calA_{AB}^{i} + c^*_A  c^B c^C c^D \calA_{BCD}^A \ .
\end{equation}
The first term corresponds to an anomaly in the symmetry of the extended action, the second term to an anomaly in the algebra of the transformations, while the third to an anomaly in the Jacobi identity. It may be difficult to get a handle on all the components (for example, for one of the reasons listed below), but it may still be possible to restrict the cohomological analysis to a particular component.

\label{page:list_of_things_to_worry_about}
Some points to worry about are:
\begin{enumerate}
\item Restoring normalization conditions,
\item Operator mixing,
\item Source terms with non-renormalizable operators.
\end{enumerate} 

If one starts with some set of normalization conditions and finds a trivial anomaly, modifying the action by $- \hbar \frac{1}{2} \calB_n$ will change the original normalization conditions. The question then is whether it is possible to modify the independent parameters of the original action by terms proportional to $\hbar$ in order to restore the original normalization conditions. For an argument that this is possible, at least in the context of gauge theories, see \cite{Baulieu:1983tg}.
 
Operator mixing refers to the fact that operators of the same dimension can mix under renormalization. So if there is a source term in the action, say $\phi^*_i R^i_0$, then in the perturbative evaluation one will get corrections
\begin{equation}
R^i_0 \rightarrow R^i_0 +\hbar R^i_1 + \cdots \ ,
\end{equation}
where $R^i_1$ is an expression involving all the operators that have the dimension of $R^i_0$. In principle this problem can be solved in the cohomological analysis by constructing the most general extended action compatible with the symmetries, but it may be difficult to do so in practice.

Non-renormalizable source terms cause similar problems as non-renormalizable couplings. That is, for a source term $\phi^*_i R^i$, where $R^i$ has negative scaling dimension, one will generally need to add an infinite number of terms to the extended action of the form
\begin{equation}
\phi^*_i \phi^*_j R^{ij} + O(\phi^{*3})
\end{equation}
in order to be able to make sense of the perturbative expansion. Unless it is possible to argue that terms non-linear in the antifields can't be generated by quantum corrections (for example, in \ref{section:spechol_discussion} we make use of Lorentz invariance to argue this), one has to find a solution to the master equation that involves all powers of antifields.


\chapter{Two-Dimensional $\sigma$-models}
\label{chapter:2d_sigma_models}


The term $\sigma$-model refers to a large class of field theories whose fields are maps from some manifold $\B$ called the \emph{base space}, to some other manifold $\M$ called the \emph{target space}. In this chapter I'll be interested in models describing the propagation of strings. In most general terms this means that the target space is a ten-dimensional Lorentzian manifold, or a submanifold thereof, and the base space a two-dimensional surface also endowed with a Lorentzian metric.  

For a start we will be even more general and consider a general $d$-dimensional surface embedded in a $D$-dimensional Lorentzian manifold.  The simplest action can be constructed when the base space $\B$ has a flat metric, and only the target space $\M$ is endowed with a general Lorentzian metric $G_{ij}$. The pullback of the target space metric to the base space is 
\begin{equation}
h_{\alpha \beta} =  G_{i j}(\phi)
\partial_{\alpha} \phi^{i}(\sig) \partial_{\beta} \phi^{j}(\sig)  \ ,
\end{equation}
where $\sig^\alpha$ denotes the base space, and $\phi^i$ the target space coordinates. The action 
\begin{equation}
\label{eq:brane_sqrt}
S = -T  \int d^d \sig \sqrt{ - \det{h_{\alpha \beta} }} 
\end{equation}
is basically the volume occupied by the embedding of $\B$ inside $\M$. The corresponding equations of motion minimize this volume and describe a $(d-1)$-dimensional extended object with tension $T$ propagating through $\M$. 

Action (\ref{eq:brane_sqrt}) can be written in a classically equivalent way if we introduce a metric  $\gamma^{\alpha \beta}$ on $\B$ \cite{Polyakov:1981rd, Brink:1976sc, Howe:1977hp}:
\begin{equation}
\label{eq:bos_sigma1}
S= - T' \int d^d \sig \sqrt{- \gamma} ( \gamma^{\alpha \beta} h_{\alpha \beta} - (d-2) c) \ ,
\end{equation}
where $\gamma$ is the determinant of  $ \gamma^{\alpha \beta}$, $c$ is a constant, and $T' = c^{\frac{d}{2} -1} T$. After eliminating $\gamma_{\alpha \beta}$ using its equations of motion, (\ref{eq:brane_sqrt}) is obtained. 

Because it doesn't contain a square root, action (\ref{eq:bos_sigma1}) is much easier to handle in a path integral than action (\ref{eq:brane_sqrt}). To fully understand such a path integral we would like to treat the base space metric as a quantum field, in which case one is coupling $h_{\alpha \beta}$ to $d$-dimensional gravity. Adding a term describing pure gravity, the action of interest is
\begin{equation}
\label{eq:bos_sigma2}
S = T'  \int d^{d} \sig \sqrt{- \gamma}  (  \gamma^{\alpha \beta} h_{\alpha \beta} - (d-2) c + R(\gamma) ) \ .
\end{equation}
The problem is that in more than two dimensions  gravity is not renormalizable, and even if we fixed the worldsheet metric and didn't treat it as a quantum field the action would not be renormalizable for general target space metrics except for $d=2$  (see (\ref{eq:renormalizability_of_sigma})). Thus, it is possible to study string theory at the quantum level in perturbation theory, but it's not possible to study action (\ref{eq:bos_sigma2}) for higher dimensional objects. A further special property of $d=2$ is that the fields $\phi^i$ have scaling dimension zero.
 
In the next section I explain the structure of the $d=2$ Lorentz group. In \ref{section:bosonic_string_theory} I give a brief introduction to bosonic string theory, in particular to the gauge fixing procedure. I make use of a bi-Hamiltonian procedure to clarify some points, and obtain an extended action describing the residual symmetries of the gauge fixed action that involves an arbitrary power of antifields. In particular, I show that antifields for the $b$ field parameterize the worldsheet metric. In \ref{section:d2_supersymmetry} and \ref{section:supersymmetric_extensions} $\sigma$-models (not coupled to $d=2$ gravity) with worldsheet supersymmetry are described. In \ref{section:sc_symmetries} the superconformal symmetries of the supersymmetric $\sigma$-model are discussed, and in \ref{section:aside_on_bi_hamiltonian} the bi-Hamiltonian formalism that features in \ref{section:bosonic_string_theory} is elaborated on. Chapters 22-24 of \cite{West:1990tg} are a useful reference for sections \ref{section:2d_spinors} and \ref{section:d2_supersymmetry}-\ref{section:sc_symmetries}.

\section{Lorentz group representations and spinors in d=2}
\label{section:2d_spinors}

The Lorentz group in two dimensions, SO(1,1), is an Abelian group, since there is just one boost generator along the single space coordinate. Therefore all its irreducible representations are one-dimensional. This is manifest when working in light-cone coordinates,
\begin{equation}
\label{eq:light_cone_coords}
\xp = \sigma^{0} + \sig^{1} \mathrm{ \ \ \ and \ \ \ } \xm = \sig^{0} - \sig^{1} \ ,
\end{equation}
where $\sig^0$ and $\sig^1$ are the coordinates in which the metric takes the form
\begin{equation}
\gamma_{\alpha \beta} = \left( \begin{array}{cc}
-1 & 0\\
0 & 1
\end{array} \right) \ .
\end{equation}
The reason for using double pluses and minuses for components, as will become clear shortly, is that there are two types of spinors labeled by $+$ and $-$ indices, and vector representations of these transform as $d=2$ scalars or vectors. Derivatives with respect to the light-cone coordinates are defined by demanding that $\partial_{++} \xp = 1$, etc.:
\begin{align}
& \partial_{++}\equiv \frac{\partial}{\partial \xp} = \frac{1}{2} \left( \frac{\partial}{\partial \sig^0} + \frac{\partial}{\partial \sig^1} \right) \equiv \frac{1}{2}( \partial_0 + \partial_1 ) \ ,  \\ \nonumber
& \partial_{--} = \frac{1}{2} (\partial_0 - \partial_1) \ .
\end{align}

The invariant line element is given by
\begin{equation}
d^2 \sig \equiv -(d \sig^0)^2 + (d \sig^1)^2 = -  d \xp d \xm \ .
\end{equation}
The metric and its inverse are given by
\begin{equation}
\label{eq:2d_metric}
\gamma_{\alpha \beta} = \left( \begin{array}{cc}
0 & -\frac{1}{2}\\
-\frac{1}{2} & 0
\end{array} \right)
\textrm{\ \ \ and \  \ \ }
\gamma^{\alpha \beta} = \left( \begin{array}{cc}
0 & -2\\
-2 & 0
\end{array} \right) \
\end{equation}
in light-cone coordinates. The two components transform independently under Lorentz transformations as
\begin{equation}
\label{eq:2d_lorentz_trans_on_coord}
\delta \xp = \omega \xp \ \ \ \mathrm{and} \ \ \  \delta \xm = - \omega \xm \ .
\end{equation}
From the transformation properties of $d \xp$ and $d \xm$ it follows that there are two types of vectors, $U^{++}$ and $U^{--}$, that transform as
\begin{equation}
\label{eq:lorentz_trans_of_2d_vectors}
U^{++} \rightarrow \omega U^{++} \mathrm{ \ \ \ and \ \ \ }
U^{--} \rightarrow - \omega U^{--}
\end{equation}
under infinitesimal Lorentz transformations. A two component vector can be written as $U_{\alpha} = \left( \begin{array}{c}
U_{--} \\
U_{++}
\end{array} \right)$. Raising the indices using  (\ref{eq:2d_metric}) yields
\begin{equation}
U^{\alpha} = \left( \begin{array}{c}
U^{--} \\
U^{++}
\end{array} \right) = - 2
\left( \begin{array}{c}
U_{++} \\
U_{--}
\end{array} \right) \ .
\end{equation}
So a lower $--$ index transforms like an upper $++$ index, and vice versa.
The dot product between vectors is
\begin{equation}
U^{\alpha}V_{\alpha} = U^{++}V_{++} + U^{--}V_{--} =
-\frac{1}{2} (U^{++}V^{--} + V^{++}U^{--}) \ .
\end{equation}
with \emph{each} of the terms on the right hand side transforming like a scalar.

In $d$-dimensional Minkowski space the Clifford algebra is generated by $d$ matrices $\gamma^{\mu}$ obeying
\begin{equation}
\gamma^\alpha  \gamma^\beta + \gamma^\alpha  \gamma^\beta = 2 \eta^{\alpha \beta} \ ,
\end{equation}
 where $\eta^{\alpha \beta}$ is the Minkowski metric. The spinor representation of the Lorentz group is generated by
 \begin{align}
 &  \delta \psi = \frac{1}{2}\omega_{\alpha \beta} \gamma^{\alpha \beta} \psi  \ , \\ \nonumber
 & \gamma^{\alpha \beta}  := \frac{1}{2}(\gamma^\alpha \gamma^\beta - \gamma^\alpha  \gamma^\beta)  \ ,
 \end{align}
 for some spinor $\psi$. Spinors in two-dimensional Minkowski space hold the representation of the double cover of $SO(1,1)$, and the single generator is proportional to $\gamma^{3}$. The irreducible representations are therefore Weyl representations.

It's clearly useful to work in a representation with $\gamma^3$ diagonal. The $\gamma^{\mu}$ matrices can be chosen as  
\begin{equation}
\gamma^{0 \boldsymbol{\beta}}_{\boldsymbol{\alpha}} = \left( \begin{array}{cc}
0 &  -1 \\
1  & 0
\end{array} \right) \ \ \ \mathrm{and} \ \ \
\gamma^{1 \boldsymbol{\beta}}_{\boldsymbol{\alpha}} = \left( \begin{array}{cc}
0 &  1 \\
1  & 0
\end{array} \right) \ ,
\end{equation}
where bold font is being used for the spinor indices. A Dirac spinor is written as 
\begin{equation}
\label{eq:2d_Dirac_spinor}
\psi_{ \boldsymbol{\alpha}} = \left( \begin{array}{c}
\psi_{-} \\
\psi_{+}
\end{array} \right) \ ,
\end{equation}
with $\psi_{+}$ and $\psi_{-}$ complex and transforming by phases under the Lorentz group:
\begin{equation}
\label{eq:2d_lorentz_trans_on_spinors}
\delta \psi_- =  \frac{1}{2} \omega \psi_- \ \ \ \mathrm{and} \ \ \  \delta \psi_+ =  -\frac{1}{2} \omega \psi_+ \ .
\end{equation}
Comparing with (\ref{eq:lorentz_trans_of_2d_vectors}) one can see that the product of two Weyl spinors of the same type does indeed transform like a vector. Similarly, the product of a $+$ and a $-$ spinor transforms like a scalar.

The Dirac conjugate is 
\begin{equation}
\overline{\psi}^{\boldsymbol{\alpha}} = (\psi^*)^T \gamma^0 = ((\psi_+)^*, -(\psi_-)^* ) \ ,
\end{equation}
where $*$ indicates complex conjugation. So indices are raised as $\overline{\psi}^+ = -(\psi_-)^*$ and $\overline{\psi}^- = (\psi_+)^*$. An object with an upstairs $(+)$ index transforms in the same way as an object with a downstairs $(-)$ index, and vice versa. The Majorana condition requires that the components be real: $(\psi_{+})^{*} = \psi_{+}$ and  $(\psi)^{*}_{-} = \psi_{-}$.  It can clearly be imposed together with the Weyl condition, and the invariant matrices that perform the raising an lowering of Majorana spinor indices are
\begin{equation}
\epsilon^{\boldsymbol{\alpha \beta}} = \left( \begin{array}{cc}
0 &  1 \\
-1  & 0
\end{array} \right) 
 \ \ \ 
 \epsilon_{\boldsymbol{\alpha \beta}} = \left( \begin{array}{cc}
0 &  -1 \\
1  & 0
\end{array} \right)  \ .
\end{equation}

The Dirac Lagrangian is decomposed in terms of Majorana-Weyl spinors as follows,
\begin{align}
& \overline{\psi}^{\boldsymbol{\alpha}}  \gamma^\beta \partial_\beta \psi^+_{\boldsymbol{\alpha}}
= - (\psi_+)^* \partial_0 \psi_+  - (\psi_-)^* \partial_0 \psi_-  + (\psi_+)^* \partial_1 \psi_+  - (\psi_-)^* \partial_1 \psi_- \\ \nonumber
& = - 2 \left[ (\psi_+)^* \dem \psi_+ + (\psi_-)^* \dep \psi_-  \right] \\ \nonumber
& = - 2\left[  \psi^1_+ \dem \psi^1_+ +  \psi_-^1 \dep \psi_-^1 +  \psi^2_+ \dem \psi^2_+ +  \psi_-^2 \dep \psi_-^2 \right] \ ,
\end{align}
where in the last line I've used  $\psi_+ = \psi^1_+ + i \psi^2_+$ and $+ \rightarrow -$.

\section{Bosonic string theory}
\label{section:bosonic_string_theory}

A string worldsheet coupled to $d=2$ gravity is described  by the action
\begin{equation}
 \label{eq:bosonic_string_action}
 S = \int d^2 \sig  \sqrt{-\gamma} \gamma^{\alpha \beta } G_{i j }  \partial_{\alpha} \phi^i \partial_{\beta} \phi^j  \ .
 \end{equation}
More accurately, the above action only describes a non-interacting string. The $n$-loop vacuum diagram is calculated by the above path integral over a Riemann surface\footnote{The worldsheet becomes a Riemann surface after Wick rotation.} with $n$-holes. The scattering amplitude of $n$ strings is given by
\begin{align}
\label{eq:string_interactions}
S_{j_1 ... j_n} = \sum_{\mathrm{Riemann} \ \mathrm{surfaces}} \int [d X d \gamma]\frac{1}{\G} \exp{i(S)} \prod_{i} \int d \sig_i \sqrt{-\gamma} V_{j_i} (k_i, \sig_i) \ ,
\end{align}
 where the vertex   operator $V_{j_i} (k_i, \sig_i)$ describes a particle of momentum $k_i$ at the point $\sig_i$ on the worldsheet; note the integral over the worldsheet to enforce the independence of the amplitude on the insertion point.\footnote{The string coupling constant has not been explicitly given in (\ref{eq:string_interactions}). The most natural way to introduce it is by adding the Ricci scalar term to the action, like in (\ref{eq:bos_sigma2}), which in $d=2$ is proportional to the Euler number topological invariant of the worldsheet. The reader is referred to \cite{Green:1987mn} or \cite{Polchinski:1998rr} for a detailed account of string interactions.} The $1/ \G$ factor indicates schematically that there are local symmetries that must be gauge fixed. In this section I will describe the gauge fixing procedure.

There are two local symmetries of (\ref{eq:bosonic_string_action}):  diffeomorphism  invariance, which in its infinitesimal form reads
\begin{equation}
\delta_{diff} \gamma_{\alpha \beta} = - \partial_{\alpha} \epsilon_{(d)\beta}  - \partial_{\beta} \epsilon_{(d) \alpha} + 2 \Gamma^{\delta}_{ \ \alpha \beta} \epsilon_{(d) \delta}  \ \ \ ,  \ \ \ \delta_{diff} \phi = \epsilon^{ \alpha}_{(d)} \partial_{\alpha} \phi \ ,
\end{equation}
and invariance under local rescalings of the worldsheet metric,  referred to as Weyl invariance
\begin{equation}
\label{eq:weyl_invariance}
\gamma_{\alpha \beta} \rightarrow \Lambda(\sig) \gamma_{\alpha \beta}  \ .
\end{equation}
The latter is specific to two dimensions. Roughly, one can see that the two-dimensional metric, which is parameterized by three independent quantities, has no degrees of freedom  since diffeomorphism invariance can be used to fix two of these, and Weyl invariance the other.

One can obtain (\ref{eq:bosonic_string_action}) by gauging the conformal symmetries of
\begin{equation}
\label{eq:bos_action}
S =  \int d^2 \sig  G_{ij} {\sdep} \phi^i \sdem \phi^j  \ ,
\end{equation}
where I've chosen to work in light-cone coordinates (\ref{eq:light_cone_coords}). Since I'll be dealing with bosonic theories for a while there will be no spinorial objects, and I'll write single pluses and minuses when strictly they should be double. Action (\ref{eq:bos_action}) is invariant under infinitesimal conformal symmetries generated by
 \begin{equation}
 \label{eq:conf_trans}
 \epsilon^\alpha R^i_\alpha =  \delta_{+} \phi^i + \delta_{-} \phi^i = \epsilon^{+} \sdep \phi^i  +  \epsilon^{-} \sdem \phi^i  \ .
 \end{equation}
That is,
\begin{equation}
\delta_{+} S = - \int d^2 \sig \sdem \epsilon^+ T_{++} \equiv  - \int d^2 \sig \sdem \epsilon^+ G_{ij} \sdep \phi^i \sdep \phi^j \ \ \ \mathrm{and} \ \ \ + \leftrightarrow -  \ ,
 \end{equation}
so (\ref{eq:conf_trans}) is a symmetry, provided the parameters  obey 
\begin{equation}
\label{eq:conf_param_condition}
\sdem \epsilon^{+} = \sdep \epsilon^{-} = 0 \ .
\end{equation}
 Strictly speaking, it is only after a Wick rotation this condition translates to the symmetries being conformal, when the two parameters become holomorphic and antiholomorphic functions over the worldsheet.  So there is a slight abuse of terminology when referring to (\ref{eq:conf_trans}) as conformal transformations. The currents $T_{++}$ and $T_{--}$ are conserved, in the sense that $\sdem T_{++} = \sdep T_{--}=0$ on-shell.

In spirit of Chapter \ref{ch:antifield_formalism} we will work with fermionic ghosts $a^{+}$ and $a^{-}$ in place of the parameters (see also Appendix \ref{app:ghost_commutators}). The commutator of conformal symmetries in the same sector is
 \begin{equation}
 \label{eq:conformal_algebra}
 a^\alpha R^i_{\alpha,k} a^\beta R^k_\beta  = a^{+}
  \sdep a^{+} \sdep \phi^i +  (+ \rightarrow -)  \ ,
 \end{equation} 
while the commutator between the $(+)$ and $(-)$ sectors vanishes, provided that (\ref{eq:conf_param_condition}) holds. The extended action expressing the conformal symmetries is
 \begin{align}
 S_{\ext}  &=   \int  d^2 \sig    \left[ \  G_{ij} {\sdep} \phi^i \sdem \phi^j + \phi^*_i ( a^{+} \sdep \phi^i     +    a^{-} \sdem \phi^i ) \right. \\ \nonumber
  &  \ \ \ \ \ \ + \left.  a^*_{+}  a^{+} \sdep a^{+}  +  a^*_{-}  a^{-} \sdep a^{-}  \ \right] \ ,
 \end{align}
where $\phi^*_i$ is a fermionic antifield, and $a^*_{+}$, $a^*_{-}$ bosonic ones. The latter depend only on half the worldsheet coordinates. 

The first step to gauging action (\ref{eq:bos_action}) is to promote the conformal ghosts $a^+$ and $a^-$ to local ghosts $c^+$ and $c^-$.  For a single sector, say the $(+)$ sector, the gauging is performed by adding the term $\int d^2 \sig \hp T_{++}$ to the action and letting $\hp$  to transform as
\begin{equation}
\delta_{+} \hp = \sdem c^{+} - \hp \sdep c^{+} +  c^{+} \sdep \hp \ .
\end{equation}
That is, the minimal solution
\begin{align}
\label{eq:chiral_conf_ext_action}
S_{\min} = & \int d^2 \sig \left[  G_{ij} {\sdep} \phi^i \sdem \phi^j - \hp G_{ij} \sdep \phi^i \sdep  \phi^j + \phi^*_i c^{+} \sdep \phi^i  \right. \\ \nonumber 
& \left. + \ahp (\sdem c^{+} - \hp \sdep c^{+} +  c^{+} \sdep \hp)  + c^*_{+}  c^{+} \sdep c^{+}   \right]
\end{align}
satisfies the master equation, with $\ahp$ a fermionic antifield for the gauge field $\hp$. 

When trying to gauge both sectors two gauge fields, $h^{++}$ and $h^{--}$, need to be introduced.  It is then necessary to  keep adding more and more terms non-linear in the gauge fields in order to close the algebra. A clever way to understand this infinite sum is to work in a covariant first order formalism \cite{Schoutens:1990ja, Hull:1993kf}. This involves introducing two momenta, $\pip{i}$  and  $\pim{i}$, and for obvious reasons I will also refer to this formalism as \emph{bi-Hamiltonian}. The reason this formalism is natural is because $d=2$ light-cone coordinates enter action (\ref{eq:bos_action}) symmetrically, and one can equally pick either as the time coordinate in a Hamiltonian procedure, i.e.:
 \begin{equation}
 \pip{i} = G_{ij} \sdep \phi^{j}    \ \ \ , \ \ \   \pim{i} =  G_{ij} \sdem \phi^{j} \ .
 \end{equation}
This enables us to write down actions of the form
\begin{equation}
\label{eq:pi_action_special}
S_{\pi} = \int d^2 \sig  \left[  G_{ij} \sdep \phi^i  \sdem \phi^j + 2 G^{ij} \pip{i} \pim{j}  -2  \pip{i} \sdem \phi^i -2  \pim{i}  \sdep \phi^i \right]  \ ,
\end{equation}
that are equivalent to (\ref{eq:bos_action}) after eliminating the momenta using their equations of motion. One can choose the terms in (\ref{eq:pi_action_special}) in different proportions and still obtain (\ref{eq:bos_action}). The particular action chosen above is invariant under the conformal symmetries
\begin{align}
\label{eq:conf_spec}
& \delta_{(+)} \phi^i = a^{+} G^{ij} \pip{i} \ \ \ , \ \ \   \delta_{(+)} \pip{i} = \sdep ( a^{+} \pip{i}) \ , \\ \nonumber
& \delta_{(+)} \pi_{-j} = - a^+ \Gamma^p_{ \ ij} \sdem \phi^i \pi_{+p} - a^+ G^{ik}_{ \ \ \ ,j} \pi_{+i} \pi_{-j} \ ,
\end{align}
and $+ \rightarrow -$. It is special in the sense that parts not containing derivatives depend only on momenta. Between the sectors the transformations acting on $\phi^i$ commute, while the transformations of momenta commute only up to the equations of motion. Since the momenta are ultimately going to be integrated out, we will be free to ignore the complicated transformations of momenta, at least for the purposes of this section.

For a general action of the form (\ref{eq:pi_action_special}), even in flat space the conformal symmetry transformations would depend on $\phi^i$  and the two sectors would commute only up to the equations of motion. The upshot is that in flat space (\ref{eq:pi_action_special}) can be gauged in closed form, by adding $\int d^2 \sig \hp \eta^{ij} \pip{i} \pip{j}$ and $\int d^2 \sig \hm \eta^{ij} \pim{i} \pim{j}$, and letting the gauge fields transform as
\begin{equation}
\delta_{+} \hp = \frac{1}{2} \sdem c^+  + c^+ \sdep \hp  - \hp \sdep c^+ \ \ \ \mathrm{and}  \ \ \ \ + \leftrightarrow - \ .
\end{equation}

The gauged system is described by the minimal solution
\begin{align}
\label{eq:conf_spec_extended}
S_{\min} =  & S_{\pi}  + \int d^2 \sig  \left[   2 \hp \eta^{ij} \pip{i} \pip{j} +  2 \hm \eta^{ij} \pim{i} \pim{j} \right. \\ \nonumber
& \phi^*_i \eta^{ij} (c^+ \pip{j} + c^- \pim{j})   - \sdep \apip{i} c^+ \pip{i}  - \sdem \apim{i} c^- \pim{i}  \\ \nonumber
& +\ahp (\frac{1}{2} \sdem c^+ + c^+ \sdep \hp - \hp \sdep c^+)  \\ \nonumber
 & \left. +\ahm( \sdep c^- + c^- \sdem \hm - \hm \sdem c^-) - c^*_+ c^+ \sdep c^+ - c^*_- c^- \sdem c^-  \right] \ .
\end{align}
Integrating out $\pip{i}$ and $\pim{i}$, we obtain
\begin{align}
\label{eq:conf_spec_extended_no_pi}
S_{\min} =&  \int d^2 \sig   \frac{1}{1- 4 \hp \hm} \left[  \vphantom{\frac{1}{2}}- (1-4\hp\hm)\eta_{ij} {\sdep} \phi^i \sdem \phi^j  + 2 \hp T_{++} + 2 \hm T_{--} \right.  \\ \nonumber
 & + \phi^*_i( c^+ \sdep \phi^i +  c^- \sdem \phi^i - 2  \hm c^+ \sdem \phi^i - 2 \hp c^- \sdep \phi^i ) \\ \nonumber
 & \sdep \apip{i}( - c^+ \sdep \phi^i + 2 \hm c^+ \sdem \phi^i)  + \sdem \apim{i} (- c^- \sdem \phi^i + 2 \hp c^- \sdep \phi^i) \\ \nonumber
 & - \left. \frac{1}{2} c^- c^+ \eta_{ij} (  \sdep \apip{i} \phi^j  -   \sdem \apim{i} \phi^j  +   \sdem \apim{i} \sdep \apip{j} ) \right] + O(h^*) + O(c^*) \ ,
\end{align}
where  $O(h^*) + O(c^*)$ stands for the terms proportional to gauge and ghost antifields, which remain the same as in (\ref{eq:conf_spec_extended}). 

It is not possible to gauge the bi-Hamiltonian action in closed form in curved space, when the transformations containing target space derivatives in (\ref{eq:conf_spec}) are non-zero. However, one can imagine constructing a solution term by term, and because all the terms involving target space derivatives contain momenta and their antifields it is clear that after setting these fields to zero (\ref{eq:conf_spec_extended_no_pi}) remains a solution to the master equation even in curved space.

The first line of (\ref{eq:conf_spec_extended_no_pi}) is equivalent to (\ref{eq:bosonic_string_action}), with the  metric parameterized as
\begin{equation}
\label{eq:gf_metric}
\gamma_{\alpha \beta} = 
\Lambda'( \sig) \left( \begin{array}{cc}
2 \hm &  \frac{1}{2} ( 1 + 4 \hm \hp)  \\
 \frac{1}{2} ( 1 + 4 \hm \hp)  & 2 \hp
\end{array} \right) \ .
\end{equation}
The parameter $\Lambda'(\sig)$ drops out of (\ref{eq:conf_spec_extended_no_pi}) due to Weyl invariance (\ref{eq:weyl_invariance}). The metric has a negative determinant,
\begin{equation}
\gamma= - \frac{1}{4} (1 - 4 \hm \hp)^2 \ ,
\end{equation}
and is thus explicitly Lorentzian, as it must be. In conclusion, 
\begin{align}
\label{eq:conf_spec_extended_no_pi2}
S_{\min} =&  \int d^2 \sig   \frac{1}{1- 4 \hp \hm} \left[  \vphantom{\frac{1}{2}}- (1-4\hp\hm)G_{ij} {\sdep} \phi^i \sdem \phi^j  + 2 \hp T_{++} + 2 \hm T_{--} \right.  \\ \nonumber
 & + \left. \phi^*_i( c^+ \sdep \phi^i +  c^- \sdem \phi^i - 2  \hm c^+ \sdem \phi^i - 2 \hp c^- \sdep \phi^i )
 \vphantom{\frac{1}{2}} \right]  \\ \nonumber
&  + \int d^2 \sig \left[ \ahm( \sdep c^- + c^- \sdem \hm - \hm \sdem c^-) - c^*_+ c^+ \sdep c^+ - c^*_- c^- \sdem c^-  \right] \ 
\end{align}
 is a minimal solution to the master equation describing the symmetries of (\ref{eq:bosonic_string_action}), which can easily be checked directly.

Next we want to obtain a gauge fixed action starting from (\ref{eq:conf_spec_extended_no_pi2}), and for this we need to introduce auxiliary fields (see \ref{section:BRST_quantization} and \ref{section:BV}).  For historical reasons, in most string theory and conformal field theory literature what I referred to as the antighost $\Oc$ in Chapter \ref{ch:antifield_formalism} is called the $b$ field, while what I've called the  $b$ field is referred to as $\lambda$. Not to be in conflict with the literature I will use the stringy notation, even though it's confusing in relation to Chapter \ref{ch:antifield_formalism} (particularly in relation to the $Sp(2)$ invariant quantization mentioned in \ref{section:BRST_quantization}, in which $c$ and $\Oc$ are naturally treated on the same footing). 

In any case, after introducing trivial pairs $b_{++}, \lambda_{++}$ and $+ \rightarrow -$,  which enter the extended action via 
\begin{equation}
\int d^2 \sig (b^{++}_* \lambda_{++} + b^{++}_* \lambda_{++}) \ , 
\end{equation}
a gauge fixed action is obtained by employing the gauge fixing fermion
\begin{align}
\label{eq:string_gf_fermion}
\Psi = & b_{++} ( \hp- \hat{h}^{++}) + b_{--} ( \hm- \hat{h}^{--}) + \phi^*_i \phi^i +c^*_+ c^+  + c^*_- c^- \ .
\end{align}
Gauge fixing is done by replacing the $\phi^*_i$s with $ \frac{\dr \Psi}{\delta \phi^i}$ and $h^*$ by $ \frac{\dr \Psi}{\delta h}$ (see (\ref{eq:canonical_trans_gauge_fixing}) and (\ref{eq:canonical_transformation1})),
\begin{align}
\label{eq:fixing_2nd_order_string}
S_{\ext} =&   \int d^2 \sig \frac{1}{1- 4 \hp \hm}  \left[  \vphantom{\frac{1}{2}}- (1-4\hp\hm)S_0  \right. \\ \nonumber 
& + 2 \hp T_{++} + 2 \hm T_{--}  + b_{++} \left( \frac{1}{2} \sdem c^+ + c^+ \sdep \hp - \hp \sdep c^+ \right)  \\ \nonumber
 & + b_{--} \left(\frac{1}{2}  \sdep c^- + c^- \sdem \hm - \hm \sdem c^- \right)  \\ \nonumber 
 & \left. + (h^{++} - \hat{h}^{++}) \lambda_{++} + (h^{--} - \hat{h}^{--}) \lambda_{--}  \vphantom{\frac{1}{2}} \right] + O(\phi^*) + O(c^*)  \ ,
\end{align}
where the terms proportional to ghost antifields and $\phi^*_i$ are the same as in (\ref{eq:conf_spec_extended_no_pi2}). Integrating over $\lambda_{++}$ and $\lambda_{--}$ fixes the $h^{\alpha \beta}$ fields to the background $\hat{h}^{\alpha \beta}$.

However, after integrating over  $\lambda_{\alpha \beta}$ the extended action \emph{does not} tell us what the BRST symmetry is. This is because in setting  the gauge fields to a background we've ignored that they were supposed to be quantum fields, albeit  auxiliary ones. The information is still there because the BRST/BV transformation for the $b_\alpha$ field is obtained by replacing $\lambda_{\alpha \beta}$ in $\delta b_{\alpha \beta} =  \lambda_{\alpha \beta}$ by using the equations of motion for $h^{\alpha \beta}$.  

When the background is set to $\hat{h}^{\alpha \beta} = 0$ the BRST transformation in the $(+)$ sector is
\begin{align}
\label{eq:string_BRST_free_background}
& \delta_{\BRST} \phi^i = c^+ \sdep \phi^i + c^- \sdem \phi^i \ , \\ \nonumber
& \delta_{\BRST} c^+ = - c^+ \sdep c^+  \ ,\\ \nonumber
& \delta_{\BRST} b_{++} = 2 G_{ij} \sdep \phi^i \sdep \phi^j - \sdep b_{++} c^+ - 2 b_{++} \sdep c^+   \ ,
\end{align}
and similarly in the $(-)$ sector. The two sectors commute. When the background fields are set to something non-trivial it is harder to  describe the situation nicely. One can of course still derive the BRST transformations, which will depend in a complicated way on the background fields $\hat{h}^{\alpha \beta}$. 

However, it is possible to find a more pleasing solution.  The easiest to see it is by gauge fixing the action containing momenta, (\ref{eq:pi_action_special}), and then integrating them out. The gauge fixing fermion is again (\ref{eq:string_gf_fermion}), and we obtain:
\begin{align}
S_{\ext} =  & S_{\pi}  + \int d^2 \sig  \left[  \vphantom{\frac{1}{2}} 2 \hp G^{ij} \pip{i} \pip{j} +  2 \hm  G^{ij} \pim{i} \pim{j} \right. \\ \nonumber
& + b_{++} \left( \frac{1}{2} \sdem c^+ + c^+ \sdep \hp - \hp \sdep a^+ \right) \\ \nonumber 
& + b_{--} \left( \frac{1}{2} \sdem c^- + c^- \sdem \hm - \hm \sdem a^- \right) \\ \nonumber 
& \left. + (\hp - \hat{h}^{++}) \lambda_{++} +  (\hm - \hat{h}^{--}) \lambda_{--} \right] + O(\mathrm{antifields}) \ .
\end{align}
Again, the above is strictly a solution to the master equation only in flat space, but will remain a solution in curved space once the momenta are integrated out and the momenta antifields set to zero. 

By using the equations of motion for $\lambda_{\alpha \beta}$ we find out what the BRST transformations are. The most elegant solution to the problem is obtained by noticing  that $-\hat{h}^{\alpha \beta}$ plays precisely the role of the sources for the BRST transformations of $b_{\alpha \beta}$.\footnote{After  some integration by parts.} Therefore, the gauge fixed action is written most simply as:
\begin{align}
\label{eq:guaged_fixed_bihamiltonian}
S_{\ext} =  & S_{\pi}  + \int d^2 \sig  \left[   2 \hp G^{ij} \pip{i} \pip{j} +  2 \hm  G^{ij} \pim{i} \pim{j} \right. \\ \nonumber
&+ b_*^{++} (-2 G^{ij} \pip{i} \pip{j} - a^+ \sdep b_{++} +2 b_{++} \sdep a^+ ) \\ \nonumber
& + b_*^{--} (-2 G^{ij} \pim{i} \pim{j} - a^+ \sdem b_{--} +2 b_{--} \sdep a^- ) \\ \nonumber
& + \phi^*_i G^{ij} (c^+ \pip{j} + c^- \pim{j})   - \sdep \apip{i} c^+ \pip{i}  - \sdem \apim{i} c^- \pim{i}  \\ \nonumber
& \left. - c^*_+ c^+ \sdep c^+ - c^*_- c^- \sdem c^-  \right] + O(\pi_*) \ .
\end{align}
Integrating out the momenta gives (\ref{eq:conf_spec_extended_no_pi2}) with the replacement $h^{\alpha \beta} \rightarrow-b_*^{\alpha \beta}$, and it can explicitly be checked that it does obey the master equation.

To summarize the main point of this section:
\begin{itemize}
\item Action(\ref{eq:conf_spec_extended_no_pi2}) can be gauge fixed by making the following replacement
\begin{equation}
\label{eq:gauge_fixing_recipe}
h^*_{\alpha \beta} \rightarrow b_{\alpha \beta}, h^{\alpha \beta} \rightarrow-b_*^{\alpha \beta} \ .
\end{equation}
\end{itemize}

There are two important observations regarding the gauge fixing of the bosonic string:
\begin{enumerate}
\item If the second order action (\ref{eq:fixing_2nd_order_string}) was  fixed to a general background not using (\ref{eq:gauge_fixing_recipe}), but by deriving the BRST transformation on the $b^{\alpha \beta}$ field from the equations of motion of the momenta, one would obtain a complicated expression in the $\hat{h}^{\alpha \beta}$ fields, but the quantization would still be BRST. The recipe above gives us a solution with all powers of antifields, and is thus inherently BV.
\item Neither of the  gauge fixing methods  quite follow the standard BV/BRST procedure as outlined in Chapter \ref{ch:antifield_formalism}. This is clearly true for the recipe (\ref{eq:gauge_fixing_recipe}). But also in the BRST case, one needs to find the transformation of $\lambda_{\alpha \beta}$ by appealing to the $h^{\alpha \beta}$ equations of motion. This does not happen for an action with a dynamic gauge field; one can generally write down the minimal solution, gauge fix and simply read off the BRST/BV transformations from the antifield terms.
\end{enumerate}

It is easiest to evaluate the path integral when the $h^{\alpha \beta}$ fields are fixed to zero. Then the part of the action independent of antifields is given by
\begin{equation}
\label{eq:gauge_fixed_bosonic_string_flat_ws}
S = \int d^2 \sig \left[ G_{ij} \sdep \phi^i \sdem \phi^j - \frac{1}{2} (b_{++} \sdem c^+ - b_{--} \sdep c^-)  \right] \ ,
\end{equation}
and there is no need to worry about a non-trivial background metric on the worldsheet. The part of the action involving ghosts is referred to as a $b$-$c$ system, and is itself a theory with a conformal symmetry. That is, 
\begin{equation}
S = \int d^2 \sig  \frac{1}{2} b_{++} \sdem c^+ 
\end{equation}
  is invariant under
\begin{equation}
\delta b_{++} = 2 \sdep \epsilon^+ b_{++} + 3 \epsilon^+ \sdep b_{++} \ \ \ , \ \ \  \delta c^+ = \sdep \epsilon^+ + 3 \epsilon^+ \sdep c^+ \ ,
\end{equation}
if  $\sdem \epsilon^+=0$, and similarly for $+ \leftrightarrow -$. The conserved current $T^{(bc)}_{++}$ associated with this symmetry features in the BRST transformation of $b_{++}$ (\ref{eq:string_BRST_free_background}):
\begin{equation}
\delta_{\BRST} b_{++} = T^{(\phi)}_{++} + T^{(bc)}_{++} \ .
\end{equation}

To understand when the BRST transformation has a quantum anomaly one can appeal to the abstract conformal field theory formalism (see for example \cite{Lust:1989tj, Ginsparg:1988ui, Polchinski:1998rq}).  In rough terms, each conformal field theory has a quantity called the central charge $c$ associated to its energy momentum tensor that dominates the short distance behavior of the expectation value of the energy momentum tensor $T$  with itself: $\Bra{0} T(\sig) T(0) \Ket{0} \propto c/ \sig^4 $. It turns out that the short distance behavior can be translated to the algebra of symmetry transformations, and that central charge deforms the classical conformal algebra.   In our case $T^{(\phi)}_{++}$ and $T^{(\phi)}_{--}$ are the components of the energy momentum tensor for the conformal theory defined by the action $\int d^2 \sig G_{ij} \sdep \phi^i \sdem \phi^j$, and $T^{(bc)}_{++}$ is the energy-momentum tensor for the theory defined by the action   $\int d^2 \sig  \frac{1}{2} b_{++} \sdem c^+ $ (and similarly for $T^{(bc)}_{--}$). Precisely because the energy-momentm tensors feature in the BRST transformations, the central charge enters as an obstruction to their classical nilpotency being maintained at the quantum level. Only when the central charge of $T^{(\phi)}_{++} + T^{(bc)}_{++}$ (and $+ \rightarrow -$) vanishes is the BRST symmetry truly nilpotent. It turns out that the central charge of $T^{(bc)}_{++}$ is -26. For a trivial target space metric the central charge of $T^{(\phi)}_{++}$ is equal to the dimension of the target space, and thus the critical dimension for bosonic string theory is 26.

The bosonic action can be generalized to include terms related to other target space tensors. The ones important for the bosonic string are the $b$ field, which is a two form related to a charge held by the fundamental string itself,  and the scalar  dilaton field $\Phi$, which is related to the string coupling. The most general bosonic action is
\begin{equation}
\label{eq:string_in_gen_back}
S = \int d^2 \sig \left\{ ( \sqrt{\gamma} \gamma^{\mu \nu}G_{ij} +
\alpha \epsilon^{\mu \nu} b_{ij} ) \partial_{\mu} \phi^{i} \partial_{\nu}\phi^{j} + \alpha' R \Phi(\phi) \right\} \ .
\end{equation}

To evaluate the gauge fixed path integral for a non-trivial target space metric $\phi^i$ must be split into a background and a quantum part. Many of the textbooks use a non-covariant split, which leads to non-covariant Green functions. The \emph{background field method} involves a non-linear but covariant split, which gives us a much better handle on the theory. It is worth consulting the original literature \cite{Fradkin:1984pq, Hull:1985rc}; see also \cite{Howe:1986vm, Blasi:1988sh}, where renormalizability in the background field method is proven.

\section{$d=2$ supersymmetry and superfields}
\label{section:d2_supersymmetry}

To promote the bosonic $\sigma$-model action (\ref{eq:bos_action}) to one with worldsheet supersymmetry\footnote{For a thorough discussion of supersymmetry and supergravity the reader is referred to one of the books \cite{West:1990tg, Wess:1992cp}, or the shorter review papers \cite{Lindstrom:2002ph, Figueroa-O'Farrill:2001tr}. For a more mathematical exposition in terms of supermanifolds see  \cite{Cartier:2002zp}. } fermionic worldsheet coordinates must be introduced. Since there are two types of fermions in two dimensions, $\theta^+$ and $\theta^-$ (see discussion around (\ref{eq:2d_Dirac_spinor})), there are two types of supersymmetric worldsheets with a single fermionic direction.  Maps from these worldsheets to the target space, which is still just an ordinary manifold, are either $(1,0)$ bosonic superfields
\begin{equation}
X^i(\sig, \theta^+) = \phi^i(\sig) + \theta^+ \psi_+^i(\sig) \ ,
\end{equation}
or $(0,1)$ superfields
\begin{equation}
X^i(\sig, \theta^-) = \phi^i(\sig) + \theta^- \psi_-^i(\sig) \ .
\end{equation}
One can expand a $(1,0)$ superfield $X^i$ in terms of the bosonic function $\phi^i(\sig)$ and a fermionic function $\psi_+^i(\sig)$  simply because $(\theta^+)^2$. In general, one can have $(n,m)$ superfields that have $n$ $\theta^+$ type fermionic coordinates and $m$ $\theta^-$ type coordinates.
  
In the context of this thesis I'll mainly work with $(1,1)$ superfields, 
\begin{equation}
X^i(\sig, \theta^+ \theta^-) = \phi^i(\sig) + \theta^+ \psi_+^i(\sig) + \theta^- \psi_-^i(\sig) + \theta^+ \theta^- F^i(\sig) \ ,
\end{equation}
which are maps from a worldsheet extended by two fermionic coordinates,  $\theta^+$ and $\theta^-$, to some target space manifold. Even though the target space superfields $X^i(\sigma, \theta^+)$ are bosonic, it is really only the leading components $\phi^i$ that are coordinates of the target space manifold. One can't think of eg. $\theta^+ \psi_+^i(\sig)$ as standing for a particular point on $\M$. Rather, these have to be thought of as slightly mysterious fudge factors that are a consequence of worldsheet supersymmetry. However, $\psi_+^i(\sig)$ are \emph{not} part of a superspace extension of the target space since they carry target space vector indices.

Differentiation with respect to the fermionic coordinates is defined by demanding that
\begin{equation}
\frac{\vec{\partial}}{\partial \theta^+} \theta^+ =1 \ .
\end{equation}
The difference to ordinary differentiation is that the fermionic derivative picks up a minus sign as it goes past fermionic objects, since $\frac{\vec{\partial}}{\partial \theta^+} (\theta^+ \theta^- ) =  -\frac{\vec{\partial}}{\partial \theta^+} (\theta^- \theta^+ ) $. It is therefore also necessary to specify a direction. Integration for fermionic coordinates is defined to be the same thing as differentiation, which is the only possible definition if we make the plausible assumption that 'exterior derivatives' of the fermionic coordinates, $d \theta^+$ and $d \theta^-$, are also fermionic. In this case $\int d \theta^+ d \theta ^+ = 0$, and hence $\int d \theta^+ = 0$. Definite integration over fermionic variables doesn't make sense, so the simplest non-zero possibility, $\int d \theta^+ \theta ^+$,  can be normalized to be equal to one.

Thus, for any $(1,1)$ superfield $\Phi(\sig, \theta^+, \theta^-)$, integration over Grassmann coordinates simply amounts to projecting out the $\theta^+ \theta^-$ component. Furthermore, we can 'integrate by parts' because
\begin{equation}
\label{eq:ferm_int_by_parts}
\int d\theta^+ d\theta^-  \frac{\vec{\partial} \Phi}{\partial \theta^+} =  \int d\theta^+ d\theta^-  \frac{\vec{\partial} \Phi}{\partial \theta^-} = 0 \ .
\end{equation}

The generators of supersymmetry transformations are given by
\begin{equation}
\label{eq:Q_def}
Q_+ \equiv   \frac{\vec{\partial}}{\partial \theta^+} - i\theta^+ \partial_{++} \ \ \ , \ \ \ Q_- \equiv  \frac{\vec{\partial}}{\partial \theta^-}   - i \theta^- \partial_{--}  \ .
\end{equation}
They are among the most basic transformations that leave the measure
\begin{equation}
\int d \sig^{++} d \sig^{--} d \theta^+ d \theta^-
\end{equation}
invariant (other symmetries include the $d=2$ Lorentz transformations, (\ref{eq:2d_lorentz_trans_on_coord}) and (\ref{eq:2d_lorentz_trans_on_spinors}), as well as translations of the bosonic and fermionic coordinates). 
The supersymmetry algebra is given by 
\begin{equation}
\frac{1}{2} \{ Q_+, Q_+ \} = Q_+^2 = -i \dep  = P_{++} \ \ \ , \ \ \  Q_-^2 = -i \dep  = P_{--} \ \ \ , \ \ \ \{ Q_+, Q_- \} =0
\end{equation}
It forms an extension of the $d=2$ Poincar\'{e} algebra,
\begin{equation}
[ Q_+, J ] = \frac{i}{2} Q_+ \ \ \ , \ \ \  [Q_+, P_{++} ] = 0 \ \ \ , \  \ \ [P_{++}, J ] = iP_{++} \ ,
\end{equation}
and similarly in the $(-)$ sector, except that the terms containing $i$ have relative minus signs. The action of supersymmetry on the coordinates of superspace is given by
\begin{equation}
\delta \xp = -i \varepsilon^+ \theta^+  \ \ \ , \ \ \  \delta \theta^+ = \varepsilon^+ \ \ \ , \ \  \  \mathrm{and} \ + \rightarrow - \ ,
\end{equation}
where $\varepsilon^+$ and $\varepsilon^-$ are fermionic parameters, and on the fields by
\begin{equation}
\label{eq:1_1_supersymmetry}
\delta X^{i} = -i (\varepsilon^+ Q_{+} + \varepsilon^-  Q_{-}) X^{i} \ .
\end{equation}
In components this  reads:
\begin{align}
\label{eq:1_1_ssm_component}
&  \delta \phi^{i} = -i \varepsilon^{+} \psi_{+}^{i} - i \varepsilon^{-} \psi_{-}^{i} \ , \\ \nonumber
& \delta \psi_+^i = \varepsilon^+ \dep \phi^i + i \varepsilon^- F^i \ \ \  ,  \ \ \  
\delta \psi_-^i = \varepsilon^- \dem \phi^i - i \varepsilon^+ F^i \ , \\ \nonumber
& \delta F^{i} = - \varepsilon^+ \dep \psi_i^i + \varepsilon^- \dem \psi_+^i \ .
\end{align}

Superfields have the following important properties: 
\begin{enumerate}
\item
A product of any two superfields $\Phi^1$ and $\Phi^2$ is itself a superfield: 
\begin{equation}
\delta (\Phi^1 \Phi^2) = -i (\varepsilon^+ Q_{+} + \varepsilon^-  Q_{-}) (\Phi^1 \Phi^2) \ .
\end{equation}
\item
The integral over superspace of any supersymmetry variation vanishes:
\begin{equation}
\label{eq:inv_uner_ssm}
\int d \sig^{++} d \sig^{--} d \theta^+ d \theta^- Q_{\pm} (\Phi) = 0 \ .
\end{equation}
The contribution due to the fermionic derivative in (\ref{eq:Q_def}) vanishes due to (\ref{eq:ferm_int_by_parts}), while the contribution due to the bosonic derivative is a total derivative in the usual sense.
\end{enumerate}
The first property enables us to take products when constructing superfield Lagrangians, while from the second property we know that an action constructed by integrating the superfield Lagrangian over all coordinates of superspace will be manifestly invariant under supersymmetry.

To construct  supersymmetric actions containing derivatives one needs to construct differential operators   that commute with the supersymmetry  generators. These are the \emph{supercovariant derivatives},
\begin{equation}
D_+ \equiv   \frac{\vec{\partial}}{\partial \theta^+} + i\theta^+ \partial_{++} \ \ \ , \ \ \ D_- \equiv  \frac{\vec{\partial}}{\partial \theta^-}   + i \theta^- \partial_{--}  \ ,
\end{equation}
and have the following properties,
\begin{align}
& \{ Q_+, D_- \} = \{Q_+, D_+ \} = 0 \ \ \ \mathrm{and } \  + \leftrightarrow - \ , \\ \nonumber
& D_- D_- = i \partial_{--}  \ \ \ , \ \  \  D_+ D_+ = i \partial_{++} \ \ \ , \ \ \  \{ D_+ , D_- \} = 0 \ .
\end{align}

\section{The $(1,1)$  $\sigma$-model}
\label{section:supersymmetric_extensions}

The action of the non-linear  $(1,1)$ supersymmetric $\sigma$-model with torsion is
\begin{equation}
 \label{eq:ssm_action2}
S = \int d^2 z (G_{ij} + b_{ij}) D_+ X^i D_- X^j (z) \ ,
\end{equation}
where 
\begin{equation}
z := \{ \sig^{++},  \sig^{--} , \theta^+, \theta^-  \} \mathrm{ \ \ \  and \ \ \  } 
d^2 z = d \sig^{++} d \sig^{--} d\theta^+ d \theta^- \ .
\end{equation}
  In general I'll use lower-case letters from the end of the alphabet to stand for all the coordinates of superspace.

The equations of motion are
\begin{equation}
\frac{ \delta S} {\delta X^k(z)}   = -2 G_{kj} D_+ D_- X^j(z) - 2 ( \Gamma_{kij} - \frac{1}{2} H_{kij}) D_+ X^i D_- X^j (z) \ ,
\end{equation}
where $\Gamma^k_{ij}$ are the Christoffel symbols for the Levi-Civita connection (\ref{eq:LC_coefficients}), and $H_{kij}$ is the totally-antisymmetric torsion tensor (\ref{eq:torsion}),
\begin{equation}
H_{ijk} = (db)_{ijk} = b_{ij,k} + b_{jk.i} + b_{ki,j} \ ,
\end{equation}
which is obviously closed. These are written in a more covariant fashion as
\begin{equation}
\label{eq:covariant_1_1_eom}
\nabla^{(+)}_+ D_- X^i = \nabla^{(-)}_- D_+ X^i = 0 \ .
\end{equation}
The upper plus and minus stand for the two natural torsionfull connections on the target space whose connection coefficients are \cite{Gates:1984nk,Howe:1985pm}
\begin{equation}
\label{eq:connection_coefficients}
\Gamma^{k(\pm)}_{ \ ij} = \Gamma^k_{ \ ij} \pm \frac{1}{2} H^k_{\ ij} \ .
\end{equation}
The lower pluses and minuses in (\ref{eq:covariant_1_1_eom})  stand for the spin that the covariant derivative carries.

It is perhaps useful to see in detail how the equations of motion are obtained for $b_{ij}=0$:
\begin{align}
\label{eq:deriving_superfield_eom}
\frac{\delta S}{\delta X^k(y)} =&  G_{ij,k} D_{+} X^{i} D_{-} X^{j}(y)  \\ \nonumber
& + \int d^2 z \left( \underbrace{G_{ij} D_{-} (\delta^i_k) D_{+} X^{j}(z) }_{a_k} 
+ \underbrace{G_{ij} D_{-} X^{i} D_{+}( \delta^{j}_k) (z) }_{b_k}  \right) \ ,
\end{align}
where $\delta^i_j \equiv \delta^i_j \delta(z-y)$. Integration by parts is performed as 
\begin{equation}
\int d^2 z D(B) A = (-1)^{\epsilon_B+1} \int d^2 z B D (A) 
\end{equation}
for any two superfields $A$ and $B$ (see (\ref{eq:ferm_int_by_parts})), therefore
\begin{align}
& a_k = D_{-}(G_{ik}D_{+} X^{i}) (y)
= D_{-} X^{l} G_{ik,l} D_{+} X^{i}(y)
+ G_{ik} D_{-} D_{+} X^{i}(y)  \ , \\ \nonumber
& b_k = D_{+}(G_{ki} D_{-} X^{i})(y)
=  D_{+} X^{l} G_{ki,l} D_{-} X^{i}(y) 
+ G_{ki} D_{+} D_{-} X^{i}(y) \ .
\end{align}
The equations of motion then follow after relabeling a few indices. 

The component expansion of the action is obtained by integrating over the Grassmann variables,
\begin{equation}
\int d \theta^{\pm} = \frac{ \vec{\partial}}{\partial \theta^{\pm}} \simeq D_{\pm} \ ,
\end{equation}
where $\simeq$ means that the equality only holds up to a total derivative. The components of $X^i(z)$ can be projected out as follows: 
\begin{equation}
\phi^i = \Phi^{i} |_{\theta = 0} 
\mathrm{ \ \  \ , \ \ \ } \psi^i_{\pm} = D_{\pm} \Phi^{i} |_{\theta=0} 
\mathrm{ \ \ \ , \  \ \ } F_i = D_{+} D_{-} \Phi^{i} |_{\theta=0}  \ .
\end{equation}
In relation to the various functions of $X^i$ that feature in the manifestly supersymmetric action and the equations of motion, I note that any function $\mathcal{F}(X)$ of $X^i(z)$ has the expansion
\begin{align}
\label{eq:function_of_superfields_exp}
 \mathcal{F}(X) =& \mathcal{F}(\phi) + \theta^+ \mathcal{F(\phi)}_{,l} \psi_+^l + \theta^- \mathcal{F(\phi)}_{,l}\psi_-^l + \theta^+ \theta^-( \mathcal{F(\phi)}_{,lm} \psi^l_- \psi^m_+ + \mathcal{F(\phi)}_{,l} F^l ) \ ,
 \end{align}
which can easily be verified by Taylor expanding $\mathcal{F}(x)$. Unlike in a context of a superfield expressions, here the subscript  "$,l$" indicates a derivative with respect to $\phi^l$. For example, the component action with $b_{ij} = 0$ is given by
\begin{align}
S  = & \int d^2 \sig D_{+} D_{-} (G_{ij} D_{+} X^{i} D_{-} X^{j}) |_{\theta=0} \nonumber \\
& =  \int d^2 \sig \left( G_{ij} \dep \phi^{i} \dem \phi^{i} - iG_{ij}\nabla_{-}(\psi^{i}_{+}) \psi^{j}_{+} 
- iG_{ij}\nabla_{+}(\psi^{i}_{-}) \psi^{j}_{-}  \right. \nonumber \\
&  \left. + G_{ij} F^{i}F^{j}  + 2 \Gamma_{jik} \psi^{k}_{+}\psi^{i}_{-} F^j 
+ G_{ij,kl} \psi^i_- \psi^j_+ \psi^k_- \psi^l_+ \right) \ ,
\end{align}
where the $\nabla$s are the covariant derivatives living in ordinary $d=2$ flat space:
\begin{equation}
\label{eq:cov_derivative_on_fermion}
\nabla_{+}  \psi_-^i= \dep \psi^{i}_{-} + \Gamma^i_{jk} \dep \phi^j \psi_- ^k \ \ \  \mathrm{and} \ \ \ + \leftrightarrow - \ .
\end{equation}
They should not be confused with the covariant derivatives in  (\ref{eq:covariant_1_1_eom}) that live in superspace; again, the same notation is used since it's clear from the context which derivative is meant.

The $F^i$ fields can be eliminated using their equations of motion, $F^{i} = \Gamma^{i}_{ \ kl} \psi^{l}_{+} \psi^{k}_{-} $. Then the action reads
\begin{align}
S  = &  \int d^2 \sig \left(  \vphantom{\frac{1}{2}} G_{ij} \dep \phi^{i} \dem \phi^{i} - iG_{ij}\nabla_{-}(\psi^{i}_{+}) \psi^{j}_{+} 
- iG_{ij}\nabla_{+}(\psi^{i}_{-}) \psi^{j}_{-}  \right. \nonumber \\
& \left. + \frac{1}{2} R_{ijkl} \psi^i_- \psi^j_- \psi^k_+ \psi^l_+ \right) \ ,
\end{align}
where $R_{ijkl}$ is the Riemann tensor (\ref{eq:riemann_tensor_def_compon}).

In the case of non-zero torsion the action takes the form
\begin{align}
S  = &  \int d^2 \sig \left(  \vphantom{\frac{1}{2}} (G_{ij}+b_{ij}) \dep \phi^{i} \dem \phi^{i} - iG_{ij}\nabla_{-}^{(+)}(\psi^{i}_{+}) \psi^{j}_{+} 
- iG_{ij}\nabla_{+}^{(-)} (\psi^{i}_{-}) \psi^{j}_{-}  \right. \nonumber \\
& \left. + \frac{1}{2} R^{(-)}_{ijkl} \psi^i_- \psi^j_- \psi^k_+ \psi^l_+ \right)
\end{align}
after the elimination of the auxiliary $F^i$ fields. Now the curvature tensor is defined using the $(-)$ connection coefficients (\ref{eq:connection_coefficients}), explicitly:
\begin{align}
\label{eq:curvature_def}
R^{i (\pm)}_{ \ jkl} = \Gamma^{i (\pm)}_{ \ lj,k} -  \Gamma^{i (\pm)}_{ \ kj,l} + \Gamma^{m (\pm)}_{\ lj} \Gamma^{i(\pm)}_{\ km} 
-  \Gamma^{m (\pm)}_{\ kj} \Gamma^{i(\pm)}_{\ lm}  \ .
\end{align}

\section{Superconformal symmetries}
\label{section:sc_symmetries}

Action (\ref{eq:ssm_action2}) is conformally invariant, as well as invariant under worldsheet supersymmetry (\ref{eq:1_1_supersymmetry}). It is possible to combine these two symmetries into \emph{superconformal transformations},
\begin{equation}
\label{eq:G_transformation}
\delta_G X^i(z) \equiv  \int  d^2 y  a^G(y) R^i_G(z,y)  =  a^G \partial_{++} X^k(z) + \frac{i}{2} D_+ a^G D_+ X^k(z) \ ,
\end{equation}
where the parameter ghost $a^G = a^{G++}$ has odd parity.\footnote{The algebra will be calculated in the sense of the antifield formalism, and therefore I'm using ghost parameters. For a quick reference see Appendix \ref{app:ghost_commutators}.} The use of the letter $G$  is meant to remind that the metric is the covariantly constant tensor related to the superconformal transformations.
 (\ref{eq:G_transformation}) is a symmetry of the action (\ref{eq:ssm_action2}) if $D_- a^G = 0$:
\begin{align}
\label{sc_variation}
 \delta_G S \equiv  \int d^2 z  \ \frac{\delta S}{\delta X^k (z)} \delta_G X^i (z) =  \int d^2z \ D_- a^G T_G (z) \ .
\end{align}
The conserved current is given by
\begin{align}
\label{eq:superconformal_current_plus}
 T_{G}(z) & : = G_{kj} D_+ X^k \partial_{++} X^j (z) - \frac{i}{6} H_{kij} D_+ X^{kij} (z) \ ,
\end{align}
where $D_{+} X^{p_1 p_2 ...p_l} := D_{+} X^{p_1} D_+ X^{p_2} \cdots D_+ X^{p_l} $. 
Conservation is understood in the sense that $D_- T_{G} = 0$ on-shell:
\begin{align}
\label{eq:TG_conservation}
D_- (T_G) = \frac{1}{2} \frac{ \delta S} {\delta X^i} \dep X^l 
- \frac{i}{2} G_{kj} D_+ X^k D_+ \left( \frac{\delta S}{\delta X^l} G^{lj} \right) \ .
\end{align} 
In deriving this relation one has to make use the Bianchi identity
\begin{align}
R^{(-)}_{a[ijk]} \equiv -\frac{1}{3} \nabla^{(-)}_a H_{ijk}  \ .
\end{align}

The components of $T_G$ are
\begin{align}
T_{G}(z) |_{\theta = 0} =& G_{ij} \dep \phi^i \psi_+^j( \sig) - \frac{i}{6} H_{ijk} \psi^i_+ \psi^j_+ \psi^k_+( \sig) \ , \\ \nonumber
D_+T_{G}(z) |_{\theta = 0}  =&  G_{ij} \nabla_+ ( \psi_+^i) \psi_+^j ( \sig )+ i G_{ij} \dep \phi^i \dep \phi^j( \sig) \ .
\end{align}
The $\theta^-$ and $\theta^+ \theta^-$ components are proportional to the equations of motions due to (\ref{eq:TG_conservation}).  

An unrestricted ghost superfield $a^{G++}(z)$ is expanded as,
\begin{equation}
\label{eq:sc_ghost_expansion}
a^{G++}(z) = \wa^{G++}(\sig) + \theta^+ \wa^{G+}(\sig) + \theta^- \wa^{G+++}(\sig) + \theta^+ \theta^- \wA^{G++}(\sig) \ ,
\end{equation}
where I've used the tilde to distinguish the component ghosts from $a^{G++}(z)$. The condition $D_-  a^{G++}=0$ amounts to the requirement that
 \begin{equation}
 \label{eq:condition_D_-a}
 \wA^{G++} = \wa^{G+++} = \dem \wa^{G++} = \dem \wa^{G+} =0 \ .
 \end{equation}
For an unrestricted ghost superfield, in components the superconformal transformation (\ref{eq:G_transformation}) reads
\begin{align}
\delta_G X^i(z) = & \wa^{G++} \dep \phi^i + \frac{i}{2} \wa^{G+} \psi_+^i \\ \nonumber
& \theta^+ \left( \frac{1}{2} \wa^{G+} \dep \phi^i - \wa^{G++} \dep \psi_+^i - \frac{1}{2} \dep \wa^{G++} \psi_+^i \right) \\ \nonumber
& \theta^- \left( \frac{i}{2} \wa^{G+} F^i + \frac{i}{2} \wA^{G++} \psi^i_+ - \wa^{G++} \dep \psi_-^i + \wa^{G+++} \dep \phi^i \right) \\ \nonumber
& \theta^+ \theta^- \left( \frac{1}{2} \wA^{G++}\dep \phi^i + \wa^{G++} \dep F^i + \frac{1}{2} \dep \wa^{G++} F^i  \right.\\ \nonumber
& \left. \frac{1}{2} \wa^{G+}\dep \psi^i_- - \wa^{G+++} \dep \psi_+^i - \frac{1}{2} \dep \wa^{G+++} \psi_+^i \right) \ .
\end{align}
Taking (\ref{eq:condition_D_-a}) into account we can see that there are only two symmetry transformations. However, if we made the symmetry local by gauging the superconformal symmetry (similarly to what was done for the bosonic model in \ref{section:bosonic_string_theory}), then all four symmetry transformations would play a role. The symmetry parameterized by $\wa^{G+}$ is just the supersymmetry transformation (\ref{eq:1_1_ssm_component}) in components (except that here the parameter is a ghost and is of conformal type), and the one parameterized by $a^{G++}$ is the conformal transformation (\ref{eq:conf_trans}). 

The above is mirrored in the $(-)$ sector with the conserved current given by
\begin{equation}
\label{eq:superconformal_current_minus}
G_{kj} D_- X^k \partial_{--} X^j (z) + \frac{i}{6} H_{kij} D_- X^{kij} (z) \ .
\end{equation}

The commutator of two superconformal transformations in the $(+)$ sector is
\begin{equation}
\label{eq:GGcom}
[ \delta_G,  \delta_G ] X^i(z) = \int d^2 y  \left[  a^G \partial_{++}a^G + \frac{i}{4} D_+ a^G D_+ a^G   \right] (y) R^i_G(z,y) \ ,
\end{equation}
and similarly in the $(-)$ sector. Two superconformal transformations in opposite sectors commute if $D_- a^{G++} = D_+ a^{G--} = 0$. If the parameters are taken to depend on all coordinates of superspace, this is not the case.\footnote{This is unlike the case of the $(+)$ $(-)$ commutator of conformal transformations which is zero even when the transformation parameters are taken to depend on all worldsheet coordinates. To understand the algebra when the parameters are local one needs to work in the context of a $(1,1)$ model with the superconformal symmetries in both sectors gauged.} 


\section{An aside on the bi-Hamiltonian formulation of string theory}
\label{section:aside_on_bi_hamiltonian}

In \ref{section:bosonic_string_theory} the use of $d=2$ covariant momenta fields was very handy for  formulating the bosonic string. The helpful feature of action (\ref{eq:pi_action_special}) was that the conserved currents were only functions of momenta, and so the conformal symmetries could be gauged in closed form in flat space. Other bi-Hamiltonian actions still descend to (\ref{eq:bos_action}) after integrating out the momenta, but generally they can not be gauged in closed form even for flat target space. Ultimately one would like to know how to gauge a general bi-Hamiltonian action for a general target space? In this section I will attempt to give some insights about the worldsheet implications of the problem, and will assume that the target space is flat. 

General conformally invariant actions containing momenta $\pi_+$ and $\pi_-$ can be parameterized by four numbers $A$, $B$, $C$, and $D$,
\begin{equation}
\label{eq:pi_action_general}
S_{\pi} = \int d^2 \sig  \left(  A \sdep \phi  \sdem \phi+ B \pi_+ \pi_-  +C  \pi_+ \sdem \phi +D  \pi_-  \sdep \phi \right)  \ ,
\end{equation}
where I have dropped, for simplicity, the dependence on target space indices. In this space of actions we can identify three special classes of interest: 
\begin{enumerate}
\item Actions like (\ref{eq:pi_action_special}) for which the currents depend only on momenta; these have been discussed in \ref{section:bosonic_string_theory}. 
\item  Action that are zero after eliminating the momenta.
\item First-order actions, for which $A=0$.
\end{enumerate}

Let's analyze the last case first. Due to the action being first order we can hope to come up with analogues of the usual Hamiltonian constructions. There are two Poisson brackets for each of the momenta\footnote{Properties of the Poisson bracket are given in Appendix \ref{app:poisson_antibracket}.},
 \begin{align}
 \label{eq:biham_poisson_brackets}
& \{ A, B \}_{-} = \int d^2 \sig   \left(  \frac{\delta A}{\delta \phi} \frac{\delta B}{\delta \pi_+ }
 -  \frac{\delta A}{\delta \pi_+ } \frac{\delta B}{\delta \phi}  \right)  \ ,\\ \nonumber
& \{ A, B \}_{+} = \int d^2 \sig   \left( \frac{\delta A}{\delta \phi} \frac{\delta B}{\delta \pi_-}
 -  \frac{\delta A}{\delta \pi_-} \frac{\delta B}{\delta \phi} \right) \ ,
 \end{align}
where the sign on the bracket indicates the spin that it carries. 

Currents which generate the conformal algebra via the $(-)$ bracket are of the form
\begin{equation}
\label{eq:current_plus}
 \Tpl \equiv  \Omega_+ \Omega_+   \ \ \ , \ \ \    \Omega_+ :=  \frac{1}{\sqrt{2}} \sdep \phi + k \pi_+  \ ,
\end{equation}
 for some constant $k$, with an analogous expression for $ \Tmi$. The normalizing factor of $1/ \sqrt{2}$ was chosen for convenience. It is convenient to calculate the algebra using currents smeared by ghosts,
 \begin{equation}
 \int d^2 \sig  a^{+} \Tpl \ \ \ \mathrm{and} \ \ \   \int d^2 \sig   a^{-} \Tmi \ .
 \end{equation}
That is, the Poisson bracket
 \begin{equation}
 \label{eq:PBplpl}
 \{ \int d^2 \sig  a^{+} \Tpl ,  \int d^2 \sig  a^{+} \Tpl \}_{-} = \int d^2 \sig  4k ( a^{+} \sdep a^{+} ) \Tpl  \ 
 \end{equation}
gives the same result as the commutator of conformal transformation (\ref{eq:conformal_algebra}). Similarly, the $(+)$ bracket generates the algebra in the $(-)$ sector. From now on I will take $k=-1/\sqrt{2}$; with this choice the smeared current is given by
\begin{equation}
 \int d^2 \sig  a^{+} \Tpl = \int d^2 \sig \left( \frac{1}{2} \pi_+ \pi_+ - \sdep \phi \pi_+ + \frac{1}{2}  \sdep \phi \sdep \phi \right) \ .
\end{equation}

The action that is invariant under the symmetries generated by the Poisson brackets (\ref{eq:biham_poisson_brackets}) is given by
\begin{align}
\label{eq:first_order_mom_action}
S = \int d^2 \sig \left( \pi_+ \pi_+ + \pi_+ \sdem \phi + \pi_- \sdep \phi \right) \ ,
\end{align}
i.e., 
\begin{equation}
\label{eq:poisson_bracket_on_action}
\{ S,  \int d^2 \sig  a^{+} \Tpl \}_{-} = \{ S,  \int d^2 \sig  a^{-} \Tmi \}_{+} = 0 \ ,
\end{equation}
provided that $\sdem a^+ = \sdep a^- =0$

Unlike (\ref{eq:pi_action_special}) for which the conformal currents depend only on momenta, the $(+)$ and $(-)$ sectors don't commute in the above cases. This means that it is not possible to gauge the algebra in closed form, but that one must keep adding terms with currents coupled to higher and higher powers of momenta, just like in the usual second order formalism. I have not  yet understood the nature of this sum, but it should be noted that if one writes the master equation that describes the above conformal symmetries, after eliminating the momenta one obtains the exact same action as (\ref{eq:conf_spec_extended_no_pi}) with the gauge fields set to zero. 

Actions that are zero after eliminating the momenta are of the form
 \begin{align}
 S_{\top} & =  \int d^2 \sig \   \Omega_+ \Omega_-  \equiv \int d^2 \sig   \mathscr{L}_{\top} \ .  \\ \nonumber
 \end{align}
The conformal symmetry is
 \begin{equation}
 \delta_{(+)} \phi^k = \{ \phi^k , \int d^2 \sig  a^{+} \Tpl  \}_{-} \ \ \ ,  \ \ \   \delta_{(+)} \pi_{+} =  - \{\pi_{+} , \int d^2 \sig  a^{++} \Tpl \}_{-}  \ , \label{eq:twisted_sym}
 \end{equation}
 and $+ \leftrightarrow -$. 
The minus sign compared to the usual Hamiltonian relations is crucial, and one can't write (\ref{eq:poisson_bracket_on_action}). By conformal here I just mean that $\sdem a^{+} = 0$; the commutator of these transformations within one sector vanishes, and the algebra is not that of the usual conformal transformations. The commutator between $(+)$ and $(-)$ sectors is proportional to the equations of motion. 
 
The result of gauging this symmetry is expressed by the following minimal solution
 \begin{align}
 \label{eq:topo_gauged}
S_{\min}  = & \int d^2 \sig  \frac{1}{1- 4\hp \hm} \left[ \vphantom{\frac{1}{2}} \mathscr{L}_{\top}  +  \hp \Tpl \ +  \hp \Tmi  \right. \\ \nonumber 
& +  \phi^*(2 c^{+}   +  4 c^{-} \hp)   \dempi
+  \phi^*   (2 c^{-}  + 4 c^{+} \hm)   \deppi \\ \nonumber
& +  \sdep \pi^{+}_*  \left( 2 c^{+} \dempi  +  4 c^{+} \hm \deppi \right)  +  \sdem \pi^{-}_*  \left( 2 c^{-} \deppi  +  4 c^{-} \hp \dempi \right) \\ \nonumber
 & + 4 c^{-} c^{+}  \left(  \sdep \pi^{+}_* \phi^*   -  \sdem \pi^{-}_* \phi^*  +  \sdep \pi^{+}_* \sdem \pi^{-}_* \right)  \\ \nonumber
 & \left. - h^*_{++} \sdem c^{+}   - h^*_{--} \sdep c^{-} \vphantom{\frac{1}{2}} \right] \ ,
\end{align}
from which one can also read off the commutation relations. The coupling in the part not involving antifields is that of an inverse worldsheet metric parameterized by
\begin{equation}
\gamma^{\mu \nu} = 
\left( \begin{array}{cc}
\hp & \frac{-1}{2} \\
\frac{-1}{2} & \hm
\end{array} \right)  \ .
\end{equation}
However, this metric is not positive definite, and we don't have the usual $\sqrt{\gamma}$ in the gauged action, but rather $\gamma$. Interestingly, the term proportional to $c^+ c^-$ is the same as in (\ref{eq:conf_spec_extended_no_pi}).  Integrating out the momenta sets 
\begin{equation}
S_{\min} =  - h^*_{++} \sdem c^{+}   - h^*_{--} \sdep c^{-} \ .
\end{equation}

What can one make of all this?  The action $S_{\top}$ is just a way of writing zero, and as such it seems like a starting point for a topological theory. Integrating out momenta for the gauged action doesn't quite yield zero, but a transformation of a general worldsheet metric, which is the starting point for the formulation of topological gravity \cite{Labastida:1988zb, Montano:1988dr} (see also \cite{Henneaux:1989ya}). This indicates that perhaps the entire action is describing some type of a topological $\sigma$-model coupled to topological gravity. One way to give a non-trivial interpretation to  (\ref{eq:topo_gauged}) is to treat the momenta antifields as quantum fields, and the momenta themselves as background fields. In this case the quantum fields have the wrong spin statistics, but this is a general feature of topological $\sigma$-models \cite{Witten:1988xj}.

Whether this gives an interesting theory remains to be investigated. The interesting topological $\sigma$-models are supersymmetric ones with Calabi-Yau target spaces. However, I stress that the type of twisting involved here is not like that in Witten's A and B models. For a start, one can make a comparison with the topological twisting of the bosonic $\sigma$-model described in \cite{Witten:1988xj} and  see that the theories are not the same. 

In the case of $(2,2)$  supersymmetric models written in $(1,1)$ superfields we have the following topological actions,
\begin{equation}
\label{eq:ques_a_model}
S_{A} = \int d^2 z I^{ij} \Omega_{+i} \Omega_{-j}    \ ,
\end{equation}
and
\begin{equation}
\label{eq:ques_b_model}
S_{B} = \int d^2 z G^{ij} \Omega_{+i} \Omega_{-j}  \ ,
\end{equation}
where now
\begin{equation}
 \Omega_{+i} = \pi_{+i} + I_{ik} D_{+} X^{k}  \  \ \ , \ \ \   \Omega_{-i} = \pi_{-i} + I_{ik} D_{-} X^{k} \ ,
\end{equation}
and $I^i_{\ j}$ is the complex structure (I am jumping the gun here a bit; see \ref{section:kahler_algebra}, and also \ref{subsec:almost_cpx_structures}).\footnote{I note at this point that in $(1,1)$ bi-Hamiltonian models the natural bracket is the antibracket rather than the Poisson bracket (see Appendix \ref{app:poisson_antibracket}),  since the momenta have the opposite parity to the fields, 
\begin{equation}
(A , B)_- :=  \int d^2 z \left( \frac{ \dr A} { \delta   X^{i}(z) } \frac{\dl B}{ \delta \pi_{+i}(z) } -
\frac{\dr A}{ \delta \pi_{+i}(z) }  \frac{ \dl B} { \delta   X^{i}(z) }  \right) \ .
\end{equation} 
The superconformal algebra is generated by
\begin{equation}
\int d^2 z a^{++} G^{ij} D_+ (\Omega_{+i}) \Omega_{+j} \ \ .
\end{equation} 
 } Although (\ref{eq:ques_a_model}) depends only on the symplectic while (\ref{eq:ques_b_model}) depends only on the complex structure, one can also consider any combination of the two actions, which is not possible for the A and B models. There is a topological theory that does share these properties. Other than the two more commonly known twistings of $N=4$ super Yang-Mills \cite{Yamron:1988qc, Vafa:1994tf}, there exists a third \cite{Marcus:1995mq} that has a whole space of actions associated with it, with two special cases akin to the A and B models (see \cite{Witten:langlands} for a full explanation). The topological  models above might be of that nature. 

To summarize, the most general locally invariant action containing momenta lies somewhere between the gauged topological action and  (\ref{eq:conf_spec_extended_no_pi}). For both the bosonic case and the supersymmetric case, one can speculate that the general gauged bi-Hamiltonian action describes a coupling of topological models with string theory. In such cases momenta should not be integrated out, but rather treated as antifields that describe a gauge symmetry acting on the quantum fields $\phi$, $\pi_*$, and $c$.  If one gauges any action containing momenta in a nontrivial background there is no simple way to integrate out the momenta from the extended action anyway, unless one sets $\pi^*$ to zero (the can be seen, for example, by constructing the first few additional terms in (\ref{eq:conf_spec_extended}) when the metric is not flat).    


\chapter{The Algebra of Symmetry Transformations Related to Covariantly Constant Forms}
\label{chapter:algebras_of_general_L_symmetries}


\section{Geometry and symmetries}
\label{section:geometry_and_symmetries}

In this chapter I will discuss, in a general setting, the interplay between target space geometry and the symmetries of $(1,1)$ supersymmetric $\sigma$-models. In the last chapter it was shown that the action
\begin{equation}
\label{eq:ssm_action_tmp3}
S = \int d^2 z (G_{ij} + b_{ij}) D_+ X^i D_- X^j (z)  
\end{equation}
 is invariant under superconformal symmetries (\ref{eq:G_transformation}). In this chapter I'll be concerned with the algebra of symmetry transformations of (\ref{eq:ssm_action_tmp3}) related to the existence of covariantly constant forms in the target space. Unlike the superconformal symmetry, which classically implies no constraints on the target space geometry, the existence of covariantly constant tensors does constrain the geometry. In the torsionless cases we have the Berger classification of special holonomy manifolds (see \ref{section:spechol_manifolds}). When torsion is non-zero there is no Berger classification, but  one can still consider the cases with the $G$-structures as those on the Berger list (see the end of \ref{subsection:connections_holonomy_groups}). All these cases will be discussed systematically in the next chapter.
 
The generator of a symmetry transformation corresponding to a covariantly constant form $L_{p_1 ... p_l} dx^{p_1} \wedge \cdots \wedge dx^{p_l}$ is
\begin{equation}
\label{eq:L_transformation}
\delta_{L} X^i (z)= a^{L} G^{i p_1}L_{p_1 p_2 ... p_l} D_{+} X^{p_2...p_l}(z) \ ,
\end{equation}
where $D_{+} X^{p_2 ... p_l} := D_+ X^{p_2} \cdots D_+ X^{p_l} $.\footnote{In this chapter I'll be using ghostly parameters for the symmetries, and calculating the algebra in the sense of the antifield formalism. For a quick reference to what this involves the reader is referred to Appendix \ref{app:ghost_commutators}.}  A similar expression exists in the $(-)$ sector.  In general, I will use the notation where $\{ l, m, n, \cdots \}$ will stand for the rank of the covariantly constant form $\{ L, M, N, \cdots \} $. The parity of the parameter ghost  $a^{L}$  is odd if the rank of $L$ is odd, and even otherwise. Its spin ensures that the transformation of $X^i$ is a scalar, i.e.  $a^L = a^{L+(l-1)}$, where the notation indicates that the spin is $\frac{l-1}{2}$ (see also \ref{section:2d_spinors}). 

Action (\ref{eq:ssm_action_tmp3}) is invariant under (\ref{eq:L_transformation}) if $L_{i_1 ... i_l}$ is covariantly constant with respect to the $(-)$ connection and $D_- a^{L} = 0$:
\begin{align}
 \delta_L S  =&    \frac{2}{l} (-1)^{l+1} D_- a^{L} L_{ p_1 ... p_l} D_{+} X^{p_1...p_l}(z) -   \frac{2}{l} a^{L} D_-X^j( \nabla^{(-)}_j L_{p_1 ... p_l} ) D_{+} X^{p_1...p_l}(z) \ .
\end{align}
The constraint on the parameter indicates that the symmetry is of superconformal type. As in the superconformal case (see discussion around (\ref{eq:sc_ghost_expansion})), the condition $D_- a^L = 0$ implies that there are only two non-vanishing components, corresponding to a supersymmetry and a conformal-type symmetry. In the $(-)$ sector $L$ has to be covariantly constant with respect to the $(+)$ connection.

The corresponding conserved current is
\begin{equation}
\label{eq:conserved_L_current}
T_L(z) := L_{ p_1 ... p_l} D_{+} X^{p_1...p_l}(z) \ .
\end{equation}
As consistency demands,  $D_- T_L $ vanishes on-shell only if $\nabla^{(-)} L =0$:
 \begin{equation}
 \label{eq:Dminus_TL}
 D_- (L_{i_1 ... i_l} D_+ X^{i_1 ... i_l} )= \nabla^{(-)}_a L_{i_1 ... i_l} D_- X^a D_+X^{i_1 ... i_l} + \frac{l}{2} \frac{\delta S}{\delta X^a} L^a_{ \ i_2 ... i_l} D_+ X^{i_2 ... i_l} \ .
 \end{equation} 
The components of the current are given by
\begin{align}
T_+(z) |_{\theta=0} & = L_{i_1 ... i_l} \psi_+^{i_1 ... i_l} \ , \\ \nonumber
D_+ T_+(z) |_{\theta = 0} & = i l L_{i_1 ... i_l} \dep \phi^{i_1} \psi_+^{i_2 ... i_l} + L_{i_1 ... i_l} \psi_+^{ i_1 ... i_l} \ ,
\end{align}
where $\psi_+^{i_1 ... i_n} := \psi_+^{i_1} ...  \psi_+^{i_n}$. The $\theta^-$ and $\theta^+ \theta^-$ components are zero on-shell due to (\ref{eq:Dminus_TL}). 
 
 A transformation of the form (\ref{eq:L_transformation}), but  involving a covariantly constant tensor that is not obtained by raising an index of an $l$-form, is not generally a symmetry. 

\section{Commutation relations for general $L$-type symmetries}

A superconformal transformation in the $(+)$ sector and an $L$-type transformation in the $(-)$ sector commute, and vice versa, wheras two $L$-type transformations in different sectors commute only up to the equations of motion:
\begin{align}
& [ \delta_{L^{+}},  \delta_{M^-} ]  X^i(z) = \\ \nonumber
& (-1)^{m+l}  \frac{1}{2}(m-1)(l-1) a^{M(-)} a^{L(+)} \left[ \ M^{i(-)}_{ \ k j_3 ... j_m} L^{ks (+)}_{ \ \ i_3 ... i_l} \right. \\ \nonumber
& \left.  - L^{i(+)}_{ \ k i_3 ...i_l} M^{ks(-)}_{ \ \ j_3 ... j_m} \ \right]  D_+ X^{i_3 ... i_l} D_- X^{j_3 ... j_m} (z)  \frac{\delta S}{\delta X^s(z)}  \ .
\end{align}
Similarly to the commutator of superconformal transformations in different sectors (see the end of \ref{section:sc_symmetries}), to close the algebra it must be assumed that the parameter ghosts depend only on half the coordinates of superspace. The algebra of local transformations can only be understood in the context of an action obtained by gauging $L$-type symmetries in both $(+)$ and $(-)$ sectors, which brings us into the realm of non-chiral $W$-strings (I will not discuss these; see Chapter \ref{chapter:W-strings} for a discussion of \emph{chiral} $W$-strings).

This exhausts the commutators between two sectors. In what follows I will calculate the commutators within the $(+)$ sector, implying that analogous expressions hold in the $(-)$ sector.

The commutator of a superconformal transformation and an $L$-type transformation closes to an $L$-type transformation:
\begin{align}
& [ \delta _G,   \delta_ L ] X^i(z)=  \\ \nonumber
& \int dy  \left[ \frac{1}{2} (l-1)\partial_{++} a^G  a^L + a^G \partial_{++} a^L + \frac{i}{2} D_+ a^G D_+ a^L \right](y) R^i_L(z,y) \ .
\end{align}

Evaluating the commutator of two $L$-type transformations within the $(+)$ sector yields
\begin{equation}
\label{eq:L_commutator}
[ \delta_L,  \delta_M ] X^i(z)= (\delta_{1} + \delta_{2} + \delta_{3})X^i(z) \ ,
\end{equation}
where
\begin{align}
 \label{eq:com_algebraic}
\delta_{1}X^i(z)  = &   (-1)^{lm} \left[  (l-1) a^L D_+ a^M  L^i_{ \ k i_3 ... i_l} M^k_{ \ j_2 ... j_m} 
 D_+ X^{i_3 ... i_l j_2 ... j_m} (z)  \right. \\ \nonumber
&+ \left. (m-1) a^M D_+ a^L \  M^i_{ \ k i_3... i_m} L^k_{ \  j_2 ... j_l} D_+ X^{i_3 ... i_m j_2 ... j_l} (z)  \right]  \ ,
\end{align}
\begin{align}
\label{eq:com_withdel}
 \delta_{2}X^i(z) =&    (-1)^{l + m(1+l)} (l-1)(m-1)  i   a^L a^M \left[    
L^i_{ \ k i_3 ... i_l} M^k_{ \ p j_3 ... j_m} \right. \\ \nonumber
&+ \left. M^i_{ \ k j_3 ... j_m} L^k_{ \ p i_3 ... i_l}  \right]  \partial_{++} X^p D_{+} X^{i_3...i_l j_3...j_m}  (z) \ ,
\end{align}
and
\begin{align}
\label{eq:com_nijenhuis}
\delta_{3}X^i(z)= &   (-1)^{m(l-1)} a^L a^M \mathcal{N}^i_{ \ i_2 ... i_l j_2 ... j_m} D_+ X^{i_2...i_lj_2...j_m} (z) \ ,
\end{align}
where
\begin{align}
\label{eq:com_nijenhuis2}
\mathcal{N}^i_{ \ i_2 ... i_l j_2 ... j_m} :=  &   L^i_{  \ [i_2 ... i_l,|k|} M^k_{ \ j_2 ... j_m]}  - M^i_{  \ [j_2 ... j_m, |k|} L^k_{ \ i_2 ... i_l]} \\ \nonumber
& +   (l-1) L^i_{ \ k [ i_3...i_l} M^k_{ \ j_2...j_m,i_2]}  - (m-1)  M^i_{  \ k[ j_3 ... j_m} L^k_{ \  i_2 ... i_l, j_2]}  \ .
\end{align}

The transformations $\delta_1$ and $\delta_2$ involve covariantly constant tensors constructed by combining $L$ and $M$ algebraically. Despite involving partial derivatives $\mathcal{N}$ is also a tensor, which can be seen by rewriting it in terms of torsion, using the covariant constancy of $L$ and $M$ with respect to the $(-)$ connection:
\begin{align}
\label{eq:com_nijenhuis3}
\delta_{3}X^i(z)= &  (-1)^{m(l-1)} a^L a^M \left[  \vphantom{\frac{1}{2}}  -H^i_{ \ ab} M^a_{ \ j_2 ... j_m} L^b_{ \ i_2 ... i_l}  \right. \\ \nonumber
& + (m-1) H^a_{ \ j_2 b} M^i_{ \ a j_3 .. j_m } L^b_{ \ i_2 ... i_l} - (l-1) H^a_{ \ i_2 b} L^i_{ \ a i_3 ... i_l} M^a_{ \ j_2 ... j_m}  \\ \nonumber
& + \frac{1}{2} (l-1)(m-1) \left( H^a_{ \ i_2 j_2} L^i_{ \ k i_3 ... i_l} M^k_{ \ a j_3 ... j_m} + H^a_{ \ i_2 j_2} M^i_{ \ k j_3 ... j_m} L^k_{ \ a i_3 ... i_l} \right)  \\ \nonumber
&\left.  \vphantom{\frac{1}{2}}   \right] D_+ X^{i_2...i_lj_2...j_m} (z) \ .
\end{align}
The tensors involved in each term are clearly covariantly constant. In the case when $T = M = I $ is  an almost complex structure, (\ref{eq:com_nijenhuis2}) is the Nijenhuis tensor, the vanishing of which implies that the almost complex structure is in fact complex. However, (\ref{eq:com_nijenhuis2}) is not the correct generalization of the Nijenhuis tensor, in the sense that it is not covariantly constant. 

In general none of the above transformations are symmetries by themselves. To disentangle what the symmetry transformations are, I will first figure out what the conserved currents are and then reconstruct the symmetry transformations from these. Varying the action with the commutator one obtains that
\begin{align}
\label{eq:com_variation}
& \int d^2 z  \frac{\delta S}{\delta X^k (z)} [\delta_L,  \delta_M ] X^i (z)  =\\ \nonumber
& \int d^2 z  \left[  (  D_- ( (-1)^{m+1} D_+a^M a^L + a^M D_+ a^L \ )  T_P(z) \right. \\ \nonumber
&   \ \ \ \ \ \ + (-1)^{l+1} D_- (a^M a^L)  T_D(x) + \left.  i D_- (a^M a^L) T_{GU}(z)   \right] \ ,
\end{align}
given that
\begin{equation}
D_- (T_P(z) + T_D(z) + T_{\mathrm{del}}(z) ) = 0
\end{equation}
on-shell. Defining these objects,
\begin{equation}
\label{eq:TP}
T_P := P_{i_2 ... i_l j_2 ...j_m} D_+ X^{i_2 ... i_l j_2 ... j_m} (z) \ ,
\end{equation}
where the $(l+m-2)$-form $P$  is given by
\begin{equation}
P_{[i_2 ... i_l j_2 ...j_m]} = L_{ a [ i_2 ... i_l} M^a_{ \ j_2 ... j_m ] }  \ .
\end{equation}
$T_D$  involves the exterior derivatives of $L$ and $M$,
\begin{align}
\label{eq:TD}
T_D := & D_{p i_2 ... i_l j_2 ... j_m}  D_{+}X^{p i_2 ...i_l j_2 ... j_m}(z) \ ,
\end{align}
where the $(l+m-1)$-form $D$ is given by
\begin{align}
\label{eq:D_tensor}
D_{[p i_2 ... i_l j_2 ... j_m]} = \frac{1}{l} (dL)_{[p i_2 ... i_l |b|} M^b_{  \ j_2 ... j_m]} - \frac{1}{m} (dM)_{[p j_2 ... j_m |b|} L^b_{ \  i_2 ... i_l]}  \ .
\end{align}
$T_{\mathrm{del}}$ is given by
\begin{align}
\label{eq:Tdel}
T_{\mathrm{del}} : = & (-1)^{l+1} \left[  (l-1) L^a_{ \ i_3 ... i_l p } M_{ r a j_3 ... j_m} \right.  \\ \nonumber
& \left. +  \ (m-1) L^a_{ \ i_3 ... i_l r} M_{ p a j_3 ... j_m}  \right] \partial_{++} X^p D_+ X^r D_+X^{i_3 ... i_l j_3 .. j_m} (z) \ .
\end{align}

$P$ is covariantly constant since $L$ and $M$ are, and $T_P$ is clearly a conserved current due to an $L$-type symmetry transformation. However, the other two terms are not separately conserved. Writing $T_D$ in terms of torsion one obtains:
\begin{align}
\label{eq:TD2}
T_D  = &  \left[ 2 H_{abp} L^a_{ \ i_2 ... i_l} M^b_{ \ j_2 ... j_m} - \frac{1}{2} (m-1) H^a_{ \ j_2 p } L_{b i_2 ... i_l} M^b_{ \ a j_3 ... j_m} \right. \\ \nonumber
& \left. + \frac{1}{2} (l-1) H^a_{ \ i_2 p} L_{ba i_3 ... i_l} M^b_{ \ j_2 ... j_m}   \right] D_+ X^{p i_2 ... i_l j_2 ... j_m}(z) \ .
\end{align}

By examining $T_{\mathrm{del}}$ one can see that $D$ can not in general be a covariantly constant form. If we write (\ref{eq:Tdel}) as a sum terms, one symmetric and one antisymmetric in the $p$ and $r$ indices, we can see that the antisymmetric term is equal to $P$. Integrating by parts:
\begin{align}
\label{eq:Tdel_int_by_parts}
&  \int d^2 z i D_- (a^M a^L) (-1)^{l+1}\left[  (l-m) L^a_{ \ i_3 ... i_l [ p } M_{ r ] a j_3 ... j_m}    \right] \partial_{++} X^p  D_+X^{ri_3 ... i_l j_3 .. j_m} (z) \\ \nonumber
& =  \int d^2 z (-1)^{m+1} D_- D_+ (a^M a^L) \left[ \frac{l-m}{l+m-2} P_{i_2 ... i_l j_2 ... j_m} \right]  D_+ X^{i_2 ... i_l j_2 ... j_m}(z) \\ \nonumber
& + \int d^2 z (-1)^{l+1} D_-(a^M a^L) \left[ \frac{l-m}{l+m-2} dP_{[ i_2 ... i_l j_2 ... j_m s ]} \right] D_+ X^{s i_2 ... i_l j_2 ... j_m} (z) \ .
\end{align}
The first term will combine with the term proportional to $T_P$ in (\ref{eq:com_variation}), and I'll show shortly that the second term will be cancelled by a part of $T_D$. 

One would expect the part of $T_{\mathrm{del}}$ symmetric in $p$ and $r$ to involve the metric,
\begin{align}
\label{eq:TGU}
& T_{G^0 U}  := (l+m -2) L^a_{ \ i_3 ... i_l ( p } M_{ r ) a j_3 ... j_m}  \partial_{++} X^p  D_+X^{ri_3 ... i_l j_3 .. j_m} (z)  \\ \nonumber
& ? = G_{pr} \partial_{++} X^p D_+ X^r  U_{i_3 ... i_l j_3 ... j_m} D_+ X^{i_3 ... i_l j_3 ... j_m}(z) = T_{G^0} T_{U}  \ ,
\end{align}
where $U$ is some covariantly constant form of rank $(l+m - 4)$. The relation 
between the tensors on the two sides of the above equation, 
\begin{align}
\label{eq:TGU_relation}
 (l+m -2) L^a_{ \ [i_3 ... i_l | p |} M_{ r  a j_3 ... j_m]} + (r \leftrightarrow p) = G_{p[r} U_{i_3 ... i_l j_3 ... j_m]} \ ,
\end{align}
can be understood as a generalization of the condition $I^i_{ \ j } I_{i k } = G_{j k }$, which requires that the metric on an (almost) complex manifold be (almost) Hermitian (see \ref{subsection:hermitian_manfiolds}).
In order to convince oneself that indeed $T_{G^0 U} = T_{G^0} T_{U} $,  it's enough to notice that the  expression for the conservation of $T_{G^0 U}$ is
\begin{align}
D_-  (T_{GU})   = &  -\frac{i}{2}   R^{b (-)}_{ \ fvs} \left[    (m-1) L^a_{ \ r i_3 ... i_l (r} M_{b) a j_3 ... j_m} \right. \\ \nonumber 
& \left.  + (l-1) L^a_{ \  i_3 ... i_l (b} M_{r) a j_3 ... j_m}  \right] D_- X^f D_+ X^{srv i_3 ... i_l j_3 ... j_m} \ ,
\end{align}
evaluated on-shell, where  $R^{b (-)}_{ \ fvs}$ is the curvature tensor of the $(-)$ connection (\ref{eq:curvature_def}). In the absence of torsion $T_{GU}$ must be conserved, since $T_D$ vanishes and $T_P$ is conserved. It is then clear that the $b$ index on the Riemann tensor must be lowered and contracted with a $D_+ X^i$, since then  the above expression vanishes due to $R^i_{ \ [ jkl ] } = 0$. This can only be true if (\ref{eq:TGU_relation}) holds. Since  (\ref{eq:TGU_relation}) is an algebraic relation it must remain valid in the presence of torsion, so indeed we have $T_{G^0 U} = T_{G^0} T_{U} $. In the presence of torsion, by using either 
\begin{equation}
\label{eq:cov_derivative_of_torsion}
R^{(-)}_{ \ a [ jkl ] } = \frac{1}{3} \nabla^{-}_a H_{jkl} \ ,
\end{equation}
or
\begin{equation}
D_- T_{G^0} = \frac{i}{6}\nabla^{(-)}_f ( H_{srv} ) D_- X^f D_+ X^{srv} \ ,
\end{equation}
we see that
\begin{align}
D_-  (T_{G^0 U}) = \frac{i}{6}   \nabla^{(-)}_f ( H_{srv}  U_{i_3 .. i_l j_3 ... j_m } )  D_- X^f D_+ X^{srv i_3 .. i_l j_3 ... j_m} \ ,
\end{align}
so $T_G T_U$ is indeed the conserved current. 

Therefore, to obtain the conserved current $T_D$ must be modified by both $H \wedge U$ and the term proportional to $dP$ in (\ref{eq:Tdel_int_by_parts}). Rewriting $T_D$ by using (\ref{eq:TGU_relation}) in terms of $U$ and $P$,
\begin{align}
\label{eq:TD3}
T_D  = &  \left[  2 H_{abp} L^a_{ \ i_2 ... i_l} M^b_{ \ j_2 ... j_m} - \frac{l-m}{l+m-2} (dP)_{i_2 ...i_l j_2 ... j_m p}  \right. \\ \nonumber
& \left. \ \ \  - \frac{1}{2} H_{p i_2 j_2} U_{i_3...i_lj_3...j_m}  \right] D_+ X^{p i_2 ... i_l j_2 ... j_m}(z) \ ,
\end{align}
we see that the $dP$ parts cancel and that the conserved current is
\begin{equation}
T_N = N_{p i_2...i_l j_2...j_m}   D_+ X^{p i_2 ... i_l j_2 ... j_m}(z) \ .
\end{equation}
The covariantly constant $(l+m-1)$-form $N$, which I'll refer to as the \emph{generalized Nijenhuis form}, is given by:
\begin{equation}
\label{eq:N}
N_{[p i_2...i_l j_2...j_m]} := 2 H_{ab[p} L^a_{ \ i_2 ... i_l} M^b_{ \ j_2 ... j_m]} - \frac{1}{3} H_{[p i_2 j_2} U_{i_3 ... i_l j_3 ... j_m]} \ .
\end{equation}
The covariant constancy can be checked explicitly as follows. Taking the covariant derivative of the first term, using (\ref{eq:cov_derivative_of_torsion}), and writing down the antisymmetrization in $\{ a,b,p \}$,  we obtain two terms that vanish due to (\ref{eq:general_holonomy_expression}), and a third term that can be rewritten using (\ref{eq:TGU_relation}) and cancels the covariant derivative of the second term in (\ref{eq:N}).

At this point we have figured out that the commutator of two $L$-type symmetries  (\ref{eq:L_transformation}) closes to three symmetry transformations whose conserved currents are $T_{GU}$, $T_P$, and $T_N$, and that  (\ref{eq:com_variation}) can be rewritten in terms of these as
\begin{align}
\label{eq:com_variation2}
& \int d^2 z  \ \frac{\delta S}{\delta X^k (z)} [\delta_L, \delta_M ] X^i (z)  =\\ \nonumber
& \int d^2 z \left[  \left(  D_- ( (-1)^{m+1} D_+a^M a^L + a^M D_+ a^L  + (-1)^{m+1} \frac{l-m}{p} D_{+}(a^M a^L)  \right)   T_P(z) \right. \\ \nonumber
&  \ \ \ \ \  + (-1)^{l+1} D_- (a^M a^L)  T_N(z) + \left.  i D_- (a^M a^L) T_{GU}(z) \vphantom{\frac{1}{2}}   \right] \ .
\end{align}

Since it is clear that  the conservation of $T_N$ and $T_P$ is related to $L$-type symmetries constructed out of the forms $N$ and $P$, the only question left is what kind of symmetry transformation has $T_{GU}$ as its conserved current. In the absence of torsion the transformation
\begin{align}
\label{eq:GU}
\delta_{G^0 U} X^i(z) := -u \int d^2 y  [ a^{G U} T_{G^0} ](y)  R^i_U(z,y) +2 \int d^2 y   [ a^{G U} T_U ] (y)   R^i_{G}(z,y)
\end{align}
is a symmetry of the action (\ref{eq:ssm_action2}), with the conserved current $T_{G^0} T_U$. The second transformation on the right side is a superconformal transformation (\ref{eq:G_transformation}) parameterized by $a^{G U} T_U$, while the first is an $L$-type symmetry (\ref{eq:L_transformation}) obtained from the form $U$, parameterized by  $a^{G U} T_{G^0}$. Neither are symmetry transformations separately since $D_-T_U$ and $D_-T_{G}$ vanish only on-shell, but in the absence of torsion they form a symmetry together. This is no longer true in the presence of torsion. Instead,
\begin{align}
\delta_{G^0 U} S = (-1)^u \int d^2z   \left[ \frac{i}{3}   a^{GU} D_- T_U H_{kij} D_+ X^{kij}(z) +  2 D_- a^{GU} T_{G} T_U(z) \right] \ .
\end{align}
Since $T_U$ is conserved, $D_- T_U$ is proportional to the equations of motion (see (\ref{eq:Dminus_TL})), so one can read off the transformation that will compensate for the unwanted first term in the above expression. It follows that the symmetry transformation having $T_{GU}$  as its conserved current is
\begin{equation}
\label{eq:GUprime}
\delta_{GU} X^i(z) =  \delta_{G^0 U} X^i(z) -  u \frac{ i}{6}  \ a^{GU}  U^i_{ \ i_2 ... i_u} H_{kjm} D_{+} X^{kjm i_2 ... i_u }(z) \ .
\end{equation}

In summary, the commutator of two $L$-type symmetries closes to three symmetry transformations, $\delta_P$, $\delta_N$, and $\delta_{G^0 U}$,  that have, respectively, the conserved currents $T_P$, $T_N$, and $T_{GU}$. Thus, (\ref{eq:L_commutator}) can be rewritten as:
\begin{align}
\label{eq:L_commutator2}
& [ \delta_L,  \delta_M ] X^i(z)= \\ \nonumber
& \frac{p}{2} (-1)^{l+m}  \int d^2 y  \left[ a^M D_+ a^L + (-1)^{m+1} D_+ a^M a^L  + (-1)^{m+1} \frac{l-m}{p} D_{+}(a^M a^L)  \right] (y)  R^i_P(z,y)  \\ \nonumber
&+  \frac{n}{2} (-1)^{ml} \int d^2y  [a^M a^L ](y)   R^i_N(z,y) +(-1)^{u+1} \frac{i}{2} \int d^2y  [a^M a^L ](y)  R^i_{GU}(z,y) \ .
\end{align}

\section{The algebra of $L$-type transformation via antibrackets}
\label{sec:algebras_via_antibrackets}

The superconformal and $L$-type symmetries can be generated using the $(-)$ antibracket,
\begin{equation}
\label{eq:min_antibracket}
(A , B)_- : = \int d^2 z  \left\{ \frac{ \dr A} { \delta   X^{i}(z) } \frac{\dl B}{ \delta \pi_{+i}(z) } -
\frac{\dl A}{ \delta \pi_{+i}(z) }  \frac{ \dr B} { \delta   X^{i} (z) } \right\} \ ,
\end{equation}
where 
\begin{equation}
\label{eq:momentum_def_replace}
\pi_{+i}(z) \equiv G_{ij} D_+X^j (z) \ .
\end{equation}
The $-$ subscript indicates the spin, and is also used to distinguish it from the BV antibracket (\ref{eq:antibracket_def}). One can think of (\ref{eq:min_antibracket}) as the first step in a Hamiltonian formalism for the action (\ref{eq:ssm_action_tmp3}) without torsion, where $D_-$  is chosen as the 'time' derivative, and momenta are given by $G_{ij} D_+X^j$. The important supersymmetric  feature is that the momenta corresponding to $X^i$ are fermionic, since $D_-$ is a fermionic derivative. If the smeared currents involve integration over all the coordinates of superspace the relevant bracket in the Hamiltonian formalism is the antibracket and not the Poisson bracket. Alternatively, one can define the smeared currents by integrating only over the bosonic coordinates and $\theta^+$, in which case the relevant bracket would be a Poisson bracket. However, it is not possible to write commutators or the variation of an action naturally in terms of such a Poisson bracket, and in this sense choosing to work with (\ref{eq:min_antibracket}) comes closer to the usual Hamiltonian formalism. 
 
The main aim of this section is to facilitate the calculation of commutation relations in Chapter \ref{chapter:algebras_on_manifolds_of_sh}. For this purpose it will not be necessary to introduce momenta as separate fields. Rather, in all expressions in this section momenta will be substituted by their definition, as has been indicated in (\ref{eq:momentum_def_replace}).  In \ref{section:aside_on_bi_hamiltonian}  I used antibrackets with momenta treated as separate fields to calculate conformal and superconformal algebras; it may be helpful to consult this in the context of what is being done here. When torsion is turned off these two approaches for calculating algebras are exactly the same, so indeed one doesn't need to introduce momenta as separate fields. When torsion is turned on I will show that there is a breakdown of the expected relation between the $(-)$ antibracket and the commutator, and will briefly discuss the implications.

Superconformal transformations are generated as\footnote{The dependence of $a^G T_G$  on $G_{ij} D_+ X^j$ can be seen by writing 
\begin{equation}
G_{kj}  \partial_{++} X^k D_+ X^j \equiv -i D_{+} ( G_{kj} D_{+} X^k ) D_{+} X^j  \ .
\end{equation}
}\label{pg:momentum_dep}
\begin{equation}
\label{eq:G_from_current}
\delta_G X^k(z) = - \frac{1}{2} (  X^k (z),  \int d^2 y   a^G T_G   )_- \ , 
\end{equation}
while $L$-type transformations are generated by 
\begin{equation}
\label{eq:L_from_current}
\delta_L X^k(z) =  \ \frac{(-1)^l}{l} (  X^k(z) ,  \int d^2y  a^L T_L )_- \ .
\end{equation}
 One needs the $(+)$ antibracket, obtained by choosing $D_+$ as the 'time' derivative, to relate the $(-)$ transformations to their currents.\footnote{It should be noted that the smeared currents $\int d^2 z  a^G T_G$ and  $\int d^2 z  a^L T_L$ actually vanish on-shell, since the integration involves both coordinates of superspace, and $D_-T_G =0$ on-shell. This is of no immediate consequence, since functional variations of smeared currents are involved in generating the transformations, and these don't vanish. If  we chose instead to work with Poisson brackets integration would be only over one of the fermionic coordinates, and the smeared currents would not vanish on-shell.}
 
Products of currents generate transformations of the form
\begin{align}
\label{eq:composite_from_current}
 & (  X^k(z) ,  \int d^2 y  a^{GLM} T_G T_L  T_M( y))_- =  m (-1)^m  \int d^2 y    [a^{GLM} T_G T_L](y)  R^k_M(z,y)  \\ \nonumber
 &  \ \ \ \ \ \ +  l (-1)^{l+ml}   \int d^2 y [a^{GLM} T_G T_M](y)  R^k_L (z,y)  \\ \nonumber
&   \ \ \ \ \ \ + 2 (-1)^{m+l+1} \int d^2 y [a^{GLM} T_L T_M](y)  R^k_G (z,y)  \ ,
 \end{align}
where $T_L$ and $T_M$ generate $L$-type transformations. When the product of currents doesn't involve a $T_G$ factor, the $(-)$ antibracket generates an $L$-type symmetry irrespective of whether torsion is turned on. If a $T_G$ factor is present things are not so simple. Taking as an example the symmetry $\delta_{GU}$ (\ref{eq:GUprime}) that emerges out of the commutator of two $L$-type symmetries, we can see that
\begin{align}
\label{eq:_GUprime_from_anti}
  \delta_{GU}X^i(z)   = & (-1)^{u+1}  (X^i (z),  \int d^2 y  a^{GU} T_G T_U )_{-}  \\ \nonumber
  & + (-1)^u  u \frac{ i}{6}  a^{GU}   U^i_{ \ i_2 ... i_u} H_{kjm} D_{+} X^{kjm i_2 ... i_u }(z) \ .
\end{align}
Thus, in the presence of torsion the relation is not as expected, and in fact there is no current that can generate the second term. When currents have derivatives acting on them the antibracket generates the correct symmetry transformations, but again only in the absence of torsion.

Writing out the commutator of two $L$-type transformations using (\ref{eq:L_from_current}) one obtains
 \begin{align}
 \label{eq:com_from_anti}
 &(-1)^{m+l} lm  [\delta_L, \delta_M ] X^k(z)  = 
  (X^k (z) ,   (  \int d^2 y  a^L T_L , \int d^2 y  a^M T_M)_- )_-   \\ \nonumber
  & \ \ \ \ \  +    \int d^2y  \left( \frac{ \dr (   X^k,   a^L T_L )_-} { \delta \pi_{+i}(y)  } \frac{\dl a^M T_M}{ \delta X^i(y) } + (L \leftrightarrow M) \right)  \ ,
  \end{align}
where the Jacobi identity for antibrackets has been used (see Appendix \ref{app:poisson_antibracket}). Since momenta are not treated as separate fields, this is actually a naive assumption; it turns out that the $(-)$ antibracket does not generally obey the Jacobi identity when torsion is involved. In the \emph{absence of torsion} the second term on the right hand side of (\ref{eq:com_from_anti}) is equal to 
 \begin{equation}
 (-1)^{m+l+1} lm  [\delta_L, \delta_M ] X^k(z) \ ,
 \end{equation}
and then the relation between the $(-)$ antibracket and the commutator is as expected:
 \begin{align}
 \label{eq:com_from_anti2}
 & [\delta_L, \delta_M ] X^i(z) = 
 \frac{(-1)^{m+l}}{2lm} (  X^i(z) ,  (  \int d^2 y a^L T_L ,  \int d^2 y  a^M T_M)_-  )_- \ .
 \end{align}
 Given two composite currents, $C_1$ and $C_2$,  the commutator is obtained from an antibracket  as
 \begin{align}
 \label{eq:com_from_anti3}
 & [\delta_{C_1}, \delta_{C_2} ] X^i(z) = 
 F (  X^i(z) ,   (  C_1 , C_2 )_-  )_- \ ,
 \end{align}
where the constant $F$ is a product of the following factors: $\frac{(-1)^l}{l}$ for each $L$-type current, and  $1/2$ for each $T_G$ current (this is regardless whether the basic currents that $C_1$ and $C_2$ are composed of have derivatives acting on them).

The $(-)$ antibracket between the smeared currents $\int d^2 z a^L T_L$ and $\int d^2 z a^M T_M$ is
\begin{align}
& \frac{1}{lm} (\int d^2 z  a^L T_L, \int d^2 z a^M T_M )_- = \\ \nonumber
&  \ \ \ \ \  (-1)^{l} \int d^2 z   \left[  (-1)^{m+1}a^M D_+a^L + D_+ a^M a^L  \right] (z)  T_P (z) \\ \nonumber
&  \ \ \ \ \  + (-1)^m \int d^2 z    [  a^M a^L  ](z) T_D(z)    +  i (-1)^{1+m}  \int d^2 x  [  a^M a^L  ](z)   T_{\mathrm{del}}(z) \ .
\end{align}
Integrating $ \int d^2 z   [ a^M a^L ](z)  T_{\mathrm{del}}(z)$ by parts, as in (\ref{eq:Tdel_int_by_parts}), and using  (\ref{eq:TGU_relation}), this can be rewritten as:
\begin{align}
& \frac{1}{lm} (\int d^2 z  a^L T_L, \int d^2 z    a^M T_M )_- =  \\ \nonumber 
&  \ \ \ \ \  (-1)^{l} \int d^2 z  \left[  (-1)^{m+1}a^M D_+a^L +  D_+ a^M a^L + \frac{l-m}{p} D_+ (a^M a^L)  \right](z) T_P(z) \\ \nonumber
& \ \ \ \ \  + \int d^2 z (-1)^m [  a^M a^L  ](z) T_{N}(z)   +  i (-1)^{1+m}  \int d^2 z   [ a^M a^L  ](z)  T_{G U}(z) \ .
\end{align}
The antibracket closes to the conserved currents, but these are not generators of their symmetry transformations unless torsion vanishes, since the transformation generated by $T_{G U}$ is not a symmetry. In the case of vanishing  torsion $\delta_{GU}X^i$ is generated by $T_{G^0 U} = T_{GU}$, and it can easily be checked that the commutator (\ref{eq:L_commutator2}) is generated by the $(-)$ antibracket via (\ref{eq:com_from_anti2}).
 
The antibracket between two smeared superconformal currents is 
\begin{align}
\label{eq:1_1_antibracket0}
& (\int d^2 z  a^G T_G, \int d^2 z   a^G T_G )_- =
  \int d^2 z   \left[  8 a^G  \dep a^G +2i D_+ a^G D_+ a^L  \right] (z)  T_G(z) \ ,
 \end{align}
and the antibracket between $\int d^2 z a^G T_G$ and an $L$-type current,  $\int d^2 z a^L T_L$, is
\begin{align}
\label{eq:1_L_antibracket}
& (\int d^2 z  a^G T_G, \int d^2 z a^L T_L )_- = \\ \nonumber
 &  \int d^2 z \left[ (-2l +2) \dep a^G a^L + 4 a^G \dep a^L +2i D_+ a^G D_+ a^L  \right] (z)  T_L (z) \ .
 \end{align}
 In fact, any superconformal primary current of spin $l$ will transform in the above manner under the superconformal current, regardless of whether it is an $L$-type current or of some other form.\footnote{See Chapters 25 and 26 of \cite{West:1990tg} for a definition of a superconformal primary that is useful in the current context. Of course, being purely classical, the antibrackets don't tell us anything about the quantum corrections, which for free theories are obtained by taking double and higher contractions in the operator product expansion. So a current that transforms as above classically may have quantum corrections in the OPE that prevents it from being a true superconformal primary.} 
 
We have seen that in the presence of torsion the antibracket can't be used reliably to calculate the algebra. It would seem that in the presence of torsion one should take 
\begin{equation}
\label{eq:momentum_def_replace2}
\pi_{+i}(z) \equiv (G_{ij}+ b_{ij}) D_+X^j (z) \ .
\end{equation}
However, one can't consistently take functional derivatives of terms involving the torsion tensor $H_{ijk}$ with respect to $b_{ij} D_+X^j$, whereas it is possible to do so with respect to $G_{ij} D_+X^j$ for all  terms containing derivatives of the metric (see, for example, the footnote on page \pageref{pg:momentum_dep}). To treat the torsionfull case via antibrackets it seems that one has to resort to a formalism where momenta are treated as separate fields. I have not investigated this in full detail, but  it turns out that the antibracket needs to be modified by a term involving torsion. As we'll see in the next chapter, the antibracket is useful primarily because one can exploit its special properties (see Appendix \ref{app:poisson_antibracket}). These no longer hold when it is modified by such a torsion term, and one needs to do more work to find a way of calculating algebras that involve composite currents quickly. 

 
\chapter{Special holonomy algebras}
\label{chapter:algebras_on_manifolds_of_sh}


In this chapter the extended algebras of $\sigma$-models on special holonomy  manifolds  are analyzed (see \ref{section:spechol_manifolds}). These are super $W$-algebras, consisting of the superconformal transformations together with $L$-type transformations due to the covariantly constant tensors that are present on special holonomy manifolds.  In Chapter \ref{chapter:algebras_of_general_L_symmetries} the algebras related to general covariantly constant tensors were worked out. Here we wish to look at all the special holonomy cases one by one. At the end we look at some torsionfull cases when the generalized Nijenhuis form (\ref{eq:N}) doesn't vanish.

The first observation of the relation between target space geometry and symmetries in the supersymmetric $\sigma$-model was that $N=2$ supersymmtery in two dimensions requires the target space to be a K\"ahler manifold \cite{Zumino:1979et}.  The more general observation, that for every covariantly constant form on the target space there is symmetry in the $\sigma$-model, was made in \cite{Delius:1989fy}. The algebra of these symmetries on Calabi-Yau target spaces was first worked out using conformal field theory methods, specifically in the context of two \cite{Eguchi:1988vr} and three \cite{Odake:1988bh} complex dimensions. The currents related to the holomorphic forms were understood to be generators of spectral flow of the $N=2$ algebra (\ref{eq:N2_abstract_algebra}). The cases of exceptional holonomy, $G_2$ and $Spin(7)$, were studied in the context of conformal field theory in \cite{Shatashvili:1994zw} and \cite{Figueroa-O'Farrill:1996hm}, and more recently in topological models \cite{deBoer:2005pt}. Classical special holonomy algebras were written down for all the cases in \cite{Howe:1991ic} and \cite{Howe:1991im}. Higher spin algebras of $\sigma$-models on on symmetric spaces (see \ref{section:spechol_manifolds}) have been studied in \cite{Evans:2005zd, Evans:2004mu}.

In \ref{sec-global_symmetries} it was argued that it's not possible to write down cohomological equations for potential anomalies of conformal-type algebras that close in a field dependent manner, due to the  parameter ghosts (which depend only on half of the worldsheet coordinates) transforming into quantum fields. Instead, we have to treat all the composite currents as separate symmetry generators and attempt to calculate the full Lie algebra if we are to avoid working with non-local expressions. Another potential problem in this case is related to Jacobi identities not closing at the level of structure functions, which prevents the derivation of naive Ward identities (see discussion around (\ref{eq:jacobi_brst_problem})).

It is also possible to gauge the original symmetries and attempt to treat the algebra in a field dependent manner, which will be studied in detail in Chapter \ref{chapter:W-strings}. Treating the gauge fields as background fields enables us to study the current algebras, whereas treating them as quantum fields defines a completely new theory, a type of $W$-superstring, in which all the currents are treated as constraints. For the former case, as shown in \ref{sec-global_symmetries}, we still have the problem that the Ward identities involve an explicit evaluation of composite operators. For a cohomological analysis it is therefore still necessary to gauge the entire Lie algebras.  It is not immediately obvious that the Jacobi identities not closing at the level of structure functions should cause problems, since now the non-closure functions can be absorbed into equation of motion terms. But as we'll see in Chapter \ref{chapter:W-strings}, the problems pop up via the closure of the transformations of the gauge fields.

With these motivations, the main aim of this chapter is to find out if it's possible to make sense of the full Lie algebras, and to find out how the Jacobi identities behave. I will be working out the algebras using smeared currents and $(-)$ antibrackets (see \ref{sec:algebras_via_antibrackets});  these are related to the commutators via (\ref{eq:com_from_anti2}) and (\ref{eq:com_from_anti3}). 

At the end I'll discuss in detail the implications for calculating anomalies.

\section{K\"ahler manifolds}
\label{section:kahler_algebra}

On a K\"ahler manifold the action of the $(1,1)$ $\sigma$-model (\ref{eq:ssm_action2}) can be written as
\begin{equation}
\label{eq:kahler_action}
S = \int d^2 z  G_{\alpha \bbeta }  D_+ X^{( \alpha } D_- X^{\bbeta )} \ .
\end{equation}

The superconformal current is
\begin{align}
T_G = & G_{\alpha \bbeta} D_+ X^\alpha \partial_{++} X^\bbeta +  G_{\balpha \beta} D_+ X^{\balpha}
\partial_{++} X^{\beta} \\ \nonumber
=: & \frac{1}{2} ( T_g + T_{\bg} ) \ .
\end{align}
As the notation indicates, $T_g$ and $T_{\bg}$ are related by complex conjugation. On a K\"ahler manifold  $T_g$ and $T_{\bg}$ generate two separate symmetries via the $(-)$ antibracket (\ref{eq:min_antibracket}):
\begin{align}
\label{eq:cpx_conf_symmetries}
& \delta_{g} X^{\alpha} = a^g \dep X^\alpha + i D_+ a^g D_+ X^\alpha \ \ \ , \ \ \ \delta_{g} X^{\balpha} = a^g \dep X^{\balpha} \ , \\ \nonumber
& \delta_{\bg} X^{\alpha} = a^\bg \dep X^\alpha  \ \ \ ,  \ \ \  \delta_{\bg} X^{\balpha} = a^\bg \dep X^{\balpha} + i D_+ a^\bg D_+ X^{\balpha} \ .
\end{align}
The transformations are not real, since the complex conjugate fields $X^\alpha$ and $X^{\balpha}$ don't transform as complex conjugates, but never the less they are symmetries of (\ref{eq:kahler_action}).
 
The current related to the complex structure is manifestly real
\begin{equation}
T_I = 2i G_{\alpha \bbeta} D_+ X^{\alpha} D_{+} X^{\bbeta} \ .
\end{equation}
In a non-holomorphic basis we would have $T_I = I_{ij} D_+X^{ij}$, but in the holomorphic basis the K\"ahler form is basically given by the metric (see (\ref{eq:def_kahler_form})). The associated symmetry transformations are:
 \begin{equation}
 \label{eq:delta_I}
 \delta_I X^{\alpha} = i a^I D_+ X^{\alpha} \ \ \ , \ \ \  \delta_I X^{\balpha} = -i a^I D_+ X^{\balpha}  \ .
 \end{equation}
 
The $(-)$ antibracket of $T_g$ with itself is
\begin{align}
\label{eq:g_g}
(\int d^2 z  a^g T_g,  \int d^2 z a^{g} T_g )_- = \int d^2z 8 \left[ a^g \partial_{++} a^g \right] T_g \ ,
\end{align}
and similarly for the $(-)$ antibracket of  $T_{\bg}$ with itself. From now on I will omit the integral signs when writing down the $(-)$ antibrackets.
 
The  antibracket of $T_g$ with $T_{\bg}$ is
\begin{align}
\label{eq:g_bg}
(a^g T_g,  a^{\bg} T_{\bg} )_- = &   \left[ 2 \partial_{++} a^g D_+ a^{\bg} - 2 D_{+} a^g \partial_{++} a^{\bg} \right] T_I \\ \nonumber
 & + \left[ -4 \partial_{++} a^g a^{\bg} + 2i D_+ a^g D_+ a^{\bg} \right] T_g  + \left[ 4 a^g \partial_{++} a^{\bg} + 2i D_+ a^g D_+ a^{\bg} \right] T_{\bg} \ .
\end{align}
 
The antibracket of $T_g$ and  $T_I$ is
\begin{align}
\label{eq:g_I}
& (a^I T_I, a^g T_g)_- =  \left[ 4 \partial_{++} a^I a^g \right] T_I - \left[ 4i a^I D_+ a^g \right] T_g  \ ,
\end{align}
while the antibracket of $T_{\bg}$ with $T_I$ is the complex conjugate of the above, since $T_I$ is real:
\begin{align}
\label{eq:I_bg}
& (a^I T_I, a^{\bg} T_{\bg} )_- =  \left[ 4 \partial_{++} a^I a^{\bg} \right] T_I + \left[ 4i a^I D_+ a^{\bg} \right]  T_{\bg}  \ .
\end{align}
Also, since $T_I$ is real, we have
\begin{equation}
\label{eq:anti_TI_TI}
(a^I T_I, a^I T_I)_- = - \left[ 8i a^I a^I \right] T_G \equiv - \left[ 4i a^I a^I \right] (T_g + T_\bg) \ .
\end{equation}
 
The closure of the currents  (\ref{eq:g_bg}),  (\ref{eq:g_I}), and (\ref{eq:I_bg}) is not unique because of the following relation
 \begin{equation}
 \label{eq:DI_relation}
 D_+ T_I = T_g - T_{\bg} \ ,
 \end{equation}
so we can integrate by parts to obtain a range of possible structure functions. If we want to write down an extended action that describes this reducible algebra, it turns out that the structure functions have to be chosen carefully, since in general the Jacobi identities will prevent us from constructing a solution to the master equation. This is slightly surprising, since transformations (\ref{eq:cpx_conf_symmetries}) are linear. There are, however, many possible choices of structure functions that do avoid the Jacobi problem.
 
The usual $N=2$ algebra is obtained by calculating antibrackets involving the real currents $T_I$ and $T_G$. In addition to (\ref{eq:anti_TI_TI}), we have:
\begin{align}
& (a^G T_G, a^G T_G)_- = \left[ 8 a^G \partial_{++} a^G + 2i D_+ a^G D_+ a^G \right] T_G \ , \\
& (a^G T_G, a^I T_I)_- = \left[ 4\partial_{++} a^G - 2 a^I \partial_{++} a^G + 2i D_+ a^I D_+ a^I \right] T_I \ .
\end{align}
 
\section{Calabi-Yau $W$-algebras}
\label{sec:cy_algebras_in_general}
 
In addition to being K\"ahler, a Calabi-Yau manifold of $D$ complex dimensions (referred to from now on as CY$D$) admits covariantly constant $(D,0)$ and $(0,D)$ forms. We can always find an atlas in which this form  is given by the Levi-Civita alternating symbol in each chart (see \ref{subsection:kah_and_CY}). Therefore, we can work with the conserved currents
 \begin{equation}
 T_M = \frac{2}{D} \epsilon_{\alpha_1 ... \alpha_D} D_+X^{\alpha_1 ... \alpha_D} \ \ \ \ \mathrm{and} \ \ \ \  
 T_{\bM} =  \frac{2}{D} \epsilon_{\balpha_{1} ... \balpha_D} D_+ X^{\balpha_{1} ... \balpha_D} \ .
 \end{equation}
The spin of the ghosts $a^M$ and $a^{\bM}$ with which the currents are smeared is $(+(D-1))$. They have odd parity in odd complex dimension, and even otherwise.
 
The antibrackets of $T_M$ and $T_{\bM}$ with $T_g$ and $T_{\bg}$ are
\begin{equation}
\label{eq:cy_tmp_comm1}
(a^g T_g, a^M T_M)_- = \left[ -4(D-1) \partial_{++} a^g a^M + 4 a^g \partial_{++} a^M \right] T_M 
\end{equation}
and 
\begin{equation}
\label{eq:cy_tmp_comm2}
(a^{\bg} T_{\bg}, a^M T_M)_- = \left[ 4i D_+ a^{\bg} D_+ a^M + 4  a^{\bg}  \dep a^M \right] T_M \ .
\end{equation}
The complex conjugate antibrackets are obtained by simply making the replacements $\bg \leftrightarrow g$ and $M \leftrightarrow \bM$ in the above expressions.
 
The antibrackets with $T_I$ are
\begin{align}
\label{eq:I_M_antibracket}
& (a^I T_I, a^M T_M )_- = \left[ - 4i a^I D_+ a^M + 4(D-1)i D_+ a^I a^M \right] T_M \ , \\ \nonumber
& (a^I T_I, a^{\bM} T_{\bM} )_- = \left[  4i a^I D_+ a^{\bM}  -  4(D-1)i D_+ a^I a^{\bM} \right] T_{\bM}  \ .
\end{align}
Unlike (\ref{eq:cy_tmp_comm1}) and (\ref{eq:cy_tmp_comm2}), the structure coefficients of these are complex conjugates of each other, since $T_I$ is real while $T_M$ and $T_{\bM}$ are not. 

The antibracket of $T_M$ with itself is zero, because the number of $D_+X^{\alpha}$ factors in the expressions is always larger than $D$. An analogous remark holds for $T_{\bM}$, so the only non-zero antibracket involving $T_M$ and $T_{\bM}$ is $(a^M T_M, a^{\bM} T_{\bM} )$. The interesting feature is that its form depends on whether the complex dimension is odd or even. For $D$ odd
\begin{align}
(a^M T_M, a^{\bM} T_{\bM} )_- =&   -(D-1)! \frac{4}{2^{D-1}} \left[ D_+ a^{\bM} a^M + a^{\bM} D_+ a^M \right] (T_I)^{D-1}  \\ \nonumber
& - (D-1)! (D-1) \frac{4}{2^{D-2}} \left[ a^{\bM} a^M \right] T_{G} (T_I)^{D-2} \ ,
\end{align}
while for $D$ even
\begin{align}
(a^M T_M, a^{\bM} T_{\bM} )_- =&   -(D-1)! \frac{4i}{2^{D-1}} \left[ D_+ a^{\bM} a^M - a^{\bM} D_+ a^M \right] (T_I)^{D-1}  \\ \nonumber
& - (D-1)! (D-1) \frac{4i}{2^{D-2}} \left[ a^{\bM} a^M \right] T_{G} (T_I)^{D-2} \ .
\end{align}
Clearly the algebra involving the complex currents closes in a field dependent sense, i.e. as a $W$-algebra. 

Next I want to examine the closure of the algebra involving only real combinations of currents. These would be $T_G$, $T_I$, and  the real and complex  parts of $T_M$:
\begin{equation}
\label{eq:TL_TM_def}
T_L := T_M + T_{\bM} \ \ \ \mathrm{and} \ \ \  T_{\bL} := i(T_M - T_{\bM}) \ .
\end{equation}
 
The antibracket of $T_G$ with $T_L$ is
\begin{equation}
\label{eq:CY_G_L_antibracket}
( a^G T_G, a^L T_L)_- = \left[ -2(D-1) \dep a^G a^L + 4 a^G \dep a^L + 2i D_+ a^G D_+ a^L \right] T_L \ ,
\end{equation}
with an analogous expression for  $( a^G T_G, a^{\bL} T_{\bL})_-$.  The antibrackets of $T_I$ with $T_L$ and $T_{\bL}$ are
\begin{equation}
( a^I T_I, a^L T_L)_- = \left[ -4 a^I D_+ a^L + 4(D-1) D_+ a^I a^L  \right] T_{\bL}
\end{equation}
and
\begin{equation}
( a^I T_I, a^L T_{\bL})_- = \left[ 4 a^I D_+ a^{\bL} - 4(D-1) D_+ a^I a^{\bL}  \right] T_{L} \ .
\end{equation}
 
In odd complex dimension we have 
\begin{align}
\label{eq:LLodd}
& (a^L T_L, a^L T_L)_- = - (D-1)! \frac{16}{2^{D-1}} \left[ a^L D_+ a^L \right] (T_I)^{D-1} \ , \\ \nonumber
& (a^L T_L, a^\bL T_\bL)_- = - (D-1)! (D-1) \frac{8i}{2^{D-2}} \left[ a^L a^\bL \right] T_G (T_I)^{D-2} \ ,
\end{align}
while in even complex dimension
\begin{align}
\label{eq:LLeven}
& (a^L T_L, a^L T_L)_- = - (D-1)! (D-1) \frac{4i}{2^{D-2}} \left[ a^L a^L \right] T_G (T_I)^{D-2}  \ ,\\ \nonumber 
& (a^L T_L, a^\bL T_\bL)_- =  (D-1)! \frac{8}{2^{D-1}} \left[ a^\bL D_+ a^L - D_+ a^\bL a^L \right] (T_I)^{D-1} \ .
\end{align}

The currents obey the following basic relations which can be verified easily when working in complex coordinates:
\begin{align}
\label{eq:IL_relation}
& T_I T_L = T_I T_\bL \equiv 0  \ , \\
\label{eq:GL_relation}
& T_G T_\bL \equiv -\frac{i}{2} D_+ T_I T_L \ \ \ , \ \ \ T_G T_L \equiv \frac{i}{2} D_+ T_I T_\bL \ , \\ 
\label{eq:IdL_relation}
& (T_I)^D T_G \equiv 0 \ , \\ 
\label{eq:GDIL_relation1}
&T_G D_+T_I T_L = T_G D_+T_I T_\bL \equiv 0 \ , \\
\label{eq:DLLhat_relations}
& D_+ T_L T_\bL \equiv -T_LD_+ T_\bL \ \ \ , \  \ \ \dep T_L T_\bL = (-1)^{D+1} T_L \dep T_\bL \ , \\
\label{eq:DLLhat_relation2}
&   D \ \mathrm{odd:} \  D_+ T_L D_+T_\bL \equiv 0   \ \ \ , \ \ \ D \ \mathrm{even:} \  \dep T_L \dep T_\bL \equiv 0 \ ,  \\ 
\label{eq:LLhat_relation}
& -\frac{8D!}{D^2 2^D} (i)^{D-1} (T_I)^D \equiv 
\begin{cases} T_L T_\bL \ \  D \ \mathrm{odd}  \\ T_L T_L \ \  D \ \mathrm{even} \end{cases}  \propto Vol
\ .
\end{align}
As we'll see in the next section,  some (but not all) of these relations are implied by the Jacobi identities involving the basic currents.

\section{The linearized  Calabi-Yau algebras}
\label{section:cy_algebras}
  
In this section I will attempt to understand the full Lie algebras generated by $T_L$, $T_\bL$, $T_G$, and $T_I$, as well as all the composite currents that appear via the antibrackets involving these basic currents. Because of the different nature of the antibrackets for odd (\ref{eq:LLodd}) and even dimensions (\ref{eq:LLeven}), it is clear that the algebras will be different for these two classes. It is not so obvious that there should be a strong distinctions between algebras within one of the classes. The $D=2$ case is special because it's linear, but for $D>2$ one may expect the non-linear algebras in all even (odd) dimensions to be of a similar nature. It turns out that this is not the case. The $D=4$ and $D=5$ cases are actually finite, essentially because they are non-linear enough, so that higher antibrackets are zero due to the dimensional bound (a product of more than $D$ $D_+ X^\alpha$s vanishes). The only case that is not finite is $D=3$, which is also the most relevant one physically. In what follows I'll go through all the dimensions up to $D=5$ in detail, in the order of complexity, which is $D=2, 5, 4, 3$. It will become clear that the algebras for $D>5$ are the same as $D=4$ for even dimensions, and $D=5$ for odd.
 
\subsection{D=2}
\label{subsection:cy2}
 
The unique CY2 manifold whose holonomy group is actually $SU(2)$, and not a subgroup thereof, is called K3 (see \ref{subsection:kah_and_CY}). This case is distinct because the algebra is linear, and in fact it is simply that of a hyper-K\"ahler four-fold.  $\sigma$-models on such manifolds were first discussed by Gates, Hull, and Ro\v{c}ek in 1984 \cite{Gates:1984nk}. They possess $N=4$ supersymmetry due to the existence of three complex structures on the target space, $I_r$, $r = \{1 ,2, 3 \}$, that obey the algebra of quaternions:
\begin{equation}
\label{eq:hyperkahler_cpx_structures}
I_r I_s = - \delta_{rs} + \epsilon_{rst} I_t \ .
\end{equation}
For K3 the role of $I_2$ and $I_3$ is played by $L$ and $\hat{L}$, and the algebra is given by:
\begin{align}
& (a^L T_L,  a^L T_L)_- = -\left[ 4i a^L a^L \right] T_G \ \ \ , \ \ \    
(a^\bL T_\bL,  a^\bL T_\bL)_- = - \left[ 4i a^\bL a^\bL \right] T_G \ , \\ \nonumber
&  (a^L T_L,  a^\bL T_\bL )_- = \left[ 4 a^\bL D_+ a^L - 4 D_+ a^\bL a^L \right] T_I \ , \\ \nonumber
& (a^I T_I, a^L T_L)_- = \left[ -4 a^I D_+ a^L + 4 D_+ a^I a^L \right] T_\bL \ , \\ \nonumber
& (a^I T_I, a^\bL T_\bL )_- =  \left[ 4 a^I D_+ a^\bL - 4 D_+ a^I a^\bL \right] T_L  \ .
 \end{align}
 
The Jacobi identities are satisfied without implying any relations between currents.
 
\subsection{D=5}
\label{subsection:cy5}
 
For CY5 the Jacobi identities between the basic currents imply the following relations:
\begin{itemize}
\item $ (a^L T_L,  (a^L T_L, a^L T_L )_-  )_-  \equiv 0$
implies that
\begin{align}
\label{eq:jacobi1_CY5}
(T_I)^3 T_\bL \equiv 0  \ .
\end{align}
\item
$(a^\bL T_\bL ,  (a^L T_L, a^L T_L )_-  )_-  + \mathrm{cyclic} \equiv 0$
implies that
\begin{align}
\label{eq:jacobi2_CY5}
(T_I)^3 T_L \equiv 0 \ \ \ \ \  (T_I)^2(T_G T_L- \frac{i}{2} D_+ T_I T_\bL)  \equiv 0 \ .
\end{align}
\end{itemize}
Thus for CY5 only (\ref{eq:IL_relation}) is needed to satisfy the Jacobi identities. The second relation in (\ref{eq:jacobi2_CY5}) is also proportional to (\ref{eq:GL_relation}). Analogous observations can be made for $L \leftrightarrow \bL$.
 
The antibrackets involving $T_L$ and $T_\bL$ are
\begin{equation}
(a^L T_L, a^L T_L )_- = - \left[ 24 a^L D_+ a^L \right] (T_I)^4 \ ,
\end{equation}
the same with $L \rightarrow \bL$, and
\begin{equation}
(a^L T_L, a^\bL T_\bL )_- = -\left[ 96i a^L a^\bL \right] T_G (T_I)^3 \ .
\end{equation}
 
Treating $(T_I)^4$ and  $T_G (T_I)^3$ as independent currents, we introduce the bosonic ghost $a^{I^4}$, the fermionic ghost  $a^{G I^3}$, and the appropriate smeared currents. The antibrackets of these among themselves, and with $T_L$ and $T_\bL$ are zero. In the antibrackets of $T_L$ and $T_\bL$ with the new smeared currents, and in $(a^{I^4}(T_I)^4, a^{I^4}(T_I)^4)_-$,  the identities (\ref{eq:IL_relation}) - (\ref{eq:LLhat_relation}) are more than saturated.  By this I mean that the  terms closest to being non-zero are $(T_I)^2 T_L \equiv 0$ and $(T_I)^6T_G \equiv 0 $.   $(T_G  (T_I)^3,  T_G  (T_I)^3)_-$ just saturates  (\ref{eq:IdL_relation}).
 
No further composite currents are generated. One still has to check how $(T_I)^4$ and  $T_G (T_I)^3$ transform under $T_G$ and $T_I$:
\begin{align}
& (a^G T_G, a^{G I^3} T_G (T_I)^3)_-  = \left[ 4 a^G \dep a^{G I^3} - 16 \dep a^G a^{G I^3} + 2i D_+ a^G D_+ a^{G I^3} \right]  T_G (T_I)^3 \ , \\ \nonumber
& (a^G T_G, a^{I^4}(T_I)^4 )_- =  \left[ 4 a^G \dep a^{I^4} -  14 a^{I^4} \dep a^G + 2i D_+ a^{I^4} D_+ a^G \right] (T_I)^4  \ .
\end{align}
This is as expected, since  $(T_I)^4$ and  $T_G (T_I)^3$ are superconformal primaries\footnote{Only up to quantum corrections; see (\ref{eq:1_L_antibracket}).} of spin $(+8)$ and spin $(+9)$ respectively. They also transform nicely under $N=2$:
\begin{align}
& (a^I T_I, a^{G I^3} T_G (T_I)^3)_-    = (8 \dep a^I a^{G I^3} - \frac{1}{2} a^I \dep a^{G I^3} + \frac{i}{2} D_+ a^I D_+ a^{G I^3}) (T_I)^4 \ , \\ \nonumber
& (a^I T_I, a^{I^4}(T_I)^4 )_- =  -32 i a^I a^{I^4} T_G (T_I)^3 \ .
\end{align}
 
In summary, the Lie algebra of CY5 is finite, containing only the additional currents $T_G (T_I)^3$ and $(T_I)^4$. It is worth noting that not all possible non-derivative currents are generated by the algebra. This is just as well if we want finiteness, because when $T_G T_I$ is included the algebra becomes unmanageable. As we'll see, this current is generated in the CY3 case (see \ref{section:CY3}). 
 
An important feature of the relations between the currents implied by the Jacobi identities, (\ref{eq:jacobi1_CY5}) and (\ref{eq:jacobi2_CY5}), is that they do not imply any relations between the basic symmetry transformations. For example, in the symmetry generated by $(T_I)^2T_G T_L \equiv 0$ (see (\ref{eq:composite_from_current})), 
\begin{equation}
\int d^2 y a^{X} (T_I)^2T_G(y)  R^i_L(z,y) \equiv 0 \ ,
\end{equation}
since  $(T_I)^2  R^i_L(z,y) \equiv 0$, where $a^X$ is an appropriate ghostly parameter.  All the combinations that arise from
\begin{equation}
\label{eq:insignificant_null}
\delta X^i = (X^i, a^X (T_I)^2T_G T_L)_- \equiv 0
\end{equation}
vanish \emph{separately} in a similar manner. This is not true for any of the basic relations (\ref{eq:IL_relation})-(\ref{eq:LLhat_relation}). For example,  $(T_I)  R^i_L(z,y)$, which is a part of
\begin{equation}
\label{eq:significant_null}
\delta X^i = (X^i, a^X T_I T_L)_- \equiv 0 \ ,
\end{equation}
doesn't vanish by itself. So (\ref{eq:significant_null}), as opposed to (\ref{eq:insignificant_null}), implies relations between the symmetry transformations. The Jacobi problem (see (\ref{eq:jacobi_brst_problem})) is automatically avoided for any linearized algebra with real transformations, but due to the above observation it also doesn't occur for the CY5 $W$-algebra. However, a word of caution is necessary because for a $W$-algebra one has the issue of ghosts transforming into quantum fields, so the Jacobi problem can really only be addressed in models with the algebra gauged. These will be discussed in Chapter \ref{chapter:W-strings}.
 
\subsection{D=4}
 
For the CY4 algebra Jacobi identities involving the basic currents imply the following relations,
\begin{itemize}
\item $ (a^L T_L,  (a^L T_L, a^L T_L )_-  )_-  \equiv 0$
implies that
\begin{align}
\label{eq:jacobi1_CY4}
& T_G D_+T_I T_\bL \equiv T_\bg T_g T_\bL \equiv 0  \ \ \ , \ \ \ D_+ T_G T_I T_\bL \equiv 0 \ \ \ , \ \ \   
\dep T_I T_I T_\bL \ , \\ \nonumber
& T_I (T_GT_\bL + \frac{i}{2} D_+ T_I T_L) \equiv 0 \ ,
\end{align}
from which non-trivial relations between the basic symmetry transformations follow. Various other relations that don't have an impact on the symmetry transformations  (such as those in (\ref{eq:insignificant_null})) are also implied. From now on I will only mention the former, significant ones.
\item
$(a^\bL T_\bL ,  (a^L T_L, a^L T_L )_-  )_-  + \mathrm{cyclic} \equiv 0$
implies the significant relations
\begin{align}
\label{eq:jacobi2_CY4}
& T_G D_+ T_I T_L \equiv 0 \ \ \  , \ \ \    D_+ T_G T_I T_L \equiv 0  \ \ \  , \ \ \ \dep T_I T_I T_L \equiv 0 \ , \\ \nonumber
&  T_I (T_G T_L- \frac{i}{2} D_+ T_I T_\bL)  \equiv 0 \ .
\end{align}
\end{itemize}
 
The algebra closes in a field dependent sense as:
\begin{equation}
(a^L T_L, a^L T_L )_- = - \left[ 18i a^L a^L \right] T_G (T_I)^2 \ ,
\end{equation}
the same with $L \rightarrow \bL$, and
\begin{equation}
(a^L T_L, a^\bL T_\bL )_- =  \left[ 6 a^\bL D_+ a^L - 6 D_+ a^\bL a^L \right] (T_I)^3 \ .
\end{equation}
 
Treating the currents $T_G (T_I)^2$ and $(T_I)^3$ as independent generators, we introduce the fermionic ghost $a^{G I^2}$, the bosonic ghost $a^{I^3}$, and the corresponding smeared currents. The antibrackets of these among themselves, as well as with $T_L$ and  $T_{\bL}$, vanish. In showing this one must make use of the null relations  $T_I T_L \equiv  0$  and  $(T_I)^4 T_G \equiv 0$. For most of the antibrackets these relations are just saturated. The exception is $( a^{G I^2} T_G (T_I)^2, a^{G I^2} T_G (T_I)^2)$, which is equal to zero but only after some intricate cancellations.
 
The currents $T_G (T_I)^2$ and $(T_I)^3$ transform as superconformal primaries  of spin $(+7)$ and spin $(+6)$, respectively (see (\ref{eq:1_L_antibracket})). Under $T_I$ they transform as
\begin{align}
& (a^I T_I, a^{G I^2} T_G (T_I)^2)_-   =  \left[ 4 \dep a^I a^{G I^2} - \frac{2}{3} a^I \dep a^{G I^2} + \frac{2i}{3} D_+ a^I D_+ a^{G I^2} \right] (T_I)^3 \ , \\ \nonumber
& (a^I T_I, a^{I^3}(T_I)^3 )_- =  - \left[ 24 i a^I a^{I^3} \right] T_G (T_I)^2 \ .
\end{align}
 
In conclusion, the Lie algebra of CY4 is finite, containing only the additional currents $T_G (T_I)^2$ and $(T_I)^3$. 

\subsection{D=3}
\label{section:CY3}
  
For CY3 the Jacobi identities imply only the following relations,
\begin{itemize}
\item $ (a^L T_L,  (a^L T_L, a^L T_L )_-  )_-  \equiv 0$, and $L \rightarrow \bL$, imply  (\ref{eq:IL_relation}),
\item
$(a^\bL T_\bL ,  (a^L T_L, a^L T_L )_-  )_-  + \mathrm{cyclic} \equiv 0$, and  $L \leftrightarrow \bL$, imply  (\ref{eq:GL_relation}).
\end{itemize}
all of which imply non-trivial relations between the basic symmetry transformations.

The field dependent algebra is:
 \begin{equation}
(a^L T_L, a^L T_L )_- = - \left[ 8 a^L D_+ a^L \right] (T_I)^2 \ ,
\end{equation}
the same with $L \rightarrow \bL$, and
\begin{equation}
(a^L T_L, a^\bL T_\bL )_- =  - \left[ 16i a^L a^\bL \right] T_G T_I \ .
\end{equation}

Treating the composite currents as new generators,  further composite currents are generated:
\begin{equation}
(a^L T_L , a^{I^2} (T_I)^2)_- = \left[ 32 i a^{I^2} a^L \right] T_G T_L \ ,
\end{equation}
the same taking $L \rightarrow \bL$, and
\begin{equation}
(a^{I^2} (T_I)^2 , a^{I^2} (T_I)^2)_- = - \left[ 32 i a^{I^2} a^{I^2} \right] T_G (T_I)^2 \ .
\end{equation}
In addition, there is an antibracket that generates currents containing derivatives:
\begin{align}
\label{eq:L_GI_antibracket}
(a^L T_L, a^{GI} T_G T_I )_- = &   \frac{1}{3} \left[ -16  a^{GI} D_+ a^L - 8 D_+ a^{GI} a^L \right] T_G T_{\bL}  \\ \nonumber
& - 8  a^{GI} a^L \left[ -\frac{1}{2} T_I \dep T_L + \frac{2}{3} T_G D_+ T_\bL +\frac{1}{3} D_+ T_G T_\bL \right] \ .
\end{align}
The first term on the right hand side is clearly a superconformal primary, but it is less obvious why the second term is.  It does in fact transform appropriately under (\ref{eq:1_L_antibracket}), but only when relations (\ref{eq:GL_relation}) are taken into account. If we integrate any of the terms in the second line by parts, the algebra will close to a current containing derivatives that is no longer a superconformal primary. However, it turns out that it makes no difference whether we work with superconformal primaries or not when calculating further antibrackets. One also always generates a finite number of currents by taking antibrackets with the $N=2$ generators.  
 
The fact that derivative currents are generated is significant. If the algebra only involved products of the basic currents it would have to be finite, because high enough powers of these always vanish. Since this is not the case there is no guarantee that the algebra is finitely generated.
 
Let us see what further composite currents are generated. For convenience, instead of the second line of (\ref{eq:L_GI_antibracket}) we can close the algebra to
\begin{equation}
-\frac{1}{2}T_I \dep T_L + T_G D_+ T_\bL \ . 
\end{equation}
Smearing this current with a fermionic ghost $a^X$ of spin $(+6)$ and taking the antibracket with the smeared $T_L$ current yields
\begin{align}
& (a^L T_L, a^X(-\frac{1}{2}T_I \dep T_L + T_G D_+ T_\bL ) )_- =  \\
&  \ \ \ \ \  \left[ \frac{10}{3} D_+ \dep a^X a^L - \frac{2}{3} a^X D_+ \dep a^L  - \frac{4}{3} \dep a^X D_+ a^L \right] (T_I)^3 \\  \label{eq:non_sc_primary}
& \ \ \ \ \ \  + \left[ a^X a^L \right] \left( 8 T_\bL  D_+ \dep T_L - 2 (T_I)^2  D_+ \dep T_I - 16i T_G T_I D_+ T_G \right) \ .
\end{align}
$(T_I)^3$ is clearly a superconformal primary, but the current on the second line, which I'll call $T_Y$, is not. There are superconformal primaries of this form,  given by
\begin{align}
\label{eq:sc_primary_der_example}
k_1  D_+ \dep T_I (T_I)^2 + k_2 D_+ \dep T_L T_\bL + k_3 \dep T_L D_+ T_\bL + k_4 \dep T_I D_+(T_I^2) \ ,
\end{align}
with the four parameters related by
\begin{align}
k_1 + \frac{4}{3} k_2 - k_3 + \frac{4}{3} k_4 = 0  \ \ \ , \ \ \ 
\frac{4}{3} k_2 - 4 k_3 + \frac{16}{3} k_4 = 0 \ .
\end{align}
One would think that by integrating by parts the algebra could be closed to (\ref{eq:sc_primary_der_example}), but $T_Y$ has the wrong coefficients for this to be possible. 
 
The current
\begin{equation}
8i T_\bL D_+ T_G D_+ T_G - 4i T_L D_+ T_G \dep T_I + 6i T_\bL \dep T_I \dep T_I + 16 T_\bL T_G \dep T_G
\end{equation}
is generated by taking an antibracket of $T_Y$ with $T_L$. Taking further antibrackets with $T_L$ generates currents with more and more derivatives. Although it becomes prohibitively difficult to check explicitly, it is unlikely that many of these can be closed to superconformal primaries. 

In conclusion, CY3 currents don't seem to generate a finite Lie algebra.
 
\section{\Hkah and quaternionic \kah manifolds}

The algebra of \hkah manifolds in general dimension is the same as that of the four dimensional \hkah case, which was worked out in \ref{subsection:cy2}. 
 
The quaternionic \kah holonomy group is $Sp(k)\cdot Sp(1)$. The three complex structures of the \hkah case are not globally defined, as the connection now sees the $r$ index. However, a covariantly constant $4$-form $L$ can be obtained as
 \begin{equation}
L = \sum_{r=1}^3 \omega_r \wedge \omega_r \ ,
 \end{equation}
where $\omega_r$ is the form associated to $I_r$ (\ref{eq:hyperkahler_cpx_structures}). 
The corresponding conserved current is
 \begin{equation}
 T_L = L_{ijkl} D_+ X^{ijkl} \ .
 \end{equation}

The antibracket of $T_L$ with itself is
\begin{align}
(a^L T_L, a^L T_L)_- = -\left[ 4i a^L a^L \right] T_G T_L \ .
\end{align}
Taking the current on the right hand side as a generator of a new symmetry we obtain the following antibrackets
\begin{align}
& (a^{X} (T_L)^p, a^{Y} (T_L)^q)_- = -4i (pq) a^{X}a^{Y} T_G (T_L)^{p+q-1} \ , \\ \nonumber
& (a^{X} T_G (T_L)^p, a^{Y} (T_L)^q )_- =   \\ \nonumber
&   \ \ \   \frac{1}{p+q} \left[ q(-8q+2) \dep a^{X} a^{Y} +4q(2p+1) a^{X} \dep a^{Y} +2iq D_+ a^{X} D_+a^{Y} \right] (T_L)^{p+q} \ ,\\ \nonumber
& (a^X T_G (T_L)^p, a^Y T_G (T_L)^q)_- =   \\ \nonumber
&  \ \ \  \left[ - (8q+4) \dep a^X a^Y + (8p+4)a^X \dep a^Y + 2i D_+a^X D_+ a^Y \right] T_G (T_L)^{p+q} \ ,
\end{align} 
where the parameter ghosts $a^X$ and $a^Y$ have the appropriate spins and parities (they stand for different ghosts in each bracket!).
 
Thus the algebra closes and is finite, since a high enough power of $T_L$ vanishes. The Jacobi identities don't imply any relations between the currents.
 
\section{$G_2$ manifolds}
 
A $G_2$ manifold has a covariantly constant form $\varphi$ that in an orthogonal basis looks like
\begin{equation}
\varphi = dx^{123} + dx^{145} + dx^{167} + dx^{246} - dx^{257} - dx^{356} - dx^{347} \ ,
\end{equation}
where $dx^{123} = dx^1 \wedge dx^2 \wedge dx^3$, etc.. The other covariantly constant form is the dual (see (\ref{eq:hodge_operator})):
\begin{equation}
* \varphi = dx^{4567} + dx^{2367} + dx^{2345} + dx^{1357} - dx^{1346} - dx^{1247} - dx^{1256} \ .
\end{equation}
We'll normalize the two currents associated with these forms as
\begin{equation}
T_L =  3! \varphi_{ijk} D_+ X^{ijk} \ \ \ \mathrm{and} \ \ \  T_M = 4! *\varphi_{ijkl} D_{+} X^{ijkl}   \ .
\end{equation}

The Jacobi identities imply the following relations:
\begin{itemize}
\item
$(a^M T_M ,  (a^L T_L, a^L T_L )_-  )_-  + \mathrm{cyclic} \equiv 0$ implies
\begin{equation}
\label{eq:GM_relation}
T_G T_M = - \frac{i}{3} T_L D_+ T_L  \ ,
\end{equation}
\item $ (a^M T_M,  (a^M T_M, a^M T_M )_-  )_-  \equiv 0$ implies 
\begin{align}
\dep T_M T_M \equiv 0 \ \ \ , \ \ \ D_+ T_M T_M \equiv 0 \ ,
\end{align}
\item
$(a^L T_L ,  (a^M T_M, a^M T_M )_-  )_-  + \mathrm{cyclic} \equiv 0$ implies
\begin{align}
\label{eq:G2_current_relations2}
& 4 T_M D_+ T_L = -3 D_+ T_M T_L \equiv 0  \ \ \ , \ \ \  4 T_M \dep T_L = -3 \dep T_M T_L \equiv 0 \ ,  \\
& D_+T_L D_+ T_M \equiv 0 \ .
\end{align}
\end{itemize}
All of these imply non-trivial relations between the basic symmetry transformations. Relations (\ref{eq:G2_current_relations2}) are very much like  (\ref{eq:DLLhat_relations}) and  (\ref{eq:DLLhat_relation2}) in the CY case. Also, 
\begin{equation}
T_L T_M \propto Vol \ ,
\end{equation}
analogously to (\ref{eq:LLhat_relation}) in the CY case. The difference is that the CY relations are not implied by the Jacobi identities.

The algebra closes in a field dependent manner as follows:
\begin{align}
& (a^L T_L,  a^L T_L )_- =  - \left[ 18   a^L D_+ a^L \right] T_M \ , \\ \nonumber
& (a^M T_M, a^M T_M)_- = - \left[ 128i  a^M a^M \right] T_G T_M = \frac{123}{3} \left[ a^M a^M \right] T_L D_+ T_L \ , \\ \nonumber
&  (a^L T_L, a^M T_M)_- =  \left[ 96 i   a^L a^M \right] T_G T_L \ .
\end{align}
In the second line I have made use of relation (\ref{eq:GM_relation}).
 
Taking the antibracket of the composite current $T_G T_L$ with the basic currents yields:
\begin{align}
(a^M T_M, a^{GL} T_G T_L)_- = & (a^M T_M, a^{GM} T_G T_M)_- = 0  \ , \\ \nonumber
(a^L T_L, a^{GM} T_G T_M)_- =  &
12 \left[ \dep a^{GM} a^L + a^{GM} \dep a^L - \frac{i}{2} D_+ a^{GM} D_+ a^L \right] T_M T_L \ ,\\ \nonumber
(a^L T_L, a^{GL} T_G T_L)_- = & \frac{1}{3} \left[ -20 D_+a^L a^{GL} - 16a^L D_+ a^{GL} \right] T_G T_M \\ \nonumber
& + \left[ a^{GL} a^L \right] \left(-\frac{16}{3}D_+ T_G T_M -\frac{11}{3} T_G D_+ T_M- 4 \dep T_L  T_L \right) \ .
\end{align}
In the last antibracket a current containing derivatives has been generated; it has been put into primary form.\footnote{It is necessary to make use of relation (\ref{eq:GM_relation}) in showing that this current is primary. A general primary current is of the form
\begin{equation}
A D_+ T_G T_M + B \dep T_L T_L + C T_G D_+ T_M \ ,
\end{equation}
with the three parameters obeying $10A - 6B - 8C=0$.}
 
As in the CY3 case, it is easier to integrate by parts and work with
\begin{equation}
-9 D_+ T_G T_M - 4 \dep T_L T_L \ .
\end{equation}
Taking the antibracket of this current with $T_L$ yields
\begin{align}
& (a^L T_L,  a^X (-9 D_+ T_G T_M - 4 \dep T_L T_L ) )_- =  \\ \nonumber
& \ \ \ 18 \left[ 24 D_+ a^L \dep a^X + D_+\dep a^X a^L + 8 D_+ \dep a^L a^X - 8 D_+ a^X \dep a^L \right] T_M T_L \\ \nonumber
&  \ \ \  +18 \left[ a^X a^L \right] \left( 4 \dep D_+ T_L T_M + 3 D_+ \dep T_M T_L \right) \ .
\end{align}
The current on the second line is obviously primary, but the one on the third line can't be forced into primary form.
 
So, much like the CY3 case, the $G_2$ algebra doesn't seem to be finitely generated.
 
\section{$Spin(7)$ manifolds}
\label{section:spin7}

$Spin(7)$ manifolds are characterized by the existence of a covariantly constant self-dual form $\Psi$, which in an orthonormal frame looks like
\begin{align}
\Psi = & d x^{1234} + dx^{1256} + dx^{1278} +dx^{1357} - dx^{1368} - dx^{1458} - dx^{1467}  \\ \nonumber
& - dx^{2358} - dx^{2367} - dx^{2457} + dx^{2468} + dx^{3456} + dx^{3478} + dx^{5678} \ .
\end{align}
The algebra generated by $T_G$ and the current related to this form,
\begin{equation}
T_{M} = 4! \Psi_{ijkl} D_+ X^{ijkl} \ ,
\end{equation}
closes as a $W$-algebra: 
\begin{equation}
(a^M T_M, a^M T_M)_- = - \left[ 96 i a^M a^M \right] T_G T_M \ ,
\end{equation}
and $T_M$ transforms as a superconformal primary.
 
The antibracket of the composite current $T_G T_M$ with $T_M$ is:
\begin{align}
& (a^M T_M, a^{GM} T_{G} T_{M} )_- = \\ \nonumber
&  \ \ \  \left[ -3 a^M \partial_{++} a^{GM} + 6 a^{GM} \partial_{++} a^M + i D_+ a^{GM} D_+ a^M \right] (T_{M})^2  \ .
\end{align}
$(T_{M})^2$ is an 8-form, and must therefore be proportional to the volume form. Being an $L$-type current, it transforms as a superconformal primary  (\ref{eq:1_L_antibracket}). The current $T_G T_M$ is also a superconformal primary: 
\begin{align}
& (a^G T_G, a^{GM} T_G T_M )_- =  \\ \nonumber
& \ \ \ \left[ -12 \partial_{++} a^G a^{GM} + 4 a^G \partial_{++} a^{GM} + 2i D_+ a^G D_+ a^{GM} \right] T_G T_M \ .
\end{align} 
The rest of the antibrackets vanish.
 
The Jacobi identity doesn't imply any relations between the transformations, a property which $Spin(7)$ shares with  with $SU(n), \  n \geq 5$.  A property shared with $SU(3)$ is that the linearized $Spin(7)$ algebra makes use of all the possible non-derivative currents.  
 
\section{Generalized Nijenhuis forms}
 
In this section we analyze the properties of the generalized Nijenhuis form (\ref{eq:N}), 
\begin{equation}
\label{eq:N2}
N_{[p i_2...i_l j_2...j_m]} \equiv N(L, M) := 2 H_{ab[p} L^a_{ \ i_2 ... i_l} M^b_{ \ j_2 ... j_m]} - \frac{1}{3} H_{[p i_2 j_2} U_{i_3 ... i_l j_3 ... j_m]} \ ,
\end{equation}
for all the holonomy groups that appear in the Berger list. The reader is reminded that for $l=m$ and $l$ odd $N$ vanishes. Since $N$ is covariantly constant, in general we can expect a reduction of the $G$-structure. In what follows it is shown that this only happens for the complex structure Nijenhuis form and in the $Sp(k)$ case. For $Sp(k)\cdot Sp(1)$ we show that $N$ vanishes for $D \leq 8$, and is non-zero for $D=12$. In other cases $N$ either vanishes, or is built up from the original covariantly constant forms.
 
\subsection{$U(m);\ D=2m$}

The holonomy group is associated with an almost complex structure $I$, and we have the usual Nijenhuis form. This case has been studied in detail in \cite{Delius:1989fy}. In a unitary frame basis the torsion can be decomposed into $(3,0)$ and $(2,1)$ components, together with their complex conjugates. It is easy to see that the Nijenhuis form is proportional to the $(3,0)$ plus $(0,3)$ part of $H$.\footnote{This provides another way of seeing that it is covariantly constant, as the curvature tensor is pure on its Lie algebra indices, and therefore mixed when one of them is lowered.} Although the Nihenhuis form is identically covariantly constant, it still implies a further reduction of the structure group. In particular, for $m=3$ the structure group is automatically $SU(3)$.

\subsection{$SU(m); D=2m$}
\label{subsection:SU2}

In addition to an almost complex structure, the $SU(m)$ holonomy group is associated with two real $m$-forms, $L$ and $\hL$, which are the real and imaginary parts of a complex $(m,0)$ form (see \ref{eq:TL_TM_def}). In this case we have a number of possible Nijenhuis forms, $\tN(I,L)$, $\tN(I,\hL)$, $\tN(L,L)$, $\tN(\hL,\hL)$, $\tN(L,\hL)$, as well as $\tN(I,I)$ which we will write as $N^{(I)}$. Apart from $N^{(I)}$ it turns out that the only non-vanishing ones are $\tN(I,L)$ and $\tN(I,\hL)$, and these are given in terms of $N^{(I)}$, $L$, and $\hL$. Therefore the only further reduction of the structure group is due to the presence of the $(3,0)+(0,3)$ form $N^{(I)}$.

As an example we sketch the proof that $\tN(L,L)$ vanishes for $m$ even (it is identically zero for $m$ odd). This is a $(2m-1)$-form so that it is convenient to look at its one-form dual. The $(2m-4)$-form $U$ is proportional to $I^{m-2}$ in this case, so that the dual of the second term in $\tN$ (\ref{eq:N2}) is
proportional to $I^2_{ijkl} H^{jkl}$. To evaluate the dual of the first term we use the fact that $L$ is self-dual for $m$ even to arrive at an expression of the form $L_{ij}^{ \ \ \ p_1\ldots p_{m-2}} L_{kl p_1\ldots p_{m-2}}H^{jkl}\propto I^2_{ijkl} H^{jkl}$. A careful evaluation shows that the two terms cancel, as indeed they must since $I^2_{ijkl} H^{jkl}$ is not covariantly constant in general. The vanishing of $\tN(\hL,\hL)$ and $\tN(L,\hL)$ can be verified in a similar fashion.

Now consider $\tN(I,L)$. In this case $U=0$ since it involves the double contraction of $I$ and $L$ which are of different type with respect to the almost complex structure. It is again easier to look at the dual, which is an $l$-form, with $l = m-1$. We find, for $m$ even,
\be
 *\tN_{i_1 \ldots
 i_l}=\frac{1}{m}\left(lH_{[i_1}{}^{jk}\hL_{j k i_2\ldots i_l]} + L_{k
 i_1\ldots i_l} I^{pq} H_{pq}{}^k\right)\ .
 \la{3.27}
\ee
Because $L$ and $\hL$ are both of type $(m,0)+(0,m)$, it follows that this expression can be either $(l,0)$, or $(l-1,1)$, or complex conjugates. It is not difficult to verify that the $(l,0)$ part vanishes and this implies that only the $(3,0)+(0,3)$ components of $H$ contribute. But this is just $N^{(I)}$, so we find
\be
 *\tN_{i_1\ldots i_l}=\frac {l}{4(l+1)}N^{(I)}_{[i_1}{}^{jk}\hL_{j k i_2\ldots
 i_l]}\ .
 \la{3.28}
\ee

Similar expressions can be derived for $m$ odd and for $\tN(I,\hL)$. These forms, although non-zero, are generated from the original set together with $N^{(I)}$ so that there is no further reduction of the structure group.

\subsection{$Sp(k);\ D=4k$}

Target spaces of this type could be called almost \hkah manifolds. There is a set of Nijenhuis three-forms given by
\be
 N^{rs}_{ijk}=\d^{rs}H_{ijk}-3 H_{lm[i}(I^r)^l{}_j (I^s)^m{}_{k]}\ ,
 \la{3.29}
\ee
where $\{I^r\}$ is a set of three almost complex structures obeying the algebra of the imaginary quaternions (\ref{eq:hyperkahler_cpx_structures}).

These forms do not vanish. One way of understanding their content is to write a real vector index $i=1\ldots 4k$ as a pair $i\rightarrow \a i$, where now $i=1\ldots 2k$ and $\a=1,2$. The latter index is acted on by the rigid $Sp(1)$ while the former is acted on by $Sp(k)$. In this notation a general three-form $H$ can be written
\be
 H_{ijk}\rightarrow H_{\a i\b j\c k}= H_{(\a\b\c) [ijk]} + \e_{\a\c}
 H'_{\b jki} + \e_{\b\c} H'_{\a kij}\ .
 \la{3.30}
\ee
The $H$-tensor on the right has the indicated symmetries while $H'_{ijk}$ is antisymmetric on the first two indices with the totally antisymmetric part being zero. In the Nijenhuis forms, this part of $H$ drops out and so they are determined by $H_{\a\b\c ijk}$. In detail,
\be
 N^{rs}_{\a i\b j\c k}= (\s^r)_{(\a}{}^{\d} (\s^s)_{\b}{}^{\e}
 H_{\c)\d\e ijk}\ .
 \la{3.31}
\ee

\subsection{$Sp(k)\cdot Sp(1);\ D=4k$}

Manifolds of this type can be called almost quaternionic \kah spaces. The only Nijenhuis form is the $7$-form $\tN(L,L)$ arising from the four-form $L$. In $D=8$ one can show that it is identically zero. This can be proven using the index pair notation of the previous section; for $k=2$, i.e. in dimension $8$, the dual can only be proportional to $L_{ijkl} H^{jkl}$. Since this object is not covariantly constant the constant of proportionality must be zero. For $D=12$ one can show that the dual \emph{is} covariantly constant, so there is no reason to assume that $\tN(L,L)$ vanishes in general. In higher dimensions calculations become prohibitively difficult.

\subsection{$Spin(7)$ and $G_2$}

For $Spin(7)$ the only possible Nijenhuis form, $\tN(\Psi, \Psi)$, vanishes. As in the previous example, its dual could only be $\Psi_{ijkl} H^{jkl}$ and this is not covariantly constant in general. 

For $G_2$ there is a three-form $\varphi$ and its dual four-form $*\varphi$, and the only non-zero Nijenhuis form is $\tN( \varphi, *\varphi)$. This a seven-form which is equal to the volume form multiplied by a constant times $\varphi_{ijk} H^{ijk}$. It is easy to see that this function is constant due to the structure of the $\gg_2$ Lie algebra.
 
\section{Discussion}
\label{section:spechol_discussion}
 
We have seen that all the special holonomy algebras can be linearized, except $G_2$ and $SU(3)$, which are the cases of most physical interest.  Thus we can attempt a cohomological analysis (see \ref{sec:anomalies}) in all except these two cases. There are a few other points we potentially need to worry about, listed in the setting of a general theory on page \pageref{page:list_of_things_to_worry_about}.

Regarding the restoration of normalization conditions, in the context of string theory the desired condition is that the superconformal anomaly vanishes. We know that there are relations between the superconformal and the non-linear currents for the $SU(N)$ (\ref{eq:GL_relation}) and $G_2$ (\ref{eq:GM_relation}) cases. Excluding the possibility that there is a genuine anomaly in the full algebra, one would expect the spurious anomalies in the $L$-type transformations to be related to the superconformal anomaly, in which case it would be consistent to keep the desired normalization condition. 

One would expect that anomalies can also be related in the Ricci flat cases when there are no relations between currents at the classical level; one possible mechanism would be via operator mixing. We know, for example, that it occurs for $Spin(7)$ \cite{Figueroa-O'Farrill:1996hm}: in order to close the OPE one has to consider a combination of $T_G$ and $T_L$. Since for $Spin(7)$ there are no relations between $T_G$ and $T_L$, perhaps this sort of mixing is necessary in order for the conformal anomaly to be related to anomalies in the non-linear symmetries. However, there is no reason to expect that operator mixing can affect the calculation of possible anomalies in the invariance of the action.


Regarding the issue of non-renormalizable sources, a priori one would expect that local terms nonlinear in the antifields are generated by quantum corrections. For the special holonomy $\sigma$-models this doesn't happen because Lorentz invariance forbids such terms from being generated. 

When computing the path integral of the $d=2$ $\sigma$-model it is of great benefit for the Green functions to be manifestly covariant. To achieve this a non-linear split between the background and quantum fields is necessary.  There is a non-linear symmetry related to this split, and  the Ward identities of this symmetry are anomaly free \cite{Howe:1986vm, Blasi:1988sh}. When introducing extra symmetries one should in principle check that they are compatible with the non-linear splitting Ward identity. Provided this works, one can calculate possible anomalies using the usual BV methods. Although a full analysis would involve the above steps it is nevertheless not unreasonable to look at the potential anomalies of the background field effective action with no external quantum lines \cite{Howe:2006si}. If one makes a special holonomy transformation of the background field accompanied by the appropriate transformation of the quantum field, the local action in the path integral is invariant and the change in the quantum field can be absorbed by a field redefinition in the path integral.\footnote{The statement in \cite{Howe:1986vm} that  non-linear symmetries in the original $\sigma$-model fields linearize in terms of the quantum fields holds only for symmetries related to Killing vectors in the target space. $L$-type symmetries are not of this type, and remain non-linear when written in terms of the quantum field. In fact, a generic CY manifold doesn't have any Killing vectors.}

The reason why it's so hard to get a handle on CY3 and $G_2$ manifolds remains somewhat mysterious. There are a few points that should be mentioned here. 
\begin{itemize}
\item These are the only linearized special holonomy algebras generating non-primary fields at the \emph{classical} level - other algebras do so only via quantum effects in the OPE \cite{Figueroa-O'Farrill:1996hm}.  In the CY case the generation of non-primary fields should be  related to the result that the $T_M$ and $T_{\bM}$ currents generally  map primary fields into non-primary ones (see \cite{Gato-Rivera:1995kd}). 
\item $SU(3)$ structure comes out naturally via the Nijenhuis tensor when the latter doesn't vanish (see \ref{subsection:SU2}). But if the torsion isn't constrained one has the same problems with linearizing the algebra as for CY3.\footnote{There is a classification of torsion classes \cite{chiossi-2001-2002, LopesCardoso:2002hd, Grana:2005jc}, but none of the cases simplify the algebra in a non-trivial way. The nearly-\kah case comes close, but the metric is non-zero only if we require that $dH \neq 0$. This happens for the Heterotic case, but 
\begin{equation}
dH = \alpha'  \mathrm{Tr} (R \wedge R - F \wedge F) \ ,
\end{equation}
which implies the metric itself has to be of order $\alpha'$, and so perturbative analysis breaks down.}
\item The work of Bonneau  \cite{Bonneau:1994mm} indicates that there is a potential anomaly in the rigid $N=2$ algebra only when the background is CY3.
\end{itemize}

An interesting observation is that the $CY$ algebras have linear subalgebras generated by the currents $T_G$, $T_I$, and $T_M$ (see \ref{sec:cy_algebras_in_general}). These subalgebras can clearly be analyzed in all cases, including CY3. It is  the complex currents that are related to the squares of the spectral flow operator by $\pm 1$ \cite{Gato-Rivera:1995kd}, and therefore it seems that commuting the spectral flows in two directions is related to the complications that occur for CY3.\footnote{The analogous subalgebra in the $G_2$ case is the tri-critical Ising model \cite{Shatashvili:1994zw}.}

Finally I would like to mention that $T_M$ is a chiral supercurrent, and as such one may expect that it is protected, analogously to short operators in four dimensional superconformal field theories. We are not aware at the moment of an argument that this is so in $d=2$, but it  would be consistent with results from the CFT side that the special holonomy algebras are unique.\footnote{They are determined by conformal bootstrap techniques. See  \cite{Blumenhagen:1992sa, Blumenhagen:1994wn}.}

The conclusion is that the linearized classical special holonomy algebras enable a direct application of cohomological techniques to anomalies in the invariance of the action. The calculation of potential anomalies in the algebra is more involved, with quite a few additional issues needing consideration. Expressing the OPE via an extended action involves gauging the symmetries, and will be discussed in more detail  in Chapter \ref{chapter:W-strings} (see \ref{section:W-string_discussion}).
 

\chapter{Special holonomy $\sigma$-models with boundaries}
\label{chapter:boundary}


We shall discuss two-dimensional $(1,1)$ supersymmetric $\sigma$-models with boundaries, with extra symmetries corresponding to covariantly constant tensors in the target space, focusing in particular on the cases of special holonomy (see Chapter \ref{chapter:algebras_on_manifolds_of_sh}).\footnote{This chapter has virtually the same content as the publication by Paul Howe, Ulf Lindstr\"{o}m, and myself \cite{howe-2006-0601}.} In a series of papers \cite{Albertsson:2001dv, Albertsson:2002qc, Lindstrom:2002jb, Lindstrom:2002vp, Albertsson:2003va} classical supersymmetric $\sigma$-models with boundaries have been discussed in detail and it has been shown how the fermionic boundary conditions involve a locally defined tensor $R$ which determines the geometry associated with the boundary. In particular, in the absence of torsion, one finds that there are integral submanifolds of the projector $P=\half (1+R)$ which have the interpretation of being branes where the boundary can be located. These papers considered $(1,1)$ and $(2,2)$ models and the analysis was also extended to models of this type with torsion where the intepretation of $R$ is less straightforward. The main purpose of this chapter is to further extend this analysis to include symmetries associated with certain holonomy groups or $G$-structures. We shall discuss models both with and without torsion.

Torsion-free $\sigma$-models with boundaries on manifolds with special holonomy were first considered in \cite{Becker:1996ay} where it was shown how the identification of the left and right currents on the boundary has a natural interpretation in terms of calibrations and calibrated submanifolds. Branes have also been discussed extensively in boundary CFT \cite{Schomerus:2002dc}, including the $G_2$ case \cite{Roiban:2001cp}, and in topological string theory \cite{Bershadsky:1995qy}.

The main new results in this chapter  concern boundary  $(1,1)$ models with torsion or with a gauge field on the brane. There is no analogue of Berger's list in the case of torsion but we can
nevertheless consider target spaces with specific $G$-structures which arise due to the presence of covariantly constant forms of the same type. In order to generalise the discussion from the torsion-free case we require there to be two independent $G$-structures specified by two sets of covariantly constant forms $\{\l^+,\l^-\}$ which are covariantly constant with respect to two metric connections $\{\C^+,\C^-\}$ and which have closed skew-symmetric torsion tensors $T^{\pm}=\pm H$, where $H=d b$, $b$ being the two-form potential which appears in the $\sigma$-model action (\ref{eq:ssm_action2}).  This sort of structure naturally generalises the notion of bi-Hermitian geometry which occurs in $N=2$ $\sigma$-models with torsion \cite{Gates:1984nk,Howe:1991ic} and which has been studied in the boundary $\sigma$-model context in \cite{Lindstrom:2002jb}. We shall refer to this type of structure as a bi-$G$-structure. The groups $G$ which are of most interest from the point of view of spacetime symmetry are the groups which appear on Berger's list and for this reason we use the term special holonomy. Bi-$G$-structures are closely related to the generalized structures which have appeared in the mathematical literature \cite{Hitchin:2004ut,gualtieri-2004-,Witt:2004vr}. These generalized geometries have been discussed in the $N=2$ $\sigma$-model context \cite{Lindstrom:2004iw,Kapustin:2004gv,Zabzine:2004dp}. In a recent paper they have been exploited in the context of branes and generalized calibrations.

We shall show that, in general, the geometrical conditions implied by equating the left and right currents on the boundary lead to further constraints by differentiation and that these constraints are the same as those which arise when one looks at the stability of the boundary conditions under symmetry transformations. It turns out, however, that these constraints are automatically satisfied by virtue of the target space geometry.

We then study the target space geometry of some examples, in particular bi-$G_2$, bi-$SU(3)$, and bi-$Spin(7)$ structures. Structures of this type have appeared in the supergravity literature in the context of supersymmetric solutions with flux \cite{Gauntlett:2002sc,Grana:2004bg,Grana:2005ny}.

This chapter is organized as follows: in \ref{section:review_of_basics} we review the basics of boundary $\sigma$-models, in \ref{section:L_type_sym_on_boundary} we discuss additional symmetries associated with special holonomy groups or bi-$G$-structures, in \ref{section:consistency} we examine the consistency of the boundary conditions under symmetry variations, in \ref{section:target_space_geometry} we look at the target space geometry of bi-$G$ structures, and in \ref{section:examples_of_solutions} we look at some examples of solutions of the boundary conditions for the currents defined by the covariantly constant forms.

\section{Review of basics}
\label{section:review_of_basics}

We write the $(1,1)$ supersymmetric  $\sigma$-model action (\ref{eq:ssm_action2}) as
\begin{equation}
S=\int  d^2 z  e_{ij} D_+ X^i D_- X^j \ , \label{2.1}
\end{equation}
where
\begin{equation}
e_{ij}:=G_{ij}+b_{ij} \ , \label{2.2}
\end{equation}
$b$ being a two-form potential with field strength $H=db$ on the $n$-dimensional Riemannian target space $(\M,G)$. As well as the usual Levi-Civita connection $\nab$, there are two natural metric connections $\nab^{\pm}$ with torsion (\ref{eq:connection_coefficients}),
\begin{equation}
\Gamma^{i(\pm)}_{ \ jk} = \Gamma^i_{ \ jk} \pm \frac{1}{2} H^i_{\ jk} \ . \label{2.5}
\end{equation}
The torsion tensors of the two connections are given by
\begin{equation}
T^{i(\pm)}_{ \ jk}=\pm H^i{}_{jk}\ , \label{2.6}
\end{equation}
so that the torsion is a closed three-form in either case.

In the presence of a boundary, $\del\S$, it is necessary to add additional boundary terms to the action \eq{2.2} when there is torsion \cite{Albertsson:2002qc}. The boundary action is
\begin{equation}
S_{\mathrm{bdry}}= \int_{\del\S}\, a_i \dot X^i +\frac{i}{4}b_{ij}(\psi_+^i
\psi_+^j +\psi_-^i\psi_-^j) \label{2.6.1}\ ,
\end{equation}
where $a_i$ is a gauge field  which is defined only on the submanifold where the boundary $\sigma$-model field maps takes its values. Note that the boundary here is purely bosonic so that the fields are component fields, $\psi_{\pm}^i:=D_{\pm} X^i|$, the vertical bar denoting the evaluation of a superfield at
$\th=0$).\footnote{In what follows we shall use $X^i$ to mean either the superfield or its leading component; it should be clear from the context which is meant.} The boundary term ensures that the action is unchanged if we add $d c$ to $b$ provided that we shift $a$ to $a-c$. The modified field strength $F=f +b$, where $f=da$,  is invariant under this transformation. In the absence of a $b$-field one can still have a gauge field on the boundary.

In the following we briefly summarise the approach to boundary $\sigma$-models of references
\cite{Albertsson:2001dv, Albertsson:2002qc, Lindstrom:2002jb, Lindstrom:2002vp, Albertsson:2003va}. We impose the standard boundary conditions \cite{Callan:1988wz} on the fermions,
\begin{equation}
\psi_-^i =\h R^i{}_j \psi_+^j \ \ \ , \ \ \  \h=\pm1 \ \ \  , \ \ \  {\rm on}\
\del\S  \ . \label{2.7}
\end{equation}
We shall also suppose that there are both Dirichlet and Neumann directions for the bosons. That is, we assume that there is a projection operator $Q$ such that
\begin{equation}
Q^i{}_j \d X^j = Q^i{}_j \dot X^j=0 \ , \label{2.7.2}
\end{equation}
on $\del\S$. If $F=0$, parity implies that $R^2=1$, so that $Q=\half (1-R)$, while $P:=\half (1+R)$ is the complementary projector. In general, we shall still use $P$ to denote $\half (1+R)$ and the complementary projector will be denoted by $\p,\ \p:=1-Q$. We can take $Q$ and $\p$ to be orthogonal
\begin{equation}
\p^k_i G_{kl} Q^l{}_j=0 \ . \label{2.8}
\end{equation}

Equation \eq{2.7.2} must hold for any variation along the boundary. Making a supersymmetry transformation we find
\begin{equation}
QR + Q=0\ . \label{2.9}
\end{equation}
On the other hand, the cancellation of the fermionic terms in the boundary variation (of $S+ S_{\mathrm{bdry}}$), when the bulk equations of motion are satisfied, requires
\begin{equation}
G_{ij}=G_{kl} R^k{}_i R^l{}_j \ . \label{2.18}
\end{equation}
Using this together with orthogonality one deduces the following algebraic relations,
\begin{eqnarray}
QR &=& RQ= -Q \ \ \ , \ \ \   QP= PQ =0 \ , \nn \w2 \p P&=& P \p =P \ \ \ , \ \ \
\p R= R \p \ . \label{2.10}
\end{eqnarray}

Making a supersymmetry variation of the fermionic boundary condition \eq{2.7} and using the equation of motion for the auxiliary field, $F^i:=\nab^\pl_- D_+ X^i|$, namely $F^i=0$, we find the bosonic boundary condition\footnote{The occurrence of (combinations of) field equations as boundary conditions is discussed in \cite{Lindstrom:2002mc}.}
\begin{equation}
i(\del_{--} X^i - R^i{}_j \del_{++} X^j) = (2 \tnab_j R^i{}_k
-P^i{}_l H^l_{jm}R^m{}_k) \psi_{+}^j\psi_+^k \ , \label{2.11}
\end{equation}
where $\tnab$ is defined by
\begin{equation}
\tnab_i:=P^j{}_i \nab_j\ . \label{2.12}
\end{equation}

Combining \eq{2.11} with the bosonic boundary condition arising directly from the variation we find
\begin{equation}
\hE_{ji} = \hE_{ik} R^k {}_j\ , \label{2.12.1}
\end{equation}
where
\begin{equation}
E_{ij}:= G_{ij} + F_{ij} \ ,  \label{2.12.2}
\end{equation}
and the hats denote a pullback to the brane,
\begin{equation}
\hE_{ij}:=\p^k{}_i \p^l{}_j E_{kl} \ . \label{2.12.3}
\end{equation}
From \eq{2.12.1} we find an expression for $R$,
\begin{equation}
R^i{}_j=(\hE^{-1})^{ik} \hE_{jk}- Q^i{}_j \ , \label{2.12.4}
\end{equation}
where the inverse is taken in the tangent space to the brane, i.e.
\begin{equation}
(\hE^{-1})^{ik} \hE_{kj}=\p^i{}_j\ . \label{2.12.5}
\end{equation}
We can multiply equation \eq{2.11} with $Q$ to obtain
\begin{equation}
P^l{}_{[i} P^m{}_{j]}\nab_l Q^k{}_m=0\ . \label{2.13}
\end{equation}
Using \eq{2.10} we can show that this implies  the integrability condition for $\p$,
\begin{equation}
\p^l{}_{[i} \p^m{}_{j]}\nab_l Q^k{}_m=0\ . \label{2.14}
\end{equation}

This confirms that the distribution specified by $\p$ in $T \M$ is integrable and the boundary maps to a submanifold, or brane, $B$. However, in the Lagrangian approach adopted here, this is implicit in the assumption of Dirichlet boundary conditions. When $F=0$ the derivative of $R$ along the brane is essentially the second fundamental form, $K$. Explicitly,
\begin{equation}
K^i_{jk}=P^l{}_{j} P^m{}_{k}\nab_l Q^i{}_m = P^l{}_j\tnab_k Q^i{}_l
 \ . \label{2.14.1}
\end{equation}

The left and right supercurrents are (\ref{eq:superconformal_current_plus}) and (\ref{eq:superconformal_current_minus}),
\begin{eqnarray}
T_{G+++}:&=&G_{ij} \del_{++}X^i D_+ X^j -\frac{i}{6} H_{ijk} D_{+3}
X^{ijk} \ , \w1 T_{G---}:&=&G_{ij} \del_{--}X^i D_- X^j +\frac{i}{6}
H_{ijk} D_{-3} X^{ijk} \ . \label{2.15}
\end{eqnarray}
The conservation conditions are
\begin{equation}
D_-T_{G+++}=D_+ T_{G---}=0 \ , \label{2.16}
\end{equation}
on-shell. The superpartners of the supercurrents are the left and right components of the energy-momentum tensor, $D_+ T_{G+++}$, and $D_-T_{G---}$ respectively. If one demands invariance of the total action under supersymmetry one finds that, on the boundary, the currents are related by
\begin{eqnarray}
T_{G+++} &=& \h T_{G---} \ , \label{2.16.1}\w1 D_+ T_{G+++}&=& D_- T_{G---} \ .
\label{2.17}
\end{eqnarray}
The supercurrent boundary condition has a three-fermion term which implies the vanishing of the totally antisymmetric part of
\begin{equation}
2Y_{i : jk} + P^l{}_i H_{ljm} R^m{}_k +\frac{1}{6}(H_{ijk} + H_{lmn}
R^l{}_i R^m{}_j R^n{}_k)\ , \label{2.19}
\end{equation}
where
\begin{equation}
Y_{i:jk}:= (R^{-1})_{jl} \tnab_i R^l{}_k \ , \label{2.20}
\end{equation}
and $:$ is being used to indicate that the indices to the right and left are being distinguished according to their symmetry properties.

\section{$L$-type symmetries on the boundary}
\label{section:L_type_sym_on_boundary}

A general variation of \eq{2.1}, neglecting boundary terms, gives
\begin{eqnarray}
\d S&=&\int d^2 z 2G_{ij}\d X^i \nab^{(+)}_- D_+ X^j \nn \w1
&=&-\int d^2 z\, 2G_{ij}\d X^i G_{ij} \nab^{(-)}_+ D_- X^j \ .
\label{3.1}
\end{eqnarray}

The additional symmetries we shall discuss are L-type symmetries (see Chapter \ref{chapter:algebras_of_general_L_symmetries})\footnote{In this chapter we work with conventional parameters with correct spin-statistics, \emph{not} ghostly parameters. Also, the indices are labeled slightly differently than in Chapter \ref{chapter:algebras_of_general_L_symmetries}: what was $l$ there is $l+1$ here.},
\begin{equation}
\d_{\pm} X^i= a^{\pm l} L^{i(\pm)}_{ \ j_1\ldots j_{l}}
D_{\pm l} X^{j_1\ldots j_{l}}\ ,\qquad  D_{\pm l}
X^{j_1\ldots j_{l}}:= D_{\pm} X^{j_1}\ldots D_{\pm} X^{j_{l}}\
, \label{3.2}
\end{equation}
where $L^{(\pm)}$ are vector-valued ${l}$-forms such that
\begin{equation}
\l^{\plmi}{}_{i_1\ldots i_{{l}+1}}:= G_{i_1 j} L^{j (\pm) }_{ \ i_2\ldots i_{{l}+1}} \label{3.3}
\end{equation}
are $(l+1)$-forms which are covariantly constant with respect to $\nab^{(\pm)}$. We shall be  interested to extending $L$-type symmetries to the boundary.

As shown in \ref{section:geometry_and_symmetries},  a left transformation of this type gives
\begin{eqnarray}
\d S&=&\int d^2 z  2 a^{+l}\l^{\pl}{}_{i_1\ldots i_{{l}+1}}
D_{+l}X^{i_2\ldots i_{l+1}} \nab^{(+)}_- D_+ X^{i_1} \nn \w1
&=&\int d^2 z \frac{2}{l+1}a^{+l}\l^{\pl}{}_{i_1\ldots
i_{{l}+1}} \nab^{(+)}_- D_{+(l+1)} X^{i_1\ldots i_{l+1}}
\nn \w1 &=&\int d^2 z  (-1)^{l}
D_-\left(\frac{2}{l+1}a^{+l}\l^{\pl}{}_{i_1\ldots
i_{{l}+1}} D_{+(l+1)} X^{i_1\ldots i_{l+1}}\right) \ ,
\label{3.4}
\end{eqnarray}
where the last step follows from covariant constancy of $\l^\pl$ and the chirality of the parameters,
\begin{equation}
D_- a^{+l}=D_+ a^{-l}=0\ . \label{3.5}
\end{equation}

Hence these transformations are symmetries of the $\sigma$-model without boundary. In the torsion-free case the $\l$s will be the forms which exist on the non-symmetric Riemannian manifolds on Berger's list (see \ref{section:spechol_manifolds}). There is no such list in the presence of torsion but the same forms will define reductions of the structure group to the various special holonomy groups. In order to preserve the symmetry on the boundary we must have both left and right symmetries so there must be two independent such reductions. Thus we can say that we are interested in boundary $\sigma$-models on manifolds which have bi-$G$-structures.

The $\l$-forms can be used to construct currents $L^{(\pm)}_{\pm (l +1)}$ (\ref{eq:conserved_L_current}),
\begin{equation}
L^{(\pm)}_{\pm (l +1)}:=\l^{(\pm)}{}_{i_1\ldots i_{{l}+1}}
D_{\pm(l+1)} X^{i_1\ldots i_{l+1}} \label{3.6}
\end{equation}
If we make both left and right transformations of the type \eq{3.2} we obtain
\begin{eqnarray}
\d S&=&\frac{2(-1)^{l}}{l+1}\int  d^2 \sig D_+ D_-
\left(D_-(a^{+l}L^{(+)}_{\pm (l +1)})- D_+(a^{-l}L^{-}_{- (l +1)})\right) \nn\w2
&=&\frac{i(-1)^{l+1}}{l+1}\int_{\del\S} 
\left(D_+(a^{+l}L^{(+)}_{\pm (l +1)})- D_-(a^{-l}L^{-}_{- (l +1)})\right) \ . \label{3.6.1}
\end{eqnarray}
In order for a linear combination of the left and right symmetries to be preserved in the presence of a boundary, the parameters should be related by
\begin{eqnarray}
a^{+{l}} &=& \h_L a^{-{l}}\ , \w1 D_+ a^{+{l}} &=& \h \h_L D_-a^{-{l}}\ , \label{3.9}
\end{eqnarray}
on the boundary, where $\h_L=\pm 1$.\footnote{In the case that there is one pair of $L$ tensors.} This implies that the currents and their superpartners should satisfy the boundary conditions
\begin{eqnarray}
L^{(+)}_{+({l}+1)}&=& \h\h_L L^{(-)}_{-({l}+1)}\label{3.7}
\w1 D_+ L^{(+)}_{+({l}+1)}&=& \h_L D_- L^{(-)}_{-({l}+1)}\ .
\label{3.8}
\end{eqnarray}
The boundary condition \eq{3.7} implies
\begin{equation}
\l^{(+)}{}_{i_1\ldots i_{{l}+1}}=\h_L \h^{l}
\l^{(-)}{}_{j_1\ldots j_{{l}+1}}R^{j_1}{}_{i_1}\ldots
R^{j_{{l}+1}}{}_{i_{{l}+1}}\ . \label{3.10}
\end{equation}

\section{Consistency}
\label{section:consistency}

In this section we shall examine the consistency of the boundary conditions, i.e we investigate the orbits of the boundary conditions under symmetry variations to see if further constraints arise. We shall show that the supersymmetry variation of the $L$-boundary condition \eq{3.7} and the $L$-variation of the fermion boundary condition \eq{2.7} are automatically satisfied if \eq{3.10} is. 

To see this we differentiate \eq{3.10} along $B$ to obtain
\begin{equation}
Y^{(+)}{}_{k:[i_1}{}^{m} \l^{(+)}{}_{i_2\ldots i_{{l}+1}]m}=0\ ,
\label{4.1}
\end{equation}
where
\begin{equation}
Y^{(+)}{}_{i:jk}:= (R^{-1})_{jl}( \tnab^{(+)}_i R^l{}_k - H^l{}_{im}
R^m{}_k )\ . \label{4.2}
\end{equation}
Note that we have contracted the derivative with $P$ rather than $\p$; this is permissible due to the fact that $P\p=\p P=P$. Equation \eq{4.1} says that $Y^{(+)}$, regarded as a matrix-valued one-form, takes its values in the Lie algebra of the group which leaves the form $\l^{(+)}$ invariant. The constraint corresponding to the superpartner of the $L$-current boundary condition is just the totally antisymmetric part of \eq{4.1}.

We now consider the variation of the fermionic boundary condition under $L$-transformations. We need to make both left and right transformations, which together can be written
\begin{eqnarray}
\d X^i &=& 2 a^{+{l}} P^i{}_k L^{k (+)}_{ \ j_1\ldots j_{l}}
D_{+{l}} X^{j_1\ldots j_{l}}\w1 &=& 2 a^{-{l}} P^i{}_k
L^{k (-)}_{ \ j_1\ldots j_{l}} D_{-{l}}X^{j_1\ldots j_{l}}\
. \label{4.2.1}
\end{eqnarray}
A straightforward computation yields
\begin{equation}
(2\tnab_{[k} R^i{}_{m]}-P^i{}_n H^n{}_{[k|p|} R^p{}_{m]} )
L^{k(+)}_{ \ j_1\ldots j_{l}}D_{+({l}+1)}X^{j_1\ldots
j_{l} m} =0 \ . \label{4.3}
\end{equation}

We define
\begin{equation}
Z^{(+)}_{i:jk}= (R^{-1})_{il}(2\tnab_{[j} R^l{}_{k]}+P^l{}_m
H^m{}_{n[j}R^n{}_{k]} )\ , \label{3.24}
\end{equation}
which is the term in the bracket in \eq{4.3} multiplied by $R^{-1}$. We claim that
\begin{equation}
Y^{(+)}_{i:jk}=Z^{(+)}_{i:jk}\ . \label{3.25}
\end{equation}
This can be proved using  \eq{2.19} with the aid of a little algebra. Thus we have shown that, if the boundary conditions \eq{3.10} are consistent, then the constraints following from supersymmetry variations of the $L$-constraints and from $L$-variations of the fermionic boundary condition are guaranteed to be satisfied.

If $\l^\pl=\l^\mi:=\l$ the boundary condition \eq{3.10} typically implies that $\pm R$ is an element of the group which preserves $\l$. If this is the case, then \eq{4.1} becomes an identity. However, it can happen that $R$ is not an element of the invariance group but that $R^{-1}dR$ still takes its values in the corresponding Lie algebra. For example, if $\l$ is the two-form of a $2m$-dimensional K\"ahler manifold and the sign $\h_L\h=-1$, $R$ is not an element of the unitary group but, since it must have mixed indices, it is easy to see that $R^{-1}d R$ is itself $\gu(m)$-valued.

A similar argument applies in the general case, when $\l^\pl\neq\l^\mi$. In the next section we discuss how the plus and minus forms are related by an element $V$ of the orthogonal group (see \eq{5.2.1}). Thus equation \eq{3.10} can be written
\begin{equation}
\l^{(-)}{}_{i_1\ldots i_{{l}+1}}=\h_L \h^{l}
\l^{(-)}{}_{j_1\ldots j_{{l}+1}}\hR^{j_1}{}_{i_1}\ldots
\hR^{j_{{l}+1}}{}_{i_{{l}+1}}\ , \label{3.10.1}
\end{equation}
where $\hR:=R V^{-1}$. If we differentiate $\eq{3.10.1}$ along the brane with respect to the minus connection we can then use the above argument applied to $\hR$.

\section{Target space geometry}
\label{section:target_space_geometry}

In this section we discuss the geometry of the $\sigma$-model target space in the presence of torsion when the holonomy groups of the torsionfull connections $\nab^\plmi$ are of special type, specifically $G_2,\, Spin(7)$, and $SU(3)$. We use only the data given by the $\sigma$-model and use a simple approach based on the fact that there is a transformation which takes one from one structure to the other. We begin with $G_2$ and then derive the other two cases from this by dimensional reduction and oxidation.

\subsection{$G_2$}

In this case we have a seven-dimensional Riemannian manifold $(\M,G)$, with two $G_2$-forms $\vf^\plmi$ which are covariantly constant with respect to left and right metric connections $\nab^\plmi$ such that the torsion tensor is $\pm H$. $G_2$ manifolds with torsion have been studied in the mathematical literature \cite{Friedrich:2001nh,Friedrich:2001yp} and have arisen in supergravity solutions \cite{Gauntlett:2002sc}. Bi-$G_2$-structures have also appeared in this context and have been given an interpretation in terms of generalized  $G_2$-structures \cite{Witt:2004vr}. They can be studied in terms of a pair of covariantly constant spinors from which one can construct the $G_2$-forms, as well as other forms, as bilinears. We will not make use of this approach here, preferring to use the tensors given to us
naturally by the $\sigma$-model. As noted in \cite{Gauntlett:2002sc} there is a common $SU(3)$ structure associated with the additional forms. We shall derive this from a slightly different perspective here.

In most of the literature use is made of the dilatino Killing spinor equation which restricts the form of $H$. The classical $\sigma$-model does not require this restriction as the dilaton does not appear until the one-loop level. The dilatino equation is needed in order to check that one has supersymmetric supergravity solutions but is not essential for our current purposes.

For $G_2$ there are two covariantly constant forms, the three-form $\vf$ and its dual four-form $*\vf$ (we shall drop the star when using indices). The metric can be written in terms of them. A convenient choice for $\vf$ is
\begin{equation}
\vf=\frac{1}{3!}\vf_{ijk}e^{ijk}=e^{123}-e^1(e^{47}+e^{56})
+e^2(e^{46}-e^{57})-e^3(e^{45}+e^{67})  \ .\label{2.35}
\end{equation}
This form is valid in flat space or in an orthonormal basis, the $e^i$s being basis forms. Another useful way of think about the $G_2$ three-form is to write it in a $6+1$ split. We then have
\begin{eqnarray}
\vf_{ijk}&=&\l_{ijk} \ , \nn\w1 \vf_{ij7}&=& \o_{ij} \ , \nn\w1
\vf_{ijk7}&=&-\ghl_{ijk}\ , \label{2.35.02}
\end{eqnarray}
where $i,j,k =1\ldots 6$, and $\{{\l,\ghl,\o}\}$ are the forms defining an $SU(3)$ structure in six dimensions. The three-forms $\l$ and $\ghl$ are the real and imaginary parts respectively of a complex three-form $\O$ which is of type $(3,0)$ with respect to the almost complex structure defined by $\o$.

On a $G_2$ manifold with skew-symmetric torsion, the latter is uniquely determined in terms of the Levi-Civita covariant derivative of $\vf$ \cite{Friedrich:2001nh,Friedrich:2001yp}. This follows from the covariant constancy of $\vf$ with respect to the torsionfull connection.

Now suppose we have a bi-$G_2$-structure. The two $G_2$ three-forms are related to one another by an $SO(7)$ transformation, $V$. If we start from $\vf^\mi$ this will be determined up to an element of $G_2^\mi$. So we can choose a representative to be generated by an element $w\in \gs\go(7)$ of the coset algebra with respect to $\gg_2^\mi$. This can be written
\begin{equation}
w_{ij}= \vf^\mi_{ijk} v^k \label{5.1}
\end{equation}
and $V=e^w$. The vector $v$ will be specified by a unit vector $N$ and an angle $\a$. It is straightforward to find $V$,
\begin{equation}
V^i{}_j=\cos\a \d^i{}_j + (1-\cos\a) N^i N_j + \sin\a\,\vf^{i (-)}_{ \ jk} N^k\ . \label{5.2}
\end{equation}
Using
\begin{equation}
\vf^\pl=\vf^\mi V^3\ , \label{5.2.1}
\end{equation}
where one factor of $V$ acts on each of the three indices of $\vf$, we can find the relation between the two $G_2$ forms explicitly,
\begin{equation}
\vf^\pl_{ijk}=A\vf^\mi_{ijk} + B\vf^\mi_{ijkl} N^l + 3
C\vf^\mi_{[ij}{}^l N_{k]} N_l\ , \label{5.3}
\end{equation}
where
\begin{equation}
A=\cos 3\a \ \ \ , \ \ \  B=\sin 3\a \ \ \ , \ \ \  C=1-\cos3\a \ . \label{5.4}
\end{equation}
The dual four-forms are related by
\begin{equation}
\vf^\pl_{ijkl}=(A+C)\vf^\mi_{ijkl} -4 B\vf^\mi_{[ijk} N_{l]} -4
C\vf^\mi_{[ijk}{}^m N_{l]} N_m\ . \label{5.4.1}
\end{equation}

The angle $\a$ is related to the angle between the two covariantly constant spinors. To simplify life a little we shall follow \cite{Gauntlett:2002sc} and choose these spinors to be orthogonal, which amounts to setting $\cos\frac{\a}{2}=0$. We then find
\begin{equation}
\vf^\pl_{ijk}=-\vf^\mi_{ijk} +  6\vf^\mi_{[ij}{}^l N_{k]} N_l 
\label{5.5}
\end{equation}
and
\begin{equation}
\vf^\pl_{ijkl}=\vf^\mi_{ijkl} -  8 \vf^\mi_{[ijk}{}^m N_{l]}N_m\ .
\label{5.10}
\end{equation}
We can use the vector $N$ to define an $SU(3)$ structure as above. We set
\begin{equation}
\o=i_N \vf^\mi\, \ \ \ , \ \ \  \l=\label{}\vf^\mi-\o\wedge N\, \ \ \ , \ \ \ 
\ghl=i_N*\vf^\mi\ . \label{5.10.1}
\end{equation}
The three-form $\ghl$ is the six-dimensional dual of $\l$ and the set of forms $\{\o,\l,\ghl\}$ is the usual set of forms associated with an $SU(3)$ structure in six dimensions. For the plus forms we have
\begin{eqnarray}
i_N\vf^\pl&=& \o  \ , \nn\w1 \vf^\pl-\o\wedge N&=&-\l  \ , \nn\w1
i_N*\vf^\pl&=&-\ghl\ . \label{5.10.2}
\end{eqnarray}

Thus a bi-$G_2$-structure is equivalent to a single $G_2$ structure together with a unit vector (and an angle to be more general). The unit vector $N$ then allows one to define a set of $SU(3)$ forms as
above. In \cite{Gauntlett:2002sc} it is shown that the projector onto the six-dimensional subspace is integrable, but this presupposes that the dilatino Killing spinor equation holds. Since we make no use of this equation it need not be the case that integrability holds.

It is straightforward to construct a covariant derivative $\hnab$ which preserves both $G_2$ structures. This connection has torsion but this is no longer totally antisymmetric. It is enough to show that the covariant derivatives of $N$ and $\vf^\mi$ are both zero. If we write
\begin{equation}
\hnab_i N_j=\nab^\mi_i N_j-S_{i:j}{}^k N_k\ , \label{5.10.3}
\end{equation}
where $S_{i:jk}=-S_{i:kj}$, then these conditions are fulfilled if
\begin{equation}
S_{i:jk}=\half
H_{ijk}+\frac{1}{4}H_i{}^{lm}\vf^\mi_{lmjk}-\frac{3}{2}
H_{ilm}\P^l{}_j\P^m{}_k-\frac{3}{4}H_i{}^{lm}\vf^mi_{jkln} N^n N_m\
. \label{5.10.4}
\end{equation}
Here $\P^i{}_j:=\d^i{}_j-N^i N_j$ is the projector transverse to $N$.

\subsection{$SU(3)$}

Manifolds with $SU(3)\times SU(3)$ have arisen in recent studies of supergravity solutions with flux \cite{Grana:2004bg, Grana:2005sn, Grana:2005ny}. They have also been discussed in a recent paper on generalized calibrations \cite{Koerber:2005qi}. A bi-$SU(3)$ structure on a six-dimensional manifold is given by a pair of a pair of forms $\{\o^\plmi,\O^\plmi\}$ of the above type which are compatible with the metric. If the $\sigma$-model algebra closes off-shell the complex structures will be integrable. The transformation relating the two structures can be found using a similar construction to that used in the $G_2$ case. However, we can instead derive the relations between the plus and minus forms by dimensional reduction from $G_2$. To this end we introduce a unit vector $N'$, which we can take to be in the seventh direction, and define the $SU(3)$ forms as in equation \eq{2.35.02} above. We consider only the simplified bi-$G_2$-structure and we also then take the unit vector $N$ to lie within the six-dimensional space. The unit vector $N$ now defines an $SO(6)$ transformation. The relations between the plus and minus forms are given by
\begin{eqnarray}
\o^\pl_{ij}&=& -\o^\mi_{ij} + 4 \o^\mi_{[i}{}^k N_{j]} N_k  \ , \nn \w1
\l^\pl_{ijk}&=& -\l^\mi_{ijk} +6\l_{[ij}{}^l N_{k]} N_l  \ , \nn \w1
\ghl^\pl_{ijk}&=& \ghl^\mi_{ijk}- 6\ghl^\mi_{[ij}{}^l N_{k]} N_l \ .
\label{5.4.3}
\end{eqnarray}

We can rewrite this in complex notation if we introduce the three-forms $\O^\plmi:=\l^{\plmi}+i\ghl^{\plmi}$ and split $N$ into $(1,0)$ and $(0,1)$ parts, $n,\bar n$. So,
\begin{equation}
N_i=n_i + \bar n_i\, \ \ \ , \ \ \  i\o_{ij} N^j =n_i-\bar n_i \ .
\label{5.4.4}
\end{equation}
Note that $n\cdot\bar n=\frac{1}{2}$. Then equations \eq{5.4.3} are equivalent to
\begin{eqnarray}
\o^\pl_{ij}&=& -\o^\mi_{ij} -2in_{[i} \bar n_{j]} \ ,  \nn\w1
\O^\pl_{ijk}&=& 6\bar\O^\mi_{[ij}{}^l n_{k]} n_l \ . \label{5.4.5}
\end{eqnarray}
This type of bi-$SU(3)$-structure is therefore equivalent to a single $SU(3)$ structure together with a normalised $(1,0)$-form.

\subsection{$Spin(7)$}

A $Spin(7)$ structure on an eight-dimensional Riemannian manifold is specified by a self-dual four-form $\F$ of a certain type. In a given basis its components can be constructed from those of the $G_2$ three-form. Thus
\begin{equation}
\F_{abcd}=\vf_{abcd}\, \qquad {\rm and} \qquad\F_{abc8}=\vf_{abc}\ ,
\label{5.4.6}
\end{equation}
where, in this section, $a,b,\ldots$ run from 1 to 7 and $i,j,\ldots$ run from 1 to 8.  $Spin(7)$ geometry with skew-symmetric torsion has been discussed \cite{Friedrich:2001nh,Ivanov:2001ma} and generalized $Spin(7)$ structures have also been studied \cite{Witt:2004vr}. A bi-$Spin(7)$-structure on a Riemannian manifold consists of a pair of such forms, covariantly constant with respect to $\nab^\plmi$. We can again get from the minus form to the plus form by an orthogonal transformation, but since the dimension of $SO(8)$ minus the dimension of $Spin(7)$ is seven it is described by seven parameters. In the presence of a $Spin(7)$ structure one of the chiral spinor spaces, $\D_s$, say, splits into one- and seven-dimensional subspaces, $\D_s=\bbR\oplus \D_7$. The transformation we seek will be described by a unit vector $n^a\in \D_7$ together with an angle.

It will be useful to introduce some invariant tensors for $Spin(7)$ using this decomposition of the spin space. We set
\begin{align}
\f_{ajk} =&\begin{cases} \f_{abc}=\vf_{abc} \\ \f_{ab8}=\d_{ab} \end{cases} \ ,
\label{5.4.7}
\w2
\f_{abkl}=&\begin{cases}  \f_{ab}{}^{cd}=\vf_{ab}{}^{cd}-2\d_{[ab]}^{cd}
\\ \f_{abc8}=-\vf_{abc}   \end{cases} \ , \label{5.4.8}
\end{align}
where $\vf_{abc}$ is the $G_2$ invariant. It will also be useful to define
\begin{equation}
\f_{aijkl}:=\f_{ab[ij}\f^b{}_{kl]}\ . \label{5.4.9}
\end{equation}
The $Spin(7)$ form itself can be written as
\begin{equation}
\F_{ijkl}=\f_{a[ij}\f^a{}_{kl]}\ . \label{5.4.10}
\end{equation}

The space of two-forms splits into $7+21$, and one can project onto the seven-dimensional subspace by means of $\f_{ajk}$. With these definitions we can now oxidise the $G_2$ equations relating the plus
and minus structure forms to obtain
\begin{equation}
\F^\pl_{ijkl}= -\F^\mi_{ijkl}-6 n_a n_b \f^{\mi a}{}_{[ij}\f^{\mi
b}{}_{kl]} \label{5.4.11}\ .
\end{equation}
Here the unit vector $N$ in the $G_2$ case becomes the unit spinor $n$.

\section{Examples of solutions}
\label{section:examples_of_solutions}

In this section we look at solutions to the boundary conditions for the additional symmetries which can be identified with various types of branes. We shall go briefly through the main examples, confining ourselves to $U(\frac{n}{2})$, $SU(\frac{n}{2})$, and the exceptional cases $G_2$ and $Spin(7)$.

\subsection{$U(\frac{n}{2})\equiv U(m)$}

This case corresponds to $N=2$ supersymmetry. For $H=F=0$ we assume that the supersymmetry algebra closes off-shell so that $\M$ is a K\"ahler manifold with complex structure $I$, Hermitian metric $G$, and K\"ahler form $\o$. The K\"ahler form is closed and covariantly constant. The boundary conditions for the second supersymmetry, which can be viewed as an additional symmetry with $\l=\o$, imply
\begin{equation}
\o_{ij}=\pm \o_{kl} R^k{}_i R^l{}_j\ . \label{2.28}
\end{equation}
Thus there are two possibilities, type A where $IR=-RI$ and type B where $IR=RI$ \cite{Ooguri:1996ck}. Consider type B first. In this case the brane inherits a K\"ahler structure from the target space and so has dimension $2k$. If there is a non-vanishing gauge field $F$, it must be of type $(1,1)$ with respect to this structure. The calibration form is $\o^k$.

For type A with zero $F$ field, $I$ is off-diagonal in the orthonormal basis in which $R$ takes its canonical form
\begin{equation}
R=\left(\begin{array}{cc} 1_p & 0\w2 0 & -1_q
\end{array}\right)\ ,
\label{2.29}
\end{equation}
where $p$ and $q$ denote the dimensions of $B$ and the transverse tangent space, $p+q=n$, and $1_p,\,1_q$ denote the corresponding unit matrices. The only possibility is $p=q=m$. The K\"ahler form
vanishes on both the tangent and normal bundles to the brane, so that the brane is Lagrangian.

When the $F$ field is non-zero but $H=0$ the situation is more complicated. We may take $R$ to have the same block-diagonal form as in \eq{2.29} but with $1_p$ replaced by $R_p$. From \eq{2.12.4}
\begin{equation}
R_p= (1+F)^{-1} (1-F)  \ .\label{2.29.1}
\end{equation}
The analysis of $IR=-RI$ shows that the brane is coisotropic \cite{Kapustin:2001ij}. This means that there is a $4k$-dimensional subspace in each tangent space to the brane where $I$ is non-singular, there is an $r$-dimensional subspace on which it vanishes, and the dimension of the normal bundle is also $r$. The product  $(I_p F)$ is an almost complex structure and both $I_p$ and $F$ are of type $(2,0)+(0,2)$ with respect to $(I_p F)$. For $m=3$ we can therefore only have $p=5$. For $m=4$ we can have $p=5$ but we can also have a space-filling brane with $p=8$.

$N=2$ $\sigma$-models with boundary and torsion have been discussed in \cite{Lindstrom:2002jb}; the geometry associated with the boundary conditions is related to generalized complex geometry \cite{Zabzine:2004dp,Kapustin:2004gv}.

\subsection{$SU(\frac{n}{2})\equiv SU(m)$}

In the Calabi-Yau case we have, in addition to the K\"ahler structure, a covariantly constant holomorphic $(m,0)$ form $\O$ where $m= \frac{n}{2}$. There are two independent real covariantly constant forms, $\l$ and $\ghl$, which can be taken to be the real and imaginary parts of $\O$. The corresponding $L$-tensors which define the symmetry transformations are related by
\begin{equation}
\hL^i{}_{j_1\ldots j_{m-1}}= I^i{}_k L^k{}_{j_1\ldots j_{m-1}} \ .
\label{2.33}
\end{equation}
Because there are now two currents we can introduce a phase rather than a sign in the boundary condition. Thus
\begin{equation}
\O_{i_1\ldots i_m} = e^{i\a} \O_{j_1\ldots j_m}
R^{j_1}{}_{i_1}\ldots R^{j_m}{}_{i_m} \ . \label{2.33.1}
\end{equation}

A second possibility is that $\O$ on the right-hand side is replaced by $\bar \O$. For type B branes, the displayed equation is the correct condition. The $R$-matrix is the sum of holomorphic and antiholomorphic parts, $R=\cR\oplus\bar{\cR}$, and \eq{2.33.1}
implies that
\begin{equation}
\det \cR=e^{i\a}\ . \label{2.33.2}
\end{equation}
If $F=0$ this fixes the phase, but if $F\neq 0$ it imposes a constraint on $F$ which must in any case be a $(1,1)$ form (from $IR=RI$) \cite{Kapustin:2003se}. The constraint is
\begin{equation}
\det\cR_p=e^{i\a} (-1)^{\frac{q}{2}}\ , \label{2.33.3}
\end{equation}
or
\begin{equation}
\det (1+f)= e^{i\a} (-1)^{\frac{q}{2}} \det(1-f) \ , \label{2.33.4}
\end{equation}
where $f^a{}_b=G^{a\bar c} F_{\bar c b}$, in a unitary basis.

For type A branes, from $IR=-RI$ it follows that $R$ maps holomorphic vectors to antiholomorphic ones and vice versa so that $\bar \O$ must be used in \eq{2.33.1}. In the case that $F=0$ the brane is a SLAG with $\gR\ge\O$ as the calibration form. For $F\neq 0$ we have  coisotropic branes with an additional constraint on the gauge field \cite{Kapustin:2003se}.

The geometry of the bi-$SU(m)$ case has been studied from the point of view of generalized geometry and generalized calibrations in \cite{Koerber:2005qi}.

\subsection{$G_2$}

The boundary conditions associated with the $G_2$ currents are
\begin{eqnarray}
\vf &=& \h_L \vf R^3 \ ,  \nn \w1 *\vf &=& \h_L *\vf R^4\, \det R \ .
\label{2.34}
\end{eqnarray}
We consider first $F=H=0$. From the first of these equations it follows that $(\h_L R)\in G_2$. From this it follows that the sign in the boundary condition for $*\vf$ is always positive because the sign of the determinant of $R$ is equal to $\h_L$. Thus the second constraint reduces to $*\vf=*\vf R^4$.

There are two possibilities depending on the sign of $\h_L$. If it is positive then non-zero components of $\vf$ must have an even number of normal indices, whereas if it is negative they must have an odd number of non-zero components. Since $(\h_L R)\in G_2$, and is symmetric, it can be diagonalised by a $G_2$ matrix so that we can bring $R$ to its canonical form in a $G_2$ basis. Looking at the components of $\vf$ we see that the only possibilities which are compatible with the preservation of the non-linear symmetries on the boundary are either $\h_L=1$, in which case $B$ is a three-dimensional associative cycle, or $\h_L=-1$ in which case $B$ is a four-dimensional co-associative cycle \cite{Becker:1996ay}.

Now let us turn to $F\neq 0$, but $H=0$. We shall assume that the tangent bundle of $\M$, restricted to the brane, splits into three, $TM|_B=T_1\oplus T_2\oplus N$, of dimensions $p_1,p_2$, and $q$ respectively. $N$ is the normal bundle and $R|_{T_2}=1_{p_2}$. If there is at least one normal direction we may assume that one of these is $7$ in the conventions of \eq{2.35.02}. Thus the problem is reduced to a six-dimensional one, at least algebraically. The six-dimensional boundary conditions are (where $R$ is now a $6\xz 6$ matrix),
\begin{eqnarray}
\l&=&\h_L\l R^3 \ , \nn\w1 \ghl&=&\h_L \ghl R^3 \det R \ , \nn\w1
\o&=&-\h_L\o R^2 \ , \label{2.35.1}
\end{eqnarray}
in an obvious notation. If the sign is negative the brane is type B, whereas if $\h_L=+1$  we have type A. These are the same conditions as we have just discussed in the preceding section, the only difference being that the phase is not arbitrary. The constraints on the $F$ field are therefore slightly stronger.

The last possibility is a space-filling brane in seven dimensions. Since $F$ is antisymmetric there must be at least one trivial direction for $R$ so that we can again reduce the algebra to the six-dimensional case. The only possibilty is $\h=+1$ in which case we have type B. The non-trivial dimension must be even, and since $\det{\cR}=1$ the case $p=2$ is also trivial.

Now let us consider the case with torsion. The boundary condition for the non-linear symmetries associated with the forms yield
\begin{eqnarray}
\vf^{\pl}&=& \h_L\vf^{\mi} R^3 \ , \nn\w1 *\vf^{\pl}&=&\h_L *\vf^{\mi}
R^4\det R\ . \label{5.13}
\end{eqnarray}
When the brane is normal to $N$ we find, on the six-dimensional subspace,
\begin{eqnarray}
\l&=&-\h\l R^3 \ , \nn\w1 \ghl&=&-\h \ghl R^3 \det R  \ , \nn\w1 \o&=&-\h\o
R^2 \ . \label{5.14}
\end{eqnarray}

The analysis is very similar to the case of zero torsion with $F$. One finds that $\h_L=-1$ corresponds to type B while $\h_L=+1$ is type A.  In particular, for type B there is a five-brane which corresponds to the five-brane wrapped on a three-cycle discussed in the supergravity literature \cite{Gauntlett:2001ur,Gauntlett:2002sc}.

\subsection{$Spin(7)$}

In the absence of torsion, the boundary condition associated with the conserved current is
\begin{equation}
\F=\ghh \F R^4\ , \label{2.42}
\end{equation}
for some sign factor $\ghh$. If this is negative then $\det R$ is also negative so that the dimension of $B$ must be odd. Furthermore, $\F$ must have an odd number of normal indices with respect to the decomposition of the tangent space induced by the brane. However, one can show that such a decomposition is not compatible with the algebraic properties of $\F$. Therefore the sign $\ghh$ must be
positive. It is easy to see that a four-dimensional $B$ is compatible with this, and indeed we then have the standard Cayley calibration with $\F$ pulled-back to the brane being equal to the induced volume form. On the other hand if $B$ has either two or six dimensions one can show that it is not compatible with the $Spin(7)$ structure. As one would expect, therefore, the only brane compatible with the non-linear symmetry associated with $\F$ on the boundary is the Cayley cycle \cite{Becker:1996ay}.

If $F\neq0$, but $H=0$, and if we assume that there is at least one direction normal to the brane, then the $Spin(7)$ case reduces to $G_2$ (with $F\neq0$). If the brane is space-filling but there is at least one trivial direction, then there must be at least two by symmetry and again we recover the $G_2$ case. But we can also have a space-filling brane which is non-trivial in all eight directions.

 
\chapter{$W$-superstrings on manifolds of special holonomy}
\label{chapter:W-strings}
 
 
\section{$W$-algebras and $W$-strings}
 
A non-linear algebra of conformal-type symmetries (see \ref{section:local_and_global_sym}) is called a \emph{W-algebra}.\footnote{The first $W$-algebra was written down by Zamolodchikov \cite{Zamolodchikov:1985wn}. A comprehensive review of  $W$-algebras is by  Bouwknegt and Schoutens \cite{Bouwknegt:1992wg}.} In Chapters  \ref{chapter:algebras_of_general_L_symmetries} and \ref{chapter:algebras_on_manifolds_of_sh} $W$-superalgebras related to special holonomy target spaces were discussed.\footnote{Other than in this context, $W$-superalgebras have been discussed in the context of supersymmetric  extensions of the $W_{\infty}$ algebra, where currents of arbitrary spin are included \cite{Bergshoeff:1990yd}. Romans \cite{Romans:1991wi}  constructs an extension of the $W_3$-algebra (see also \ref{subsection:W3string}).  Many types of $W$-superalgebras have been classified using conformal bootstrap methods;  see \cite{Blumenhagen:1992sa, Blumenhagen:1994wn} and references therein.}

As explained in \ref{section:bosonic_string_theory}, the bosonic string action describes a worldsheet coupled to two-dimensional gravity. We obtained it by gauging the conformally invariant $\sigma$-model (\ref{eq:bos_action}), and showed that after gauge fixing the BRST operator basically imposes the conformal currents as constraints (\ref{eq:string_BRST_free_background}). If action (\ref{eq:bos_action}) is invariant under any $W$-symmetries, they remain symmetries of the gauge fixed action but are not themselves imposed as constraints. In \emph{W-string theory}\footnote{For a detailed review of $W$-strings the reader is referred to \cite{Hull:1993kf} and \cite{Pope:1992mi}.} one imposes all the currents present in a $W$-algebra as constraints, rather than just those in the conformal subalgebra. To obtain the BRST operator the entire $W$-symmetry algebra of the conformally invariant action must be gauged. Similarly, a $W$-superstring action is obtained by gauging the entire $W$-superalgebra.\footnote{$W$-superstrings have not been explored much. Most of the work is in the context of the supersymmetric extension of the bosonic $W_{\infty}$-string \cite{Bergshoeff:1990vt}.}  

In this chapter I will explore the superstring actions obtained by gauging the extended special holonomy symmetries described in Chapter \ref{chapter:algebras_on_manifolds_of_sh}. Most of the explicit calculations will be done for the CY3 algebra, but we will be able to extend our conclusions to the other cases straightforwardly. One of the main issues is an obstruction to obtaining the BRST operator which occurs when the Jacobi identities aren't satisfied at the level of structure functions; I will refer to Jacobi identities that vanishes at the level of structure functions as \emph{well behaved} (see (\ref{eq:jacobi_brst_problem})). Such an obstruction will be analyzed in detail for the non-supersymmetric $W_{\frac{5}{2}}$-algebra\footnote{The subscripts denote the spins of the currents in the algebra other than the conformal current, although sometimes the conformal spins are included as well, in which case one would write $W_{2,\frac{5}{2}}$.} before tackling the special holonomy algebras. 

In \ref{section:bosonic_string_theory} we saw that conformal symmetries in a single chiral sector can be gauged rather simply. In this chapter I'll only be looking at chiral versions of $W$-strings. It is possible to gauge both chiralities simultaneously, but this is even more involved than in the conformal case. Again, things simplify if one works in a covariant Hamiltonian formalism (see \ref{section:bosonic_string_theory}), but for the $W$-string case it is not possible to obtain a closed expression for the second order action, analogue of  (\ref{eq:conf_spec_extended_no_pi}), after eliminating the momenta \cite{Schoutens:1990ja, Hull:1993kf}. 

In \ref{section:general_form_of_solution} I describe, in a general setting, the gauging procedure and the form of the solution to the master equation for $W$-strings, as well as possible obstructions. In in \ref{subsection:W3string} I give the classical $W_3$-string \cite{Vandoren:1993bw, Geyer:2003de} as an example of a well behaved system. In \ref{section:W52string} I illustrate the problems in attempting to define a classical $W$-string theory from the $W_{\frac{5}{2}}$-algebra \cite{Lu:1994sc,Thielemans:1995hn}, whose Jacobi identities are not well behaved.  In \ref{section:W52string} I review the analysis of this problem in \cite{Thielemans:1995hn}, and go on to show that the relations between the generators imply that the action with only the conformal symmetry gauged possesses extra gauge symmetries. In \ref{section:W_superstrings_on_CY} I perform the analogous analysis for the CY3 algebra. The closure of the gauge symmetries in the full CY3 $W$-superstring implies that the $N=2$ string action on a CY3 background possesses extra gauge symmetries.  

Furthermore, for both the CY3 and $W_{\frac{5}{2}}$ cases, I show that there are subalgebras involving \emph{complex} currents that have no relations among themselves and involve the generators of spectral flow in one direction only. Classical $W$-strings can be defined from such algebras. In  \ref{section:W-string_discussion} I discuss the implications for other special holonomy cases.

This chapter serves a double purpose, since to write down naive Ward identities related to the operator product expansion one needs to introduce source terms for the currents, and these come out naturally from the gauging procedure. Further remarks are given in \ref{section:W-string_discussion}.

\section{The general  solution}
\label{section:general_form_of_solution}

In this section I'll describe, in a general setting, what is involved in gauging a single chiral sector, and will show that there is an obstruction to closing the algebra on the gauge fields when the Jacobi identities are not well behaved. One may hope that the BV formalism somehow provides a way of dealing with this obstruction without introducing generators for new symmetries, but it turns out that there is no such mechanism.

Starting with a matter action $S_{\mathrm{mat}}$ that is invariant under conformal-type symmetries and varying it with a \emph{local} ghostly parameter yields
\begin{equation}
\delta S_{\mathrm{mat}} = T_A \sdem c^A \ ,
\end{equation}
where $T_A$ are the conserved currents, $c^A$ are the BRST/BV ghosts, and I'm temporarily using the deWitt convention (see \ref{section:local_and_global_sym}). Applying the Noether method, the action
\begin{equation}
S_0 = S_{\mathrm{mat}} + T_A h^A
\end{equation}
is invariant  if the gauge fields transform as
\begin{equation}
\label{eq:gauge_trans}
\delta_A h^B = - \sdem c^B -  h^A c^C f^B_{AC}  \ ,
\end{equation}
where $f^B_{AC}$ are structure functions (for non-linear symmetries they are generally field dependent).  The second term in (\ref{eq:gauge_trans}) arises due to the variation of  $T_A h^A$, which is given by
\begin{equation}
 \label{eq:schematic_pb}
\{ T_A h^A, T_C c^C \} =  h^A c^C f^B_{AC} T_B \ .
\end{equation}

In terms of the Poisson bracket the Jacobi identities are
\begin{equation}
[  f^D_{[AB} f_{C]D}^E + \{ f_{[AB}^E , T_{C]} \} ] T_E \equiv 0 \ . \label{eq:jacobi_poisson}
\end{equation}
If they are well behaved the object inside the square parentheses vanishes, and the symmetry transformations on the gauge fields close:
\begin{equation}
h^A c^B c^C\left[  f^D_{[AB} f_{C]D}^E + \{ f_{[AB}^E , T_{C]} \} \right] =0 \ .
\end{equation}
If they are not well behaved, the Jacobi identities imply relations between the currents,
\begin{equation}
T_A U^A_{\bB} (T) \equiv 0 \ ,
\end{equation}
labeled by the index set $\bB$, with $U^A_{\bB}$ generally functions of the currents. In this case the symmetry transformations close to new gauge symmetries that act only on the gauge fields as
\begin{equation}
\label{eq:null_symmetry1}
\delta h^A = c^{\bB} U^A_{\bB} (T) \ .
\end{equation}
 
The equations of motion for the gauge fields are simply
\begin{equation}
T_A = 0 \ .
\end{equation}
Since (\ref{eq:null_symmetry1}) describes relations between the currents, the new symmetries are symmetric rather than antisymmetric in the equations of motions (they are null symmetries; see \ref{subsection:properness}). Therefore, they can't be absorbed in the equations of motion by adding terms proportional to $(h^*)^2$ to the extended action, and  have to be treated as genuine new symmetries. The Jacobi problem is also present in terms linear in $c^*$ in the master equation, but  terms proportional to $h^* c^*$ can be added to the extended action to solve this part of the master equation. In the rest of this section we'll see how this works in detail.

The master equation (see (\ref{eq:quantum_master_eq1}), (\ref{eq:classical_master_equation}), and (\ref{eq:ward_identity_effective_action})) is
\begin{equation}
\frac{ \dr S (\Phi , \Phi^{*}) } { \delta   \Phi^{Y} } \frac{\dl S (\Phi, \Phi^{*}) }{ \delta \Phi^{*}_{Y} } = 
\frac{1}{2} (S, S) = 0 \ .
\end{equation}

In the setting of chiral $W$-(super)strings the fields are divided between gauge, $h^A$, and matter, $\phi^i$, so $\Phi^Y = \{ \phi^i, h^A \}$. The minimal solution to the master equation begins as
\begin{align}
\label{eq:chiral_gauge_action_min}
S_{\min}  = & S_0(\phi,h) + \phi^*_i c^B R^i_B(X) + h^*_A c^B D^A_B(\phi,h) + c^*_A c^{BC} N^A_{BC}(\phi) \\ \nonumber
 &  +   \phi^*_i \phi^*_j c^{AB} M^{ij}_{AB} + h^*_C h^*_D c^{AB} M^{CD}_{AB} \\ \nonumber
 &+ c^*_D \phi^*_i c^{BCD} 
M^{Di}_{ABC}  +   c^*_D h^*_F c^{ABC} M^{DF}_{ABC}(\phi) + \cdots \ ,
\end{align}
where $c^{ABC} := c^A c^B c^C$, while $\cdots$ stands for terms not relevant to the master equation expanded to first order in antifields, and also for terms that are not relevant in the context of $W$-strings. The first line contains the generators of transformations of the matter fields $R^i_B$, the generators of the gauge transformations $D^A_B$, and the structure functions $N^{A}_{BC}$.  I'll make the reasonable assumption that $R^i_B$ don't depend on the gauge fields. The objects denoted by $M$ in the second line of (\ref{eq:chiral_gauge_action_min}) correspond to various non-closure functions. 

The master equation expanded to linear order in antifields is
\begin{align}
 \label{eq:solution_to_me}
&   \frac{1}{2}  (S_{\min} , S_{\min})  =  S_{0,k} c^B R^k_B + S_{0, H} c^B D^H_B  \\ \nonumber
&  +  \phi^*_i [ c^B R^i_{B,k} c^C R^k_C + R^i_H c^{BC} N^H_{BC}+  S_{0,k} c^{AB} M^{ki}_{AB} +    
S_{0,H} c^{AB} M^{iH}_{AB}  ]  \\  \nonumber
&  +  h_A^* [ c^B D^A_{B,H} c^C D^H_C  + c^B D^A_{B,k} c^C R^k_C 
+ D^A_H c^{BC} N^H_{BC}  \\ \nonumber 
&  \ \ \ \ \  \ \ \ + S_{0,k} c^{BC} M^{kA}_{BC} + 2 S_{0,H} c^{BC} M^{HA}_{BC}  ] \\ \nonumber
&  +  c^*_A [ 2 c^C N^A_{BC} c^{DE} N^B_{DE}+  c^{BC} N^{A}_{BC,k} c^D R^k_D   S_{0,k} + c^{BCD} M^{Ak}_{BCD} + S_{0,H} c^{BCD} M^{AH}_{BCD}  ]  \ ,
\end{align}
where $S_{,k} := \frac{\dr S}{\delta \phi^k}$ and $S_{,H} := \frac{\dr S}{\delta h^H}$. The term independent of antifields  corresponds to the symmetries of the original action. The terms linear in $X^*_i$ and $h^*_A$ correspond to closure equations for transformations acting on the matter and gauge fields, respectively. The terms linear in $c^*$ are related to the Jacobi identities.

Analyzing these terms separately:
\begin{itemize}
\item The part independent of antifields is satisfied when the action is properly gauged. 
\item The term proportional to $\phi^*$ contains the commutator of the matter field transformations in the first term. Next  we have a term containing the usual structure functions of the matter algebra, $N^H_{BC}$, followed by terms proportional to the matter and gauge equations of motion. We expect to make use of the last term in the second line since the algebra of matter field transformations closes to functions of currents, so the commutator term can be absorbed either into  the structure functions, via $R^i_H c^{BC} N^H_{BC}$,  or into terms proportional to the gauge field equations of motion via $S_{0,H} c^{AB} M^{iH}_{AB}$.
\item The first two terms in the part proportional to $h^*$ describe the closure of the transformations on the gauge fields, followed by a term containing the structure functions and  terms proportional to the gauge and matter equations of motion. In particular, the functions $M^{kA}_{BC}$ that featured in the term proportional to $X^*$ features here as well. This is related to the $\sdem c^A$ term in the gauge transformation (\ref{eq:gauge_trans}). If the structure functions are chosen to be non-zero, one will get terms containing $\sdem T_A$, which is proportional to the \emph{matter} equations of motion (see, for example, (\ref{eq:Dminus_TL})). Finally we have the term antisymmetric in the gauge field equations of motion. It is not possible to absorb null symmetries into such terms, but for well behaved Jacobi identities it may be necessary to include $M^{AH}_{BCD}$ terms to the extended action. The choices of $M^{kA}_{BC}$, $M^{HA}_{BC}$, and $N^H_{BC}$ are interdependent, with different solutions related by canonical transformations (\ref{eq:canonical_transformation_def})
\item The first two terms in the part proportional to $c^*_A$ constitute the structure function part of the Jacobi identities. If the Jacobi identities aren't well behaved this doesn't vanish, and then one has to make use of the $M^{AH}_{BCD}$ term  in the extended action. Alternatively, one could work with a solution for which $N^A_{BC} = 0$.
\end{itemize}

In conclusion, a solution to the master equation can be obtained to linear order in antifields provided that the Jacobi identities are well behaved. Terms quadratic and higher in the antifields  exist  and are unique up to canonical transformations, provided that the solution is proper.\footnote{The proofs of this statement can be found in \cite{Henneaux:1992ig} and \cite{Vandoren:1993bw}.} However, it may be that an infinite number of such terms is necessary. When the Jacobi identities are not well behaved we are forced to introduce generators for the null symmetries, and it's not possible to find a proper solution. As we'll show in \ref{section:W52string} and \ref{section:W_superstrings_on_CY}, non-proper solutions are possible, but this depends on a particular realization of the algebras.

\section{$W_3$-string}
\label{subsection:W3string}

An example of a theory with well behaved Jacobi identities is the $W_3$-string \cite{Vandoren:1993bw, Geyer:2003de}. The simplest realization for a single boson $\phi$ is
\begin{equation}
S_0   =  \int d^2 \sig \left( \frac{1}{2} \parp \phi \parm \phi - \frac{1}{2}  (\parp \phi )^2 h^G - 
\frac{1}{3} (\parp \phi)^3 h^W \right) \ .
\end{equation}
A minimal solution is given by:
\begin{align}
S_{\min} = & \int d^2 \sig \left(  \vphantom{\frac{1}{2}}  S_0 +
\phi^*[c^G \parp \phi + c^W (\parp \phi)^2] \right. \\ \nonumber
& + h^*_G [ \sdem c^G - \sdep c^G h^G + c^G \sdep h^G + (c^W \sdep h^W - \sdep c^W h^W)(\sdep \phi)^2 ] \\ \nonumber
& + h^*_W [\sdem c^W + c^C \sdep h^W - 2 \sdep c^G h^W - \sdep c^W h^G + 2 c^W \sdep h^G ] \\ \nonumber
& +c^*_G [ \sdep c^G c^G + \sdep c^W c^W (\sdep \phi)^2 ]  \\ \nonumber
& \left. \vphantom{\frac{1}{2}} + c^*_W [ \sdep c^W c^G + 2 \sdep c^G c^W] + \phi^* h^*_G [ 2 \sdep \phi \sdep c^W c^W ] \right) \ .
\end{align}
This solution  is in a sense the simplest one since  the structure coefficients,
\begin{equation}
c^*_G c^{WW} N^G_{WW} = \int d^2 \sig c^*_G \parp c^W c^W (\parp \phi)^2
\end{equation}
and
\begin{equation}
\phi^* h_G^*c^{WW} M_{WW}^{\phi G} = \int d^2 \sig \phi^* h_G^* 2  \parp c^W c^W \parp \phi \ , 
\end{equation}
are chosen in the correct proportion, so that no other terms non-linear in the antifields are neccessary. If they were chosen differently one would need to introduce terms non-linear in the gauge antifields. All the solutions are related by canonical transformations (\ref{eq:canonical_transformation_def}). In the context of this simple example one can easily test in detail how the interdependence of the various structure coefficients in (\ref{eq:solution_to_me}) works.

\section{$W_{\frac{5}{2}}$-string}
\label{section:W52string}

The $W_{\frac{5}{2}}$-algebra \cite{Lu:1994sc, Thielemans:1995hn} consists of the conformal current, $T_G$, and a current of spin $5/2$, $T_W$. Abstractly it is characterized by the following algebra 
\begin{align}
\{ T_G(\sig_1 ),  T_G (\sig_2) \} & = -2 T_G(\sig_2) \dep \delta(\sig_1- \sig_2) + \dep T_G(\sig_2) \delta(\sig_1 - \sig_2) \ , \nonumber \\
\{ T_G (\sig_1), T_W(\sig_2) \} &= -\frac{5}{2} T_W (\sig_2) \dep \delta(\sig_1-\sig_2) + \dep T_W(\sig_2) \delta(\sig_1-\sig_2) \ , \nonumber \\
\{ T_W(\sig_1), T_W(\sig_2) \} &= (T_G(\sig_2) )^2  \delta(\sig_1-\sig_2) \ ,
\end{align} 
expressed using Poisson brackets. The Jacobi identity with three $T_W$ currents reveals that there is a relation between the currents given by\footnote{Since two-dimensional spinors are involved in this section, and in the rest of the chapter, from now on I'll be using the double $(++)$ and $(--)$ for vector indices, and a single plus or minus for spinors. See \ref{section:2d_spinors}.}
\begin{equation}
\label{eq:W52_current_relations}
-5 \dep T_G T_W + 4 T_G \dep T_W \equiv 0 \ .
\end{equation}

A representation is given by
\begin{equation}
T_G = - \frac{1}{2} \psi \dep \psib + \frac{1}{2} \dep \psi \psib \ \ \ , \ \ \ 
T_W = (\psi + \psib)T_G \ ,
\end{equation}
where $\psi \equiv \psi_{+}$ is a complex chiral fermion. In this representation both terms in relation (\ref{eq:W52_current_relations}) vanish separately. The action
\begin{equation}
S = \int d^2 \sig \left( \psi \dem \psib +T_G h^G + T_W h^W  \right)
\end{equation}
is invariant under a gauge symmetries that act as
\begin{align}
\label{eq:W52_matter_sym}
&  \delta_G \psi = - \frac{1}{2} \dep c^G \psi - c^G \dep \psi \ \ \ \mathrm{and \ complex \ conjugate} \ ,  \\ \nonumber
&  \delta_W \psi = - \frac{1}{2} \dep c^W  \psib \psi + c^W ( - \frac{1}{2} \psi \dep \psi
+ \psi \dep \psib - \frac{1}{2} \psib \dep \psi ) \ \ \ \mathrm{and \ c.c.} \ ,
\end{align}
on the matter fields, and as
\begin{align}
\label{eq:W52_gauge_fld_sym}
&  \delta_G h^G  = \dem c^G - \dep c^G h^G + c^G \dep h^G  \ , \\ \nonumber
& \delta_W h^G = - 4 c^W h^W T_G \ ,  \\   \nonumber 
& \delta_G h^W = - c^G \dep h^W + \frac{3}{2} \dep c^G h^W  \ ,  \\ \nonumber
&  \delta_W h^W = \dem c^W - \dep c^W h^G  + \frac{3}{2} c^W \dep h^G \ ,
\end{align}
on the gauge fields. The ghost $c^G \equiv c^{G++}$ is fermionic, and $c^W \equiv c^{W+++}$ is bosonic, while the corresponding gauge fields have opposite parity.

We can attempt to solve the master equation in the same manner as when the Jacobi identities are well behaved, and add the following terms to $S_{\min}$:
\begin{equation}
\int d^2 \sig \left\{  c^*_G ( -c^G \dep c^G + 2 c^W c^W T_G) + c^*_W ( c^T \dep  c^W - \frac{3}{2} c^W \dep c^W) \right\}  \ ,
\end{equation}
corresponding to  $c^*_A c^{BC} N^A_{BC}$, and 
\begin{equation}
\int d^2 \sig \left\{ \vphantom{\frac{1}{2}} \psi^* [ h^*_G c^{WW} \dep \psi + \dep(h^*_G c^{WW} \psi) ] \right\}  + \mathrm{c.c.}  \ ,
\end{equation}
corresponding to $c^*_D h^*_F c^{ABC} M^{DF}_{ABC}$ in (\ref{eq:chiral_gauge_action_min}).
In addition,  the following can be added to satisfy the part of the master equation proportional to  $c^*$ (\ref{eq:solution_to_me}):
\begin{align}
& \int d^2 \sig \left\{ - \frac{5}{3} c^*_G h^*_W   \dep(c^{WWW}) + 3 h^*_W \dep ( c^*_G  c^{WWW}) \right. \nonumber \\
& \left. \ \ \  - \frac{4}{3} c^*_W h^*_G \dep(c^{WWW}) - 3 h^*_G \dep(c^*_W c^{WWW} ) \right\}  \ .
\end{align}
Then the extended action doesn't satisfy the master equation solely in the part proportional to $h^*$, and we have isolated the problem due to the Jacobi identities not being well behaved. 

Next, we can introduce generators for the null symmetries:
\begin{align}
\label{eq:W52_new_sym1}
& \delta_X h^W  = 4 \dep c^X T_G + 9 c^X \dep T_G \ , \\ \nonumber
& \delta_X h^G =  9 c^X \dep T_W + 5 \dep c^X T_W \ .
\end{align}
This is a symmetry due to (\ref{eq:W52_current_relations}), and obviously has no additional conserved currents associated with it. Writing $\delta_X$ as we've done above is not unique, since it can be recast in various ways using terms graded antisymmetric in the equations of motion. For example, we can add
\begin{equation}
\pm \int d^2 \sig \left\{ h^*_W h^*_G 4 (c^{WW} \dep h^W + 8 c^W \dep c^W  h^W) -9 h^*_G \dep(h^*_W c^{WW} h^W) \right\}
\end{equation}
to the minimal solution, and then only one of the transformations in (\ref{eq:W52_new_sym1}) has to be introduced: for the plus sign we only need to introduce $\delta_X h^T$, and for the minus sign $\delta_X h^W$.

Supposing that  only $\delta_X h^W$ is kept, then one can actually close the algebra and find a \emph{non-proper} solution to the master equation after introducing also the following null symmetries:
\begin{equation}
\label{eq:W52_new_sym2}
\delta_Y h^G = 2 c^Y (T_G)^2 \ \ \ , \ \ \ \delta_Z h^W  = 2 c^Z \dep T_W \ ,
\end{equation}
and choosing appropriate additional structure and non-closure functions. There are two points about this solution. First of all, it depends on using the single fermion model. If a more complicated representation was used, for example one that simply included more fermions, further null symmetries would be generated by the algebra. Furthermore, the solution is not proper since the null symmetries vanish on shell and are thus infinitely reducible. Properness requires that for on-shell symmetries any $v^A$ in $R^i_A v^A$ be treated as a reducibility relation, which makes things apparently unmanageable. However, because the gauge fields are non-propagating, it is possible to ignore properness and come up with some other requirement that makes obtaining a solution to the master equation manageable. These issues were discussed in a general setting in \ref{subsection:properness}.
 
In \cite{Thielemans:1995hn} a rigorous approach to constructing a solution to the master equation is attempted that involves requiring the nilpotence and acyclicity of the Koszul-Tate differential rather than properness. The theory is still infinitely reducible, but in a more easily controllable manner, with a finite number of ghosts needing to be introduced at each reducibility level. A problem with this approach is that one often needs to introduce null symmetries in theories that would otherwise be well behaved. I'll illustrate this using the action for a single real fermion $\psi$,
\begin{equation}
S^{(1\mathrm{ferm})} = \int d^2 \sigma \psi \dem \psi  \ ,
\end{equation}
which is invariant under the conformal symmetry,
\begin{equation}
\delta \psi = a \dep \psi + \frac{1}{2} \dep a \psi \ ,
\end{equation}
for a ghostly parameter $a$.\footnote{This example is taken from \cite{Vandoren:1996ku}.} Gauging this involves promoting the conformal-type ghost to a local one, $a \rightarrow c$, adding  
\begin{equation}
\int d^2 \sig h \psi \dep \psi
\end{equation}
to the action, and letting  the gauge field $h$ transform as
\begin{equation}
\delta h = \dem c + \dep c h - c \dep h \ .
\end{equation}
There is a relation $T \psi = 0$ simply due to the fermionic nature of $\psi$, and by the Koszul-Tate approach taken in \cite{Thielemans:1995hn, Vandoren:1996ku} one must introduce the generators for
\begin{equation}
\label{eq:1ferm_null_sym}
\delta h =  c^x \psi \ \ \mathrm{and} \ \  \delta h =  c^y \dep \psi
\end{equation}
in the master equation, with $c^x$ and $c^y$ both bosonic ghosts. Doing this  introduces the troubles of infinite reducibility, although in a manageable way, whereas ignoring them yields a proper solution to the master equation:
\begin{align}
S_{\min}^{(1 \mathrm{ferm})} = & S^{(1\mathrm{ferm})} + \int d^2 \sig \left( \psi \dem \psi + \psi \dep \psi h + \psi^* (c\dep \psi + \frac{1}{2} \dep c \psi ) \right. \\ \nonumber 
& \left. + h^{*} ( \dem c + \dep c h - c \dep h ) + c^* (c \dep c ) \vphantom{\frac{1}{2}} \right) \ .
\end{align}

One can also look for a solution that needs the least number of extra ghosts, ignoring both properness and the Koszul-Tate requirements. In the $W_{\frac{5}{2}}$ case one such solution can be obtained by introducing $\delta_Y$ and $\delta_Z$ (\ref{eq:W52_new_sym1}). In the context of the one fermion model, for example
\begin{align}
S_{\min} = & S_{\min}^{(1\mathrm{ferm})} +  \int d^2 \sig \left\{ h^* (c^x \psi c^y + \dep \psi ) + c_x^* ( - c \dep c^x + \frac{3}{2} \dep c c^x \right. \\ \nonumber 
&\left. + \frac{1}{2} c^y \dep \dep c ) + c_y^* ( - c \dep c^y + \frac {5}{2} \dep c c^y ) \right\}
\end{align}
is a non-proper solution.

In conclusion, it seems quite evident that null symmetries must be ignored if possible. The problem with the $W_{\frac{5}{2}}$-string is that closure  imposes the introduction of (\ref{eq:W52_new_sym1}).
We can try to make sense of this in one of the following ways: 
\begin{itemize}
\item Work with a non-proper solution,
\item Try to make sense of the infinite reducibility by requiring the acyclicity of the Koszul-Tate differential,
\item Work with a non-trivial subset of conserved currents among which there are no reducibility relations.
\end{itemize}
In seems to me that the first two possibilities are ill advised; the former because it's too arbitrary, and the second because it requires introducing null symmetries in many situations in which they can be safely ignored. I will propose an interesting resolution of the third kind at the end of this section, that generalizes to all the reducible special holonomy algebras and is independent of a particular representation.

Beforehand, I wish to make a further point about  the null symmetries of $W_{\frac{5}{2}}$. Namely, if only $\delta_X h^G$  in (\ref{eq:W52_new_sym2}) is kept, we have effectively closed the algebra to a symmetry that lives in the model with only $\delta_G$  gauged. That is, the action
\begin{equation}
\label{eq:W52_sub_action}
S =  \int d^2 \sigma ( \psi \dem \psib +T_G h^G )
\end{equation}
is invariant not only under the gauged conformal symmetry but also under the $\delta_X$ symmetry.  Taking the commutator of $\delta_X$ with $\delta_G$, closure demands that we introduce the symmetry that expresses the null relation
\begin{equation}
\label{eq:W52_new_null} 
T_G T_W \equiv 0 \ .
\end{equation}
This relation is not related to the Jacobi identities, but the algebra does seem to know about it. It is not immediately obvious how to close this algebra, but we can still assert that the full algebra involves infinite reducibility, and as such makes the quantum interpretation very problematic. Never the less, it is true that the $W_{\frac{5}{2}}$-algebra implies a particular null symmetry for (\ref{eq:W52_sub_action}). This point is more meaningful when the target space is a curved manifold, since then one can't pick and choose null symmetries, but must relate them to target space tensors (see (\ref{eq:N_2_null_algebra})).

There exists a subalgebra of $W_{\frac{5}{2}}$ for which a $W$-string can be formulated at the classical level - the caveat is that it involves complex currents and transformations. Each of the terms in the $T_W$,
\begin{equation}
T_W = T_w + T_\bw  \ \ \ , \ \ \ T_w:= \psib T_G \ \ \ , \ \ \ T_\bw := \psi T_G \ ,
\end{equation}
is separately a generator of a symmetry:
\begin{align}
& \delta_w \psi  =  - \frac{1}{2} \dep c^w \psib \psi + c^w \psi \dep \psib 
- \frac{1}{2} c^w \psib \dep \psi  \ , \nonumber \\
& \delta_w \psib   =  - \frac{1}{2} c^w \psib \dep \psib \  \ \ \mathrm{and \ c.c.} \ .
\end{align}
A similar observation is true about the conformal currents,
\begin{equation}
T_G = T_g+ T_\bg \ \ \ , \ \ \ T_g = \frac{1}{2} \psi \dep \psib  \ \ \ , \ \ \ T_\bg = \frac{1}{2} \psib \dep \psi \ ,
\end{equation}
where both $T_g$ and $T_\bg$ generate a symmetry:
\begin{align}
\delta_g \psi = \frac{1}{2} \dep(a^g \psi) \ \ \ , \ \ \ \delta_g \psib = \frac{1}{2} a^g \dep \psib \ \ \ \mathrm{and \ c.c.} \ .
\end{align}
The commutator of $\delta_g$ with itself closes to $\delta_g$, and similarly for $\delta_\bg$. However,  the commutator of $\delta_g$ and $\delta_\bg$ closes to the $U(1)$ symmetry. This is expected, since it is only due to the representation being complex that this splitting of the currents is possible.

Let us consider the $W_{1, \frac{5}{2}}$-algebra, given by the $W_{\frac{5}{2}}$-algebra enlarged by the $U(1)$ current,
\begin{equation}
T_I = \frac{i}{2} \psi \psib \ .
\end{equation}
It is possible to gauge various  subalgebras involving complex currents, so that the Jacobi identities don't imply any reducibility relations. The possibilities are $\{ T_G, T_I,  T_w \}$, $\{ T_g, T_I, T_w \}$, and the complex conjugates. If $T_w$ and $T_\bw$ are included as separate generators one has the same problems as when attempting to construct the $W_{\frac{5}{2}}$ string.

The minimal solution for the $\{ T_g, T_I,  T_w \}$ combination is
\begin{align}
S_{\min}  =&\int d^2 \sig \left\{  \vphantom{\frac{1}{2}} \psi \dem \psi + T_g h^g +  T_I h^I + T_w h^w  \right. \\ \nonumber
&+ \psi^*\left(  \dep( c^g \psi) + i c^I \psi + c^w \psi \del \psib - \dep(c^w \psib \psi) \right) \\ \nonumber
& + \psib^* \left( a^g \dep \psib -i c^I \psib -c^w \psib \dep \psib \right) \\ \nonumber
& +  c^*_g c^g \dep c^g + \half c^*_I c^g \del c^I + c^*_w(ic^w c^i + c^g \dep c^w - c^w \dep c^g) \\ \nonumber
&+  h^*_g \left( 2 \dem c^g + h^g \dep c^g - \dep h^g c^g   \right) +  h^*_I \left( -2 \dep c^I - h^g \dep c^I - \dep h^I c^g \right) \\ \nonumber
& \left. +  h^*_w \left( - 2 \dep c^w - h^g \dep c^w + \dep h^g c^w + h^w \dep c^g - \dep h^w c^g - i h^w c^I - i h^I c^w \right) \vphantom{\frac{1}{2}} \right\} \ .
\end{align}
A solution containing the $T_w$ and $T_g$ subalgebra of $W_{\frac{5}{2}}$ is obtained by setting the $U(1)$ related terms to zero. It is  possible to include $T_g$ and $T_\bg$ as separate currents, but then one needs to take care of the reducibility relation between these two currents and $T_I$:
\begin{equation}
\dep T_I = i ( T_g - T_\bg) \ .
\end{equation}
 
\section{$W$-superstrings on Calabi-Yau 3-folds}
\label{section:W_superstrings_on_CY}

This example is related to extended algebras of the $(1,1)$ supersymmetric $\sigma$-model on a six-dimensional Calabi-Yau target space related to the covariantly constant $(3,0)$ and $(0,3)$ forms (see \ref{section:kahler_algebra},  \ref{section:cy_algebras}, and in particular  \ref{section:CY3}). The action of the (1,1) model (\ref{eq:kahler_action}) is
\begin{equation}
S_{\mathrm{mat}} =  \int d^2 z G_{\mu \nub} D_{-} X^{(\mu} D_{+} X^{\nub )} \ ,
\end{equation}
and the superconformal symmetry (\ref{eq:G_transformation}) is generated by
\begin{equation}
a^G R_G^\mu = a^G \dep X^{\mu} + \frac{i}{2} D_+ a^G D_+ X^{\mu} \ \ \ \mathrm{and \ c.c.} \ ,
\end{equation}
where $a^G$ is a fermionic parameter ghost of spin $-1$.  Because the manifold is K\"{a}hler\footnote{Characterized by $G_{\mub \nu, \beta} = G_{\mub \beta, \nu}$ and  $G_{\mu \nub, \betab} = G_{\mu \betab, \nub}$ (see  (\ref{eq:kah_metric_relations})).} there is an additional supersymmetry (\ref{eq:delta_I}),
\begin{equation}
a^I R^\mu_{I} = i a^I D_+ X^\mu \ \ \ , \ \ \  a^I R^{\mub}_{I} = -i a^I  D_+ X^{\mub} \ ,
\end{equation}
which is also of conformal-type.  Furthermore, when the manifold is Calabi-Yau,  $S_{\mathrm{mat}}$ is invariant under the following symmetries associated with the $(3,0)$ and $(0,3)$ forms (see \ref{sec:cy_algebras_in_general} and \ref{section:cy_algebras}):
\begin{align}
& a^{L} R^{\mu}_L = a^L G^{\mu \nub} \epsilon_{\nub \alphab \betab} D_{+} X^{\alphab \betab}
\ \ \ \mathrm{and \ c.c.}  \ , \\ \nonumber
& a^{\bL} R^{\mu}_{\bL} = -i a^{\bL} G^{\mu \nub} \epsilon_{\nub \alphab \betab} D_{+} X^{\alphab \betab}
\ \ \ \mathrm{and \ c.c.} \ .
\end{align}
To close the algebra both the $L$ and $\bL$ symmetries must be introduced.

The locally invariant action is given by
\begin{equation}
\label{eq:CYsting_bad}
S_0 = S_{\mathrm{mat}} + \int d^2 z ( T_G h^G + T_I h^I + T_L h^L + T_{\bL} h^{\bL} ) \ ,
\end{equation}
where the conserved currents are given by
\begin{align}
\label{eq:reducible_currents}
& T_G = G_{\mu \nub} \dep X^{(\mu} D_+ X^{\nub)} \ \ \ T_I = 2i G_{\mu \nub} D_+ X^{\mu \nub} \ , \\ \nonumber
 & T_L = \frac{2}{3}(\epsilon_{\alpha \mu \nu} D_{+} X^{\alpha \mu\nu} + \epsilon_{\alphab \mub \nub} D_{+} X^{\alphab \mub\nub} ) \ ,  \\  \nonumber
 & T_{\bL} = \frac{2}{3}(i \epsilon_{\alpha \mu \nu} D_{+} X^{\alpha \mu\nu} - i \epsilon_{\alphab \mub \nub} D_{+} X^{\alphab \mub\nub} ) \ .
\end{align}
The gauge fields transform as
\begin{align}
\label{eq:CY3_gauge_filed_transf}
& \delta_G h^G = D_- c^G - c^G \dep h^G + \dep c^G h^G - \frac{i}{2} D_+ c^G D_+ h^G  \ , \\ \nonumber
& \delta_I h^G= -2i c^I h^I   \ \ \  , \ \ \     \delta_L h^G = 2 i c^L h^{\bL} T_I  \ , \\ \nonumber
& \delta_I h^I= D_- c^I + \dep c^I h^G - \frac{1}{2} c^I \dep h^G + \frac{i}{2} D_{+} c^I D_{+} h^G  \ , \\ \nonumber
&  \delta_G h^I = c^G \dep h^I - \frac{1}{2} \dep c^G h^I - \frac{i}{2} D_+ c^GD_+ h^I  \ , \\ \nonumber
& \delta_L h^I = (c^L D_+ h^L + D_+ c^L h^L ) T_I + 2 i c^L h^{\bL} T_G \ ,  \\ \nonumber
& \delta_L h^L = D_- c^L - c^L \dep h^G + \dep c^L h^G - \frac{i}{2} D_+ c^L D_+ h^G \ , \\  \nonumber 
& \delta_\bL h^L= 2 c^\bL D_+ h^I + D_+ c^\bL h^I   \ \ \ , \ \ \ 
\delta_I h^L = - c^I D_+ h^{\bL} + 2 D_+ c^I h^{\bL} \ , \\  \nonumber 
& \delta_G h^L =- c^G \dep h^L + \dep c^G h^L - \frac{i}{2} D_+ c^G D_+ h^L \ ,
\end{align}
together with the transformations obtained by replacing $L$ with $\bL$ and taking complex conjugates. 

The solution  is obstructed by null relations due to the Jacobi identities. In \ref{section:CY3} it was shown that for CY3 these are only (\ref{eq:IL_relation}) and (\ref{eq:GL_relation}):
\begin{equation}
T_G T_\bL = -\frac{i}{2} D_+ T_I T_L \ \ \ , \  \ \ T_G T_L = \frac{i}{2} D_+ T_I T_\bL \ \ \ , \ \ \ T_I T_L= T_I T_\bL = 0 \ .
\end{equation}
These relations imply that the $N=2$ \emph{string}\footnote{The $N=2$ string action is obtained by gauging only $T_G$ and $T_I$. It was first studied  by Ademolo et al. \cite{Ademollo:1975an, Ademollo:1976pp}.  It gives a stringy description of self dual gravity and Yang-Mills theories, for closed and open heterotic $N=2$ strings, respectively. For its relevance to integrable models in two dimensions see \cite{Witten:1976ck, Leznov:1980tz}, and also the review by Marcus \cite{Marcus:1992wi}.}  \label{N2_string_footnote} has the extra gauge symmetries,
\begin{align}
\label{eq:N_2_null_algebra}
& \delta_X h^G = c^X T_\bL \ \ \ , \  \ \  \delta_X h^I =  + \frac{i}{2} c^X D_+ T_L \ , \\ \nonumber
& \delta_Y h^G = c^Y T_L \ \ \ , \  \ \   \delta_Y h^I = - \frac{i}{2} c^Y D_+ T_\bL  \ ,\\ \nonumber
& \delta_{Z_1} h^I = c^{Z_1} T_L \ \ \ , \  \ \ \delta_{Z_2} h^I = c^{Z_2} T_\bL \ ,
\end{align}
that act trivially on the mater fields. These are symmetries only due to the fact that the classical $N=2$ string is defined on a CY3 background.

The algebra of the transformations (\ref{eq:N_2_null_algebra}) together with the $N=2$ gauge transformations closes, but this is no longer the case when the $\delta_L$ and $\delta_\bL$ symmetries are added to the algebra. Another problem if this is done is that we are dealing with symmetries proportional to the equations of motion - since $T_L$ and $T_\bL$ currents are the equations of motions for  $h^L$ and $h^\bL$. But even when appended only to the $N=2$ string, in which case we don't have the $h^L$ and $h^\bL$ equations of motion, it is  difficult to see how this algebra can make sense, given that the  symmetries (\ref{eq:N_2_null_algebra}) are  infinitely reducible. Also, the $N=2$ string is a well defined quantum  theory that lives in four real dimensions, and it's difficult to see how these new gauge symmetries could be incorporated to try and make sense of an $N=2$ string in six dimensions.

More interestingly, we are dealing with currents that have holomorphic and an antiholomorphic parts: $T_L = T_M + T_ \bM ,   \ T_\bL = i(T_M - T_\bM)$, with
\begin{align}
T_M = \frac{2}{3} \epsilon_{\alpha \beta \gamma} D_+ X^{\alpha \beta \gamma}  \ \ \ , \ \ \ 
T_\bM = \frac{2}{3} \epsilon_{\balpha \bbeta \bgamma} D_+ X^{\balpha \bbeta \bgamma}  \ .
\end{align}
The superconformal current can also be separated into two complex conserved currents,
\begin{equation}
T_g = G_{\alpha \bbeta} D_+ X^\alpha \dep X^\bbeta \ \ \ , \ \ \ T_\bg = 2  G_{\balpha \beta} D_+ X^\balpha \dep X^\beta \ .
\end{equation}
while $T_I$ is manifestly real. These algebras have been discussed in detail in Chapter \ref{chapter:algebras_on_manifolds_of_sh}, and we can find various algebras containing a subset of these complex currents for which there are no reducibility relations implied by the Jacobi identities. One can include  $T_M$ together with $T_I$ and $T_G$,  or simply $T_M$ and $T_g$. Like in the $W_{\frac{5}{2}}$ case, the commutator of $T_g$ and $T_\bg$ necessarily includes $T_I$, and then one has to deal with the reducibility relations $D_+T_I = T_g - T_\bg$.  The difference to the $W_{\frac{5}{2}}$ case is that we can't work with $T_g$, $T_I$, and $T_M$, without having a reducible algebra, since the commutator of $\delta_I$ with itself closes to $T_G$.  It is not possible to include both $T_M$ and $T_\bM$ and avoid the Jacobi identity problem.

The minimal solution to the master equation for the $W$-string containing only $T_g$ and $T_M$ is quite simple
\begin{align}
S_{\min} = & S_{\mathrm{mat}} + \int d^2 z \left\{  T_g h^g +T_M h^M + \right. \\ \nonumber
&+  X^*_\mu ( c^g \dep X^\mu + i D_+ c^g D_+ X^\mu) + X^*_\bmu c^g \dep X^\bmu \\ \nonumber
& + c^*_g c^g \dep c^g + c^*_M( -c^g \dep c^M - 2 c^M \dep c^g) \\ \nonumber
& \left. + h^*_g(D_- c^g + \dep c^g h^g + \dep h^g c^g) + h^*_M(D_-c^M + \dep c^M h^g - 2 c^M \dep h^g) \right\} \ .
\end{align}
The solution involving $T_G$, $T_I$, and $T_M$ is more elaborate. I won't write it down in full, but will explain how to obtain it starting from the gauged system (\ref{eq:CYsting_bad}). The commutators of $\delta_M$ with $\delta_G$ are the same as those of $\delta_L$ or $\delta_\bL$ with $\delta_G$ (see (\ref{eq:1_L_antibracket}) and (\ref{eq:CY_G_L_antibracket})), while the commutator of $\delta_M$ with $\delta_I$ now closes to $\delta_M$ (\ref{eq:I_M_antibracket}), since we are allowing complex structure coefficients.

\section{Discussion}
\label{section:W-string_discussion}

It is not possible to obtain a consistent classical $W$-string action by gauging symmetries that have relations between them implied by the Jacobi identities. I've shown that attempting to construct such a theory reveals there are gauge symmetries when only a subset of the generators are gauged. At least in the CY cases these symmetries close nicely, but  are infinitely reducible and don't imply extra conserved currents, so it would seem natural to ignore them. As was discussed for the $W_{\frac{5}{2}}$ and CY3 examples, one way of avoiding the Jacobi problem is to work with complex subalgebras. Below I will also discuss some classically well behaved special holonomy strings, and show that working with composite currents could potentially avoid the Jacobi problem.

At this point it is also worth mentioning potential quantum obstructions to a $W$-string that is classically well defined. In particular, the role of anomalies in the operator product expansion (OPE) of the $W$-algebra is crucial when attempting to construct the quantum $W$-string.\footnote{This is explained nicely in chapter 9 of \cite{Hull:1993kf}. } Field dependent anomalies in the OPE will generally prevent the BRST operator from being nilpotent at the quantum level, while the central charge anomaly determines the critical dimension. One can find ways around the field dependent anomalies in very special ways for the $W_3$-string, but such methods are not applicable to $W$-strings living on special holonomy manifolds. 
 
For the $SU(5)$ algebra Jacobi identities don't imply relations between the symmetry transformations (see \ref{subsection:cy2}). Although they do imply relations between the currents, these are highly non-linear and one can absorb the null symmetries in terms antisymmetric in the equations of motion, simply because the $SU(5)$ analogues of the matter field dependent parts in $\delta_L h^G$ and $\delta_L h^I$ of (\ref{eq:CY3_gauge_filed_transf}) vanish trivially. For the $Spin(7)$ algebra the Jacobi identities imply no relations between the currents (see \ref{section:spin7}). Therefore, for these cases it is possible to write down proper solutions describing the classical $W$-strings. At the quantum level all the special holonomy algebras except $Spin(7)$ generate derivative currents in the OPE via higher order contractions \cite{Figueroa-O'Farrill:1996hm}. This means that even $SU(5)$ is likely to be anomalous, and it seems that only $Spin(7)$ has a chance of producing a well defined $W$-string that involves real, non-composite currents. 
 
The complex subalgebras of $W_{\frac{5}{2}}$ for which the Jacobi identities are well behaved have a chance of defining consistent $W$-strings, since the OPE doesn't have any field dependent anomalies, and there are no reducibility relations between the currents. The same is true for the complex CY algebras discussed in the previous section. The $G_2$ algebra has an analogous subalgebra, namely the tri-critical Ising model \cite{Shatashvili:1994zw}, for which the Jacobi identities are well behaved.
 
One can also come up with algebras that consist of composite currents, and may close in a field dependent way, but such that the Jacobi identities imply no relations between the transformations. A nice example is $T_G$,  $(T^{I})^2$, and $T_L$ (but not $T_{\bL}$!)  for CY3,  which also happens to have the correct critical dimensions. I am not sure at the moment whether the OPE causes problems in this case. Of course, in the general case of the complex special holonomy algebras one has to get  the critical dimension correct. It is actually not possible to do so with any of the cases without introducing  composite currents.
 
What interpretation could a $W$-string obtained from the CY3 algebra have, assuming that one of the proposals ultimately works at the quantum level? An important lesson may be that it's not possible to  write down a classical $W$-string action when both $\delta_M$ and $\delta_\bM$ are gauged. The currents $T_M$ and $T_\bM$ are the generators of spectral flow by $\pm 1$ \cite{Eguchi:1988vr, Odake:1988bh, Gato-Rivera:1995kd}. The naive interpretation would be that it's not possible to define a  $W$-string that projects out states invariant under the spectral flow in both directions. On the other hand, spectral flow by one unit is related to the square of the spacetime supersymmetry generator, which is in turn related to a target space diffeomorphism. So it may be that one can relate the problematic $W$-string algebras to some subalgebra of target space diffeomorphisms.
 
In the $W_{\frac{5}{2}}$ as well as the problematic special holonomy cases ($SU(3)$, $SU(4)$, $G_2$), one can gauge the entire classical algebras in a finite number of steps, but than has to deal with the solutions not being proper. This gauging is not a completely futile exercise, because in order to write down naive Ward identities related to the OPE one must introduce sources for the currents and treat the gauge fields and ghosts as background fields. However, it is only possible to obtain naive Ward identities that involve explicit insertions of composite operators in this manner. As discussed in \ref{sec-global_symmetries}, it is necessary to linearize the algebras to avoid the non-local problems that emerge when background fields transform into quantum fields. Therefore, in order to write down cohomological equations for the OPE one is forced to linearize the $W$-algebras and gauge the symmetry transformations related to composite currents. 
 
\appendix

\chapter{}
\pagestyle{fancyplain}
\lhead[\fancyplain{}{\bfseries\thepage}]{\fancyplain{}{\itshape\rightmark}}
\rhead[\fancyplain{}{\itshape
Appendices}]{\fancyplain{}{\bfseries\thepage}} \cfoot{}


\section{Conventions}
\label{app:conventions}

\begin{itemize}
\item A derivative  acts only on the first object to the right of it, unless indicated otherwise by parentheses. So for any two objects $A$ and $B$,
\begin{equation}
\overleftarrow{\partial} A B \equiv \overleftarrow{\partial} (A) B \neq \overleftarrow{\partial} (AB) \ .
\end{equation}
\item Symmetrization and antisymmetrizartion implies a symmetry factor, e.g.:
\begin{equation}
T_{[ij]} \equiv \frac{1}{2} (T_{ij} - T_{ji} ) \ .
\end{equation}
\item $a$ is used for a parameter ghost that doesn't depend on all the coordinates of (super)space, while $c$ is used for a fully local  ghost.
\end{itemize}

\section{Calculating the commutation relations in the antifield formalism}
\label{app:ghost_commutators}
 
Throughout the thesis I generally use the antifield formalism to calculate the algebras. This is a summary of what is involved.

The central objects are  the symmetry generators $R^i_\mathbf{X}$, and they enter the master equation via 
\begin{equation}
X^*_k a^{\mathbf{X}} R^k_{\mathbf{X}} \ ,
\end{equation}
where the shorthand deWitt convention is being used, so repeated indices implying integration as well as Einstein summation. The index $\mathbf{X} = \{ A, B, C, ... \}$ stands for a set of labels of the symmetries of the theory, and $R^i_A$ is the generator of the symmetry transformation labeled by $A$. deWitt notation will be used when talking about the antifield formalism in general, but not in reference to a particular theory, since this is likely to cause confusion. For particular theories I will write out the integrals. 

For example, a symmetry transformation parameterized by an infinitesimal parameter $\varepsilon^A$ is
\begin{equation}
\delta_{\varepsilon^A} X^i(x) = \int dy \varepsilon^A(y) R^i_A(x,y) \ .
\end{equation}
The object $R^i_A(x,y)$ contains  a $\delta$-function, possibly with derivatives acting on it. In the master equation generators are not contracted with proper transformation parameters but with ghosts, which can be thought of as parameters but with the 'wrong' parity. I will often write
\begin{equation}
\delta_{A} X^i(x) = \int  dy  a^A(y) R^i_A(x,y) \ ,
\end{equation}
and refer to $\delta_{A} X^i$ as the generator of a symmetry, although it is of course $\delta_{\varepsilon^A} X^i(x)$ that is the infinitesimal generator. Similarly, I will not refrain from calling $ [ \delta_A, \delta_ B] X^i(x)$ the commutator.

As discussed in Chapter \ref{ch:antifield_formalism}, the information about the commutation relations of a theory is contained in the master equation, the relevant term being
\begin{equation}
\label{eq:commutator_from_ME}
X^*_i \left( \ a^\mathbf{X} \frac{ \dr R^i_{\mathbf{X}}}{ \delta X^j} a^\mathbf{Y} R^j_\mathcal{\mathbf{Y}} 
+  (-1)^{\epsilon_{\mathbf{Z}}} R^i_{\mathbf{Z}}a^{\mathbf{X}} a^{\mathbf{Y}} N^{\mathbf{Z}}_{\mathbf{X} \mathbf{Y}}  
+  \ \mathrm{e.o.m. \ terms} \ \right) =0 \ .
\end{equation}
Up to sign conventions, $N^{\mathbf{Z}}_{\mathbf{X} \mathbf{Y}}$, which enter the extended action via $a^*_{\mathbf{Z}} a^{\mathbf{X}} a^{\mathbf{Y}} N^{\mathbf{Z}}_{\mathbf{X} \mathbf{Y}}$, are the same structure functions one would obtain from the usual commutators between the infinitesimal transformations $\delta_{\varepsilon^\mathbf{X}}$ and $\delta_{\varepsilon^{\mathbf{Y}}}$. 

The algebra is calculated by evaluating
\begin{align}
 \label{eq:commutator}
 & [ \delta_A,  \delta_ B] X^i(x) :=  \\ \nonumber 
 & \int  dz  \left\{ \frac{ \dr \int dy a^A(y) R^i_A(x,y) } { \delta X^{j} (z) }  \int  dy  a^B(y) R^j_B(z,y) + (A \leftrightarrow B) \right\}  \ .
\end{align}
As well  as being necessary in order to solve the master equation and analyze the symmetries at the quantum level, there are computational advantages to calculating the algebra in this way. This flip of parity introduces sign differences compared to the usual commutators. Importantly, in the 'wrong sign' case there is no need to introduce two separate parameters when calculating the commutator of a transformation with itself, and the commutation relations of an entire algebra are written as a  simple sum over all the transformations (\ref{eq:commutator_from_ME}) without having to worry whether the correct bracket is a commutator or an anticommutator.

\section{Properties of the Poisson bracket and the antibracket}
\label{app:poisson_antibracket}

The Poisson bracket is characterized by the properties:
\begin{align}
\epsilon \{ A, B \} & =  \epsilon_A + \epsilon_B \ , \\ 
\{A, B\} & = - (-1)^{ \epsilon_A  \epsilon_B } \{ B, A \}  \ , \\ 
\{ A + B, C \} & =  \{ A, C \} + \{ B, C \} \ ,\\ 
(-1)^{\epsilon_A \epsilon_C} \{ A, \{ B, C \} \} + \mathrm{CYCLIC} &= 0 \ ,\\ 
\{ AB, C \} & =  A \{ B, C \} + (-1)^{\epsilon_B \epsilon_C} \{ A,C \} B \ .
\end{align}

The antibracket is characterized by the properties:
\begin{align}
\epsilon ( A, B ) & =  \epsilon_A + \epsilon_B \ , \\
( A, B ) & =  - (-1)^{ ( \epsilon_A +1)( \epsilon_B +1) } ( B, A )  \ , \\
( A + B, C ) & =  ( A, C ) + ( B, C )  \ ,\\
(-1)^{(\epsilon_A+1)( \epsilon_C+1)} ( A, ( B, C ) ) + \mathrm{CYCLIC} &= 0 \ , \\  \label{eq:ab_jacobi}
( AB, C ) & =  A ( B, C ) + (-1)^{\epsilon_B (\epsilon_C+1)} ( A,C ) B \ .
\end{align}

\newpage
\pagestyle{fancyplain}
\renewcommand{\chaptermark}[1]{\markboth{#1}{}}
\renewcommand{\sectionmark}[1]{\markright{\thesection\ #1}}
\lhead[\fancyplain{}{\bfseries\thepage}]{\fancyplain{}{\itshape\rightmark}}
\rhead[\fancyplain{}{\itshape\leftmark}]{\fancyplain{}{\bfseries\thepage}}
\cfoot{}

\bibliographystyle{custom_bibliography2}
\bibliography{bibliography}

\newpage
\pagestyle{plain} \addcontentsline{toc}{chapter}{Acknowledegments}
\begin{center}\LARGE\bfseries{\sc Acknowledgements}\end{center}

In first place I would like to thank my parents, without whose help I wouldn't  have had the opportunity of doing this PhD.

It was a pleasure working under the supervision of Paul Howe. I am grateful to him for providing an interesting problem that has forced me to learn some difficult and important ideas, which have opened up many possibilities for future research. I especially thank him for his patience with many of my questions and comments.

Many thanks to Ulf Lindstr\"om for beeing very welcoming during my stay in Sweden, and helpful during and after the work on our paper.

I am grateful to Andreas Recknagel, Nathan Berkovits, Thomas Quella, and Sylvain Ribault for their help with the CFT approach to to the extended algebras.

Thanks to Christian S\"{a}mann for letting me use his \LaTeX   \ code.

Finally, I'd like to acknowledge the great people of Room 102 in Drury Lane, and of course Rachel who has been very nice to me, and amazingly understanding, during the last year of this PhD.

\end{document}